\documentclass[a4paper,11pt]{article}
\usepackage{Mypos}
\usepackage{hyperref}
\usepackage{amsmath,mathtools}
\usepackage{physics}
\usepackage{wrapfig}
\usepackage{subdepth}
\usepackage{float}
\usepackage[version=4]{mhchem}
\usepackage{cellspace}
\usepackage{setspace}
\makeatletter
\renewcommand{\tableofcontents}{%
  \@starttoc{toc}
}
\makeatother

\long\def\exclude#1{}

\title{Cosmology of axion dark matter}
\ShortTitle{Axion cosmology}

\author*{Ciaran A.~J.~O'Hare}

\affiliation{ARC Centre of Excellence for Dark Matter Particle Physics, The University of Sydney, School of Physics, NSW 2006, Australia}

\emailAdd{ciaran.ohare@sydney.edu.au}

\abstract{I present an introduction and topical review on axions as a dark matter candidate. Emphasis is placed on issues surrounding the cosmology of axion dark matter that are relevant for present-day searches, including: early-Universe production mechanisms, predictions of the axion mass, bounds on axion properties derived from cosmological data, as well as the direct and indirect detection of relic axion populations.}

\FullConference{%
  First Training School COST Action {\em COSMIC WISPers} (CA21106)\\
  11--14 September 2023 \\
  Rettorato Universit\'a del Salento, Lecce, Italy
}

\begin{document}

\maketitle

{\setstretch{1.66}
\setlength{\parskip}{-10pt}\tableofcontents}

\section{Introduction}
\label{sec:introduction}

Of the particle physicist's many imaginary friends, the axion might be one of the current favourites. If the number of papers people write has any bearing on the nature of our Universe (which it does not), then you may get the impression that axions are a sure bet for our next major discovery. It is unclear if this will happen on any reasonable timescale, or at all, but we will never know if they are real unless we try to find them. As it stands, we have barely scratched the surface when it comes to ruling them out. Thankfully, axions turn out to be quite fascinating things, and so that gives us plenty to think about, and plenty more papers to write until we find them and/or move on to our next imaginary friend (see Refs.~\cite{Antypas:2022asj, Antel:2023hkf} for a selection of alternatives).

One of the fields that has something important to say about the axion is cosmology. The main reason behind this is that cosmology tells us that there seems to be a whopping great amount of dark matter around and we cannot make sense of galaxies or the Universe at large without it. But cosmology provides more than just a suggestion that dark matter exists. Cosmology in the 21st century is a precision science---any hypothesis we come up with for the identity of dark matter is immediately confronted with a large amount of exquisite data.

So what is dark matter? While there are no answers to be found in the zoo of known particles, there are some suspicious elements of the theory underlying their interactions that offer us some clues. In particular, a solution to one of the Standard Model's long-standing internal puzzles known as the Strong CP problem inspires us to invent a new particle which we call the axion---a particle which seems to fit the bill nicely when it comes to simultaneously explaining a Universe filled with invisible matter.

It is safe to say there is no evidence currently that we have hit upon the correct dark matter candidate in the axion. However, cosmology still has a lot to tell us. Are there models for the axion which let them work as a dark matter candidate, and those which do not? When in cosmic history did the axion need to be created for the Universe to remain consistent with those exquisite observations of it? What sort of imprints would relic populations of axions leave in astrophysical data? Can we assist terrestrial experiments seeking to detect the axion by telling them where to look first? These are some of the questions being addressed in axion cosmology right now, and this review will showcase some of what we are learning.

That said, this review is merely an introduction to this aspect of axion physics---the aim being to assist students and newcomers to the field, in particular those who are not aiming to become cosmologists and just want to understand the important ideas. I especially want to get across some of the guidance that cosmology provides to the people who are on the ground, hunting for axions. I encourage those who are looking for something more technically substantial to seek out some of the major reviews of this subject, like that of Marsh~\cite{Marsh:2015xka} or Sikivie~\cite{Sikivie:2006ni} on axion cosmology, as well as Hui~\cite{Hui:2021tkt}, Niemeyer~\cite{Niemeyer:2019aqm} or Ferreira~\cite{Ferreira:2020fam} for reviews focusing on ultralight axions and wave-like dark matter. I will also skip over a lot of axion theory and experimental techniques, and I will only cover astrophysical searches for axions specifically as a dark matter candidate. For an introduction to those other topics, there are many other excellent reviews out there (here's a few:~\cite{Irastorza:2018dyq, Hook:2018dlk, DiLuzio:2020wdo, Sikivie:2020zpn, Semertzidis:2021rxs, Safdi:2023wsk, Adams:2022pbo, ParticleDataGroup:2022pth}), including in the very series this work is a part of: Refs.~\cite{Choi:2024ome, Caputo:2024oqc}. 
\\~\\
Throughout I will use natural units where $c=\hbar=k_B = 1$.

\section{Cosmology}
Before getting stuck into the axion stuff, let us recap a few of the basics. You will find this and much more in any good cosmology textbook, so if you know your cosmology already, you can skip it. I will just focus on the definitions and concepts that we will use later. 

\subsection{The definitions and concepts we will use later}
What is physical cosmology? It is the study of the Universe at large as a dynamical system. The axio\textbf{m} upon which we begin thinking about cosmology is the statement that the Universe looks essentially the same in all directions (it looks isotropic) and that it would look like that no matter where you observed it from (it is homogeneous). Of course, the Universe has stuff in it, but when you zoom out to the multi-billion-parsec scale, galaxies look like little specks of dust floating in space. Accordingly, it is appropriate to describe spacetime on these scales as being smooth, flat\footnote{By the way, the assertion of flatness is inspired by observations and not theoretical bias. In textbooks on cosmology, you will see a factor of $(1-kr^2)^{-1}$ with the radial part of the metric which describes the possibility that the Universe could have an overall large-scale positive or negative curvature. Analyses of the cosmic microwave background constrain the contribution of curvature to the overall energy density of the Universe to be close to zero, so our Universe seems to be geometrically flat as far as we can tell~\cite{Boomerang:2000efg}.}, and Minkowski-like,
\begin{equation}
    \mathrm{d} s^2=-\mathrm{d} t^2+a^2(t)\mathrm{d}\mathbf{x}^2 \, .
\end{equation}
Except it's not quite Minkowski, we have a dimensionless and time-dependent function $a(t)$, multiplying the space part. This is a scaling factor that encodes one of the most important discoveries of the 20th century that kicked off the field of cosmology: that the Universe is expanding. This scale factor is an increasing function of time, and since we usually talk about comparisons in the Universe's size from one moment to another, it can be useful to normalise it to $a({\rm today}) = 1$.

This spacetime is called FRW or FLRW depending on whether or not you remember to recognise the work of Georges Lemaître. We can use it to extract from Einstein's equations a few useful formulae for the expansion rate ($\mathrm{d}a/\mathrm{d}t = \dot{a}$) and acceleration ($\mathrm{d}^2a/\mathrm{d}t^2 = \ddot{a}$):
\begin{equation}
    H^2 \equiv\left(\frac{\dot{a}}{a}\right)^2=\frac{1}{3 M^2_{\rm Pl}} \sum_i \rho_i \, ,
\end{equation}
\begin{equation}
    \frac{\ddot{a}}{a}=-\frac{1}{6 M_{\rm Pl}^2}\sum_i(\rho_i+3 P_i) \, .
\end{equation}
I am using \mbox{$M_{\rm Pl} = 1/\sqrt{8\pi G_N} = 2.435\times 10^{18}$~GeV} for the reduced Planck mass, and $H(t)$ is called the Hubble parameter. The different $\rho_i$ and $P_i$ are the large-scale average energy densities and pressures of the different types of stuff in the Universe. We model each component as a perfect fluid with energy-momentum tensor $T^\mu_\nu = \text{diag}(-\rho, P, P, P)$. The relationship between the density and pressure of one of these fluids is called its equation of state, $P_i = w \rho_i$. The ones we will need are ``matter'' which has no pressure $w = 0$; ``radiation'' which has $w = 1/3$; and the ``cosmological constant'' which has negative pressure $w=-1$. For the purposes of this, I will be conflating the cosmological constant with the term ``dark energy''.

The expressions above can be combined to reveal the continuity equation (which can also be derived by enforcing the conservation of energy and momentum):
\begin{equation}
    \dot{\rho}+3 H(\rho+P) = 0 \, .
\end{equation}
The continuity equation is solved by a perfect fluid whose density evolves with expansion as $\rho\propto a^{-3(1+w)}$. This implies that matter dilutes with the Universe's volume $\rho_m \propto a^{-3}$, dark energy always has a constant energy density $\rho_\Lambda = $\emph{const.}~and radiation is diluted by an extra factor of $a$: $\rho_r \propto a^{-4}$. 

The intuition behind the extra factor of $a$ in the dilution of radiation is that it also gets stretched out and loses energy due to the expansion of space. We see this directly when we observe distant galaxies. Photons emitted by a source billions of years ago get shifted to longer wavelengths by the same factor as how much space has expanded over that time. To an observer this looks the same as if the galaxy emitting the light was flying off away from us, so we use the word redshift to describe it, as in the Doppler effect. This, of course, is precisely how the discovery of cosmic expansion was made in the first place by Hubble. To have a closer connection with observations, we like to relate the redshift of light emitted by a distant galaxy, $z = \Delta \lambda/\lambda$, to the scale factor via $a = 1/(1+z)$. This allows us to use redshift as a sort of combined time and distance coordinate, with $z=0$ referring to here and now, and $z = {\it large}$, as far away and long ago.

The Hubble parameter, $H(t)$, was first measured by plotting the redshift of a bunch of galaxies against their distances from us. Its value today (cosmologically speaking) is called $H_0$ and is typically expressed in the units reflecting how it is measured: km/s/Mpc. Physically, $H(t)$ is a rate, and so $H^{-1}(t)$ can be interpreted in natural units as either a time or a distance. We will use $H^{-1}(t)$ frequently as an estimate of the cosmic horizon---the size of the observable Universe at a particular time. 

Measuring the Hubble constant today is important because its role in the Friedmann equation means we can use it to calculate the average density of all the stuff in the Universe right now,
\begin{equation}
    \sum_i \rho_i({\rm today}) \equiv \rho_{\rm tot} = 3 H_0^2 M_{\rm Pl}^2 \, .
\end{equation}
One thing that we will be doing a lot of is comparing the density of stuff now with the density of stuff at some other time in the past. This means that $H_0$ will often lurk around in our equations. Unfortunately, $H_0$ can be a tricky quantity to measure, so it is customary to just keep track of it by writing it in terms of a dimensionless parameter, $h$:
\begin{equation}
    H_0 = 100 h \, \mathrm{km\,s}^{-1} \mathrm{Mpc}^{-1} \equiv 2.13\times 10^{-33} \, h \, {\rm eV} \, .
\end{equation}
There are several ways to measure $h$. The most direct approach involves plotting the redshifts of galaxies against their distances obtained using the distance ladder. These local measurements yield $h\approx 0.73\pm 0.01$~\cite{Riess:2021jrx, Riess:2023egm}. Another approach is to infer how fast our Universe should be expanding based on our best-fit cosmological model called $\Lambda$CDM. Strangely, this model-dependent approach yields a Hubble constant of $h=0.674 \pm 0.005$~\cite{Planck:2018vyg}---a value that is significantly slower (by something like 5$\sigma$ now) than the distance ladder approach, which makes people think maybe the model is wrong. However, for pretty much every other bit of cosmological data\footnote{One other example of where $\Lambda$CDM has a potential crack in it is the so-called S8 or $\sigma_8$ tension, reviewed in Ref.~\cite{Abdalla:2022yfr}.}, $\Lambda$CDM works stunningly well. This is the famous \textit{Hubble tension}~\cite{DiValentino:2021izs}. Could axions be the solution? Maybe, but not in any immediately obvious way, e.g.~\cite{DEramo:2018vss, Alexander:2019rsc, Berghaus:2019cls, Ivanov:2020ril, Mawas:2021tdx}.

That issue notwithstanding, since we measure $H_0$ and therefore $\rho_{\rm tot}$, we simplify things and just express each component of the cosmic density as a fraction, $\Omega_i = \rho_i({\rm today})/\rho_{\rm tot}$. We can anchor these densities to their values today because we know how the different cosmological substances should scale as the Universe expands. Assuming we have accounted for everything (i.e.~matter, radiation and dark energy are all there is), the entire cosmic history of $H(t)$ is then fully described:
\begin{equation}\label{eq:hubbleparameter}
    H^2(t)=H_0^2\left[\Omega_r a(t)^{-4}+\Omega_m a(t)^{-3}+\Omega_{\Lambda}\right] \, ,
\end{equation}
recalling the solution $\rho\propto a^{-3(1+w)}$ mentioned above. 

\subsection{Thermodynamics of the early Universe}\label{sec:thermodynamics}
Matter, radiation and dark energy are the three major ingredients of our standard model of cosmology. The best-fit $\Lambda$CDM parameters right now tell us that around $\Omega_m = 31$\% is matter, $\Omega_\Lambda = 68$\% is in the form of dark energy, with less than 0.01\% in radiation. For most of the Universe's history, it has been dominated by matter. Only in the last billion years or so has $\rho_m$ diluted away sufficiently for $\rho_\Lambda$ to become the dominant component of Eq.(\ref{eq:hubbleparameter}). Because $\rho_\Lambda = const.$, a dark-energy dominated universe expands with $\ddot{a}>0$. In other words, we observe the recent Universe undergoing accelerated expansion. 

That's not what we are interested in right now though. If we roll the clock all the way back to a mere 50,000 years after the Big Bang, there was a time when it was \textit{radiation} that was the dominant of the three components, rather than matter. The point at which the two densities crossed over is called matter-radiation equality, which occurs at a redshift satisfying $\Omega_r(1+z_{\rm eq})^4 = \Omega_m(1+z_{\rm eq})^3$, where $z_{\rm eq} \approx 3400$.

In cosmology, all regular matter---including nuclei and electrons---are lumped together and called baryons. As we will see by the end of this section, the majority of the matter going into $\Omega_m$ is not baryons, but another type of matter whose defining feature is that it only seems to be coupled gravitationally. We are going to be trying to get axions to explain what this ``dark'' matter component is, so here is our first constraint: we want axions to exist and be exerting their gravitational influence on the Universe in the form of matter by $z_{\rm eq}$ at the latest.

So when we talk about axions as dark matter, most of our discussion is going to take place prior to $z_{\rm eq}$---during radiation domination. Going this far back requires us to understand how the Universe behaved when it was nothing but a swarm of particles and radiation, held (most of them anyway) in equilibrium. The temperature of this thermal bath will start out exceedingly hot, hotter than a few MeV at the very very least, but likely much hotter, and will then cool down over time as $T \sim a^{-1}\sim (1+z)$ roughly. 

Digging up a bit of statistical mechanics, we recall that the macroscopic properties of an ensemble of particles in thermal equilibrium can be obtained by doing the relevant integrals over their distribution functions. If we want the average energy density for example, we take the particle energy $E = \sqrt{\mathbf{p}^2 + m^2}$, where $\mathbf{p}$ is momentum and $m$ is the particle mass, and do:
\begin{equation}\label{eq:energydensity_gas}
    \rho = g \int \frac{{\rm d}^3 p}{(2 \pi)^3} E f(\mathbf{p}) \, ,
\end{equation}
where $g$ counts up the spin states for the particles. We have two choices for the distribution function,
\begin{equation}
    f(\mathbf{p})=\frac{1}{\exp[E(\mathbf{p})/T] \pm 1} \, ,
\end{equation}
where `$+$' is used for fermions and `$-$' for bosons.

Evaluating these integrals in the relativistic limit, $p\gg m$, the relationship between the energy density of a species and the temperature of the cosmic thermal bath, $T$, is revealed to be,
\begin{align}\label{eq:rho_vs_T}
   \text{relativistic bosons}: \rho &= \frac{\pi^2}{30} g T^4 \, , \\ 
   \text{relativistic fermions}: \rho & =\frac{7}{8} \frac{\pi^2}{30} g T^4 \, . \nonumber
\end{align}
Since fermions scale with temperature in the same way as bosons just with a factor of 7/8, we can conveniently express the density-temperature relation of all the species at once by defining an ``effective'' number of relativistic species,
\begin{equation}
    g_{\star}(T)  =\sum_{\text {bosons }} g_i\left(\frac{T_i}{T}\right)^4+\sum_{\text {fermions }} \frac{7}{8} g_i\left(\frac{T_i}{T}\right)^4  \, .
\end{equation}
With this single effective species, the Friedmann equation during radiation domination can be written in terms of the temperature as,
\begin{equation}\label{eq:Hubble_raddom}
    3 H^2(T) M_{\rm Pl}^2 = \frac{\pi^2}{30} g_{\star}(T) T^4 \, ,
\end{equation}
which is a useful formula that we will use later. 

The reason for giving each species its own temperature $T_i$ which could deviate from the average $T$ is because the thermal equilibrium may not be maintained for all particles over all timescales. For example, if some particle is kept in thermal equilibrium by a process where it scatters with other particles, then the frequency of those scattering events needs to be sufficient, even as the Universe gets bigger and those particles get spread further apart. 

As an example where this won't be the case: if a particle is equilibrated by a feeble process with a small probability of occurring, then there may come a time when the thermal bath has cooled down too much for that equilibrium to be kept up. When this occurs, the particles will no longer have their temperature maintained by the surrounding bath and the species is said to kinetically decouple. Similarly, if some particle is kept in equilibrium by production and annihilation and the temperature (i.e.~typical energy of particles in the bath) drops below the mass of that particle, then annihilation is no longer reversible and they are said to chemically decouple. Once the temperature falls below the mass scale of the particle they will also become non-relativistic. For non-relativistic particles, Eq.(\ref{eq:rho_vs_T}) gets a Boltzmann suppression factor $e^{-m/T}$, which causes $\rho$ to drop exponentially as the temperature cools. Depending on the mass and interaction strengths of a certain species, decoupling can occur when the particle is either still relativistic or after it has gone non-relativistic. In the standard picture of thermal (non-axion) dark matter production for example, a weakly interacting massive particle first chemically decouples, becomes non-relativistic, and then kinetically decouples, leaving its comoving abundance at some frozen amount that depends on its interaction strengths.

Another ingredient we will need is entropy. Without getting into the thermodynamical weeds, you can show by applying the second law of thermodynamics, that the entropy \textit{density} of the Universe can be expressed as,
\begin{equation}
    s \equiv \frac{S}{V}=\frac{\rho+P}{T} \, .
\end{equation}
Taking radiation as the dominant contributor to the Universe's entropy ($\rho = P/3$), we see that the entropy density has a similar form to the energy density, $s = 2\pi^2/45 g T^3$ for relativistic bosons (and $\times 7/8$ for fermions). Recall that $T \sim a^{-1}$, so the $T^3$ there hints towards an important fact: the \textit{comoving entropy density of the Universe is conserved}. ``Comoving'' means we imagine the entropy in a box that expands at the same rate as the Universe does (i.e.~$s a^3$). The conservation of comoving entropy density can be derived by applying the first law of thermodynamics to a highly relativistic system of particles and holds even when there are multiple species in the bath falling out of equilibrium at different times. 

It will prove useful to make use of the fact that $sa^3 = $ \emph{const}.~by keeping track of what $s$ is doing. Using the same logic of modelling the thermal bath of the Universe in terms of a single substance that has an effective temperature-varying number of degrees of freedom, we write down,
\begin{equation}\label{eq:entropydensity}
s = \frac{2\pi^2}{45} g_{\star,s} T^3    \, ,
\end{equation}
where the effective number of \textit{entropic} degrees of freedom, $g_{\star,s}$, is very similar to $g_\star$ but has a temperature dependence lowered by one factor of $T$,
\begin{equation}
g_{\star,s} = \sum_{\text{bosons}} g_i\left(\frac{T_i}{T}\right)^3+\sum_{\text{fermions}} \frac{7}{8} g_i\left(\frac{T_i}{T}\right)^3 \, .
\end{equation}
So how can the comoving entropy density stay constant when particles are falling out of equilibrium and changing $g_{\star,s}$? The way this works is that when a species of particle decouples, that entropy is transferred to the particles still in equilibrium, which causes their temperature to cool slightly more slowly. Only in regimes where $g_{\star,s}$ remains stable with temperature does the bath cool simply as the Universe expands like $T\sim a^{-1}$.

The values of $g_\star(T)$ and $g_{\star,s}(T)$ for standard cosmology follow each other very closely with the exception of one period when the neutrinos decouple from the bath ($T \sim 1$~MeV) and then electrons and positrons become non-relativistic and annihilate away shortly after ($T \sim m_e \sim 0.5$~MeV). The bath has $g_\star = 2$ (just photons) after $e^+e^-$ annihilation, but had $g_\star = 2 + (7/8)\times(2+2) = 11/2$ beforehand. So, by the conservation of $s a^{3}$, the photon bath gets heated by a factor of $(11/4)^{1/3}$ compared to the neutrinos. The fact that the photons receive this extra injection but the neutrinos didn't because they had decoupled means that $g_{\star,s}(T<m_e)=3.91$ and $g_{\star}(T<m_e)=3.35$ are ever so slightly different. It also means that the temperature of the left-over ``relic'' neutrinos is this same factor cooler than the temperature of the relic photons, as we will see later.

Going back to the fact that the comoving entropy density is conserved, this means that $g_{\star,s}(T) T^3 a^3 = const.$ and so we can use this as a trick to relate the temperatures and scale factors at two different times. As a pertinent example, if we apply this trick to the energy density of something that scales like matter between temperatures $T_1$ and $T_0$, we can find,
\begin{equation}\label{eq:densities_vs_T}
    \rho_m\left(T_0\right)=\rho_m\left(T_1\right) \frac{g_{\star,s}\left(T_0\right)}{g_{\star,s}\left(T_1\right)}\left(\frac{T_0}{T_1}\right)^3
\end{equation}
where we use the fact that matter scales like $\rho_m \propto a^{-3}$ to get rid of $a$. This is useful because we have measured the cosmological density of matter in the Universe today, and so we can use formulae like this to answer questions such as: at what temperature did a population of axions have to be created for them to exist in the right abundance to explain dark matter today?

\subsection{A brief history}
\begin{figure*}[t]
    \centering
    \includegraphics[width=0.99\textwidth]{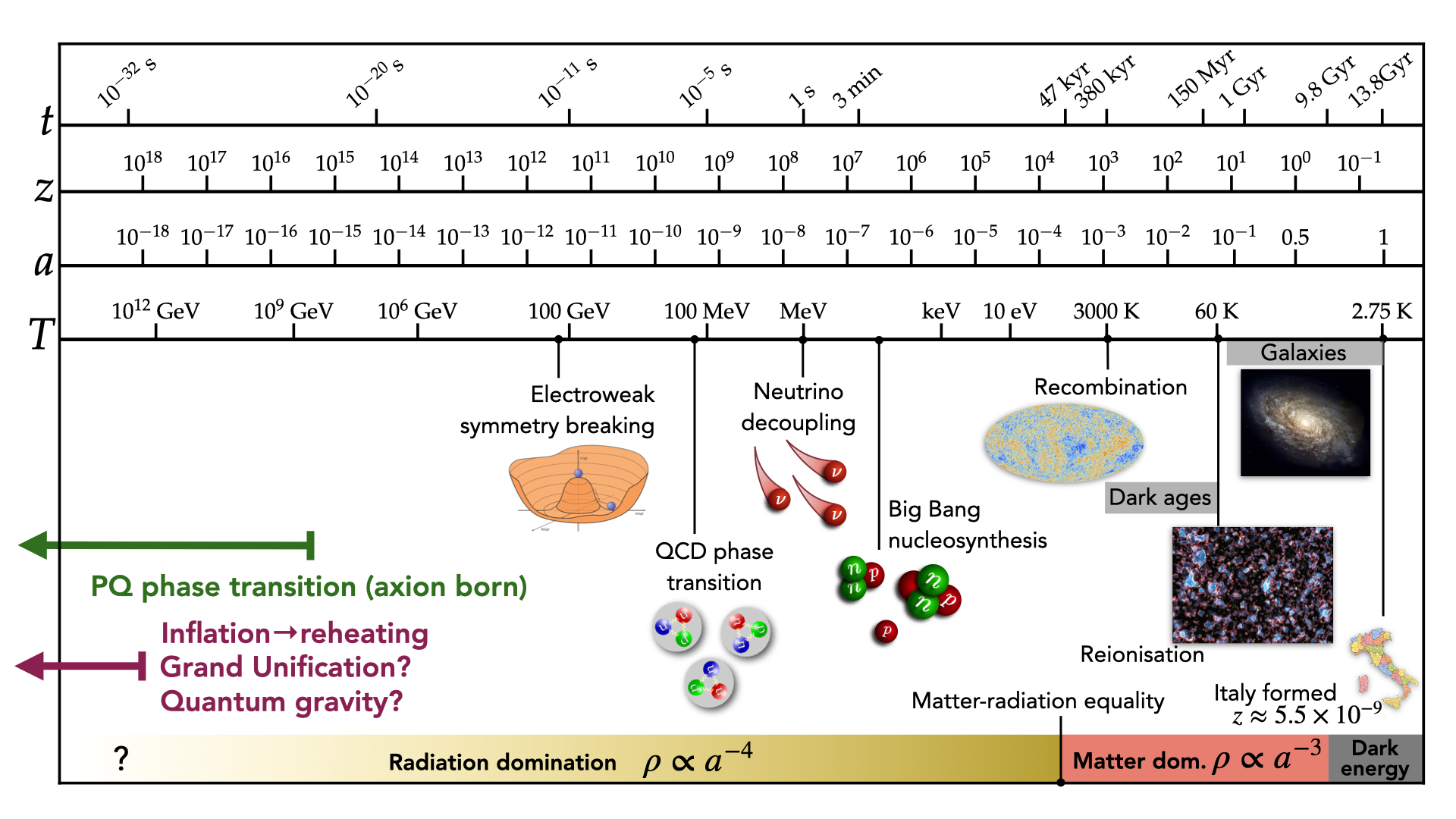}
\caption{A timeline of the early Universe, with scales for time ($t$), redshift ($z$), scale factor ($a$), and photon temperature ($T$). If the QCD axion exists, then astrophysical bounds suggest they were produced when the Universe was hotter than $T\gtrsim 10^8$~GeV.}
    \label{fig:timeline}
\end{figure*}

Armed with the basic physics of the thermodynamics of the early Universe, let us now go through the 13.8 billion years of cosmic history. As a visual reference, see Fig.~\ref{fig:timeline}. 

\begin{itemize}
    \item \textbf{Inflation}. We suppose the Universe began with a period of astonishingly rapid accelerated expansion, $a\propto e^{H t}$, driven by some field with a negative equation of state. Inflation explains why the Universe is homogeneous and spatially flat, but also explains why it has structure at all. The Universe during inflation is described as de Sitter space with a constant Hubble scale $H(t) = H_I$, where $H_I$ could be anything from $10^{14}$~GeV to $10^{-15}$~eV. The nearly scale-invariant quantum fluctuations in the field driving inflation were also inflated to scales well beyond the horizon. These perturbations in curvature are subsequently imprinted on the densities of matter and radiation---the seeds of all structures.
    \item \textbf{Electroweak symmetry breaking} ($\sim 100$ GeV). Below this temperature, the electroweak force is broken down to the electromagnetic and weak nuclear forces. The four massless electroweak bosons become the photon, W$^{\pm}$ and Z bosons, with the latter three acquiring masses through the Higgs mechanism. We know this must have occurred thanks to the monumental experimental success of the Standard Model of Particle Physics, for which the electroweak theory is a cornerstone. Nonetheless, particle colliders on Earth probe only up to the TeV scale, and so any other supposed unifications---like grand unification between the electroweak and strong interactions, or theories incorporating gravity---are entirely speculative.
    \item \textbf{QCD cross-over} ($\sim$200 MeV). Below this temperature, the quarks become confined into hadrons like the proton and neutron, bound together by the gluons---the gauge bosons of the strong interaction. The so-called QCD axion, which solves the strong CP problem, is designed to interact with these gluons in a very particular way, and so this era will be important in the story of the axion if they are produced in the great abundance needed to explain dark matter.
    \item \textbf{Neutrino decoupling} (1 MeV). Neutrinos interact solely via the weak force and so they are the first particles to decouple from the bath. Since the temperature is very hot and neutrinos very light, neutrinos decouple while they are still relativistic. The massive population of neutrinos that decouple at this moment still linger around for the next 13.8 billion years doing little else but lose energy through redshift. A density on the order of $\sim$340 cm$^{-3}$ of these relic neutrinos exists around us today, but because they have undetectably small energies $\lesssim$meV, there is only a dim hope that we might be able to use them as messengers of the Universe when its temperature was around an MeV.
    \item \textbf{Electrons become non-relativistic} (0.5 MeV). Here the population of electrons is being exponentially depleted by $e^+ e^-$ annihilation, which the thermal bath is now no longer energetic enough to reverse.
    \item \textbf{Nucleons freeze-out} (0.8 MeV). Nucleons do not decouple until temperatures below the proton-neutron mass difference $T<m_n-m_p=1.3$~MeV, which is when the bath is no longer hot enough to keep Weak-nuclear processes like $n+\nu_e \to p+e^-$ in equilibrium. This temperature is smaller than the nucleon mass, $\sim$GeV, and hence they are already non-relativistic when they freeze out.
    \item \textbf{Big-Bang Nucleosynthesis} (0.1 MeV): from here, protons and neutrons will be fusing to form the first nuclei. Detailed calculations can precisely predict the abundance of the light primordial nuclei such as helium, deuterium, and lithium~\cite{Cyburt:2015mya, Burns:2023sgx}, and so BBN represents our earliest true observational handle on cosmic history. From here, everything is much more tightly constrained.
    \item \textbf{Matter-radiation equality} (0.8 eV). This is the point when the Universe stops being dominated by radiation and instead matter takes over as the dominant influencer in the Universe's evolution. A very important moment for constraining dark matter models, so it is useful to remember the redshift at which this occurs: $z_{\rm eq} = 3400$.
    \item \textbf{Recombination} (0.26 eV). There is a long gap of several hundred thousand years after the first nuclei form.  The plasma of non-relativistic nuclei and electrons, as well as photons, cools down until the temperature drops below $\mathcal{O}$(10 eV), the typical ionisation energy of Hydrogen. At this point, the electrons can be captured by the nuclei to form neutral atoms. However, because there were $10^{10}$ photons for every baryon, the photon bath actually has to cool much more than this before atoms can really stick around as stable entities. The period when the Universe neutralises itself at last is called recombination, and happens at a redshift of $z_{\rm re}\approx 1100$. At some point around this time, each photon will enjoy its final scattering off of the last remaining charged particles before they, too, decouple. Like the neutrinos before them, these photons then simply dilute away in the form of a population of relic photons, which by the present day will have redshifted down to the microwave end of the electromagnetic spectrum.
    \item \textbf{Reionisation} (60 K). A very long period then passes while the Universe remains neutral and dark. The atoms and dark matter clump together into gravitationally bound structures called halos. Eventually though, over a period lasting 600--800 million years, the majority of the baryons in the Universe turn from neutral hydrogen clouds into an ionised plasma. The culprits behind this period of reionisation are believed to be the first generation of massive but very short-lived stars that formed from dense knots of primordial gas. From here, the Universe is lit up and ready for us to observe.
    \item \textbf{Hierarchical structure growth}. The next several billion years of cosmic history are dominated by matter. The dark matter halos that the baryons have fallen into exist across many different scales. Galaxies for example are also grouped together into galaxy clusters and superclusters. Within these environments, several successive generations of massive stars are born and then die in spectacular explosions called supernovae which enrich the Universe with chemical elements. Stars explode, galaxies merge, clusters collide and interact, and the entire Universe becomes an interconnected web of galaxies surrounded by dark-matter halos and separated by vast cosmic voids.
    \item \textbf{Dark-energy domination} (1 billion years ago). The universe keeps on expanding. Eventually, the cosmic voids have expanded so much that matter is diluted to the point where dark energy---which had been lurking in the background all this time with its constant energy density---starts to become the dominant component of $\rho$. From here, it takes hold of the evolution, and its negative pressure causes the expansion rate to start accelerating. This is where we are now.
\end{itemize}

\subsection{Cosmological observations}\label{sec:cosmologicalobservations}
Glancing through the timeline above, we see there are several messengers that we could use to learn about the history of the Universe, and indeed this story was written thanks to those messengers. Much can be learned from the stars and the galaxies of course, which I will come to soon, but the most direct messenger of the hot primordial state of the early Universe are those particles that were emitted by it directly: the relic photons and neutrinos.

The energy densities of the relic population of photons compared to neutrinos today are very similar, yet notably different,
\begin{equation}
    \rho_\nu=\left[3 \times \frac{7}{8} \times\left(\frac{4}{11}\right)^{4 / 3}\right] \rho_\gamma=0.68 \rho_\gamma \, ,
\end{equation}
where the factor of 3 comes from the three flavours of neutrino, the 7/8 because they are fermions, and the $(4/11)^{4/3}$ comes from the fact that the photons feel the injection of entropy from $e^+e^-$ annihilation but the neutrinos do not, as discussed earlier. The number density of photons today is around 411~cm$^{-3}$ compared to relic neutrinos at around 340~cm$^{-3}$, and they are slightly hotter at 2.73~K, compared to 1.9~K. The main difference though is that we can actually detect the photons~
\cite{Scott:2024rwc}.\footnote{Directly at least, we have \textit{indirectly} detected the cosmic neutrino background because of the fact that neutrinos still contribute a decent amount towards the total energy density of the Universe, even at recombination. Indeed, to get a good fit to the CMB we require around 3 flavours of neutrinos to be present in the early Universe (more on this in Sec.~\ref{sec:thermalaxions})~\cite{Planck:2018vyg}. In the future, it may also be possible to detect them gravitationally too. Unlike photons, neutrinos are massive, so eventually at least 2 of the mass states will redshift to non-relativistic speeds and start behaving as a form of `hot dark matter', which has a small but potentially detectable impact on the distribution of structure on large scales. Future datasets will hopefully provide the first measurement of the quantity that affects this---the sum of the three neutrino masses $\sum m_\nu$~\cite{Euclid:2022qde}.}

This relic population of photons is called the cosmic microwave background (CMB). It was discovered in the 1960s, a short time after its first theoretical prediction, and we have been making ever more precise maps of it across the sky ever since. This background is really little more than some thermal radiation redshifted down to $T = 2.73$~K. In fact, it is the most perfect thermal Black Body spectrum ever observed in nature. But what it represents is much more significant---the CMB is a photo of the early Universe, so it unsurprisingly provides a wealth of information we can use to understand its contents at that time.

\begin{figure}
        \centering
        \includegraphics[width=0.99\textwidth]{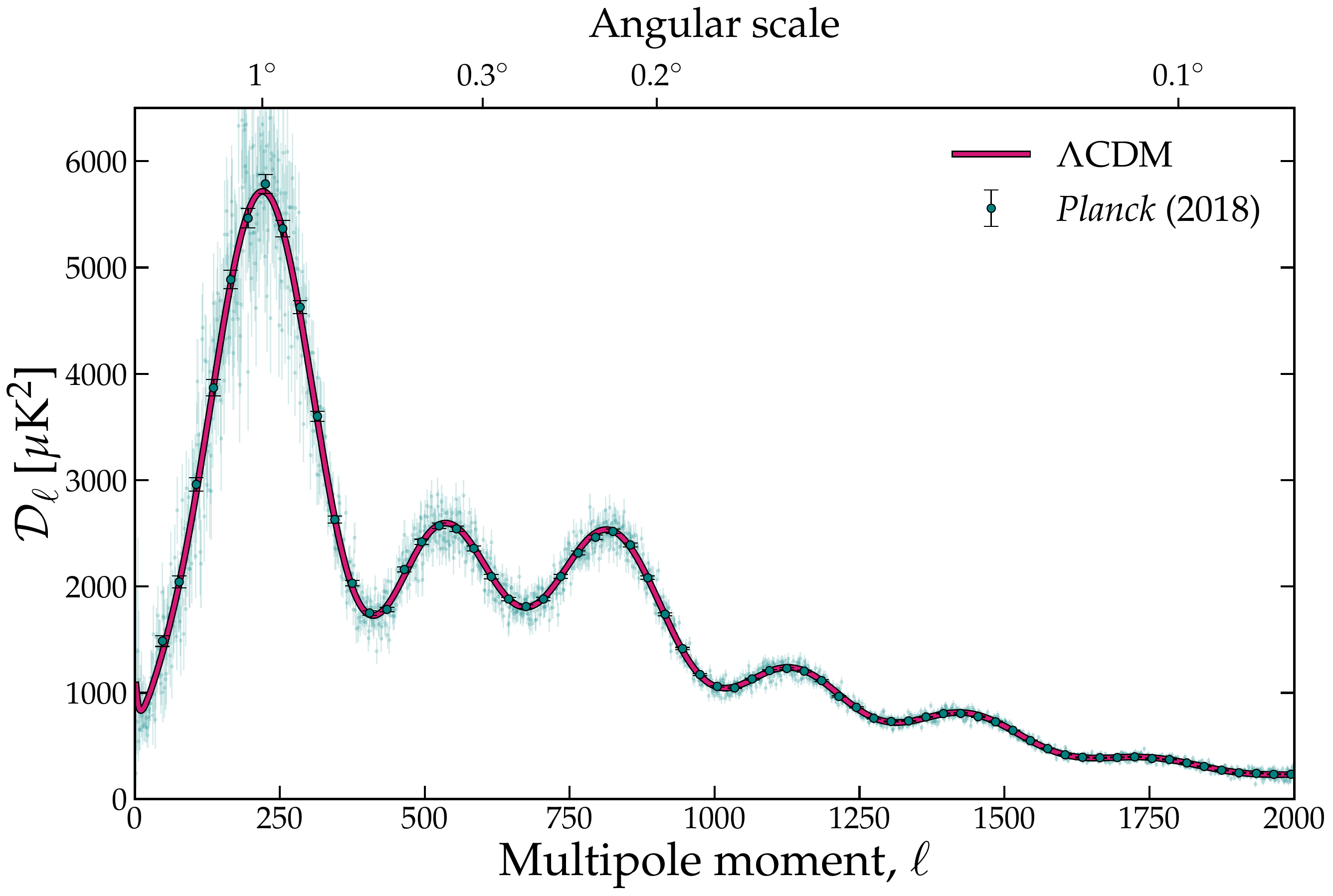}
        \caption{The CMB power spectrum as measured by \emph{Planck}~\cite{Planck:2019nip}, and compared against the best-fit 6-parameter $\Lambda$CDM model~\cite{Planck:2018vyg}. The temperature anisotropies versus multipole number are related to the spherical harmonic expansion coefficients via $\mathcal{D}_\ell = \ell(\ell +1)C^{TT}_\ell$. The little semi-transparent dots are the measurements at each value of $\ell$ individually, whereas the points with error bars are the binned values, which is what is usually shown.}
        \label{fig:CMB}
\end{figure}

But while the CMB is \textit{almost} spectrally featureless, and \textit{almost} perfectly isotropic, its most important features are nevertheless the deviations from perfection. We quantify these by trying to correlate the change in the photon temperature from one direction ($\hat{\mathbf{n}}$) to another ($\hat{\mathbf{n}}^\prime$). Since the CMB stretches across the whole of the sky, the resulting correlation functions are then decomposed into spherical harmonics (or equivalently, Legendre polynomials, $P_\ell$):
\begin{equation}
\left\langle\delta T(\hat{\mathbf{n}}) \delta T\left(\hat{\mathbf{n}}^{\prime}\right)\right\rangle=\sum_{\ell} \frac{2 \ell+1}{4 \pi} C_{\ell}^{T T} P_{\ell}\left(\hat{\mathbf{n}} \cdot \hat{\mathbf{n}}^{\prime}\right) \, .
\end{equation}
The temperature fluctuations are at the level of $\delta T/T\sim 10^{-5}$. When showing CMB data, what is usually plotted are those expansion coefficients, $C^{TT}_\ell$, as a function of the multipole moment, where larger $\ell$ values correspond to smaller angular scales on the sky. This ``power spectrum'' is shown in Fig.~\ref{fig:CMB}, with data from the \emph{Planck} satellite plotted alongside a 6-parameter best-fit model, called $\Lambda$CDM. The model fits better than you could ever ask for, despite the fact that two of its major components, $\Lambda$ and cold dark matter (CDM), are a mystery.

There are some important features in there that have physical significance. The power spectrum has a major peak for fluctuations on angular scales of about a degree, and then a series of peaks at smaller angular scales with decaying amplitudes. These are the `acoustic peaks'---evidence of sound waves rippling across the primordial plasma of the Universe. The oscillations exist because of two things: the primordial density fluctuations that set them in motion, and gravitational instability that sustained them.

These oscillations involve the tightly-coupled baryon-photon plasma sloshing in and out, while the dark matter just gravitationally collapses in the background. This state of affairs proceeds until recombination when the plasma turns to gas and the photons are released. Recombination effectively switches off the coupling between the photons and baryons and stops the oscillations. From that point on, baryons decouple from the photons and join dark matter for the gravitational ride---the unstoppable collapse that it had been undergoing since the beginning.

The scales where there are peaks in the CMB power spectrum correspond to the modes of the photon-baryon oscillations which were at maximum compression or rarefaction when recombination put a stop to them. The largest of those scales, the first peak in the CMB, is the mode with the longest period where this is true---the one which only had time to complete half an oscillation by recombination. The angular size of the first peak of the CMB therefore tells us the size of the sound horizon: the maximum distance that a ripple can travel between the Big Bang and recombination. Because we know the time of recombination as well as the speed of sound in a fluid of relativistic particles (about $c_s \approx c/\sqrt{3}$ for the early Universe), the sound horizon represents a very well-defined ruler. By comparing the size this ruler should have with its apparent size when projected on the sky, we can effectively tell how `far away' the CMB is, and in the process, measure how much the Universe has expanded from recombination until now. This is how the CMB lets us infer a value of $H_0$ for a given cosmological model.\footnote{Note that this inferred value depends on the other parameters in the model fit and since it is not a direct measurement of the current expansion rate, whether or not it is correct depends on whether our cosmological model has incorporated everything that influences the expansion history of the Universe. Many of the resolutions to the Hubble tension address it in precisely this way, by inventing something that can change the size of the sound horizon on the sky~\cite{Schoneberg:2021qvd}.}

The fact that baryons were tightly coupled to the photons before recombination, whereas the dark matter was not, is what allows us to extract the density of baryons and dark matter separately from the CMB. This information is stored in the heights of the various higher peaks: the ratios of odd to even peaks, as well as their decay towards higher $\ell$\footnote{This effect is called Silk damping and is caused by the fact that photons can diffuse out of ripples smaller than their mean-free path.}. $\Lambda$CDM is the simplest model that reproduces all of the peaks. The latest all-sky measurement of the CMB made by the \emph{Planck} satellite leads to best-fit density ratio parameters for the baryons and dark matter of~\cite{Planck:2018vyg}:
\begin{equation}
\begin{aligned}
\Omega_{b} h^2 & =0.02237 \pm 0.00015 \, ,\\
\Omega_{\mathrm{DM}} h^2 & =0.1200 \pm 0.0012 \, . \\
\end{aligned}    
\end{equation}
So we see only 16\% of all the matter that exists is in a form that the photons were able to see, a striking statement about our Universe that the CMB makes quite unambiguously. If \textit{all} of the matter was in a form that was coupled to the photons then the amplitude of the acoustic peaks would be huge, so many $\sigma$ in excess of the data that it is absurd to even talk about this as a possibility. In fact, any scenario in which baryons dominate the matter content is also impossible because of how precisely the ratios of the odd to even peaks are measured\footnote{Here is a fun app you can play with to see this for yourself: \url{http://chrisnorth.github.io/planckapps/Simulator/}.}. For what it's worth, the key indicator of the particular balance of dark matter to baryons in our Universe is the fact that the third peak is very close in height to the second. In baryon-dominated universes, the peaks would decay sequentially. %But since dark matter dominates in our universe, its growing gravitational wells pull on the baryons, which suppresses the modes which end on a rarefaction at recombination (the even peaks), relative to those that end on a compression (odd peaks).

If the spectacular agreement between $\Lambda$CDM and the CMB weren't enough, yet more evidence in favour of a dark matter-dominated universe comes from considering how this conclusion aligns with other sets of cosmological data. One of those datasets is the matter power spectrum---how we measure the distribution of structure on large scales.

We arrive at the matter power spectrum theoretically by doing perturbation theory in terms of small initial deviations in density away from the average: $\delta = \delta \rho/\bar{\rho}$. At early times, when these perturbations are small and before any structures have properly formed, their equation of motion is linear and each mode in Fourier space evolves independently. We find it convenient to follow the Fourier-transformed $\delta$ as a function of wavenumber $k$, where the power spectrum, $P_m(k)$ is a measure of their correlations:
\begin{equation}
    \left\langle\delta_m(\mathbf{k}) \delta_m(\mathbf{k}^{\prime})\right\rangle=(2 \pi)^3 \delta\left(\mathbf{k}-\mathbf{k}^{\prime}\right)P_m(k) \, .
\end{equation}
The two things which influence how structure appears in the Universe today are the same two things that influenced the way the CMB looks: the random primordial fluctuations in density inherited from inflation that set the initial sound waves in motion, and then the prescriptive evolution of those fluctuations under gravitational instability. The matter power spectrum is therefore divided into two parts that reflect these two ingredients:
\begin{equation}
    P_m(k, t)=\frac{2 \pi^2}{k^3} \mathcal{P}(k) T^2(k, t) \, ,
\end{equation}
where $\mathcal{P}$ is the power spectrum of primordial fluctuations and $T$ is the transfer function describing their subsequent evolution. The primordial part of the matter power spectrum is usually written in the form,
\begin{equation}\label{eq:primordial_scalar}
    \mathcal{P}(k)=A_{\mathrm{s}}\left(\frac{k}{k_{\star}}\right)^{n_s - 1} \, ,
\end{equation}
where $A_s = 2.2 \times 10^{-9}$ sets the amplitude and $k_\star = 0.05$~Mpc$^{-1}$ is just a reference wavenumber called the pivot scale. \emph{Planck} finds that the spectrum of primordial fluctuations was almost, but not quite, scale-invariant: \mbox{$n_s=0.9649 \pm 0.0042$}~\cite{Planck:2018jri}. We believe these primordial fluctuations originate in the inflaton, and so these numbers are what any good theory of inflation must reproduce.

The transfer function, on the other hand, is independent of anything to do with inflation but does require knowledge about the nature of dark matter, baryons, and any relationship between them, because it encodes the growth and evolution of those initial primordial ripples. The main piece of physics that it must capture is the fact that the baryons remained tightly coupled to the photons right up until $z=1100$, after which they decouple and the baryons are left to fall into the gravitational wells formed by dark matter. Dark matter's perturbations had been growing all throughout those early times, but they really got going after matter-radiation equality---which marks an important moment in the story of gravitational collapse.

To understand the evolution of a particular Fourier mode of a density perturbation more quantitatively we have to compare its physical size with the size of the horizon at a particular time. Since we have modes existing across all scales, there will be short subhorizon ones $k>aH$ as well as large superhorizon ones $k<aH$ which evolve differently. An important moment in the life of any mode is when it `enters' the horizon and becomes subhorizon, and we also need to know whether that moment occurs before or after matter-radiation equality. This is because, during radiation domination, perturbations in the dark matter grow only logarithmically, whereas, in a matter-dominated universe, they grow linearly. The different evolution of modes that entered the horizon before or after matter-radiation equality gives the resulting power spectrum a distinctive peaked shape. The power spectrum ascends $P \propto k^{n_s}$ for large-scale modes that entered the horizon after equality; it reaches a peak at $k\sim k_{\rm eq} = a_{\rm eq} H(t_{\rm eq}) \sim 0.01\,{\rm Mpc}^{-1}$; and then descends $P \propto k^{-3} \ln \left(k/k_{\mathrm{eq}}\right) k^{n_s-1}$ for small-scale modes that entered the horizon before equality. Once in the matter-dominated era, the shape of the power spectrum does not change---it merely amplifies itself across all scales $\propto a^2$.

All of that is true for dark matter, whose story is one of simple gravitational collapse. Baryons, as we know, are tightly coupled to the photons at early times and so they slosh about, unable to collapse until the photons finally leave them alone at recombination. The extremely late arrival of the baryons to the gravitational collapse party is why the matter power spectrum only has the slightest hint of these oscillations imprinted onto it---the CMB, on the other hand, since it directly traces the baryons, is nothing \textit{but} oscillations. Nevertheless, while the imprint of baryon-acoustic-oscillations (BAOs) is small, it is still measurable. This is yet another catastrophe for those seeking to explain our Universe without inventing dark matter~\cite{Dodelson:2011qv}. The baryon-only version of our Universe would be one in which those oscillations are painted clearly across the sky in the form of giant well-defined rings of galaxies. As it is, we have to try quite hard to pick out the rings.

\begin{figure}
    \centering
    \includegraphics[width=0.9\linewidth]{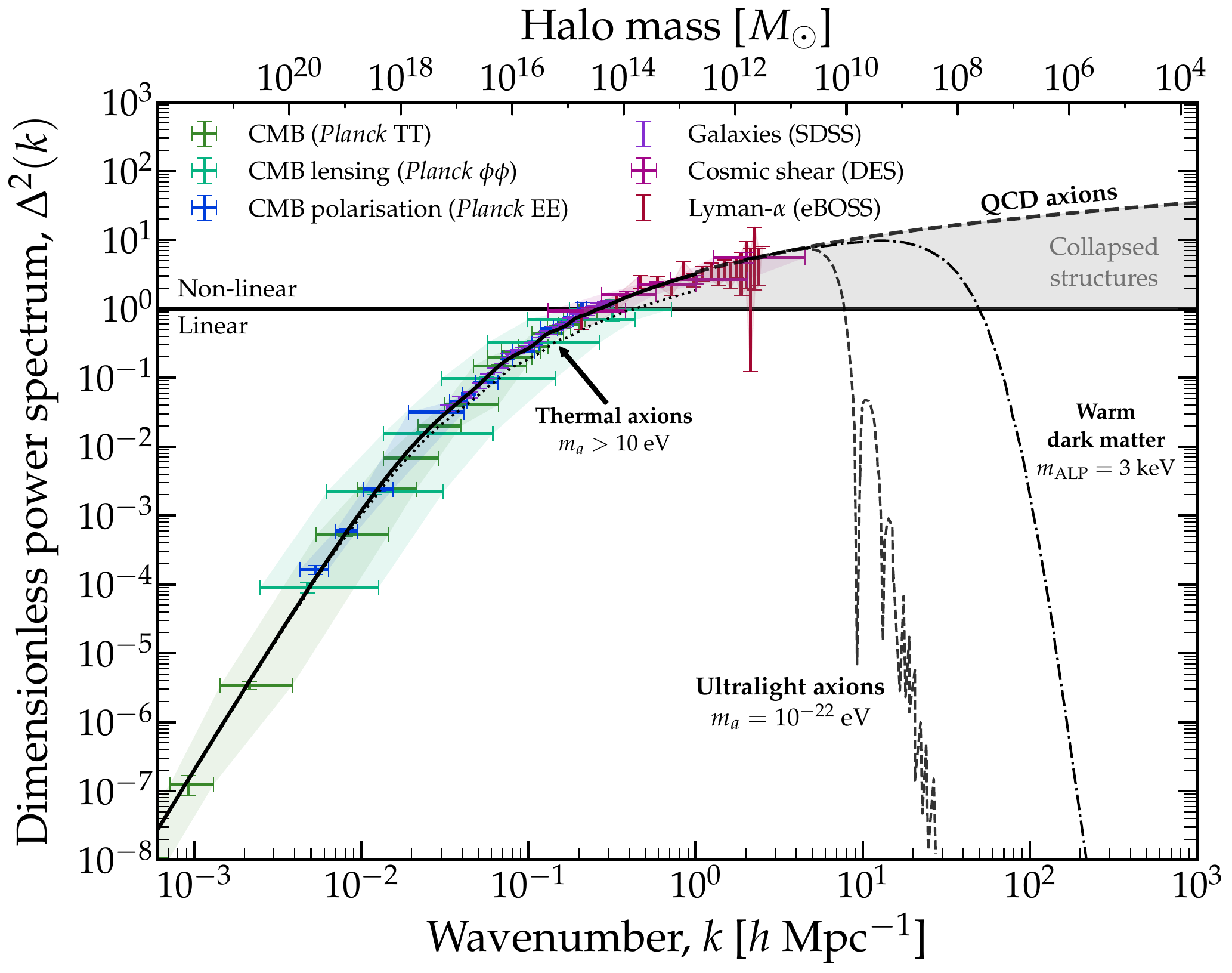}
    \caption{The dimensionless linear matter power spectrum at $z=0$, as a function of the wavenumber of matter fluctuations, $k$, or equivalently the halo mass within that scale. The power spectrum measurements at large scales come from a recasting of CMB correlations in temperature (TT), E-mode polarisation (EE), and lensing ($\phi \phi$) measured by \emph{Planck}; clustering of luminous red galaxies from the Sloan Digital Sky Survey (SDSS); cosmic shear in the Dark-Energy Survey (DES); and at the smallest scales, measurements of the 1D line-of-sight clustering of matter from the Lyman-$\alpha$ forest. I also show three examples of axion-related models that would lead to departures from the expected power spectrum under CDM (solid black line going into the dashed line for the extrapolation at high $k$). These are described at the end of Sec.~\ref{sec:cosmologicalobservations}. This plot was inspired by Refs.~\cite{Kuhlen:2012ft, Bechtol:2022koa} and uses data compiled at Ref.~\cite{Chabanier:2019eai}.}
    \label{fig:DimensionlessPowerSpectrum}
\end{figure}
I show an example of the matter power spectrum in Fig.~\ref{fig:DimensionlessPowerSpectrum}, where the BAOs are just about visible if you zoom in around $k\sim 0.1 h$/Mpc. Notice that in this plot, rather than showing the dimensionful power spectrum, I am showing the dimensionless quantity,
\begin{equation}
    \Delta^2(k) = \frac{k^3}{2\pi^2} P_m(k)\, ,
\end{equation}
which is helpful because scales where $\Delta^2(k)>1$ indicate where linear perturbation theory breaks down and perturbations evolve non-linearly. The state-of-the-art in predicting the \textit{non-linear} matter power spectrum is to use extensive N-body simulations---I am showing here only the linear one. To get a more tangible sense of what each part of the power spectrum corresponds to, we can convert the physical scale of a perturbation $R \sim \pi/k$ to the mass contained within a sphere of that radius, 
\begin{equation}
M =\frac{4 \pi}{3} R^3\bar{\rho}_m \sim 10^{10} M_{\odot}\left(\frac{k}{10 \,\mathrm{Mpc}^{-1}}\right)^{-3} \, ,
\end{equation}
where $\bar{\rho}_m = \rho_{\rm tot}(\Omega_b + \Omega_{\rm DM})$. This is where the auxiliary halo-mass axis in Fig.~\ref{fig:DimensionlessPowerSpectrum} comes from. Galaxy clusters exist below $M\lesssim 10^{16}~M_\odot$, individual galaxy halos for $M\lesssim10^{13}~M_\odot$, and sub-galactic halos, ``sub-halos'', do not kick in until masses smaller than $M\lesssim 10^9~M_\odot$.

As Fig.~\ref{fig:DimensionlessPowerSpectrum} shows, we have observed the matter power spectrum on a wide range of scales. On the largest scales, we can recast the CMB fluctuations into it~\cite{Tegmark:2002cy}, whereas various types of astronomical surveys can map it at smaller scales. Examples of the latter include the clustering of galaxies (e.g.~\cite{Reid2010}), maps of neutral Hydrogen clouds constructed using the so-called Lyman-$\alpha$ forest (e.g.~\cite{SDSS2017}), as well as reconstructed maps of dark matter halos created using the way they subtly gravitationally lens fields of galaxies (e.g.~\cite{DES:2017qwj}). Connecting up all of these probes results in a matter power spectrum that is marvellously consistent with $\Lambda$CDM and a Universe dominated by dark matter.

The matter power spectrum also reveals why our best cosmological model is specifically $\Lambda$CDM and not simply $\Lambda$DM. Dark matter is \textit{cold}. In the lingo, this means that whatever dark matter is, it was produced with initial velocities that were significantly smaller than the typical velocities it accelerates to when falling into its halos. Departures from this go under the name ``hot dark matter'' (HDM) for the extreme case where dark matter starts out with relativistic velocities $T_i \gg m_{\rm DM}$, or ``warm dark matter'' (WDM) if it starts with only mildly relativistic velocities $T_i > m_{\rm DM}$. The various types of dark matter are distinguished because of something called the free-streaming scale, $\lambda_{\rm fs}$. This is the typical distance over which a population of particles will spread out due to their random individual velocities until redshift slows them down to non-relativistic speeds.

The point is that any would-be structures smaller than the free-streaming scale (i.e.~$k \gtrsim \pi/\lambda_{\rm fs}$) are washed away because the dark matter is fast enough to escape out of them. Observations of the matter power spectrum show that dark matter forms structures down to all scales that we have been able to probe so far. If dark matter did have a substantial initial velocity then $P_m(k)$ would get suppressed above some value of $k$ depending on how far the particles could free-stream during the initial growth of structure. Since we have decent knowledge of the power spectrum down to $k\sim 10$~Mpc$^{-1}$, this means that dark matter must have already been non-relativistic when modes of that size crossed the horizon, $k\sim a H$. This corresponds to a redshift of $z\sim 10^7$ or equivalently when the temperature of the Universe was around a keV---if dark matter was produced in thermal equilibrium, then it must be heavier than this. In fact, even a small sub-population of HDM present in the Universe is also excluded---halos in this scenario would be larger and puffier, incompatible with the observed granularity in structure down to sub-Mpc-scales~\cite{Bode:2000gq, Schneider:2011yu}. This turns out to be a relevant statement when it comes to some axions models, where we do expect a small thermally-produced population, as I discuss in Sec.~\ref{sec:thermalaxions}. 

WDM, however, posits that the temperature dark matter was produced at may leave the particles mildly relativistic, and so 100\% of DM in the form of WDM with masses around a couple of keV is excluded because these particles free-stream a bit too much, but those with slightly heavier masses, around 3--7 keV, would have less time to free-stream until they became non-relativistic and so would generate a suppression to $P(k)$ only on scales just below the resolution of our finest maps of structure. See e.g.~Refs.~\cite{Irsic:2023equ, DES:2020fxi, Dekker:2021scf} for recent bounds on the WDM mass and Ref.~\cite{Bechtol:2022koa} for a discussion on the prospects to push this bound further upwards by mapping structure down to even smaller scales. There are some crude similarities between the effects of WDM, and axions with masses in the ultralight regime, $m_a\sim 10^{-22}$~eV, which will be described in Sec.~\ref{sec:ultralight}.

CDM as a model states that the dark matter has negligible initial velocities and is firmly non-relativistic during structure formation. However, many candidates labelled as CDM are not absolutely cold in the ideal sense. For example, if a weakly-interacting massive particle (WIMP) is thermally produced it will have some finite initial velocity spread even if it decoupled from the Standard Model bath while non-relativistic. For a WIMP with $\mathcal{O}(100\,{\rm GeV})$ mass, structures would only grow down to the Earth mass~$10^{-6}\,M_\odot$~\cite{Green:2003un, Anderhalden:2013wd,Ishiyama:2014uoa}\footnote{See also Refs.~\cite{Delos:2019mxl, Delos:2022bhp, White:2022yoc, Delos:2022yhn, Zheng:2023myp, Wang:2019ftp} for more recent work on understanding the small-scale limits of CDM halos.}. Interestingly, axions produced via the misalignment mechanism are even colder than this, as we will see in the next section.

So it is clear that any deviations from CDM, like what would be expected if dark matter is not \textit{absolutely} non-relativistic---but also if it were not totally collisionless or non-interacting~\cite{Boehm:2000gq}---must therefore emerge only on smaller scales. These scales represent the frontier of our knowledge: if dark matter possesses some kind of physics that will become available to cosmologists in the near future, this is where it will be hiding. I have shown a few axion-related examples in the plot. Just to be concrete (I will cover them in detail later) the examples are: the imprint of a small relic population of thermally-produced hot-dark-matter axions (dotted), the high-$k$ extrapolation of pure CDM expected for misalignment-produced QCD axions (thick dashed), and the suppressed power spectra of warm dark matter and ultralight-aka-fuzzy dark matter (thin dashed). As for their compatibility with our Universe: Pure CDM is frustratingly consistent with all data. The warm and ultralight DM models I'm showing are around the level of what can be tested right now on non-linear scales, whereas the thermal axion mass I've assumed here is firmly ruled out.

\section{Axions as cold dark matter}\label{sec:axionCDM}
Having revealed just how unambiguously our observations of the Universe lead us down the path of cold dark matter, let us now make the leap and assume that what we are seeing are the imprints of a large relic population of some very light and feebly-coupled scalar particle---the axion.

We already know that everything is consistent with the CDM paradigm in which dark matter is born effectively non-relativistic, allowing it to collapse into sub-galaxy-sized halos, so let us see what this implies about the nature of axion dark matter. Take the dark matter around us in the galaxy for example. The local density is measured to be around $\rho_{\rm DM,\, local} = 0.4$~GeV~cm$^{-3}$ and the dark matter particles should be moving with speeds similar to the stars they are orbiting alongside, around 300~km/s\footnote{There will be a bit more on these numbers in Sec.~\ref{sec:directdetection} when I discuss direct detection.}. We can then try and reconcile these numbers with the expected properties of the axion. One of these properties is the mass, whose value we do not know, except that it is constrained to be very, very light, e.g.~$m_a \lesssim$~meV for the ``QCD axion"---the version of this model that is able to solve the strong CP problem. This is so light, in fact, that the \textit{number} densities of axion dark matter inside galactic halos must be macroscopically huge.

An interesting comparison to make is between the number density of dark matter particles, and the de Broglie `volume' which is sort of like the space taken up by that particle's quantum state--- multiplying them will give us an order-of-magnitude estimate of how many particles will need to occupy each of those states to make up the density. Taking a typical\footnote{Getting a little ahead of ourselves here, the reason why this is typical will be explained in the next section.} dark-matter axion mass of $m_a = 100~\upmu$eV we get,
\begin{equation}
    \lambda_{\mathrm{dB}} \equiv \frac{2 \pi}{m_a v} = 12.4 \,\mathrm{m}\left(\frac{100\,\upmu\mathrm{eV}}{m_a}\right)\left(\frac{300 \mathrm{~km} / \mathrm{s}}{v}\right) \, ,
\end{equation}
for the de Broglie wavelength, and
\begin{equation}
    \mathcal{N} \sim \frac{\rho_{\rm DM,\,local}}{m_a} \times \frac{1}{\lambda_{\rm dB}^3} \sim 10^{22} \left(\frac{100\,\upmu \mathrm{eV}}{m_a}\right)^4\left(\frac{300 \mathrm{~km} / \mathrm{s}}{v}\right)^3 \, ,
\end{equation}
for the number of particles per de Broglie volume---a rather large number indeed. When we have a vast number (a macroscopically vast number) of particles collectively occupying a single state, what we really have on our hands is something classical. There's nothing strange about this, it's perfectly correct to model some photons as classical electromagnetic waves when you have loads of them. So this is how we prefer to treat light bosonic dark matter, rather than talking about discrete particles, and we give it an appropriate name: \textit{wave-like dark matter}~\cite{Sin:1992bg}.\footnote{Wave-like dark matter applies strictly to bosonic dark matter candidates. For fermions, it is impossible to make this description due to the Pauli exclusion principle, hence why all dark matter candidates with masses below $\sim$0.16 keV must be bosonic~\cite{Alvey:2020xsk}. This is the Tremaine-Gunn bound---a more sophisticated version of the argument I have hand-waved here applied to known dark-matter-rich environments like dwarf galaxies~\cite{Tremaine:1979we}.}

A classical description of the macroscopic behaviour of axions is best obtained by computing the coherent state of its quantum field---any quantum fluctuations in such a highly occupied system will be negligible. This essentially boils down in the end to a classical field that we can assign the required energy $E = m_a + \frac{1}{2} m_\phi v^2$ and momentum $\mathbf{p} = m_\phi \mathbf{v}$. Something like,
\begin{equation}
    \phi(x,t) \approx \frac{\sqrt{2 \rho_{\rm DM,local}}}{m_\phi} \sin( m_a t + m_a \frac{1}{2} v^2t + \mathbf{p}\cdot \mathbf{x}) \, .
\end{equation}
Once I have introduced the necessary formula (Eq.~\ref{eq:rho_scalar}) you can check that the time-averaged energy density for a scalar field with this amplitude will give you back the required $\rho_{\rm DM,\, local}$.

This hypothesis turns out to work in explaining the cosmological behaviour of dark matter when we go to very large scales too. However, you would be forgiven for thinking that an oscillating classical field doesn't sound very matter-y. So our first goal is to understand how an oscillating classical field is consistent with the cosmological behaviour that we assign to matter. To get there, we need a production mechanism---a reason why the oscillations got started in the first place. The most popular one that the vast majority of axion dark matter cosmology is centred around is called the \textit{misalignment mechanism}.

\subsection{The misalignment mechanism}
The behaviour and abundance of axion dark matter is where we want to get to, but there are a few extra pieces of baggage associated with the axion that it is less instructive to include from the beginning. So, to get across the main essence of the physics of misalignment, I will begin with the mechanism applied to a generic real scalar field described solely by a mass $m_\phi$. Much of this will carry over into the axion, up to a few modifications, and in any case, these results can be used for scalar dark matter and axion-like particles.

\subsubsection{Scalar misalignment}\label{sec:misalignment}
The gist of the misalignment mechanism is in the name. The field starts its oscillations because it began life at some value that was misaligned from the potential minimum that it would eventually settle to. The reason for this misalignment could be that at some high temperature, a previously preserved symmetry was broken during a phase transition, causing the field's potential to acquire a minimum at a location unrelated to where it happened to be sitting beforehand, but the details of that are not essential right now. The only important ingredients are that we have a scalar field with potential $V(\phi) = \frac{1}{2}m_\phi^2 \phi^2$ sitting at some arbitrary initial value $\phi_i \neq 0$.

\begin{figure}
        \centering
        \includegraphics[width=0.49\textwidth]{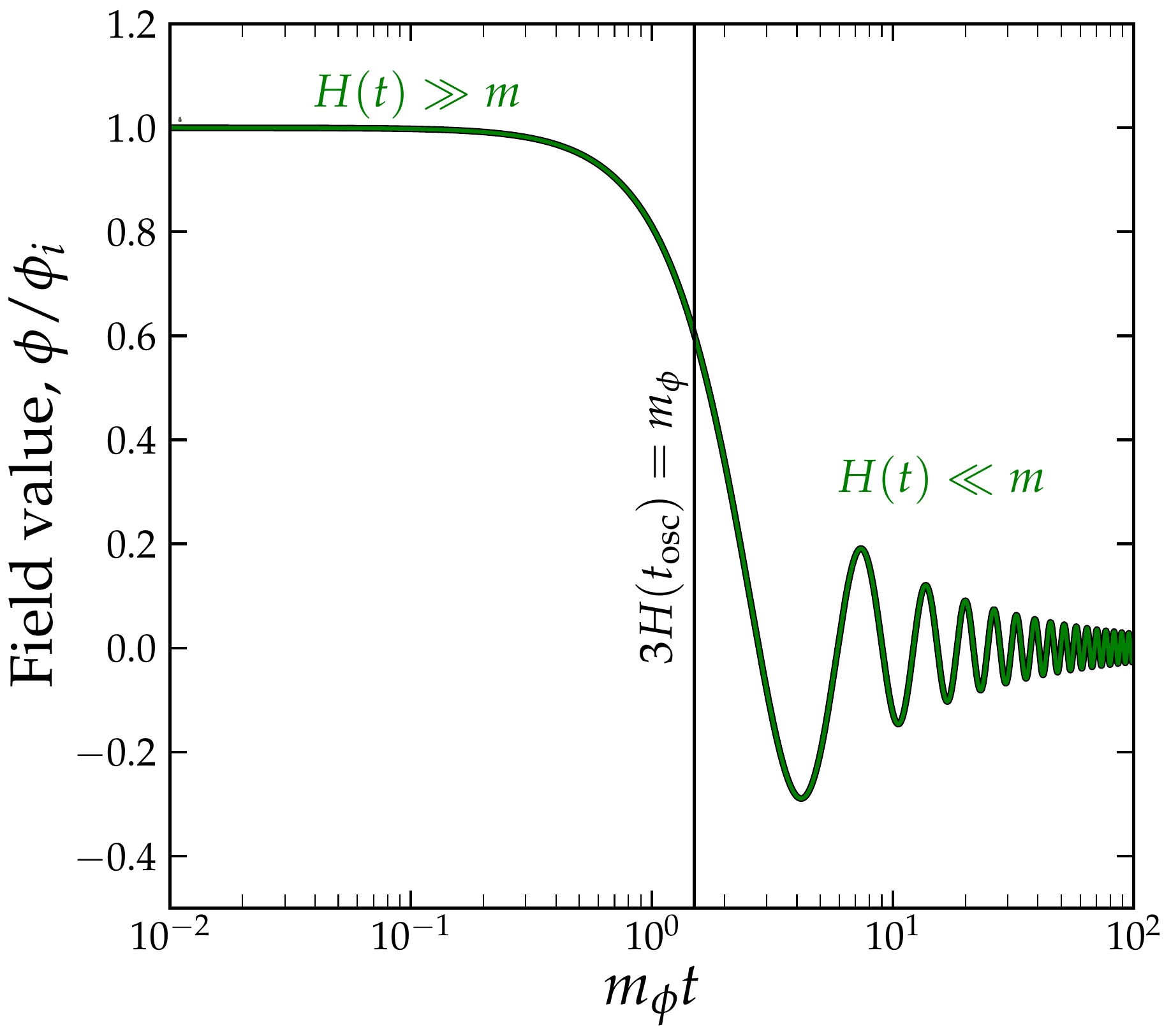}
        \includegraphics[width=0.49\textwidth]{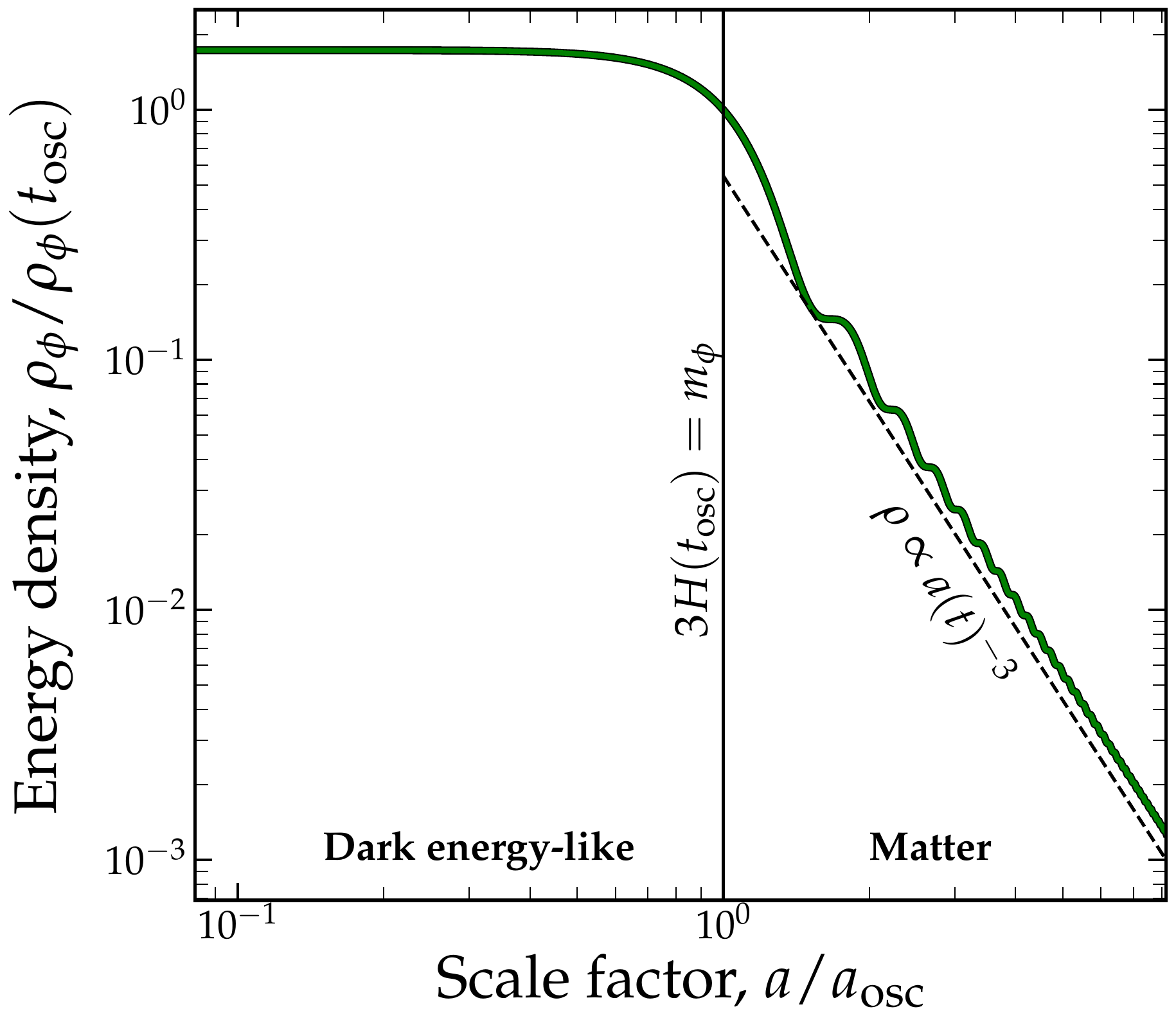}
        \caption{Cosmological evolution of a scalar field rolling down a quadratic potential during radiation domination. The left-hand panel shows the field value, $\phi(t)$, relative to its initial value $ \phi(0) = \phi_i$. The right-hand panel shows the energy density in the field, defined in Eq.(\ref{eq:rho_scalar}). The time $t_{\rm osc}$ is defined here as when $3H(t_{\rm osc}) = m_\phi$ is satisfied---it roughly corresponds to the time when the system transitions from an over-damped to a damped harmonic oscillator. This is also the moment when the cosmological scaling of the energy density as a function of scale factor, $a$, transitions from dark-energy-like ($\rho_\phi \approx const.$), to matter-like behaviour. In other words, damped oscillations in a scalar field can be thought of as dark matter.}
        \label{fig:phi}
\end{figure}

We start with the dynamics. The action for this system is the following,
\begin{equation}
    S=\int \mathrm{d}^4 x \sqrt{-g}\left[-\frac{1}{2}\left(\partial_\mu \phi \partial^\mu \phi\right)-\frac{1}{2} m^2 \phi^2\right] \, ,
\end{equation}
where the quantity in the square brackets is the Lagrangian $\mathcal{L}$, and $g$ is the determinant of the metric, for which we will use the FRW one from earlier: $g_{\mu \nu}=\operatorname{diag}\left(-1, a^2, a^2, a^2\right)$.

The equation of motion for the field we find by varying the action
\begin{equation}
    \frac{\partial \sqrt{-g} \mathcal{L}}{\partial \phi} - \partial_\mu \frac{\partial \sqrt{-g} \mathcal{L}}{\partial\left(\partial_\mu \phi\right)} = 0 \, ,
\end{equation}
resulting in the Klein-Gordon equation,
\begin{equation}\label{eq:KGeq}
    \square \phi-m_\phi^2 \phi=0 \, .
\end{equation}
For the time being, we are just going to be thinking big-picture: we first want to check that the behaviour of a scalar field can be thought of as dark matter with a large-scale density that scales like $\bar{\rho} \sim a^{-3}$. So let us assume that we are following only the homogeneous zero-momentum mode of the solution to Eq.(\ref{eq:KGeq}), and bring in any higher modes/perturbations later if it looks like we need to. The upshot of that is simply that when we write out the d'Alembertian ($\square$) we will ignore the spatial derivative,
\begin{equation}
    \square=-\partial_t^2-3 H \partial_t \, .
\end{equation}
Plugging this in we get the equation of motion we need to solve,
\begin{equation}\label{eq:scalar_eom}
    \ddot{\phi}+3 H(t) \dot{\phi}+m^2 \phi=0 \, .
\end{equation}
Since all of this physics will have to happen in advance of matter-radiation equality for any of this to make sense for dark matter, we assume we will be in a radiation-dominated background where $H = 1/2t$. The equation of motion then has a very familiar form: just a harmonic oscillator with a (decaying) damping term.

Now to solve it, we plug in our initial conditions, for which we will assume\footnote{Note that this \textit{is} just an assumption, one that you can play with if you don't get the answer you want for some set of model parameters. People have considered, for instance, the situation where the field starts with a very large initial misaligned value~\cite{Arvanitaki:2019rax}, as well as the case where it starts with some initial velocity, $\dot{\phi}(0)>0$~\cite{Co:2019jts}. I will go into elaborations on the misalignment mechanism in Sec.~\ref{sec:ALPs}.} the field starts from some arbitrary value misaligned away from zero ($\phi(0) = \phi_i \neq 0$) and with no derivative ($\dot{\phi}(0) = 0$). The solution is then,
\begin{equation}
    \phi=\phi_i\left(\frac{2}{m_\phi t}\right)^{\frac{1}{4}} \Gamma\left(\frac{5}{4}\right) J_{\frac{1}{4}}(m_\phi t) \, ,
\end{equation}
where $J_n(x)$ is Bessel's function and $\Gamma(x)$ the gamma function. This solution is shown in the left-hand panel of Fig.~\ref{fig:phi}. Notice there are two regimes for this evolution, which depend on how important the damping term ($3 H(t)\dot{\phi}$) is compared with the mass term ($m_\phi^2 \phi$). When the Hubble parameter is large, the system is overdamped, but once the damping term has decayed away, the field's mass drives its dynamics and so it starts oscillating.

We're interested in the late time behaviour, so it's instructive now to take the limit where $mt$ is large. You can find that $J_{1/4}(x) \approx (\pi x/2)^{-1/2} \cos(x - \pi/3)$ for large $x$, so the late-time solution essentially involves an envelope that decays slowly multiplied by quickly oscillating part,
\begin{equation}
    \phi(t) \approx \phi_{\mathrm{env}}(t) \cos m_\phi t  \, .
\end{equation}
Now that we know how the field is behaving at late times, we can go and look at the energy density stored in it. The energy density of our scalar field can be read off from its energy-momentum tensor, which I won't do here, but the result is,
\begin{equation}\label{eq:rho_scalar}
    \rho_\phi=\frac{1}{2} \dot{\phi}^2+\frac{1}{2} m^2 \phi^2 \, .
\end{equation}
Working this out during radiation domination ($a \propto t^{1/2}$) and putting everything in terms of $a$, we see that,
\begin{equation}
    \rho_\phi \propto a^{-3} \, ,
\end{equation}
on timescales much longer than the timescale of the oscillations. This dilution with the volume of space is exactly how the energy density of matter scales. 

The full behaviour can be seen in the right-hand panel of Fig.~\ref{fig:phi}. Note that this scaling like matter takes hold only once the field starts oscillating. Evaluating the energy density at early times when the field is still slowly rolling down its potential, you will find that $\rho_\phi = \text{const}$, which is the behaviour we attribute to dark energy. Indeed, a slowly rolling scalar field is a decent candidate for dark energy, as well as the driver of inflation which seems to require the same sort of behaviour.

So we have shown that the average density of an oscillating scalar field scales like dark matter---that was our first task. Now let us try to make sure we get the right \textit{amount} of dark matter.

As mentioned in the previous section, we have measured the large-scale average dark matter density in our Universe to percent-level precision, and in a simple world, one candidate would explain all of it. So let us try to engineer our free parameters to align such that the following equation is true:
\begin{equation}\label{eq:Omegaphi}
    \Omega_\phi h^2 \equiv \frac{\rho_\phi({\rm today})}{3 H_0^2 M_{\mathrm{Pl}}^2} h^2 = 0.12 \, .
\end{equation}
Since our setup only has two unknowns, $m_\phi$ and $\phi_i$, enforcing the expression above will tell us how those parameters have to be related to each other if our scalar field is to add up to all of the dark matter.

So now what we need is the energy density in the field today, $\rho_\phi({\rm today})$, which we can find by rolling the clock forwards from the density in the scalar field at the point when it started scaling like matter. We already defined this time to be $t_{\rm osc}$, so the density today will just be the density at that time, diluted by how much space has expanded from then until now:
\begin{equation}
    \rho_\phi({\rm today})= \rho_\phi (t_{\rm osc}) \left(\frac{a_{\text {today }}}{a_{\text {osc }}}\right)^{-3} \, .
\end{equation}
where $a_{\rm osc} = a(t_{\rm osc})$ is the scale factor when the field started behaving like matter. As a simplifying approximation, let us also assume the field value at this time was $\phi(t_{\rm osc}) \approx \phi_i$, which shouldn't be too far off since $\phi$ doesn't evolve much in the overdamped period. Because $\dot{\phi}(t_{\rm osc})\approx 0$, this means that we can write $\rho_\phi(t_{\rm osc}) = \frac{1}{2} m_\phi^2 \phi_i^2$.

Now we can bring back Eq.(\ref{eq:densities_vs_T}) and write this in terms of the temperatures of the Universe now ($T_0 = 2.35\times 10^{-4}$~eV) and at $t_{\rm osc}$:
\begin{equation}
        \rho_\phi({\rm today}) = \rho_\phi\left(T_0\right)=\rho\left(T_{\rm osc}\right) \frac{g_{* s}\left(T_0\right)}{g_{* s}\left(T_{\rm osc}\right)}\left(\frac{T_0}{T_{\rm osc}}\right)^3 \, .
\end{equation}
The reason for doing this is so that we can find the temperature at $t_{\rm osc}$ using the Friedmann equation,
\begin{equation}\label{eq:3Hm}
    3 H(T_{\rm osc})^2 M_{\mathrm{Pl}}^2=\frac{\pi^2}{30} g_*(T_{\rm osc}) T_{\rm osc}^4 \, ,
\end{equation}
where we already know the Hubble parameter when the field began oscillating because that was how we defined it in the first place: $3 H(T_{\rm osc}) = m_\phi$. Plugging these expressions back into Eq.(\ref{eq:Omegaphi}), and using the fact that $3 H_0^2 M^2_{\rm Pl} = 8.07\times10^{-11}~h^2$~eV$^4$, we have numerical values for everything other than $m_\phi$ and $\phi_i$, leaving us with,
\begin{equation}
    \Omega_\phi h^2=0.12\left(\frac{\phi_i}{4.7 \times 10^{16}\, \mathrm{GeV}}\right)^2\left(\frac{m_\phi}{10^{-21} \mathrm{eV}}\right)^{\frac{1}{2}} \, .
\end{equation}
To get this I have assumed $g_\star = 3.4$ and $g_{\star,s}(T_0)\approx g_{\star,s}(T_{\rm osc})$, which will only hold if the number of degrees of freedom are not changing much between $t_{\rm osc}$ and now. The ramification of this choice is that it only applies when $t_{\rm osc}$ is relatively \textit{late}, or equivalently when the scalar mass is small. To see this, inspect Eq.(\ref{eq:3Hm}) and enforce the condition $3H(T_{\rm osc}) = m_\phi$. This reveals the relationship $T_{\rm osc}\propto m_\phi^{1/2}$, i.e.~the heavier the scalar's mass, the higher the temperature when it starts oscillating and hence the \textit{earlier} it starts oscillating. At the same, if we try to make the scalar mass \textit{too} light we will run into problems. We need our field to be behaving like dark matter by matter-radiation equality at the latest. Enforcing $T_{\rm osc}>T_{\rm eq}$ implies there will be a bound on the scalar mass $m_\phi > T_{\rm eq}^2/M_{\rm Pl}\sqrt{\pi^2 g_\star/10}\sim 10^{-28}$~eV. However, this turns out to be far from the most competitive lower limit it is possible to draw on the mass of dark matter, as I will discuss in Sec.~\ref{sec:ultralight}.

\subsubsection{Axion misalignment}\label{sec:axionmisalignment}

The previous section went through the misalignment mechanism for a generic scalar field. We will now modify the setup so that we can describe \textit{axion} dark matter. To avoid bringing along any unnecessary theoretical baggage, let's start with a very minimal definition of the QCD axion that will get us where we need to go: the axion is a pseudo-Nambu Goldstone boson that emerges when some $U(1)$ symmetry is spontaneously broken at a high energy scale labelled $f_a$, and then explicitly broken at a temperature $T_{\rm QCD}$.

\begin{figure}
        \centering
        \includegraphics[width=0.99\textwidth]{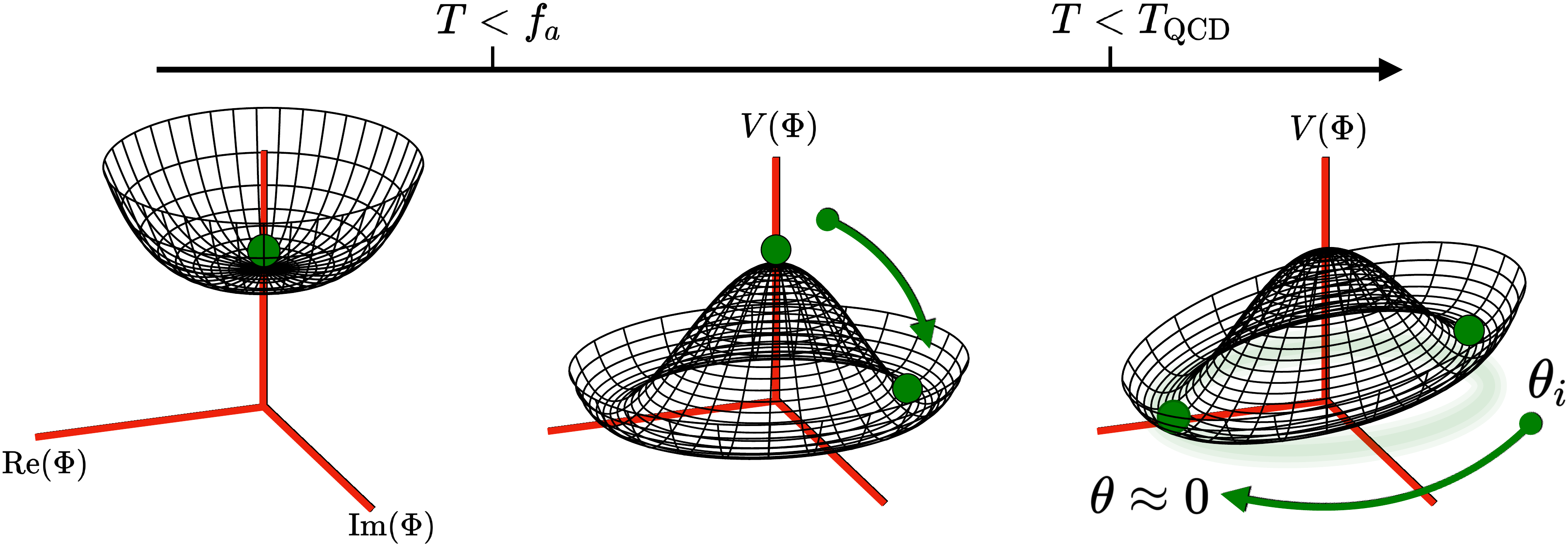}
        \caption{A cartoon (i.e.~not to scale) illustration of the evolution of the PQ complex scalar's tilted-wine-bottle potential, Eq.(\ref{eq:fullpotential}). The axion appears as a massless degree of freedom when the PQ symmetry is spontaneously broken at $T \lesssim f_a$. It then acquires a mass when the PQ symmetry is \textit{explicitly} broken at $T\lesssim T_{\rm QCD}$. The initial angle, $\theta_i$, that is chosen at $T\sim f_a$ is therefore expected to be misaligned from $\theta = 0$ where the axion field eventually resides today.} 
        \label{fig:Axion_PQPotential}
\end{figure}

What does this entail? Firstly the Goldstone mode of a field with this broken $U(1)$ symmetry can be expressed as the angular component of a complex scalar, $\Phi$, governed by the potential,
\begin{equation}
    V_{\rm PQ}(|\Phi|) = \frac{\lambda}{8}(|\Phi|^2 - f^2_a)^2 \, ,
\end{equation}
This equation describes the potential at temperatures below $f_a$ but above $T_{\rm QCD}$. PQ stands for Peccei-Quinn, who came up with this mechanism~\cite{Peccei:1977ur, Peccei:1977hh} and the $U(1)$ symmetry I will often call the PQ symmetry. We will write out the complex scalar in terms of a radial mode and a phase: $\Phi = |\Phi| e^{i\theta}$, where the phase $\theta$ we interpret as the axion. The `pseudo' part of pseudo-Nambu-Goldstone boson suggests that the shift symmetry enjoyed by the axion under the potential above is broken somehow. For the QCD axion, this breaking occurs explicitly at the Lagrangian level, and is by design. We do this by coupling the axion to the gluons in such a way that the potential for the full field becomes,
\begin{equation}\label{eq:fullpotential}
    V(\Phi)=V_{\mathrm{PQ}}(|\Phi|)+V_{\mathrm{QCD}}(\theta)=\frac{\lambda}{8}\left(|\Phi|^2 -f_a^2\right)^2+\chi(T)(1-\cos \theta) \, ,
\end{equation}
around and below the QCD scale, $T \sim \mathcal{O}(100\,{\rm MeV)}$. The temperature-dependent function $\chi(T)$ comes from QCD and is called the topological susceptibility. This function is zero at very high temperatures, grows as the temperature of the Universe cools, and plateaus below the confinement scale $T<T_{\rm QCD}$. A cartoon of how the full PQ potential evolves through these different stages is shown in Fig.~\ref{fig:Axion_PQPotential}. 

The topological susceptibility encapsulates the quantum-mechanical effects that give rise to the puzzle for which the axion is a solution. As the formula above suggests, this function governs the axion's dynamics, so we should understand where it comes from. As you will learn about in reviews on axion theory (e.g.~\cite{Choi:2024ome}), the axion was invented to solve the `strong-CP problem'---an apparent fine-tuning of a fundamental parameter of the Standard Model to the value that happens to preserve charge-parity (CP) symmetry in the strong interaction. This parameter shows up as the coefficient of a term in the QCD Lagrangian involving the gluons, and even though the energy-minimising value of that coefficient is zero~\cite{Vafa:1984xg}, because it is merely a parameter, it has no dynamics. By creating a new field that couples to the gluons in the same way, the combined CP-violating coefficient is a field+constant which \textit{is} dynamical. So the reason CP is preserved is then simply a result of the field settling to that same energy-minimising value~\cite{Peccei:1977ur, Peccei:1977hh, Weinberg:1977ma, Wilczek:1977pj}. 

The upshot of all of that is that the potential for this new field, the axion, is generated by the same physical processes that gave rise to the original CP problem. To give it a name, these physical processes are a phenomenon known as instantons, which, in a nutshell, are tunnelling solutions that connect the landscape of topologically inequivalent vacua of QCD. These instantons govern the potential that the axion feels, but their effects are temperature dependent and so this is what the function $\chi(T)$ captures.

Notice that if we expand out $V(\theta)$ to second order in the field, $\chi(T)$ essentially plays the role of a mass squared. Except the axion, as we've written it, is just an angle---the canonically normalised field that we are coupling to QCD is $\phi = \theta f_a$. So if we want to express the topological susceptibility in terms of a `temperature-dependent axion mass', as in $V(\phi) = \frac{1}{2}m_a^2 \phi^2$, then we should express it as $\chi(T) = m^2_a(T) f_a^2$, which is what we will do.

\begin{figure}
    \centering
    \includegraphics[width=0.99\linewidth]{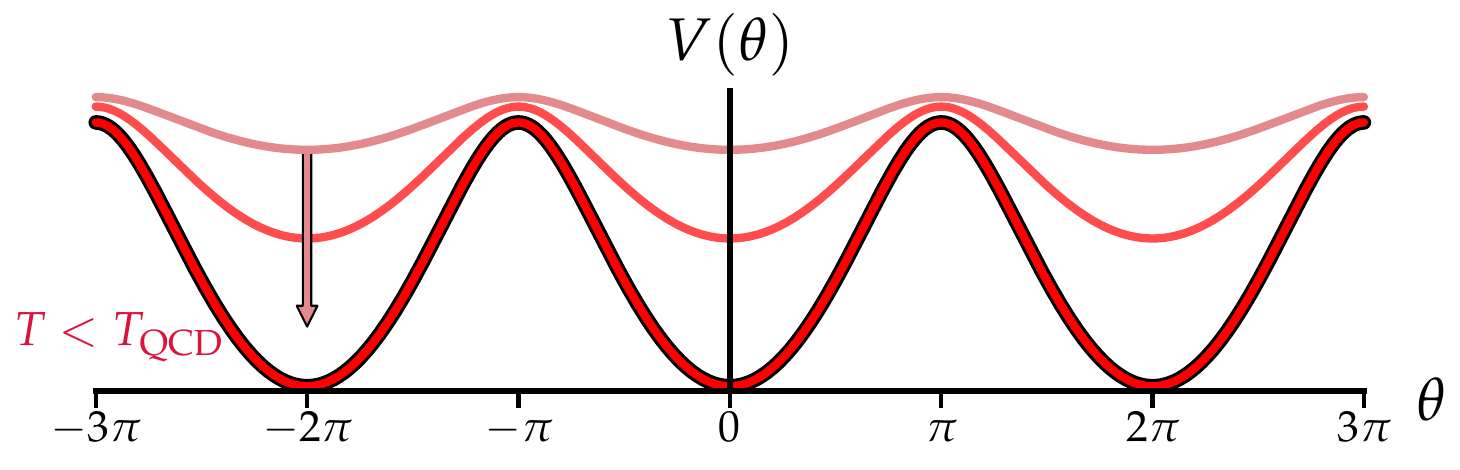}
    \caption{Evolution of the axion's potential as the temperature of the Universe cools below the QCD confinement scale. When the axion is produced at high temperatures, it has a flat potential and hence no mass. However because of the instanton temperature-dependence encapsulated in the topological susceptibility (Eq.~\ref{eq:topologicalsusceptibility}), the axion mass grows like $m_a\sim T^{-4}$ as the temperature cools, until it acquires its present-day mass at $T\lesssim 150$~MeV.}
    \label{fig:TopologicalSusceptibility}
\end{figure}

We can now go through the same exercise as before, and find the equation of motion for the field $\phi$ and rewrite it in terms of $\theta$ to remove the $f_a$'s:
\begin{equation}\label{eq:axioneom}
    \ddot{\theta}+3 H \dot{\theta}+m_a(T)^2 \sin (\theta)=0 \, .
\end{equation}
This is very similar to the scalar case from before, Eq.(\ref{eq:scalar_eom}); however, there are some important differences. One of these differences is that in the last term, we have a $\sin(\theta)$ instead of just a $\theta$. If our initial condition is $\theta_i\ll 1$ then this distinction won't matter much and we will approximately have the same dynamics as for the generic scalar case, but for initial values approaching $\theta_i \approx \pi$, we will have to correct for the fact our oscillator is not exactly harmonic but is sitting in a cosine-like potential.\footnote{I'm also brushing over another detail here which is that the axion potential written down in Eq.(\ref{eq:fullpotential}) is also an approximation. The low-temperature potential (i.e.~below the confinement scale) can be obtained from chiral perturbation theory and is actually more like $V(\theta) \propto \sqrt{1-A \sin^2(\theta/2)}$ where $A$ is a constant involving the up and down quark masses. See Eq.(54) of Ref.~\cite{DiLuzio:2020wdo} for the full expression and a derivation. This shape is what I am plotting for the $T<T_{\rm QCD}$ line in Fig.~\ref{fig:TopologicalSusceptibility}. The difference can be spotted if you pay close attention to the peaks at $|\theta| \sim n\pi$---they are a little sharper than a cosine. It turns out that the simple one-instanton potential $V \propto -\cos{\theta}$ is more accurate at $T\gtrsim$~GeV anyway, which is the typical temperatures we are working at right now in this calculation, hence the use of this approximation.} I'll mention this again a little later on, it is a relatively simple correction.

The more major change from the scalar case is that we now have a temperature-dependent mass. This makes finding the solution for $\theta(t)$ a bit more complicated. Firstly, let us at least make a start by writing down a functional form for how $m_a(T)$ could behave:
\begin{equation}\label{eq:topologicalsusceptibility}
    m_a(T)^2 = \begin{cases}
        m_a^2 \left(\frac{T}{T_{\rm QCD}}\right)^{-n} \, &\text{for} \, T>T_{\rm QCD} \, , \\
        m_a^2 \quad &\text{for} \, T<T_{\rm QCD} \, .
    \end{cases}
\end{equation}
The full temperature-dependence of the axion mass has been computed using analytic techniques with differing regimes of validity, such as the dilute instanton gas approximation~\cite{Gross:1980br} and the interacting instanton-liquid model~\cite{Wantz:2009it}. It has also received a full numerical treatment using lattice QCD~\cite{Borsanyi:2016ksw}. For the calculation I will present now, I am just going to use the simple power-law approximation written above with $T_{\rm QCD} \approx 150$~MeV and $n \approx 8$, which is a reasonably good fit to the results of Ref.~\cite{Borsanyi:2016ksw}. Nevertheless, the parameters appearing here ultimately reflect an underlying theoretical uncertainty, so we will keep them around in our expressions to see where they end up.

We have all the ingredients now to go and compute the full solution to $\theta(t)$ over some cosmological background. You could do this numerically if you wanted, but there is a faster way to get an answer for the average density in axions as a function of temperature, $\rho_a(T)$, which is the only thing we need. The reason this was easy to get previously was that we assumed the comoving energy density for a constant-mass scalar was conserved, i.e.~$\rho_\phi a^3 = const$. This was fine in that case, but strictly it is not the energy density that is comovingly conserved: the \textit{number} density of particles is our adiabatic invariant. This means we should start by writing down,
\begin{equation}\label{eq:na_step}
    n_a(T_0) = n_a(T_{\rm osc}) \left( \frac{a_{\rm today}}{a_{\rm osc}} \right)^{-3} \, ,
\end{equation}
The number density around the time the oscillations began can be approximated in the same way as before: by assuming the field starts at an initial value $f_a\theta_i$ that hasn't changed much by $t_{\rm osc}$, and that $\dot{\theta}_i = 0$. This means we can write down,
\begin{equation}
    n_a(T_{\rm osc}) = \frac{\rho_a(T_{\rm osc})}{m_a(T_{\rm osc})} \approx \frac{1}{2} m_a(T_{\rm osc}) f_a^2 \theta_i^2 \, .
\end{equation}
Combining this with Eq.(\ref{eq:na_step}) and using the conservation of entropy density as before, we get, 
\begin{equation}
\begin{aligned}
        \rho_a(T_0) &= \frac{1}{2}\theta_i^2 f_a^2 m_a m_a(T_{\rm osc}) \left( \frac{a_{\rm today}}{a_{\rm osc}} \right)^{-3} \\ 
         &= \frac{1}{2}\theta_i^2 f_a^2 m_a m_a(T_{\rm osc}) \left(\frac{g_{\star,s}(T_0) T^3_0}{g_{\star,s}(T_{\rm osc}) T^3_{\rm osc}} \right) \, .
\end{aligned}
\end{equation}
To evaluate this we will need to know $T_{\rm osc}$. Let us again assume this is the temperature when the condition $3H(T_{\rm osc}) = m_a(T_{\rm osc})$ is true. I must emphasise that the `3' there is chosen as a decent approximation to the full result~\cite{Wantz:2009it}---in general, we should be solving $\theta(t)$ numerically if we want a precise answer. 

Rearranging the condition that defines $T_{\rm osc}$, we find,
\begin{equation}\label{eq:Tosc}
        T_{\rm osc} = \left(\sqrt{\frac{10}{\pi^2 g_\star(T_{\rm osc})}} m_a M_{\rm Pl}  \right)^\frac{2}{n+4} \, T_{\rm QCD}^\frac{n}{n+4}
\end{equation}
which again requires a solution involving the temperature dependence of $g_\star(T)$. Nonetheless, we can check the numbers to see that $T_{\rm osc} \approx 1$~GeV for $m_a = 10\,\upmu$eV, $n = 8$ and $T_{\rm QCD} = 150$~MeV. So the oscillations will start sometime between the electroweak (EW) and QCD eras, where the effective number of relativistic degrees of freedom is roughly,
\begin{equation}
    g_\star(T_{\rm EW}>T>T_{\rm QCD}) \approx g_{\star,s}(T_{\rm EW}>T>T_{\rm QCD}) \approx 18+\frac{7}{8} \times 50=61.75 \, .
\end{equation}
We are also going to need the entropic degrees of freedom today, which is just $g_{\star,s}(T_0) = 3.91$.

The number we wish to calculate is then simply the present-day axion energy density $\rho_a(T_0)$ as a fraction of the total cosmic energy density: $\rho_{\rm tot} = 3 H_0^2 M_{\rm Pl}^2 = 8.07\times 10^{-11}\,h^2$~eV$^4$,
\begin{equation}\label{eq:Omega_a_step}
    \Omega_a h^2 = \frac{\theta_i^2 f_a^2 m_a m_a(T_{\rm osc})}{6 H_0^2 M_{\rm Pl}^2 } \left(\frac{g_{\star,s}(T_0) T^3_0}{g_{\star,s}(T_{\rm osc}) T^3_{\rm osc}} \right) \, .
\end{equation}
Before we put this into numbers, we should pause to notice that the expression Eq.(\ref{eq:Omega_a_step}) looks a little more complicated than our previous case where we had only two unknowns $\phi_i$ and $m_\phi$. It seems like we now have three unknowns $m_a$, $f_a$ and $\theta_i$. However, the magic of axion dark matter is that everything (up to details) boils down to a single number. 

First of all, $\theta_i$ is just an angle, so it can only span the circle as opposed to $\phi_i$ which was in principle unbounded. Secondly, and most importantly, is the fact that the axion's low-temperature mass has a simple and fixed relationship to the PQ symmetry-breaking scale $f_a$. Roughly speaking, the axion's parameters are connected to the pion's mass and decay constant through $m_\pi f_\pi = m_a f_a$, but more precise calculations leveraging chiral perturbation theory at next-to-leading order as well as lattice QCD now pin down this relationship to be the following~\cite{GrillidiCortona:2015jxo}:
\begin{equation}\label{eq:axionmass}
    m_{a}= (5.70\pm 0.007)\,\upmu \mathrm{eV}\left(\frac{10^{12} \,\mathrm{GeV}}{f_{a}}\right) \, .
\end{equation}
So given the fact that we can remove $f_a$ from the equation and guess what value $\theta_i$ could take, Eq.(\ref{eq:Omega_a_step}) is solved. In other words, \textit{we can predict the mass that the axion has to be for $\Omega_a h^2 = 0.12$ to be true.}

Let us now do that. We start by seeing how $\Omega_a$ depends on the parameters. Rewriting everything in terms of $m_a$ and $\theta_i$ we see that,
\begin{equation}
    \Omega_a h^2 \propto \theta_i^2 m_a^{-\frac{n+6}{n+4}}
\end{equation}
But of course what we want is a number; so choosing now $n=8$, we can find that $\Omega_a h^2 = 0.12$ is satisfied when,
\begin{equation}
\Omega_a h^2 \approx 0.12 \, \theta_i^2\left(\frac{4.7 \, \upmu \mathrm{eV}}{m_a}\right)^{\frac{7}{6}} \approx 0.12 \left(\frac{\theta_i}{2.155}\right)^2\left(\frac{9.0 \, \upmu \mathrm{eV}}{m_a}\right)^{\frac{7}{6}} \, ,
\end{equation}
A more careful numerical calculation that includes an accurate temperature dependence for the axion mass and correctly tracks the temperature dependence of $g_{\star,s}$ etc.~yields a slightly higher value~\cite{Borsanyi:2016ksw}
\begin{equation}\label{eq:Omega_a_final}
    \Omega_a h^2 = 0.12 \left(\frac{\theta_i}{2.155}\right)^2\left(\frac{28 \, \upmu \mathrm{eV}}{m_a}\right)^{1.16} \, .
\end{equation}
where the choice $\theta_i = 2.155$ is explained below.

The scaling with $m_a$ here, $\Omega_a \propto m_a^{-7/6}$ seems at first glance a little counter-intuitive---why do we get a smaller dark matter density when the axion mass is heavier? The thing to remember is that it is the \textit{number} density that gets set at $T_{\rm osc}$, and this just gets diluted by the expansion until today. So if $T_{\rm osc}$ occurs later on, then the axions get diluted less and we are left with a higher density of them today. Because $T_{\rm osc}$ is defined when the Hubble scale drops below the mass, \textit{lighter} axions start oscillating later and so they are the ones that get diluted the least.

So the conclusion is that it looks very much like a mass of $\mathcal{O}(10\,\upmu$eV) is natural for the QCD axion to explain dark matter. By no coincidence at all, this prediction lands within reach of existing haloscope experiments and represents the ``classic'' QCD axion window.

So what about the initial angle? Although this is indeed just an angle, if we want a better estimate for $m_a$ we need to decide what we should pick for it. In principle, an angle squared could be anything between 0 and $\pi^2$, and if we were to draw it truly at random, on average it would have a value, 
\begin{equation}
    \langle \theta_i^2 \rangle = \frac{1}{2\pi} \int_{-\pi}^{+\pi} \theta^2 \textrm{d}\theta = \left(\frac{\pi}{\sqrt{3}}\right)^2 = (1.81)^2 \, .
\end{equation}
However, the axion abundance is not exactly quadratically dependent on $\theta_i$ because the axion's potential is not exactly quadratic, but is a cosine. Accounting for these anharmonic corrections, a value of $\langle\theta_i\rangle^2 = (2.155)^2$ turns out to be more representative~\cite{GrillidiCortona:2015jxo}. The reason this corrected value is higher than $1.81^2$ is because the gradient of a cosine potential is shallower compared to a harmonic potential at large $\theta$---therefore the onset of oscillations is delayed when the field starts up there, and so we get more axions out in those cases than we would in the harmonic approximation.

\begin{figure}
    \centering
    \includegraphics[width=0.99\linewidth]{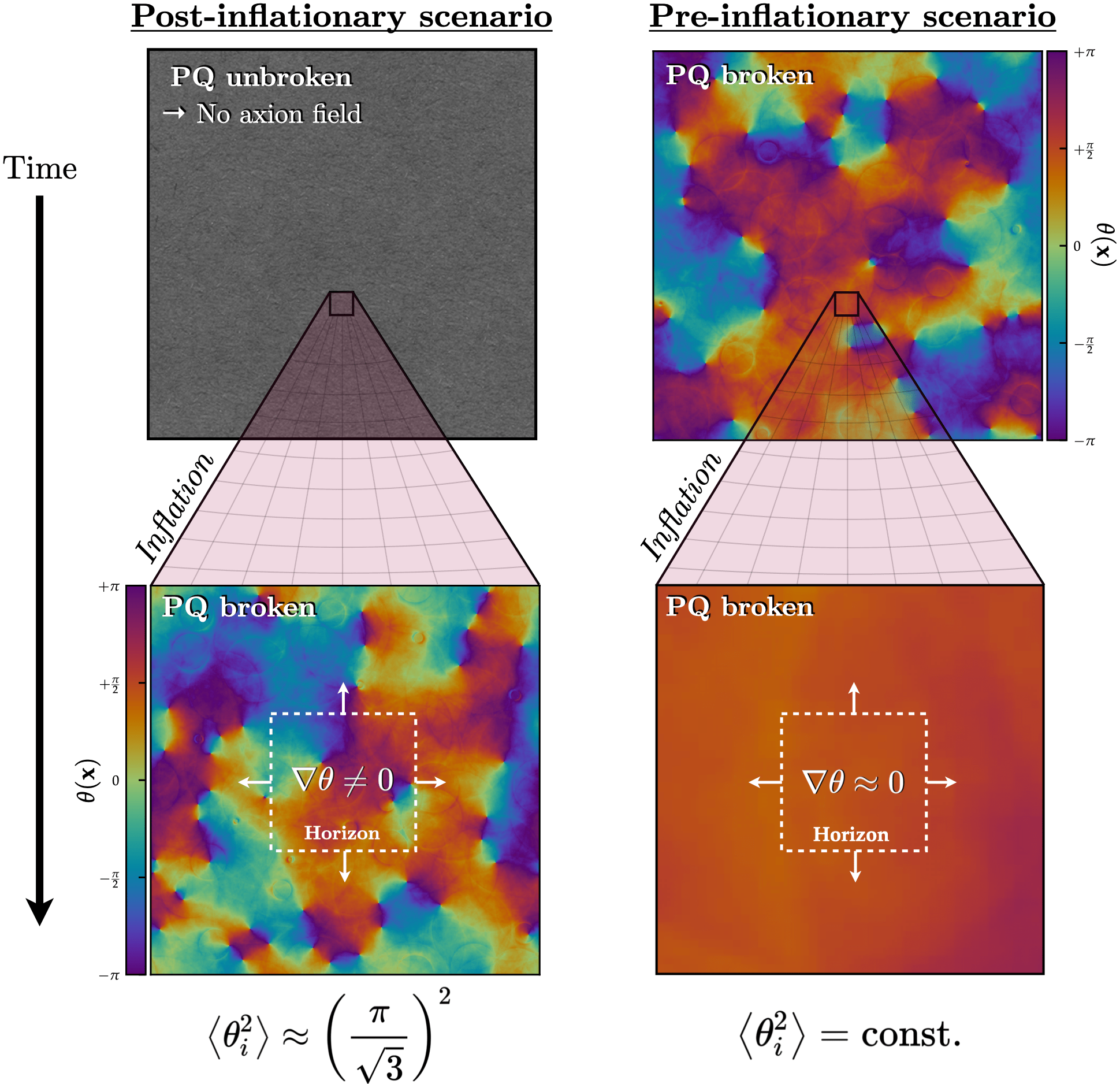}
    \caption{Diagram depicting the different implications of the pre and post-inflationary PQ breaking scenarios. In the post-inflationary scenario (left), there is no need to choose $\theta_i$ because a random ensemble of values fills the horizon. However we do run into a complication in that there are gradients in the field which can have a serious impact on the calculated dark matter abundance. On the other hand, the pre-inflationary scenario (right) has no field gradients because a single initial field value fills the horizon, but as a consequence, we then lack a way to predict what mass the axion should have because of the undetermined nature of $\theta_i$.}
    \label{fig:PrePostInflation}
\end{figure}

But should we be assuming that the axion density is just set by some stochastic average over all possible misalignment angles? It may seem like this is the natural thing to assume---the axion was born as a massless Goldstone boson and so it is reasonable to think it would have taken on a random value at every causally unconnected patch of the Universe. However, there is a very important era in cosmology that may change this picture completely: \textit{inflation}. 

What I've mentioned so far is the correct line of thinking if the PQ symmetry is broken \textit{after} inflation has already ended. However, we do not know the PQ scale and we only have a bound on the scale of inflation, so it is quite possible that the PQ symmetry is broken before or during inflation, and not restored afterwards. In that scenario, we shouldn't be talking about a stochastic ensemble of $\theta_i$ because the thing that inflation does above all is homogenise the Universe. If the axion was born before inflation ended, then the field across, and even far beyond, the observable Universe should take on a single initial value. In this case, thinking about \textit{variations} in $\theta_i$ from one causally-disconnected patch to another is wrong because inflation has inflated those regions to scales way beyond our horizon today.

So we now have a fork in the road: the QCD axion could have been produced in a \textit{pre-inflationary} scenario, or a \textit{post-inflationary} one.\footnote{I remark in passing here that people have recently thought more about the intermediate scenario where the PQ breaking occurs during inflation but is restored partially by inflationary fluctuations. The topological defects (that will be described in Sec.~\ref{sec:postinflation}) in this case are not fully inflated away and instead some re-enter the horizon later on~\cite{Redi:2022llj, Harigaya:2022pjd, Gorghetto:2023vqu}.} A visual depiction of these two scenarios is shown in Fig.~\ref{fig:PrePostInflation}. The pre-inflation scenario is in some sense simple, but it is also the less predictive of the two: the $\theta_i$ that our bit of the Universe got stuck with could really have been anything. The post-inflationary scenario, on the other hand, seems a bit more restricted: since all $\theta_i$ values will eventually enter our horizon we should just take the average. However, as we will see, the post-inflation scenario does have further complications: the production of topological defects.

\subsection{Pre-inflationary scenario}\label{sec:preinflation}
I will now discuss the issues arising in each scenario separately, beginning with the pre-inflation case. The issues to cover here are two-fold: the choice of the initial misalignment angle and how this informs our predictions for $m_a$, and what are the consequences of the axion field existing during inflation. It is worth pointing out that the issues associated with the pre-inflationary scenario will also apply to the case where there never was a PQ symmetry to begin with. This is the case for string-theory-inspired scenarios where the axion does not have to be a pseudo-Nambu Goldstone boson but instead could emerge from a 4d compactification of a higher-dimensional gauge field. In this review I am remaining agnostic about the high-scale origins of the axion, whatever they may be, so I will not get much further into this issue. I refer you to these excellent notes by Reece~\cite{Reece:2023czb} for more discussion.

\subsubsection{Initial misalignment angle}
\begin{figure*}
    \centering
    \includegraphics[width=0.99\textwidth]{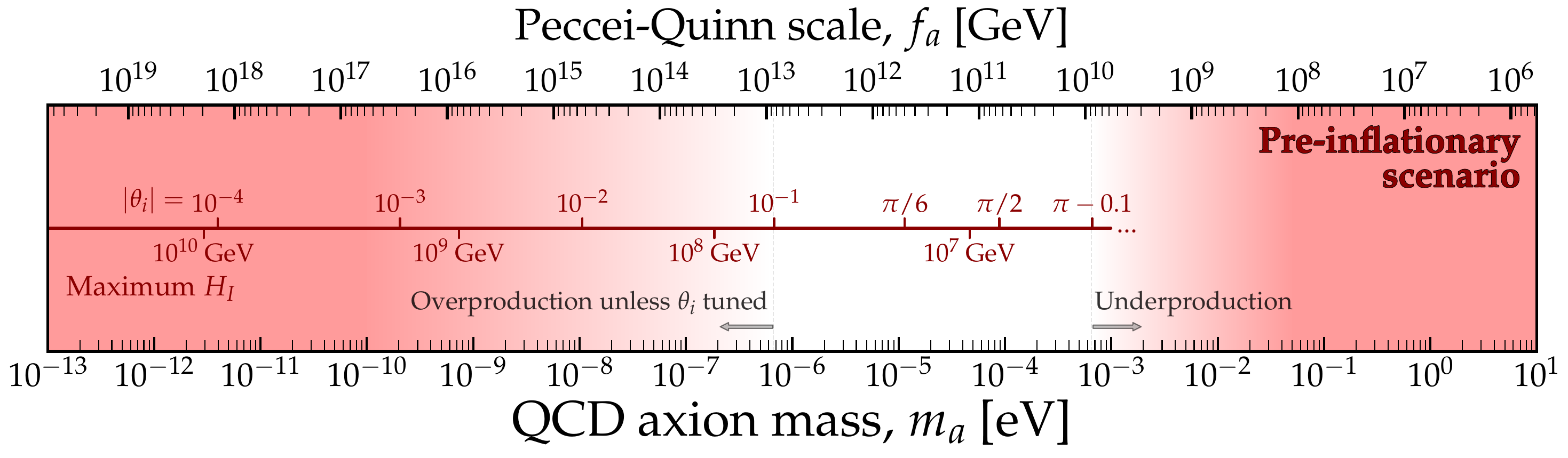}
    \caption{Constraints on the QCD axion mass in the pre-inflationary scenario. The axis through the middle of the diagram shows the value of misalignment angle $\theta_i$ required for a QCD axion of that mass to produce the correct abundance of dark matter, $\Omega_a h^2 = 0.12$, as well as the maximum allowed value for the Hubble scale of inflation, $H_I$, for axion dark matter to not produce isocurvature in excess of the \emph{Planck} bound. Models with small axion masses require fine-tuned values of $\theta_i$ to not overproduce the dark matter, whereas models with large masses cannot produce enough dark matter. The range 1--100 $\upmu$eV can be considered in some sense the ``natural'' QCD axion window for this reason.}
    \label{fig:PreInflation}
\end{figure*}
The axion abundance in the pre-inflationary scenario~\cite{Dine:1982ah} is described simply in terms of a single, but undetermined, initial angle. Choosing ``typical'' values---i.e.~order-1 numbers---leads to an axion mass in the range $\sim$1-100$~\upmu$eV. From inspecting Eq.(\ref{eq:Omega_a_final}) and insisting that $\theta_i\sim 1$, we can see that smaller values of axion mass than this then drastically overproduce the amount of dark matter, while larger values will result in underproduction\footnote{Keep in mind that we want to avoid the former scenario much more than the latter because we can always find another dark matter candidate to supplement whatever is missing to make up $\Omega_{\rm DM} h^2 = 0.12$}. See Fig.~\ref{fig:PreInflation} for a graphical depiction. A consideration of these statements in a Bayesian framework, folding in existing constraints on the axion's couplings, can be found in Ref.~\cite{Hoof:2018ieb}. A 95\% credible interval is given as \mbox{$0.12~\upmu$eV~$< m_a < 0.15$~meV} assuming a uniform prior on $\theta^2_i$.

Then again, there is nothing to say that $\theta^2_i$ had to be order-1. If you really wanted, you could imagine that our Universe, just by luck of the draw, happened to be given an exceedingly small value $\theta_i\ll1$, meaning that even an axion with a very small mass would not overproduce dark matter. The problem with this argument is that it feels like a bit of a fine-tuning issue. Still, some might argue that it is less uncomfortable than other instances of fine-tuning in physics because you and I are a \textit{consequence} of a Universe containing structurally and chemically rich galaxies that emerge inside the deep gravitational wells of a Universe dominated by dark matter. Axion masses down to below an neV that require such tuning have been dubbed ``anthropic axions'' for this reason~\cite{Wilczek:2004cr, Tegmark:2005dy, Hertzberg:2008wr}. This would be a way for axions to be produced around the scale of Grand Unification, $f_a \sim 10^{15}$~GeV, which is a feature of some models that people have cooked up, see e.g.~Refs.~\cite{Wise:1981ry, Ernst:2018bib, DiLuzio:2018gqe}.\footnote{Interestingly, a subset of axions and axion-like particles with photon couplings higher than the canonical values for the QCD axion are actually incompatible with many generic grand unified theories, as shown using an anomaly matching argument in Ref.~\cite{Agrawal:2022lsp}.} It has also been shown that in low-scale inflation models the axion field can become distributed around small values of $\theta_i$, which avoids the need for tuning to obtain the correct dark matter abundance~\cite{Graham:2018jyp, Takahashi:2018tdu}. Small masses for the QCD axion should therefore not be discounted as a possibility.

You could in principle tune the other way too~\cite{Wantz:2009it}: drive $\theta$ up to the limit of $\pi$, which could allow for large values of the axion mass, perhaps up to an meV at most. In Fig.~\ref{fig:PreInflation} I have included a scale showing the value of $\theta_i$ required for the corresponding axion mass to make up 100\% of the dark matter---the upper end of this line incorporates the necessary correction function to account for the anharmonic dependence of the abundance on initial angles close to $\pi$~\cite{Visinelli:2009zm}.

However, if we are now saying that the axion is rolling down from a point very close to the peak of its potential $\theta\to \pi$, then we need to be sure we are assuming the correct shape of this potential beyond the harmonic approximation to get the correct abundance~\cite{Turner:1985si, Lyth:1991ub, Strobl:1994wk, Bae:2008ue}. There are further consequences of doing this as well. One is that there is the threat of producing too much isocurvature~\cite{Kobayashi:2013nva}, as I will discuss in the next section. The other can be appreciated by expanding the axion's cosine potential beyond the mass term, where we discover that the axion has $(\phi/f_a)^4$ attractive self-interactions. These could have important effects if we decide to tune the field value to be as large as possible. The interplay between these interactions and gravity can become important later in this story, see e.g.~Ref.~\cite{Arvanitaki:2019rax} and the discussion of more elaborate misalignment scenarios in Sec.~\ref{sec:ALPs}.

\subsubsection{Isocurvature and the scale of inflation}\label{sec:isocurvature}
\begin{figure*}
    \centering
    \includegraphics[width=0.99\textwidth]{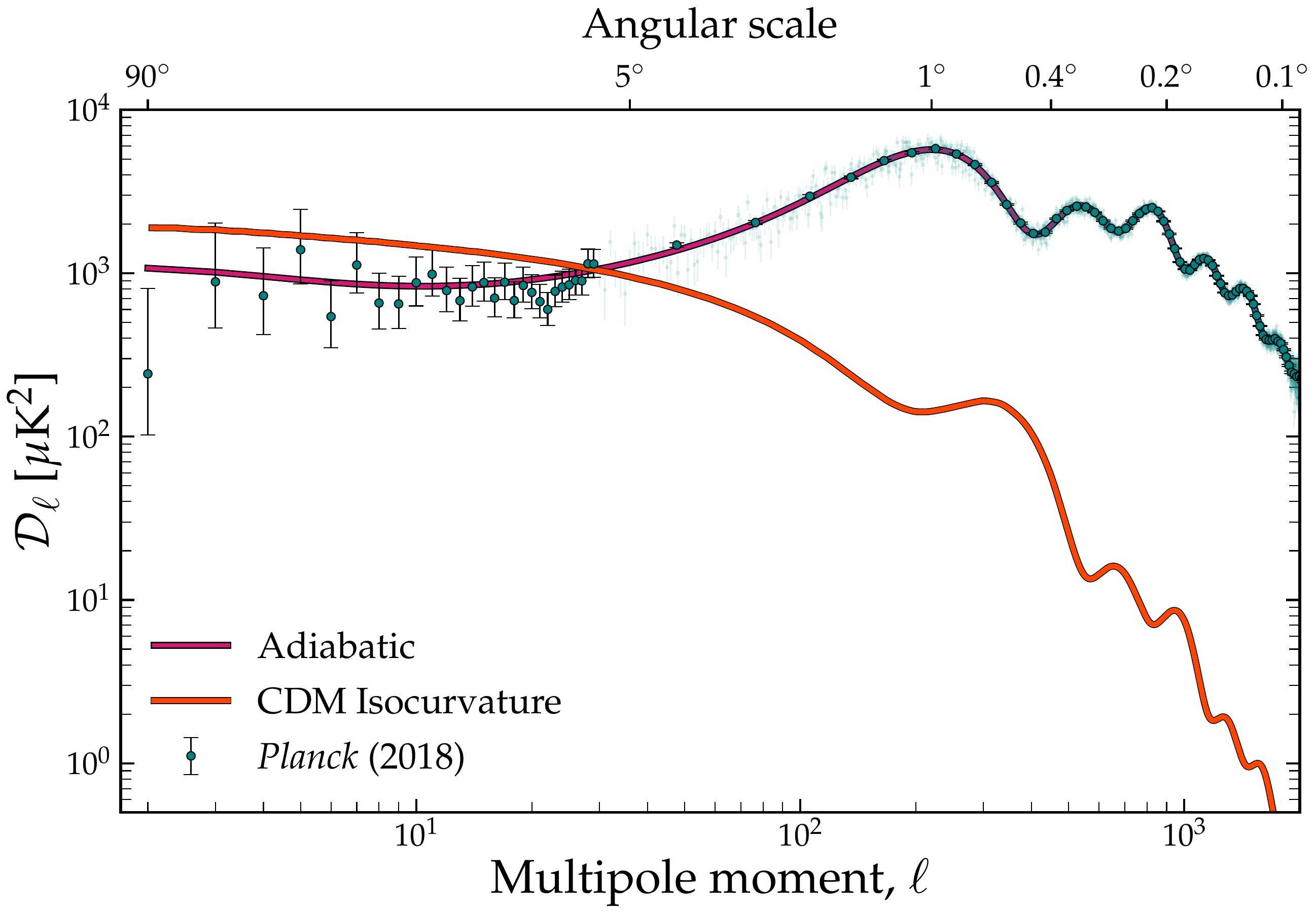}
    \caption{The CMB temperature correlation power spectrum as in Fig.~\ref{fig:CMB}, but now compared against the shape of the power spectrum produced by primordial CDM isocurvature modes. \emph{Planck} data is consistent with all of the primordial power being from adiabatic modes which are able to explain the large amplitude for the first acoustic peak (and indeed everything else). Any CDM isocurvature, for example from an axion field that existed during inflation, must therefore be constrained to a subdominant contribution: $A_{\rm iso}/A_s<0.038$.}
    \label{fig:CMB_w_Isocurvature}
\end{figure*}
Although it seems like that is all there is to say about the pre-inflationary scenario, there is an outstanding wrinkle in this story that we have to address. This relates to a potentially fatal type of fluctuation created by the axion field if it exists during inflation which goes by the name of ``isocurvature''~\cite{Seckel:1985tj, Linde:1985yf, Bucher:1999re}.

There are two types of fluctuations that can exist in the fluid of matter and radiation at very early times: curvature and isocurvature. Curvature perturbations are adiabatic---fluctuations in the density of the medium that leave the \textit{ratios} of the different components (matter and radiation) fixed with the relationship $\delta_m = \frac{3}{4} \delta_r$. The curvature perturbations are predominant in our Universe and we attribute these to the quantum fluctuations in the inflaton field that were fed into matter and radiation during the period of reheating after inflation ended. Isocurvature fluctuations, on the other hand, have no associated variation in density, but rather are fluctuations in the ratios of the components that cancel out to keep the curvature the same, i.e.~$\delta_m = -\delta_r$. The key point is that the CMB is consistent with all of the primordial perturbations being adiabatic. This is primarily because of the very steep rise up to the first acoustic peak which is a unique feature of the adiabatic mode---compare the two curves in Fig.~\ref{fig:CMB_w_Isocurvature}. Any isocurvature, if present, must be a very small contribution to what we ultimately see in the CMB.

Why is this important for axion dark matter? It is important because if the axion field already exists during inflation, then its quantum fluctuations will be inflated too~\cite{Seckel:1985tj, Linde:1985yf, Crotty:2003rz, Beltran:2006sq, Beltran:2005xd}. These are perturbations in the number density of axions, but once the axion acquires its mass they turn into perturbations in the matter density and are of the isocurvature type. Axion isocurvature perturbations are totally \textit{uncorrelated} with the curvature perturbations in the matter inherited from the inflaton field, which the CMB tells us are the majority. So we must make sure our pre-inflation axion model doesn't create too much isocurvature, and this is how we get another important bound.

In analogy with Eq.(\ref{eq:primordial_scalar}) for the primordial curvature perturbations, we can also write down a power spectrum for isocurvature perturbations with its own amplitude and spectral index,
\begin{equation}
    \mathcal{P}_{\rm iso}(k) = A_{\rm iso}\left(\frac{k}{k_\star} \right)^{n_{\rm iso}-1} \, .
\end{equation}
The amplitude of the primordial curvature perturbations was measured by \emph{Planck} to be $A_s \equiv \mathcal{P}(k_\star) = 2.1 \times 10^{-9}$, whereas the amplitude of isocurvature perturbations measured relative to the same pivot scale $k_\star = 0.05$~Mpc$^{-1}$, is constrained to be $A_{\rm iso} \lesssim 0.038 A_s$~\cite{Planck:2018jri}.

During inflation, we model the Universe as a de Sitter spacetime experiencing exponential expansion, where the value of the Hubble parameter is constant at $H_I$. All massless fields undergo quantum fluctuations in de Sitter space of size $\delta \phi = H_I/2\pi$. In the case of the axion, where $\phi = \theta f_a$, these can be interpreted as fluctuations in the initial misalignment angle: 
\begin{equation}
    \delta \theta = \frac{H_I}{2 \pi f_a} \, .
\end{equation}
Because the axion eventually turns into matter with energy density $\rho_a = \frac{1}{2} m_a^2 f_a^2 \theta_i^2$ these fluctuations in the initial angle turn into matter perturbations of size $\delta \rho_a = m_a^2 f_a^2 \theta_i \delta \theta$ and these will be uncorrelated with the adiabatic ones. The primordial amplitude $A_{\rm iso}$ of isocurvature can be related to the square of the size of the perturbations in the axion density it ends up producing,
\begin{equation}
    A_{\rm iso} = \left(\frac{\delta \rho_a}{\rho_a}\right)^2 = \left( \frac{H_I}{\pi f_a \theta_i} \right)^2 \lesssim 8.36 \times 10^{-11} \, ,
\end{equation}
where the final inequality expresses the \emph{Planck} bound on $A_{\rm iso}/A_s$. So we now have a new bound, one that connects $H_I$ to $f_a$ and $\theta_i$~\cite{Hertzberg:2008wr, Kobayashi:2013nva}. Essentially what this bound is saying is that if an axion born at a scale $f_a$ existed during inflation, then the scale of inflation has to be lower than,
\begin{equation}
H_I \lesssim 2.8 \times 10^7\, \mathrm{GeV} \, \, \theta_i \, \left(\frac{f_a}{10^{12} \mathrm{GeV}}\right) \, ,
\end{equation}
for it to not have produced isocurvature in excess of the \emph{Planck} bound.

This is interesting in its own right, but we can go further. Recall that we also have a relationship between $\theta_i$ and $f_a$ if we insist that axions make up all of the dark matter---this is expressed in Eq.(\ref{eq:Omega_a_final}). We can bring this in to derive a consistency condition between pre-inflationary axion dark matter and the scale of inflation:
\begin{equation}
    H_I \lesssim 8.8\times 10^8 \, {\rm GeV}\left( \frac{1\,{\rm neV}}{m_a}\right)^\frac{n+2}{2n + 8} \, .
\end{equation}
where the value is again given for $n = 8$. This bound on the maximum that $H_I$ can be for a given axion mass is displayed in Fig.~\ref{fig:PreInflation}.

This consistency condition will really come into play if future CMB probes \textit{measure} the scale of inflation. What this would require is a measurement of the so-called tensor-to-scalar ratio which is best extracted from so-called B-mode polarisation (attributed in this case to inflationary gravitational waves, but there are other more mundane sources of B-mode polarisation in the CMB). Let's say it was found that $H_I = 10^{11}$~GeV, then this would rule out the entire QCD axion parameter space for the pre-inflationary scenario. So far, this has not happened\footnote{In recent memory there was a short-lived discovery claim of B-mode polarisation made by the BICEP-2~\cite{BICEP2:2014owc} CMB experiment, which would have implied that pre-inflationary axions were ruled out~\cite{Visinelli:2014twa}. Unfortunately, the signal was later attributed to dust mimicking the polarisation signal expected from inflationary gravitational waves, see e.g.~\cite{Flauger:2014qra, Planck:2014dmk, Mortonson:2014bja}.}, we only have a bound on $H_I\lesssim10^{13}$~GeV~\cite{Planck:2018jri}, so not yet very helpful in narrowing down the parameter space.

\subsection{Post-inflationary scenario}\label{sec:postinflation}
In the post-inflationary scenario, we do not have the unsatisfying situation of not knowing the initial misalignment angle. As Fig.~\ref{fig:PrePostInflation} showed, our horizon expands to encompass many patches that were once causally disconnected, all of which contribute to the eventual dark matter abundance in our Universe today. In this case, it really is the correct thing to do to simply take an average when calculating $\Omega_a h^2$.

Accounting for the slightly anharmonic dependence of the dark matter abundance on $\theta_i$, the ``effective'' average misalignment angle ends up being $\theta_i \sim 2.155$~\cite{GrillidiCortona:2015jxo}, meaning the equivalent plot to Fig.~\ref{fig:PreInflation} for the post-inflation case does not span a window in $m_a$, but would rather be constrained to a single allowed value: $m_a = 28~\upmu$eV~\cite{Borsanyi:2015cka} (up to some minor theoretical uncertainties stemming from the topological susceptibility etc.). This is a very nice conclusion to give to an experimentalist because they could then go and test our grand claim by building some device that resonates at the exact frequency $\omega = m_a$, in which case the axion could be discovered tomorrow...

If only it were that simple. Unfortunately, this answer is not as solid as it may seem. The reason is that the simple calculation of $\Omega_a h^2$ presented in Sec.~\ref{sec:axionmisalignment} does not exactly apply to the post-inflationary scenario because we were only calculating the evolution of homogeneous zero-momentum mode of the classical field. In doing so, we ignored one of the terms that should have been in our equation of motion from the beginning:
\begin{equation}\label{eq:axioneom_w_laplace}
    \ddot{\theta}+3 H \dot{\theta}-\frac{1}{a^2} \nabla^2 \theta+m_a^2 \theta=0 \, .
\end{equation}
The term we were missing was the third one, which involves spatial gradients in the field. We did not need to worry about these in the pre-inflation case, precisely because inflation drove $\nabla \theta$ to zero (at least on the scales relevant for computing the abundance). Since the axion field $\theta(\mathbf{x},t)$ in the post-inflation scenario does not have its inhomogeneities inflated away, we need to understand the impact of the spatial gradient term as modes of the field enter the horizon. It turns out they do complicated things.

\begin{figure*}[t]
    \centering
    \includegraphics[width=0.99\textwidth]{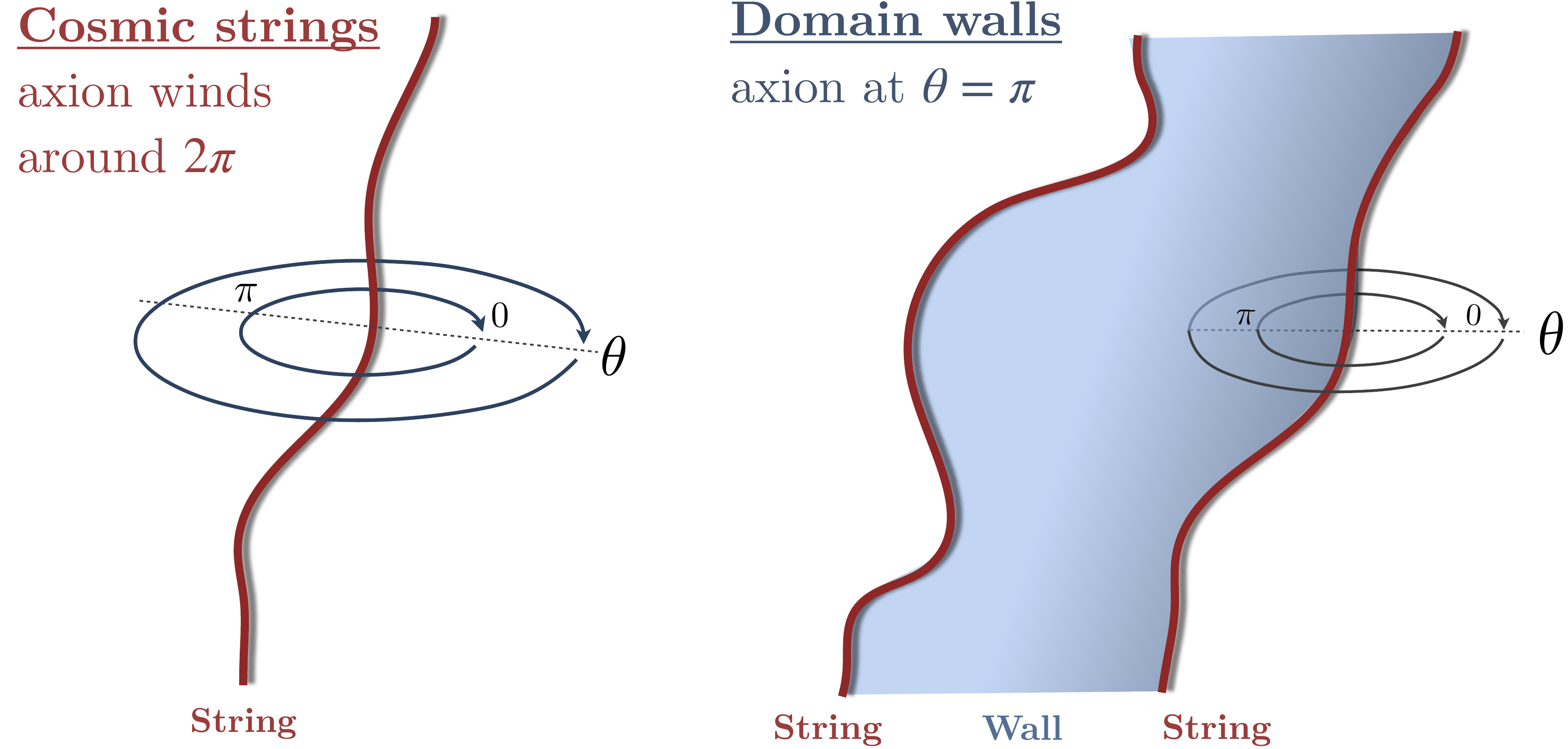}
    \caption{Diagram of the two types of topological defect appearing in the post-inflationary scenario. Cosmic strings arise along 1-dimensional lines around which the axion field wraps around its circular domain. Quasi-stable domain walls arise once the PQ symmetry is explicitly broken around the QCD era. They correspond to 2-dimensional surfaces where the axion field is stuck at $\theta = \pi$.}
    \label{fig:TopologicalDefects}
\end{figure*}

\subsubsection{Cosmic strings}
The name we give to these complicated things is `topological defects', and they arise as follows. First, notice that the axion field angle exists on the domain ($-\pi,\pi$). If we consider some continuous field of $\theta(\mathbf{x})$, with every point adopting a randomly chosen angle, you can imagine that there will be certain locations where those angles will just happen to loop around $2\pi$. If this occurs, then somewhere inside that loop there will have to be a point where the angle is undefined. The axion is the phase of a complex scalar field, and so what is occurring at this singular point is that the field is forced into the centre of the complex plane at $|\Phi| = 0$, which is effectively where the PQ symmetry is restored. This point is a potential maximum, so it ought to be unstable, but because of the winding of the field around it, the configuration turns out to be stable. In three dimensions, these singular points all connect up along a one-dimensional line, containing energy associated with the potential for the field's radial mode.

This kind of field configuration that is conjectured to appear after a cosmological phase transition is known as a topological defect~\cite{Kibble:1980mv}.\footnote{In the post-inflationary QCD axion these objects appear to be inevitable, but this is not generically true for string theory axions in which there is a qualitatively different picture~\cite{Benabou:2023npn}.} In this case, we are breaking a global $U(1)$ and so the topological defects that emerge are called global cosmic strings. The string can either stretch off beyond the horizon or connect back up to itself to form a loop. The strings are thin, but not infinitesimally so---they have a finite high-energy core set by the shape of the potential in the radial direction $\Phi$. It is useful to think in terms of the radial degree of freedom called the ``saxion'', which will have a mass,
\begin{equation}
    m_s = \sqrt{\lambda} f_a \, ,
\end{equation}
where $\lambda$ is the same constant as in Eq.(\ref{eq:fullpotential}).

To understand the properties of cosmic strings, we can construct a simple static model by imagining we have an infinite one aligned along the $z$ direction in cylindrical coordinates $(r,\varphi,z)$. The full field will have some solution: $\Phi = |\Phi(r)|e^{i\theta(\varphi)}$ where the axion field winds around the $z$-axis so we can assume $\theta(\varphi) = \varphi$. The full radial dependence can be obtained numerically by solving the equation of motion, Eq.(\ref{eq:axioneom}), but here I will just state its behaviour at the two extremes~\cite{Vaquero:2018tib}:
\begin{equation}
    |\Phi(r)| \approx \begin{cases}
0.53 \, f_a m_s \,r \quad \textrm{for}\quad r \to 0 \, ,\\
f_a \left(1- \frac{1}{2 m^2_s r^2}\right) \quad \textrm{for}\quad  r\to \infty \, .
\end{cases}
\end{equation}
We can then use this to look at the energy stored in the cosmic string. The solution is static so the Hamiltonian is just the sum of the gradient and potential energies,
\begin{equation}
    \mathcal{H} = \frac{1}{2}\left((\nabla|\Phi|)^2+|\Phi|^2(\nabla \theta)^2\right) + V_{\rm PQ}(\Phi) \, .
\end{equation}
For the time being, we are ignoring $V_{\rm QCD}(\theta)$, which will allow the string to unwind itself, so we are working at temperatures between PQ breaking and QCD. The useful quantity to calculate is the energy per unit length, also called the string tension $\mu$:
\begin{equation}
    \mu \equiv 2\pi \int_0^\infty r \textrm{d}r \left(\frac{1}{2}\left(\frac{\textrm{d}|\Phi(r)|}{\textrm{d} r}\right)^2+\frac{1}{2}\left(\frac{|\Phi(r)|}{r}\right)^2 + V_{\mathrm{PQ}}(|\Phi|)\right) \, .
\end{equation}
The first and third terms integrate just fine, however looking at the large $r$ behaviour of $\Phi(r)$ we see we need to do $\int^\infty r^{-1}\textrm{d}r$, and so the final answer is divergent. To evaluate the string tension we therefore need to impose some large-scale cutoff to the integral $r_{\rm max}$, in which case we get,
\begin{equation}
    \mu = f_a^2\left[4.5 + \pi \ln\left(\frac{m_s r_{\text{max}}}{4}\right)\right] \, .
\end{equation}
Cosmic strings may well be infinite in extent, but there is actually a physically motivated choice for this cutoff here because we have a cosmological horizon: $r_{\rm max} = H(t)^{-1}$.

So $\mu\approx f_a^2$, meaning that if cosmic strings exist in the universe they can potentially store a considerable amount of energy. As well as complicating the evolutionary history of the field, their motion through space as they straighten and intersect each other will act to stir up waves in the axion field~\cite{Davis:1986xc, Harari:1987ht, Lyth:1991bb}. Although the axion at the moment in our discussion is still massless, it won't stay massless, and so the axions radiated by the cosmic strings will eventually bestow the dark matter with some distribution of momenta over a wide range of scales.

The waves through the axion field stirred up by the motion of cosmic strings will come in a range of wavelengths. There will be long-wavelength modes coming from the large-scale motion of the long horizon-sized strings, as well as short-wavelength modes generated by the collapse and intersection of smaller closed loops of strings. This period of evolution is described by a so-called ``scaling solution'' in terms of the number of strings inside some comoving volume, $V$:
\begin{equation}
    \xi(t) = \frac{\ell_{\rm tot}(t) t^2}{V} \, .
\end{equation}
Given that there could be more than $\xi>1$ string per horizon, we should slightly shrink the long-distance cutoff in our integral from $H^{-1}$ to the typical distance to the nearest string, $(H\sqrt{\xi})^{-1}$. This inspires an effective string tension for a network:
\begin{equation}
    \mu_{\mathrm{eff}}=\pi f_a^2 \ln \left(\frac{m_s \eta}{H \sqrt{\xi}}\right) \, ,
\end{equation}
where $\eta$ is also included to parameterise the departure away from the perfectly straight string model assumed earlier.

The naive expectation is for $\xi(t)$ to stay at a roughly $\mathcal{O}(1)$ constant because the decay of smaller string loops is compensated by new lengths of string entering the horizon. This expectation is broadly what was observed in early numerical simulations~\cite{Yamaguchi:1999dy, Hiramatsu:2010yu, Klaer:2017qhr, Klaer:2017ond}. However, there has been a lot of discussion more recently about a mild violation of this scaling expectation discovered in Ref.~\cite{Gorghetto:2020qws}.\footnote{For a counter-argument see Ref.~\cite{Hindmarsh:2021zkt}.} Instead of remaining constant, the number of axion strings per Hubble volume has been confirmed in several subsequent studies~\cite{Gorghetto:2020qws, Buschmann:2021sdq, Kim:2024wku} to grow logarithmically in time as, 
\begin{equation}
    \xi(t) \approx c_0 + c_1 \ln \left(\frac{m_r}{H(t)}\right) \, .
\end{equation}
Moreover, this scaling appears to be an attractor that $\xi$ will try to converge towards, even if the system happens to start with an over or under-density of strings~\cite{Saikawa:2024bta}.

The energy density in the strings at a given time depends on how many of them there are:
\begin{equation}
    \rho_{\text {strings }} =\frac{\xi(t) \mu}{t^2} \approx \xi \times f_a^2 H^2 \log \frac{m_r}{H} \, ,
\end{equation}
and this energy has to go somewhere. The vast majority will go into axions, and a small amount into the saxion field~\cite{Gorghetto:2018myk, Saikawa:2024bta}. A very small amount of energy also goes into gravitational waves, but the present-day amplitude of this emission is only measurable in gravitational wave observatories for $f_a\gtrsim 10^{14}$~\cite{Gorghetto:2021fsn}, which is sadly well into the over-production regime for the QCD axion.\footnote{The additional freedom in non-QCD axion-like particle models could produce detectable gravitational waves from cosmic strings, whilst avoiding overproduction and other early-Universe constraints~\cite{Servant:2023mwt}.}

Strings contribute axions \emph{on top of} the ``free'' axions we get from the standard misalignment production in the regions far away from the strings.\footnote{See Ref.~\cite{Dine:2020pds} for a counterargument to this.} So if we account for the axions radiated by the strings, our predicted axion mass will need to change to match the total dark matter abundance today. Since $\Omega_a h^2 \propto m_a^{-1.167}$ from Eq.(\ref{eq:Omega_a_final}), if we get more axions out of the strings, then this means the predicted mass that matches the dark matter abundance will be shifted \textit{higher}.

Recall from the calculation in Sec.~\ref{sec:axionmisalignment} that the adiabatic invariant we used to get the abundance of dark matter in the end was the number density in the axion's zero-momentum mode at the time it became dark matter, $n_a(T_{\rm osc})$. We now have a situation where a potentially large amount of the energy contained in the system will be fed into modes of the axion field with \textit{nonzero} momentum, $k$. So the number density we get from them will depend on how they are distributed across momentum,
\begin{equation}
    n^{\rm strings}_a(t) = \int \frac{\textrm{d}k}{k} \frac{\partial \rho^{\rm strings}_a}{\partial k} \, ,
\end{equation}
where $\rho_a^{\rm strings}$ is the energy density of axions radiated by strings. Because the network is evolving, with strings entering the horizon as well as loops collapsing, we might also expect the energy spectrum to vary with time as well. It is better to keep track of the \textit{instantaneous} spectrum of string-radiated axions~\cite{Gorghetto:2018myk},
\begin{equation}
    \mathcal{F}(k,t) = \frac{1}{(f_a H(t))^2} \frac{1}{a^3(t)} \frac{\partial}{\partial t}\left(a^3(t) \frac{\partial \rho^{\rm strings}_a(k,t)}{\partial k} \right) \, ,
\end{equation}
so that the integral of this function over time and momentum is what sets the eventual number density of axions:
\begin{equation}
    n^{\rm strings}_a(t) = \frac{f^2_a}{a^3(t)} \int  \frac{\textrm{d}k}{k} \int_0^t \textrm{d}t^\prime \left[ a^3(t') H^2(t^\prime) \mathcal{F}(k,t^\prime) \right] \, .
\end{equation} 
The functional form adopted in the recent literature on axion strings is a power law in terms of a spectral index, $q$,
\begin{equation}\label{eq:stringspectrum}
    \mathcal{F}(k,t) \propto \begin{cases} k^{-q} & \text{for}\,\,k_{\rm min}(t) < k <k_{\rm max}\, , \\ 0 & \text{otherwise,}\end{cases} 
\end{equation}
which is observed to fit the simulation results well. As implied by the two cases written there, the power-law spectrum is fitted between some minimum (IR) and maximum (UV) cutoffs in momentum. Again, these are not arbitrary cutoffs, but rather come about from the two physical scales in the system. At large-scales/low-momenta we have the horizon again $k_{\rm min}(t) \sim H(t)$, beyond which the strings are essentially frozen; whereas at small-scales/high-momenta we have the size of the string cores, $k_{\rm max} \sim m_s \sim f_a$, below which the axion disappears as a degree of freedom.\footnote{Although the cutoffs themselves are physically motivated, the precise range over which the fit is obtained is more arbitrary and represents another potential source of uncertainty in the final answer~\cite{Gorghetto:2018myk, Gorghetto:2020qws, Buschmann:2021sdq, Saikawa:2024bta}.}

Since the spectrum sets $n_a$, which in turn sets $\Omega_a$, we need to know what $q$ is if we want to solve the inverse problem and find the axion mass that gives the right amount of dark matter in our Universe. To demonstrate the implications of this, imagine the spectrum happened to be tilted away from scale invariance with $q>1$, then the total energy would be distributed amongst a larger number of low-momentum axions, compared to the alternative, $q<1$ where the spectrum is tilted toward there being a smaller number of high-momentum axions. Remember that it is the number density that matters for the abundance. So our prediction for the axion mass would need to be shifted higher for $q>1$ compared to $q<1$ to make sure their total energy density today stays at $\Omega_a h^2 = 0.12$.

\begin{figure*}[t]
    \centering
    \includegraphics[width=0.99\textwidth]{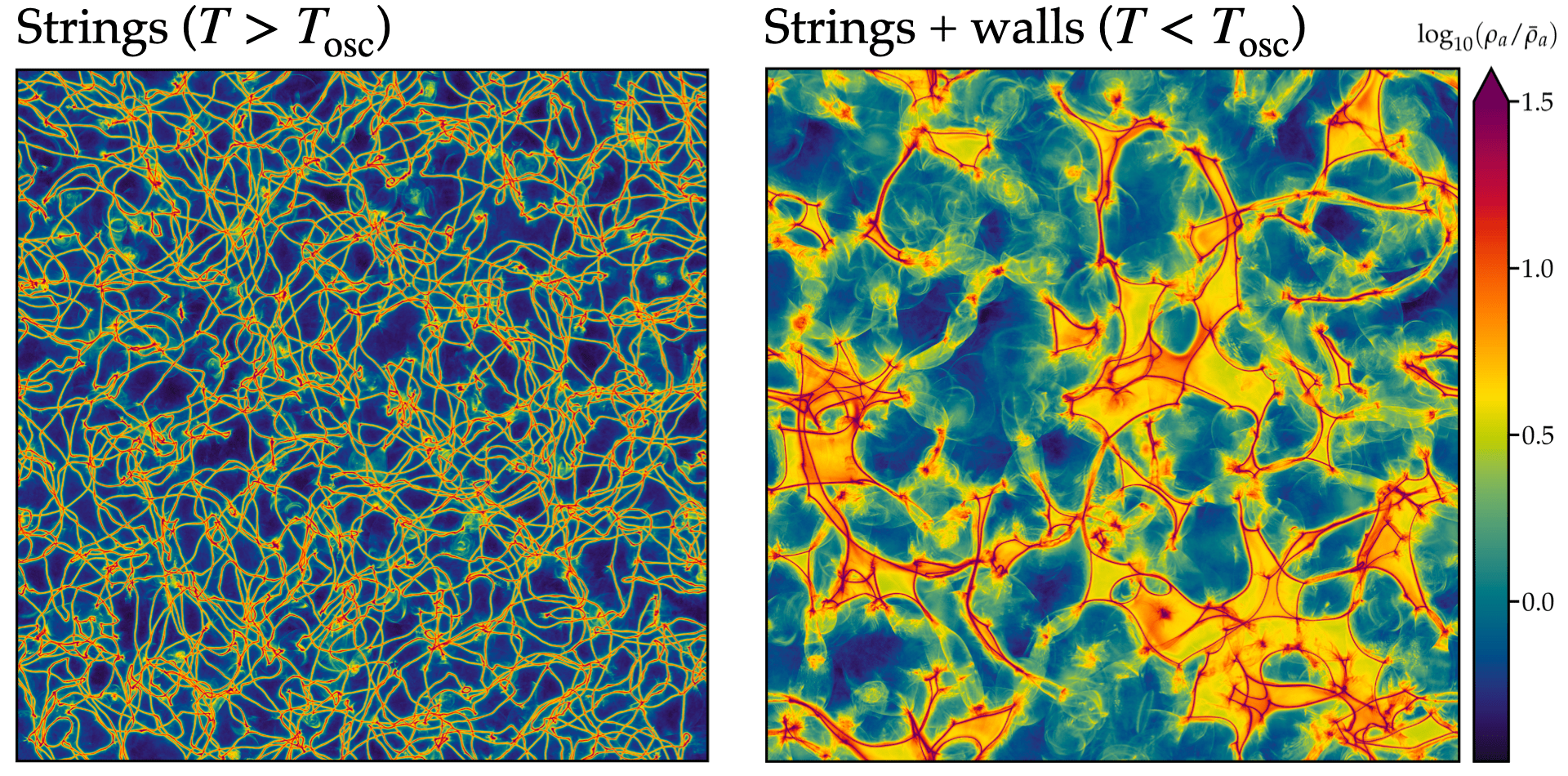}
    \caption{Visualisation of a post-inflation axion simulation before and after the axion becomes dark matter at $T_{\rm osc}$. The colour scale is logarithmic and corresponds to the axion energy density integrated along a third spatial axis extending into the page. The reddish linear objects are cosmic strings, visible in both panels, and the yellowish surfaces connected to them are domain walls, visible only in the right-hand panel which is for a time after the axion's mass has become relevant. Wavefronts in the axion string radiation are visible, especially from the cusps of the cosmic strings.}
    \label{fig:sim}
\end{figure*}
So an answer to which is correct is important, but hard to find. What is required are numerical simulations in which the string network emerges from random initial conditions once the equations of motion are evolved through cosmic time on a comoving grid. To get a sense of what these look like, Fig.~\ref{fig:sim} shows two snapshots of the axion energy in a simulation from Ref.~\cite{OHare:2021zrq}. The colour encodes the projected energy density in axions defined as,
\begin{equation}
\rho_a = \frac{1}{2} f_a^2\dot{\theta}^2+\frac{f_a^2}{2 a(t)^2}(\nabla \theta)^2 + \chi(T)\left(1-\cos \theta\right) \, .
\end{equation}
The two simulation snapshots are taken during two important eras: the time around where strings appear, which is when $T_{\rm osc}<T\lesssim f_a$, and when the temperature is $T\lesssim T_{\rm osc}$, which is when the axion mass becomes important (the features appearing in this stage are described further in the next subsection). 

Unfortunately, while simulations like Fig.~\ref{fig:sim} are based on sophisticated computer codes\footnote{Go to~\url{https://github.com/veintemillas/jaxions} for the code that the simulation shown in Fig.~\ref{fig:sim} is based on~\cite{jaxions}.} they do not fully reflect the system we are trying to understand. Axion string simulations demand a daunting dynamical range that is related to the two cutoffs appearing in Eq.(\ref{eq:stringspectrum}). We simultaneously need to simulate Hubble-scale boxes, $L \sim H^{-1}$, while resolving scales comparable to the string width $\Delta x \sim f_a^{-1}$---the ratio of those increases as a function of time and will grow up to something between $L/\Delta x \sim 10^{28}$--$10^{30}$ for the QCD axion. Since the simulations use comoving coordinates, this looks like the strings are shrinking over time, and if the string cores slip between the grid sites, then the results are no longer reliable. The largest simulations so far have been able to run up to where this ratio is around $10^4$, so quite a way off, and achieved primarily by brute force\footnote{Although see Ref.~\cite{Klaer:2017qhr} for an interesting alternative scheme for creating strings with an effective tunable tension where this issue could be circumvented.}---making the simulation boxes bigger so the strings remain resolved for longer. An alternative approach that has been explored very recently is to utilise adaptive mesh refinement to increase the resolution just around the string cores where it is needed~\cite{Buschmann:2021sdq, Drew:2019mzc, Drew:2022iqz, Drew:2023ptp,Buschmann:2024bfj}. However, the required dynamical range is still far away, which means that the value of $q$ must be obtained by extrapolating beyond whatever can be simulated---a procedure which is, of course, highly uncertain. 

There has been some lively disagreement between the results of the different groups involved in this work. Gorghetto et al.~originally found an axion spectrum with $q<1$ that was trending upwards to $q\gtrsim 1$~\cite{Gorghetto:2020qws}. The adaptive mesh technique was then successfully implemented in the simulation of Buschmann et al.~\cite{Buschmann:2021sdq} and showed evidence for scale invariance, $q=1$.\footnote{Interestingly, both of these conclusions had precedence in the literature over the decades prior to the current simulations of axion strings: see e.g.~Refs.~\cite{Davis:1986xc, Davis:1989nj, Battye:1993jv} and \cite{Harari:1987ht, Hagmann:1990mj} respectively.} The most recent suite of simulations presented by Saikawa et al.~\cite{Saikawa:2024bta} is the largest to date in terms of size---up to 11264$^3$ grid sites for their largest ones. This study highlighted and quantified several potential contributions to the disagreement in the literature---including the impact of the precise initial conditions, discretisation effects, and the handling of oscillatory features in the axion energy spectrum, among others---but the situation remains inconclusive about how $q$ should be extrapolated to the required physical values $\ln(m_s/H)\to 65$--$70$.
So while the simulations attempting to quantify the axions from strings have come a long way since the early efforts~\cite{Davis:1989nj, Battye:1993jv, Yamaguchi:1998gx, Yamaguchi:1999dy, Hagmann:2000ja}, the takeaway should be that there is not going to be an easy route to simulating the entire physical system and getting a precise prediction for the QCD axion mass. That is not to say people have not given out predictions anyway and in Sec.~\ref{sec:predictions} I will summarise them and place them in the broader context.

\subsubsection{Domain walls}
Continuing on from the axion string era, we eventually enter a new phase. As the QCD cross-over approaches, the axion stops being a massless Goldstone boson and instead starts to become dark matter. This is when the misalignment mechanism really gets going---the axion now has a preferred vacuum expectation value at $\theta = 0$ that it wants to go towards, and this preference in angle allows the strings to finally unravel themselves. What then proceeds is the collapse of the network until the axion field is oscillating around $\theta =0$ everywhere. Whilst this process happens, a different type of topological defect called an axion domain wall emerges temporarily in the system. Axion \textit{domain walls} are two-dimensional surfaces where the axion field is stuck at the saddle point in the potential at $\theta = \pi$~\cite{Kibble:1982dd, Vilenkin:1982ks}. The structure of an axion domain wall is explained graphically in the right-hand panel of Fig.~\ref{fig:TopologicalDefects} and can be seen as yellow sheets bounded by strings in the right-hand simulation snapshot of Fig.~\ref{fig:sim}. Note that in this simulation, the QCD era is brought on while the strings are still fully resolved, i.e.~$\ln(m_s/H)\ll 70$, and so the domain walls are connected to strings with unphysically small tension.

\begin{figure*}[t]
    \centering
    \includegraphics[width=0.99\textwidth]{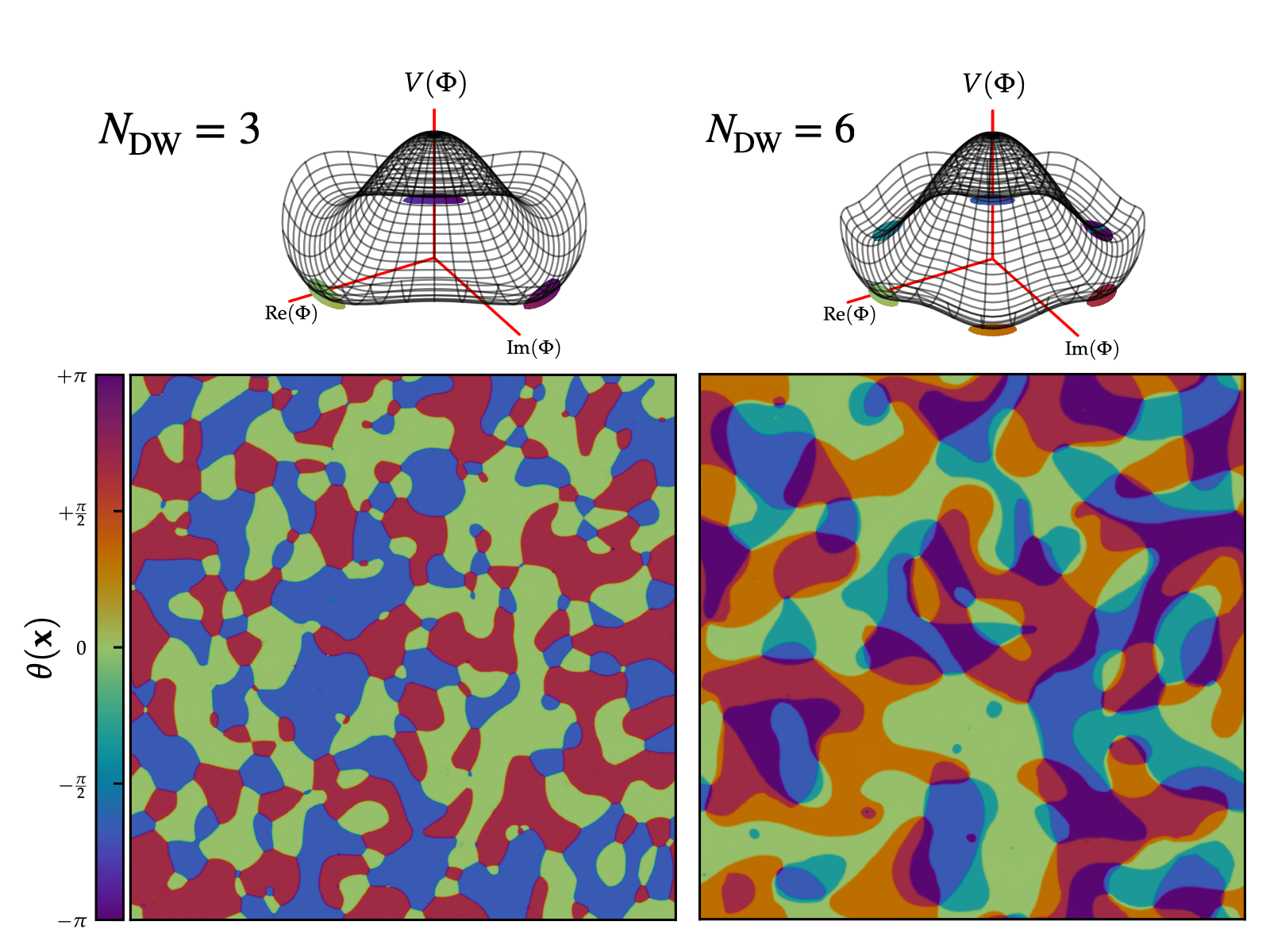}
    \caption{The PQ potential (top) and example axion field distributions (bottom) in scenarios with a domain-wall number larger than 1. The distribution of $\theta(\mathbf{x})$ shown is a slice through 3-dimensional space. Cosmic strings appear as points where the field winds around the domain $(-\pi,\pi)$. The domain walls appear as dividing lines between the regions where the field takes on values $\theta = 2\pi n/N_{\rm DW}$, for integer $n$.}
    \label{fig:DomainWalls}
\end{figure*}

Domain walls have a thickness set by the axion mass $\sim m_a^{-1}(t)$. In analogy to the linear string tension, they have a surface tension~\cite{Vilenkin:1982ks, Huang:1985tt},
\begin{equation}
    \sigma_{\rm DW}(t) = 8 m_a(t) f_a^2 \, .
\end{equation}
Recall that the axion mass is growing with temperature, so the domain walls are getting thinner and more tense as the temperature cools towards $T_{\rm QCD}$. Because the point $\theta = \pi$ is only quasi-stable, fairly quickly the network of defects will collapse, and the remaining energy contained in the radial saxion mode is converted into (now massive) axions\footnote{There is further opportunity to produce detectable gravitational waves here too, see e.g.~Refs.~\cite{Figueroa:2020lvo, ZambujalFerreira:2021cte, Chang:2021afa, Gorghetto:2021fsn, Gelmini:2021yzu,Blasi:2023sej,Ge:2023rce}. It is also possible to get closed domain walls which may collapse into black holes~\cite{Vachaspati:2017hjw,Ge:2019ihf,Gouttenoire:2023gbn,Ferreira:2024eru,Dunsky:2024zdo,Gouttenoire:2023ftk,Ge:2023rrq}, or remain stable with baryons trapped inside, in which case they are called `axion quark nuggets'~\cite{Liang:2016tqc, Ge:2019voa, Zhitnitsky:2002qa}.}.

What I described above is the general picture when the axion's potential is periodic in the range $(-\pi$  to $\pi)$, but this doesn't have to be the case. The axion potential $V(\theta)$ could cycle multiple times inside the domain $\theta \in (-\pi,\pi)$, i.e.~we could have written down $V \sim - \cos{N\theta}$ instead of $V\sim -\cos{\theta}$. This number, which I will write as $N_{\rm DW}$, is called the ``domain wall number'' because it refers to how many distinct \textit{types} of domain walls emerge in this scenario. They arise because of the discrete $Z(N_{\rm DW})$ symmetry that is spontaneously broken at the QCD phase transition. Domain walls form a network, connected at junctions by cosmic strings, but only in the $N_{\rm DW}=1$ case (the case discussed above) is the network of walls unstable to collapse. When $N_{\rm DW}=1$ there is a single true vacuum state that the axion field can always unwind itself to find. But if $N_{\rm DW}>1$ then there are $N_{\rm DW}$ \textit{equivalent} vacua with the same energy, each one separated from its neighbour by a potential barrier. Since there is no preference for the field to evolve towards any particular one---the domain walls are absolutely stable. In Fig.~\ref{fig:DomainWalls} I have shown a sketch of the potential for two possible cases---$N_{\rm DW} = 3$ and 6 (the relevance of these numbers is mentioned below)---as well as a 2D slice through the field showing the patchwork of domains and walls.

When domain walls are stable like this they will enter a scaling solution similar to cosmic strings. They want to flatten themselves and will do this by effectively repelling each other until huge flat walls are separated by scales on the order of the horizon, $L\sim H^{-1}\sim 1/t$. The energy density of the network therefore scales cosmologically like, 
\begin{equation}
 \rho_{\rm DW}(t)\sim \frac{\sigma_{\rm DW}L^2}{L^3} \sim \sigma H(t) \propto \frac{1}{t} \, ,
\end{equation}
which implies that it is possible for $\rho_{\rm DW}$ to overwhelm matter and radiation which dilute away faster than this~\cite{Sikivie:1982qv}. Let us imagine there is some time when the Universe becomes domain-wall dominated, this occurs when the condition $3H^2(t) M^2_{\rm Pl} = \rho_{\rm DW}(t)$ is satisfied. Rearranging, we can find that the temperature of the Universe when axion domain walls start dominating is,
\begin{equation}
    T_{\rm wall-dom.} = \left( \frac{640}{\pi^2} \frac{f_a^4 m_a^2}{g_\star(T) M^2_{\rm Pl}} \right)^{1/4} \approx 20 \, {\rm keV} \, \left( \frac{m_a}{100\,\upmu{\rm eV}}\right)^{-\frac{1}{2}} \, ,
\end{equation}
which is between BBN and recombination. In other words, this is not good. But it gets worse. A domain wall network possesses an equation of state $p_{\rm DW} = -2/3 \rho_{\rm DW}$, so after they come to dominate the energy density of the Universe, the scale factor will evolve like $a(t)\sim t^2$, i.e.~it has an acceleration $\ddot{a}>0$. But even though the accelerating effect this negative pressure has on expansion is reminiscent of dark energy, we clearly do not see a network of vast energetic walls spanning the cosmos. A domain wall-dominated cosmology is in catastrophic disagreement with observations~\cite{Zeldovich:1974uw}---this is the so-called ``domain wall problem'' of post-inflationary axion models with $N_{\rm DW}>1$. 

It might seem like this conclusion is telling us that we just shouldn't choose $N_{\rm DW}$ to be larger than 1. But while $N_{\rm DW}=1$ is indeed the case for one of the two major categories of UV-complete axion models, the Kim-Shifman-Vainshtein-Zakharov (KSVZ) model, for the other---the Dine-Fischler-Srednicki-Zhitnitsky (DFSZ) model---it is not. In fact, DFSZ models typically have $N_{\rm DW} = 3$ or 6 depending on details in the potential. The domain wall number is related to how many are quarks charged under the $U(1)_{\rm PQ}$ symmetry, and so emerges from the specific contents of that theory. Those contents are required in those models to make a UV-complete theory that solves the strong-CP problem. Escaping the domain wall problem adds another layer of work, for example by arranging charges so that $N_{\rm DW}$ happens to end up equal to one. For the DFSZ case, this requires the PQ symmetry to be made flavour-dependent, see the recent Ref.~\cite{Cox:2023squ} for a list of DFSZ models with no domain-wall number problem. 

The second reason why we might not want to just discount the $N_{\rm DW}>1$ cases is that there could well be effects that can drive the network to be unstable and fix this potential domain wall problem for us. In general, what is needed is some kind of bias that tips the scales to the axion preferring one of its vacua over others~\cite{Sikivie:1982qv}. There are many ways this can be achieved, see for example Refs.~\cite{Barr:1982uj, Lazarides:1982tw,Dvali:1994wv, Chang:1998bq,Rai:1992xw,Barr:2014vva, Reig:2019vqh,Caputo:2019wsd, Gelmini:1988sf, Harigaya:2018ooc}. The essential idea in many of these solutions is to engineer there to be an effect that \textit{explicitly} breaks the $Z(N_{\rm DW})$ symmetry, causing the $N_{\rm DW}$ vacua to be non-degenerate~\cite{Holdom:1982ew}. In scenarios like this, the domain wall network would collapse on some timescale that could be chosen by tuning the explicit breaking terms in the potential. Unfortunately, the required CP-violating effects work against the solution the axion was originally designed to solve, and so face observational constraints from the neutron electric dipole moment~\cite{Hiramatsu:2012sc, Beyer:2022ywc}. Some numerical simulations of this scenario were performed by Hiramatsu et al.~\cite{Hiramatsu:2012sc}. They tested mechanisms for the annihilation of the domain walls but found them highly constrained except in a small window of parameter space.\footnote{On this, I direct you also to the recent Ref.~\cite{Beyer:2022ywc} which overviews issues with certain solutions to the axion domain wall problem in the context of the DFSZ model. Another conflict highlighted recently in Ref.~\cite{Lu:2023ayc} is that QCD axion models with $N_{\rm DW}>1$ are harder to reconcile with solutions to a worrying technical issue in axion theories known as the PQ quality problem.}

These details aside, the generic expectation for $N_{\rm DW}>1$ scenarios is that we should get more axions out of the domain walls collapsing. So matching the cosmological dark matter abundance leads to mass predictions that are larger than the $N_{\rm DW}=1$ case. Any more precise statement that this really demands we specify some particular mechanism for why the network collapses, as well as full numerical simulations of that collapse, of which rather few can be found in the literature.

\subsubsection{Axion miniclusters}\label{sec:miniclusters}
\begin{figure*}[t]
    \centering
    \includegraphics[width=0.99\textwidth]{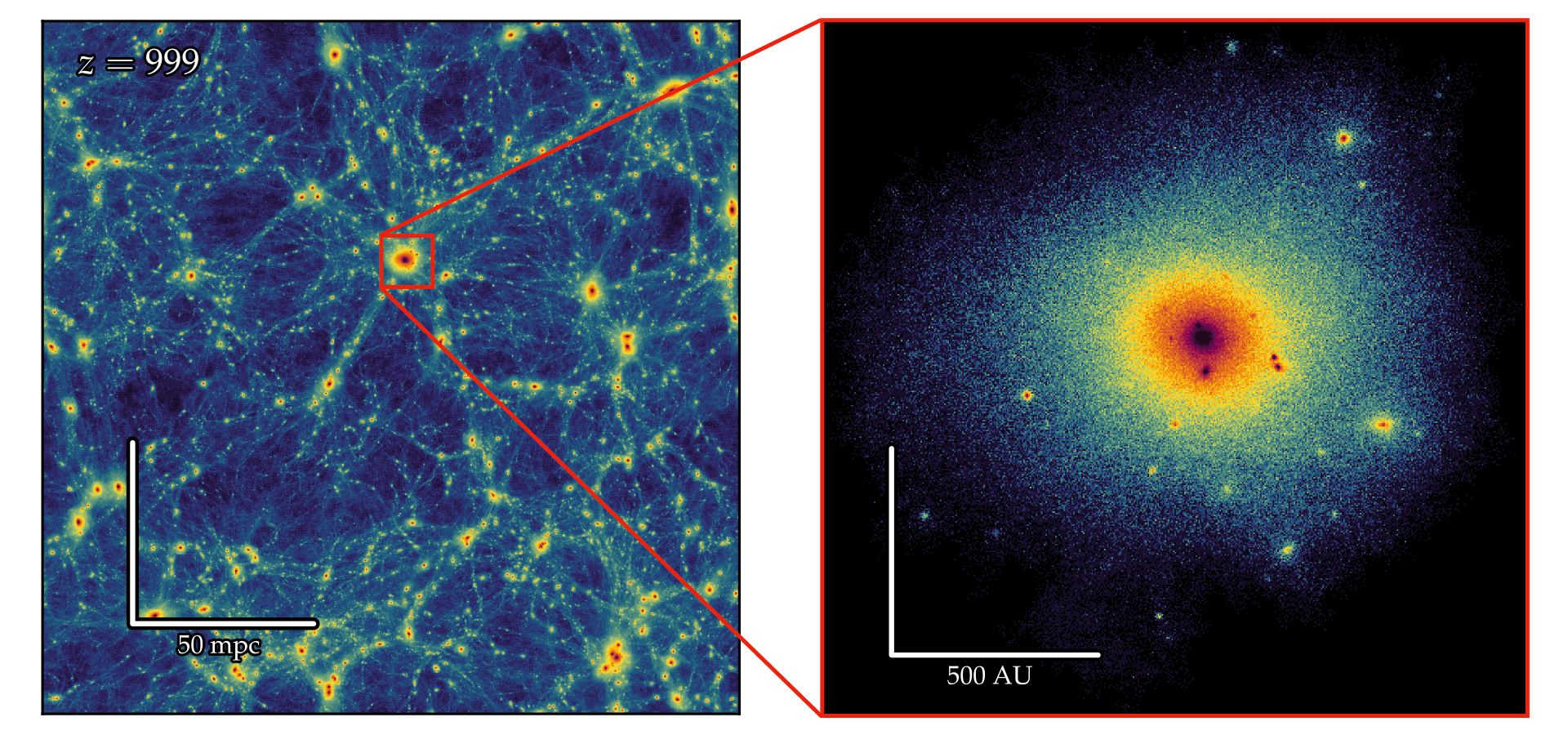}
    \caption{Snapshot of some merging axion miniclusters in the N-body simulations performed for Ref.~\cite{Eggemeier:2022hqa}. The redshift of the snapshot is $z=999$, so well before the formation of galaxies. The right-hand panel zooms in on the largest minicluster halo in the (0.2~pc)$^3$ box. Notice that the scale of this merged minicluster has grown to be significantly larger than the estimate of $R\sim $AU made in Eq.(\ref{eq:miniclusterradius}). Nonetheless, you can still see that there are unmerged AU-scale miniclusters orbiting around it that would have formed from the prompt collapse of horizon-sized overdensities around the QCD phase transition.}
    \label{fig:miniclusters}
\end{figure*}

As should be clear by now, the phenomenological consequences of the post-inflationary scenario are very rich, but we are still not done. After the string-wall network has unravelled itself, the axions will free-stream until they settle down to non-relativistic speeds and become dark matter in earnest. However, the remnants of that highly inhomogeneous distribution of $\theta_i$ and the now-destroyed topological defects will still be imprinted in the axion distribution. Now that we are somewhere after the QCD cross-over where the axion has a mass, that inhomogeneous distribution has been converted into genuine physical fluctuations in the density of dark matter. The important fact about these inhomogeneities is that they are large in amplitude, with $\delta = \delta\rho_a/\rho_a \sim \mathcal{O}(1)$, but physically small in size. So what becomes of them?

A highly inhomogeneous distribution of dark matter on small scales is yet another unique feature of axion dark matter in the post-inflationary scenario that is not generically expected in the pre-inflationary case nor indeed for many other dark matter candidates.\footnote{Interestingly, this form of perturbation that we are discussing here are of the isocurvature type---exactly what we were trying to avoid in the pre-inflationary case as discussed in Sec.~\ref{sec:isocurvature}. Fortunately, these isocurvature perturbations develop late and are only present on scales that are tiny in cosmological terms.} Usually, all of the inhomogeneities in the dark matter are the scale-invariant ones inherited ultimately from inflationary curvature perturbations. The density fluctuations in the axion field, on the other hand, emerge at temperatures around $T_{\rm QCD}$. The size of the horizon at those times was significantly smaller than the size of the horizon during matter-radiation equality which is when sub-horizon cold dark matter fluctuations begin collapsing to form halos. These early-forming structures go by various names in the literature---``minihalos'' or ``microhalos''---I will refer to them as ``axion miniclusters'', which is the name I believe to be the most common~\cite{Hogan:1988mp, Kolb:1994fi, Kolb:1995bu}.

We can derive some of the main properties of these miniclusters. Let us start with some initial overdensity of axions $\delta_i\sim \mathcal{O}(1)$ defined as the local departure away from the large-scale average, $\bar{\rho}_a$. This overdense region will begin to collapse ahead of matter-radiation equality, at a value of $a(t)$ that is a factor $1/\delta_i$ \textit{smaller} than $a_{\rm eq}$ (this is because $\rho_m/\rho_r \propto a$), i.e.~an overdensity gets a head-start on gravitational collapse. If we say the axions become dark matter at $T_{\rm osc}$, then adding up all the axions inside the horizon at that time will tell us the typical amount of mass that could end up inside a minicluster,
\begin{equation}
    M = \frac{4\pi}{3} (1+\delta_i) \, \bar{\rho}_a \, L_H^3 \, ,
\end{equation}
where $L_H(T_{\rm osc}) \sim H(T_{\rm osc})^{-1}$ is the physical size of the horizon at $T_{\rm osc}$, which in turn is related to $m_a(T_{\rm osc})$ because of the condition we wrote down: $m_a(T_{\rm osc})\approx 3H(T_{\rm osc})$, that defines $T_{\rm osc}$. 

In principle we should now plug in $\rho_a(T_{\rm osc}) = m_a(T_0) n_a(T_{\rm osc})$ for the eventual density of axions that would end up in the minicluster. However, this would be somewhat inconvenient since $n_a(T_{\rm osc})$ depends on $\theta_i$, and we only know what this is at the level of a statistical distribution. It is also inconvenient because previously we were tuning $m_a$ to get the right dark matter density, but as we saw in the last section, that number gets shifted upwards slightly by an (as yet undetermined) amount that depends on how much extra axions we get from the decay of topological defects. So a better strategy is to simply assume for now that we \textit{have} figured out the axion mass that correctly reproduces the observed abundance of dark matter, but still leave the final answer in terms of $m_a$ because technically we haven't. That approach is easy because it just entails working out $\rho_a(T_{\rm osc})$ by scaling the present-day dark matter density back in time to $a(T_{\rm osc})$:
\begin{equation}
    \rho_a(T_{\rm osc}) = \rho_{\rm DM} \left( \frac{a(T_0)}{a(T_{\rm osc})} \right)^3 \, ,
\end{equation}
where $\rho_{\rm DM} = 3H_0^2 M_{\rm Pl}^2 \Omega_{\rm DM} = 9.672\times 10^{-12}$~eV$^4$, and we take $a(T_0) =1$.

So to compute $M$ we need to know the quantity $a(T_{\rm osc})H(T_{\rm osc})$. Since we know the Universe is radiation dominated at $T_{\rm osc}$, we can assume $H^2(T_{\rm osc}) = H^2_0 \Omega_r a(T_{\rm osc})^{-4}$ from Eq.(\ref{eq:hubbleparameter}). Instead of dealing with $\Omega_r$ it is easier to write this in terms of $\Omega_m = 0.31$ and $z_{\rm eq} = 3402$, as $\Omega_r  = \Omega_m (1+z_{\rm eq})^{-1}$. 

Putting numbers in, we find the mass of axions in an overdensity of size $\delta_i$ that is going to collapse is,
\begin{equation}
    M = \frac{4\pi}{3} (1+\delta_i) \bigg(\frac{1}{3}H_0 m_a(T_{\rm osc})\bigg)^{-\frac{3}{2}} \,\Omega_m^{-\frac{3}{4}} (1+z_{\rm eq})^{\frac{3}{4}} \propto m_a^{-\frac{6}{n+4}} \, ,
\end{equation}
where the last step is just there to highlight the dependence on the temperature-dependence index of the axion mass. Putting in $n=8$ for this we get,
\begin{equation}
     M = 2.4\times 10^{-12} \, M_{\odot} \, \left(\frac{100\,\upmu{\rm eV}}{m_a} \right)^\frac{1}{2} \, ,
\end{equation}
where solar masses is $M_\odot = 1.3\times 10^{66}$~eV, and to evaluate the final number we use the temperature-dependence of the axion mass from Eq.(\ref{eq:topologicalsusceptibility}), and the expression for $T_{\rm osc}$ from Eq.(\ref{eq:Tosc}). A few pico-Suns is around the mass of the asteroid Chiron\footnote{one of the many possible misspellings of my first name.}.

So we conclude that axions in the post-inflationary scenario begin clumping just ahead of matter-radiation equality into miniclusters that have masses around that of an asteroid. Similar objects can also form on a wider range of scales if we relax some of the model assumptions and constraints on the axion mass, for example by considering non-QCD axion-like particle (ALP) models~\cite{Arvanitaki:2019rax, Hardy:2016mns} or alternative cosmological histories from the one presented here~\cite{Visinelli:2018wza, Blinov:2019jqc, Gorghetto:2023vqu}. Although the typical minicluster mass can vary substantially between different model configurations, the conclusion that heavier axions produce lighter miniclusters tends to be generic.

We know their masses now, but what is the nature of these objects? Are they compact or fluffy? To find out, let us estimate their physical sizes and densities. This can be done rigorously using semi-analytic techniques as well as numerical simulations, but we can make do with an estimate. For that, we can adopt a simple spherical collapse argument~\cite{Kolb:1994fi} which goes as follows. Gravitational collapse occurs when the gravity of some clump of matter is sufficient to decouple it from the outward pull of expansion, also called the Hubble flow. In the case of axions, as we have seen, this process will start during radiation domination and proceed through matter-radiation equality. The equation of motion for some shell of physical radius $r$ that is part of a spherical overdensity with mass $M$ inside it is,
\begin{equation}
\ddot{r}=-\frac{8 \pi G_N}{3} \rho_{r} r-\frac{G_N M}{r^2} \, ,
\end{equation}
where $\rho_r$ is the homogeneous radiation density. At early times the solution for $r(t)$ increases as the shell is dragged outwards by Hubble flow. But at some point---earlier if the overdensity is larger---the shell will halt and turn around. After turn-around, $r(t)$ will decrease as a function of time as the shell collapses inwards. See Ref.~\cite{Ellis:2020gtq} for the full solution\footnote{A nice python notebook evaluating the numerical solution to spherical collapse in radiation domination can be found here: \url{https://github.com/David-Ellis/thesis-code/blob/main/KTode/KT_spherical_collapse.ipynb}}, as well as the classic paper of \mbox{Kolb \& Tkachev} who first did the calculation of the spherical collapse of axion overdensities~\cite{Kolb:1994fi}. The upshot of this calculation is that the density of the collapsed minicluster depends on the initial overdensity parameter via~\cite{Ellis:2020gtq},
\begin{equation}
    \rho_{\rm mc} = 136 \rho_a(t_{\mathrm{eq}}) \, \delta_{i}^3\left(1+\delta_{i}\right) \approx 2.4\times 10^{6} \, {\rm GeV}\,{\rm cm}^{-3} \, \delta_i^3 (1+\delta_i) \, ,
\end{equation}
where $\rho_a(t_{\rm eq}) = \rho_{\rm DM}(1+z_{\rm eq})^3$ is the axion density at matter-radiation equality. The slightly inconvenient units I have converted this density into are so that we can see that it is much larger than the value $\rho_{\rm DM, local} \approx 0.3$~GeV~cm$^{-3}$ mentioned at the beginning of Sec.~\ref{sec:axionCDM}---the local density of dark matter in our galaxy. Making the crude assumption now that the miniclusters are spheres of radius $R_{\rm mc}$ with mass $M$ and a constant density $\rho_{\rm mc}$, we find that,
\begin{equation}\label{eq:miniclusterradius}
    R_{\rm mc} = \left(\frac{3 M}{4 \pi \rho_{\rm mc}}\right)^{\frac{1}{3}} = 0.32 \mathrm{~AU} \, \frac{1}{\delta_i}\left(\frac{100\, \upmu \mathrm{eV}}{m_a}\right)^\frac{1}{6} \, ,
\end{equation}
where an astronomical unit is $1~\text{AU} = 1.496\times 10^{13}$~cm. So we have an asteroid's worth of axions spread over a size comparable to the Earth's orbit, making them rather fluffy objects, despite the fact that the density in axions would be many orders of magnitude larger than the density of dark matter in our galaxy local to us~\cite{Read:2014qva, Evans:2018bqy, deSalas:2020hbh}.

A word of warning that when surveying the literature to find similar estimates to the one I have done here~\cite{Hogan:1988mp,   Kolb:1994fi, Kolb:1995bu, Fairbairn:2017sil, Enander:2017ogx}, you will sometimes encounter disagreements of up to orders of magnitude arising from different ways of defining various quantities like the minicluster mass and the horizon volume etc. It is not particularly enlightening to go through these disagreements; ultimately, they are just estimates. We will get a more complete picture by doing gravitational simulations~\cite{Eggemeier:2019khm, Xiao:2021nkb, Shen:2022ltx, Eggemeier:2022hqa, Pierobon:2023ozb, Eggemeier:2024fzs}, as I will get to shortly.

To give a sketch of the next steps: the continued growth of the axion density fluctuations during matter domination can be modelled for a time using linear perturbation theory. We move back into Fourier space for ease since each mode with comoving momentum $k$ evolves independently according to the equation of motion,
\begin{equation}\label{eq:delta_eom}
    \ddot{\delta}+2 H \dot{\delta}+\left(\frac{c_s^2 k^2}{a^2}-4 \pi G_N\bar{\rho}_a\right) \delta=0 \, .
\end{equation}
where physical momenta are related to comoving ones via $k_{\rm phys} = k/a$. The $k$-dependent quantity $c_s$ is the effective speed of sound, which can be derived by computing the perturbations for a cosmological fluid composed of a scalar field (e.g.~Refs.~\cite{Park:2012ru, Marsh:2015xka}),
\begin{equation}
    c_s^2 = \frac{\delta P}{\delta \rho} =\frac{k^2 / 4 m_a^2 a^2}{1+k^2 / 4 m_a^2 a^2} \approx \frac{k^2}{4 m_a^2 a^2} \, ,
\end{equation}
the approximation in the final step holds because we are considering modes with a physical size larger than the axion's Compton wavelength, i.e.~$k<m_a a$. Looking at the $k$-dependence of the third term, you may notice that there will be a value of $k$ where it swaps sign from negative to positive:
\begin{equation}
  k_J=\left(16 \pi G_N a^4 \bar{\rho}_a(a) \right)^\frac{1}{4} \, \sqrt{m_a} \approx 66.5\, {\rm mpc}^{-1} \, a^{1/4} \left(\frac{\Omega_a h^2}{0.12}\right)^\frac{1}{4} \left(\frac{m_a}{100\,\upmu{\rm eV}}\right)^\frac{1}{2}\, .
\end{equation}
For any mode with $k<k_J$ the solutions are for $\delta_k$ to grow\footnote{Or decay, but these transient solutions are not usually kept track of.}. When $k>k_J$, on the other hand, the third term in the equation of motion behaves like an outward pressure which acts against the inward pull of gravity. We can draw an analogy here with a classic model for the collapse of regular gas clouds in astrophysics---gravitational collapse takes hold above a certain length scale called the \textit{Jeans length} when there is insufficient hydrodynamical pressure to stop it. Accordingly, we call $k_J$ the axion Jeans wavenumber. Its associated physical length-scale is given by,
\begin{equation}
\lambda_J  = \frac{2\pi a}{k_J} \approx 0.04 \, {\rm AU}~ \left( \frac{1+z_{\rm eq}}{1+z} \right)^\frac{3}{4}\left(\frac{\Omega_a h^2}{0.12}\right)^{-\frac{1}{4}} \left(\frac{m_a}{100\,\upmu{\rm eV}}\right)^{-\frac{1}{2}} \, ,
\end{equation}
evaluated at matter-radiation equality. The physical origin of this scale comes from the pressure exerted when the scalar field's gradient energy is large. As might be expected, $\lambda_J$ turns out to be very similar in scale to the coherence length of the field, $\lambda_{\rm coh} = 2\pi/m_a v$, because this is related to the velocity of the field, which in turn is related to the gradient energy. The Jeans scale also implies there is a Jeans mass---a minimum mass where the mass distribution of miniclusters will be cut off, a feature that will be important for the next stage.

The behaviour of the solutions to Eq.(\ref{eq:delta_eom}) depend on the wavenumber, with modes at $k<k_J(t)$ experiencing a growing solution and those $k>k_J(t)$ an oscillatory one that does not grow. Calling back to Sec.~\ref{sec:cosmologicalobservations}, the $k$-space correlations between the density fluctuations can be written in terms of a power spectrum,
\begin{equation}
\left\langle\delta_a(\mathbf{k}) \delta_a\left(\mathbf{k}^{\prime}\right)\right\rangle=(2 \pi)^3 \delta\left(\mathbf{k}-\mathbf{k}^{\prime}\right) P_a(k) \, ,
\end{equation}
where the dimensionless variance is written as,
\begin{equation}
    \Delta_a^2(k) = \frac{k^3}{2\pi^2} P_a(k) \, ,
\end{equation}
%\begin{equation}
% \left\langle\delta\left(\mathbf{k}^{\prime}\right) \delta(\mathbf{k})\right\rangle=(2 \pi)^2 \delta\left(\mathbf{k}^{\prime} - \mathbf{k}\right) \frac{2 \pi^2}{k^3} \Delta_a^2(k) \, .
% \end{equation}
For small $k$ the axion miniclusters have no knowledge of each other, and so we expect them to follow the power spectrum associated with uncorrelated white noise $\Delta_a^2(k)\propto k^3$ (or equivalently a power spectrum $P_a(k)$ that is flat in $k$). Eggemeier et al.~\cite{Eggemeier:2019khm}, using initial conditions taken from the final stages of the string-wall lattice simulations of Vaquero et al.~\cite{Vaquero:2018tib}, find $\Delta_a^2(k) \approx  0.05 (k L_{\rm osc})^3$ well before matter-radiation equality, which has grown to $\Delta^2_a \sim 10^{-15}\,{\rm Mpc}^{-3}\times k^3/(2 \pi)^2$ by $z=99$. This can be compared to the matter power spectrum from the regular adiabatic fluctuations, which as described in Sec.~\ref{sec:cosmologicalobservations} scales like $\Delta_m^2(k) \sim \ln(k/k_{\rm eq}) k^{n_s-1}$ at high $k$. So there will come some $k$ where the two will cross over, and the fluctuations associated with miniclusters (which are of the isocurvature type) are dominant, i.e.~ $\Delta_a^2(k) \gtrsim  \Delta_m^2(k)$. Extrapolating the adiabatic power spectrum towards high $k$, this occurs for $k \gtrsim 10^6$~Mpc$^{-1}$, and so above this wavenumber, the power spectrum exhibits an enhancement due to the miniclusters. Don't bother going back to Fig.~\ref{fig:DimensionlessPowerSpectrum} to check where this is---it's way off the right-hand edge of the plot.

% This will be when: 
% \begin{equation}
%     \Delta_a^2(k) \sim  \Delta_m^2(k) \approx A_s \left(\frac{k}{k_\star}\right)^{n_s-1} \times \mathcal{T}^2(k) \, ,
% \end{equation}
% is satisfied. Recall that $A_s = 2.2\times 10^{-9}$, $n_s = 0.965$ for $k_\star = 0.05\,$ Mpc$^{-1}$. The transfer function which describes the gravitational growth of initial perturbations is,
% \begin{equation}
%     T(k, a)=\frac{12 k_{\mathrm{eq}}^2}{k^2} \ln \left(\frac{k}{8 k_{\mathrm{eq}}}\right) D(a) \, ,
% \end{equation}
% where the linear growth factor is,
% \begin{equation}
%     D(a)=1+\frac{3}{2} \frac{1+z_{\mathrm{eq}}}{1+z} \, .
% \end{equation}
% Combining these together we find that for

Beyond this value of $k$ we expect the spectrum to depart from perfectly uncorrelated white noise through several effects. Firstly, and most fundamentally, the power spectrum has to get cut off at the Jeans wavenumber because this sets the minimum length scale below which axion overdensities do not grow. Evaluated at $z=0$ the Jeans wavenumber is $k_J \sim 10^{10}$~Mpc$^{-1}$ so this will be the maximum possible $k$ for the minicluster power spectrum, however it will be a smooth rather than a sharp cut off. Another factor that will introduce features into the shape of $\Delta_a^2(k)$ is the fact that many of the initial perturbations were laid down not just by the random distribution of $\theta_i$ but also by the topological defects that were present. This brings in another length scale into the problem related to the spacing of the strings around the time when the domain walls are collapsing the network, which ultimately makes the shape of $\Delta_a^2(k)$ somewhat lumpy prior to matter-radiation equality and not well approximated by white noise towards high $k$. See Ref.~\cite{Vaquero:2018tib} for results on the shape of the power spectrum of the initial seeds to the axion miniclusters and comparisons against the semi-analytic treatment of Ref.~\cite{Enander:2017ogx}.

The shape aside, the fact that the amplitude of the power spectrum after matter-radiation equality will continue to grow $\Delta^2(k)\propto D(a)^2$, where $D(a)$ is the linear matter growth factor:
\begin{equation}
    D(a)=1+\frac{3}{2} \frac{1+z_{\mathrm{eq}}}{1+z} \, .
\end{equation}
So at some point we expect $\Delta_a^2(k)>1$ and for the miniclusters to start growing non-linearly, with larger $k$ modes becoming non-linear first. However, even as early as redshift $z\approx 20$, the power spectrum across the whole range of wavenumbers we have been discussing, i.e.~$k\gtrsim 10^6$~Mpc$^{-1}$ will all have $\Delta_a^2(k)>1$. Some headway can be made with semi-analytic techniques for propagating structure growth like the Press-Schechter or Peak-Patch methods, for which I refer you to Refs.~\cite{Davidson:2016uok, Fairbairn:2017sil, Enander:2017ogx, Blinov:2019jqc, Ellis:2020gtq, Ellis:2022grh}. However, for this next stage of evolution, a better way to get a handle on the non-linear growth of miniclusters, their mergers into \textit{minicluster halos}, as well as their internal structures, is to perform N-body simulations~\cite{Eggemeier:2019khm, Eggemeier:2022hqa, Xiao:2021nkb, Shen:2022ltx, Eggemeier:2024fzs}. \footnote{It is probably important to mention here that, although these simulations lend us a more precise insight into the detailed properties of miniclusters compared to semi-analytic treatments, the same technical problems encountered in running physically realistic simulations of cosmic strings do indeed have an effect on the eventual properties of minicluster populations, as studied recently in~\cite{Pierobon:2023ozb}. The statements I make here should be broadly and qualitatively correct but refined quantitative statements about issues to do with mass distributions and power spectra have the potential to change in the future as work continues.}

N-body simulations, as the name suggests are simulations in which the gravitational attraction of $N$ particles on each other is evolved through cosmic time. There are now a handful of studies in the literature which have performed N-body simulations of miniclusters, starting either from an initial power spectrum left over from early-Universe lattice simulations~\cite{Eggemeier:2019khm, Eggemeier:2022hqa, Pierobon:2023ozb, Eggemeier:2024fzs} or from white noise~\cite{Xiao:2021nkb}. To do the former type of simulations, fluctuations in the density of the continuous axion field are mapped into discrete particles. I have shown a snapshot of one in Fig.~\ref{fig:miniclusters} at a redshift of $z=999$.

So what do we learn from doing N-body simulations? As we might expect, the resulting miniclusters end up distributed across a wide range of masses. The mass function at the latest times it is possible to simulate using current resources is well fit by a falling power law: $\textrm{d}n/\textrm{d}\ln{M} \propto M^{-0.7}$~\cite{Eggemeier:2019khm}. Over time, the miniclusters will merge and the mass distribution is expected to creep up towards larger masses, many orders of magnitude larger than the initial seeds and potentially reaching up close to the Earth mass. Towards this high mass end it is expected for the distribution to scale like $\textrm{d}n/\textrm{d}\ln{M} \propto M^{-0.5}$ which is the naive expectation for mergers forming out of large-scale Gaussian fluctuations. 

The simulation boxes here are tiny compared to the state-of-the-art N-body simulations usually employed to study the non-linear growth of large-scale structures in cosmology. Unfortunately, this is unavoidable because of the extremely fine spatial resolution needed to even resolve the miniclusters in the first place. Furthermore, because N-body simulations impose periodic boundary conditions, only a handful of the most massive minicluster halos ever form inside the boxes and there is a finite time limit on how long they can be run until structures inside the box start to be influenced by neighbouring copies of themselves. Minicluster simulations have been extended up to $z=99$ so far for those beginning from initial minicluster seeds leftover from lattice simulations~\cite{Eggemeier:2019khm, Eggemeier:2022hqa, Pierobon:2023ozb, Eggemeier:2024fzs}, and up to $z=19$ for larger-scale/lower-resolution simulations whose aim is to model the non-linear growth from the white noise part of the power spectrum~\cite{Xiao:2021nkb}. The two approaches are complementary---the former provides insights into the statistics of the more abundant low-mass miniclusters, whereas the latter tells us about the structures of the largest minicluster halos which is where the majority of the dark matter mass is actually contained.

As for the internal structures of miniclusters, those first forming miniclusters from the prompt collapse of the initial density peaks show power-law density profiles with $\rho \propto r^\alpha$~\cite{Pierobon:2023ozb} where $\alpha \sim [-4,-2.5]$. This is similar but not identical to the power-law index expected from models of the self-similar collapse of isolated density perturbations, $\rho(r)\propto r^{-9/4}$~\cite{Bertschinger:1985pd}. On the other hand, the miniclusters that emerge as a result of many smaller mergers trend, unsurprisingly, towards the famous Navarro-Frenk-White (NFW) profile~\cite{Xiao:2021nkb},
\begin{equation}
    \rho_{\mathrm{NFW}}(r)=\frac{\rho_0}{r / r_s\left(1+r / r_s\right)^2} \, ,
\end{equation}
where $\rho_0$ is some characteristic density and $r_s$ is a scale radius where the power law transitions from its small-radii scaling $\rho \sim r^{-1}$ to its large radii scaling $\rho\sim r^{-3}$~\cite{Navarro:1995iw}. There is also interesting evidence presented very recently that the densest of the NFW miniclusters with masses above $M>10^{-12}~M_\odot$ may have even steeper inner profiles than this, with $\rho \sim r^{-2}$~\cite{Eggemeier:2024fzs}.

The total mass in an NFW profile is divergent when integrated out to infinity. This is a problem, but then again, halos are not infinite in extent. The convention for how to cut off the density profile is to enforce the integrated mass out to some virial radius, $r_v$ to be,
\begin{equation}
    M(<r_v) = \int_0^{r_v} 4\pi r^2 \rho_{\rm NFW}(r) \textrm{d}r  = 4 \pi \rho_0 r_s^3\left(\ln \left(1+c\right)-\frac{c}{1+c}\right) \, ,
\end{equation}
where we define $c = r_v/r_s$ and call it the concentration of the halo. A common convention for how to define $r_v$ is to take it to be the radius where the average internal density is 200 times the average density of the Universe.
\begin{equation}
    M(<r_v) = 200 \times \frac{4 \pi}{3} r_{\mathrm{vir}}^3 \bar{\rho} \, .
\end{equation}
Compared to the concentrations for regular galaxies that are $c\sim \mathcal{O}(10)$, miniclusters at redshifts around $z=99$ have concentrations $\mathcal{O}(100)$---a result that can be reproduced using semi-analytic techniques like the Peak-Patch formalism, as studied in Ref.~\cite{Ellis:2020gtq}. These concentrations will grow naturally over time, potentially up to values as large as $\mathcal{O}(10^4)$ when extrapolated to $z=0$. They may also get even more concentrated still if their outer layers are tidally stripped from them. Miniclusters, especially those with NFW profiles, have relatively loosely bound outer layers. In fact, this process is very likely to occur once the miniclusters fall into galactic halos and orbit alongside stars which will exert tidal forces on them.

The presence of miniclusters means that there is an intrinsic granularity to the sub-galactic dark matter distribution in the post-inflationary scenario. Their anticipated presence in galaxy halos today will therefore be important to keep in mind once we go to try and think about the possible observational signatures of axions as a dark matter candidate. I will discuss these signatures in the case of miniclusters in a dedicated subsection Sec.~\ref{sec:minicluster_signatures}. In anticipation of that discussion, one final result of the minicluster N-body simulations that is worth mentioning is exactly how much of the total population of axions actually ends up inside of these objects. Currently, signs point to this being quite a sizeable fraction, but not the entirety. Upwards of 75\%, of axions are gravitationally bound in miniclusters by the time galaxies are beginning to form~\cite{Eggemeier:2019khm}, with the mini\textit{voids}--the spaces between miniclusters---having an ambient density that is slightly smaller than $\sim$10\% of the large-scale average dark matter density~\cite{Eggemeier:2022hqa}.

\subsubsection{Axion stars}\label{sec:axionstars}
It turns out that miniclusters are not the only class of dark matter substructure that could form in the post-inflationary scenario. Another type that may be important in this story are \textit{solitons}---spatially localised field configurations. The equation of motion we use to model axion dark matter permits solutions that have solitonic behaviour, and so we must understand if they form under cosmological conditions.

The first type of soliton (and the slightly less important type in the broader context) are ``axitons'', also called oscillons. These are quasi-stable solutions of the Sine-Gordon equation with a growing mass term~\cite{Kolb:1993hw}---which is precisely what Eq.(\ref{eq:axioneom_w_laplace}) is. In the case of the axion, these appear at times well after $T_{\rm osc}$, and they consist of spherical lumps in the field of a size given by the axion's Compton wavelength $\lambda \sim 1/m_a(T)$. They are seeded towards the end of the post-inflationary simulations that include the turn-on of the axion mass which destroys the string-wall network~\cite{Vaquero:2018tib, Buschmann:2019icd}. They are a curiosity and something of a technical nuisance---they shrink because $m_a(T)$ is growing, and they also tend to proliferate in great numbers at late times when $n$ is large, see e.g.~Ref.~\cite{OHare:2021zrq} for some visualisations of this. Fortunately, because axitons simply disappear when the axion mass stops changing with time, it is unlikely they have many dramatic consequences for cosmology.

The other class of soliton that could be much more important are axion ``stars''~\cite{Tkachev:1991ka}. These are a similar type of object, but appear as solutions to the axion's equation of motion once it is also coupled to gravity. An axion star forms when the inward pull of gravity is balanced against the outward pressure from the axion gradient energy.\footnote{An analogy for this pressure can be understood within the wave-like description of axion dark matter in a similar way to how we think of quantum wave functions. Recall how Heisenberg's uncertainty principle would stop you from confining a quantum wavefunction in space without a simultaneous spread in momentum. This is why people use the term ``quantum pressure'' even though the system we are studying is entirely classical.}
Therefore, they are essentially the smallest objects that can form out of axion dark matter, and are in some sense the ground state of the system.

The properties of axion stars can be derived by finding the stable solution of the Schr\"odinger-Poisson equation. As the name suggests, this is a combination of the Schr\"odinger wave equation and Poisson's equation that together dictate the dynamics of a classical field interacting with itself gravitationally.\footnote{I will be more mathematically explicit about this when discussing ultralight axions in Sec.~\ref{sec:ultralight} where the details are a bit more important. You can also go to the reviews by Niemeyer~\cite{Niemeyer:2019aqm} or Hui~\cite{Hui:2021tkt} for much more detail on the description and behaviour of this system.} The Schr\"odinger-Poisson system in a cosmological context was first studied in the seminal paper of Schive et al.~\cite{Schive:2014dra}, which revealed stable, solitonic cores forming in the centres of self-gravitating halos of wave-like dark matter. This result has been confirmed and elaborated on in numerous studies since then, see e.g.~\cite{Schwabe:2016rze, Veltmaat:2019hou, Zagorac:2021qxq, Glennon:2022huu, Zagorac:2022xic, Mocz:2023adf, Kendall:2023kit, Painter:2024rnc} for a very incomplete sample.

Most relevant for the present discussion is the result from Ref.~\cite{Schive:2014hza} which revealed a relationship between the mass of the central axion star/soliton, $M_s$, and the mass of the host halo that surrounds it, $M_h$:
\begin{equation}
    M_s=\left(\frac{M_h}{M_{\min }(z)}\right)^\frac{1}{3} M_{\min }(z) \, .
\end{equation}
The mass-scale $M_{\rm min}(z)$ encapsulates the axion's Jeans mass mentioned in the previous section, i.e.~the minimum virial mass of a halo that can exist at redshift $z$:
\begin{equation}
M_{\min }(z) = 4.4\times 10^{-20}\,M_{\odot}\,\left(\frac{m_a}{100\,\upmu \mathrm{eV}}\right)^{-\frac{3}{2}} \left(\frac{\zeta(z)}{\zeta(0)}\right)^\frac{1}{4}(1+z)^{3 / 4} \, ,
\end{equation}
where the function $\zeta(z) =\left[ 18\pi^2+82(\Omega_m(z)-1)-39 (\Omega_m(z)-1)^2 \right]/ \Omega_m(z)$, parameterises the redshift dependence of the virial mass. The physical size of the soliton on the other hand is observed to be,
\begin{equation}\label{eq:axionstarradius}
    R_s = 1.06 \times 10^{5} \, {\rm km} \left( \frac{100\,\upmu{\rm eV}}{m_a} \right) (1+z)^{-\frac{1}{2}} \left(\frac{\zeta(z)}{\zeta(0)}\right)^{-1 / 6}\left(\frac{M_h}{10^{-10}\, M_{\odot}}\right)^{-1 / 3} \, .
\end{equation}
So because $R_s \sim M_s^{-1}$, we have the funny result that more massive axion stars are physically \textit{smaller} than lighter ones. Why is this? The point is that axion stars, by definition, are coherent lumps of the field and hence their physical size should be related to the coherence length, i.e.~$R_s \sim \lambda_{\rm coh} \sim 1/m_a v$ where $v \sim \sqrt{G_N M_s/R_s}$ is the virial velocity of axions in the star. Combining these two expressions we find, 
\begin{equation}
    R_s \sim \frac{1}{G_N M_s m_a^2} \sim \frac{1}{G_N M^{1/3}_h m_a} \, ,
\end{equation}
where in the second step we input parametrically the halo-soliton mass relation. This lends some intuition to where Eq.(\ref{eq:axionstarradius}) comes from---if an axion star is heavier, then its virial velocity is higher, and so its coherence length is smaller. Or, you could imagine the balance of forces: a heavier axion star with more gravity has to be balanced by more gradient energy, which is only possible if it is physically smaller.\footnote{If you are feeling dangerous, you could also imagine the scenario where an axion star is so massive that $R_s$ is smaller than its Schwarzschild radius~\cite{Eby:2016cnq,Chavanis:2022fvh}...}

Several later simulations of solitons forming inside of dark matter halos, and then growing as halos undergo mergers have been performed in the years since Refs.~\cite{Schive:2014dra, Schive:2014hza}, generally confirming the core-halo mass relation up to some disagreement about the exact value of the power-law index~\cite{Du:2016aik, Chan:2021bja, Nori:2020jzx}. A resolution that has been suggested is that there simply is some level of intrinsic diversity around the exact value of the index that relates somehow to the specific environment and merger history of their host halos~\cite{Zagorac:2022xic}. Reference~\cite{Du:2023jxh} used the results of their merger-tree model to advocate for the use $M_s \sim M_h^{2/5}$ instead of $M_s \sim M_h^{1/3}$ as a more representative ``average'' scaling that accounts for the intrinsic scatter.

I should perhaps not brush over the fact that these relations were derived from simulations of the Schr\"odinger-Poisson system on a radically different scale to the scale I have cast their results into---Mpc as opposed to mpc. This may seem unusual, but it turns out that there is a scaling symmetry in the Schr\"odinger-Poisson system~\cite{Guzman:2006yc, Ji:1994xh} where all quantities in the problem can be rescaled by a constant to the appropriate power and the dynamics remain the same. This means that we can extrapolate the results from Refs.~\cite{Schive:2014dra, Schive:2014hza} down to minicluster scales. Nonetheless, it is important to check this is indeed the case (and people have done or I would not be discussing it).

Axion stars are indeed seeded inside miniclusters~\cite{Levkov:2018kau, Chen:2020cef} and grow until they reach the core-halo mass relation in alignment with the expectation from simulations of the ultralight-axion/galaxy-halo regime~\cite{Eggemeier:2019jsu}. The main mechanism for this growth is attributed to wave condensation, which occurs on a timescale~\cite{Levkov:2018kau},
\begin{equation}
    \tau \sim 10^{-2} \times\left(\frac{\lambda_{\mathrm{coh}}}{R_{\rm mc}}\right)^{-3} t_{\mathrm{cr}} \, ,
\end{equation}
where $t_{\rm cr} \sim R_{\rm mc}/v \sim \sqrt{R_{\rm mc}/G_N M_{\rm mc}} \ll H^{-1}$ is the timescale for crossing the minicluster. Further mass growth is seen to slow down, but not halt entirely, once the core-halo mass relation is reached~\cite{Chavanis:2019faf, Eggemeier:2019jsu, Niemeyer:2019aqm, Chen:2020cef}. So axion stars could indeed form rapidly inside of miniclusters, and since the solitons are stable, it stands to reason that they could survive until the present day assuming there is sufficient time for them to regrow after experiencing mergers.

Interestingly, you may notice something odd about the fact that the axion-star-radius-to-minicluster-mass relation is $R_s\sim M_{\rm mc}^{-1/3}$ while the minicluster-radius-to-minicluster-mass relation is $R_{\rm mc}\sim M_{\rm mc}^{1/3}$. The opposite scaling means that at small enough miniclumpster masses, those two relations will intersect where $R_s>R_{\rm mc}$, i.e.~the axion star is larger than its host. This poses even more questions about the abundance of axion stars in miniclusters, their survival, and observational signatures---some of which I will discuss more in Sec.~\ref{sec:minicluster_signatures}.

There is a lot more I could have mentioned about axion stars---what is written here is what is relevant for cosmology and signatures of axion dark matter. The complete phenomenology of solitonic objects formed out of the axion field goes much deeper. This is particularly true in regimes in which axion-axion self-interactions become important. I have not included these in the discussion but they should indeed be there at some level. For example, the QCD axion has attractive $\phi^4$ self-interactions that will be pronounced at the large field excursions associated with axion stars. So it is possible for other classes of solutions to exist where the objects are stabilised by some balance of gravity, gradient energy and self-interactions~\cite{Chavanis_soliton, Chavanis_soliton2, Chavanis:2017loo,Amin:2019ums, Braaten:2015eeu}. See, for example, Ref.~\cite{Visinelli:2017ooc} on the difference between so-called dilute and dense axion stars and how the various mass-radius relations that emerge connect to the axion's properties. For more general discussion I also point you to Ref.~\cite{Eby:2019ntd} for a review on QCD axion stars and Ref.~\cite{Visinelli:2021uve} for a review on boson stars\footnote{Boson stars are solitons of a gravitationally-coupled \textit{complex} scalar field.} and other similar objects.

\subsection{Axion-like particles as dark matter}\label{sec:ALPs}

\begin{figure*}[t]
    \centering
    \includegraphics[width=0.99\textwidth]{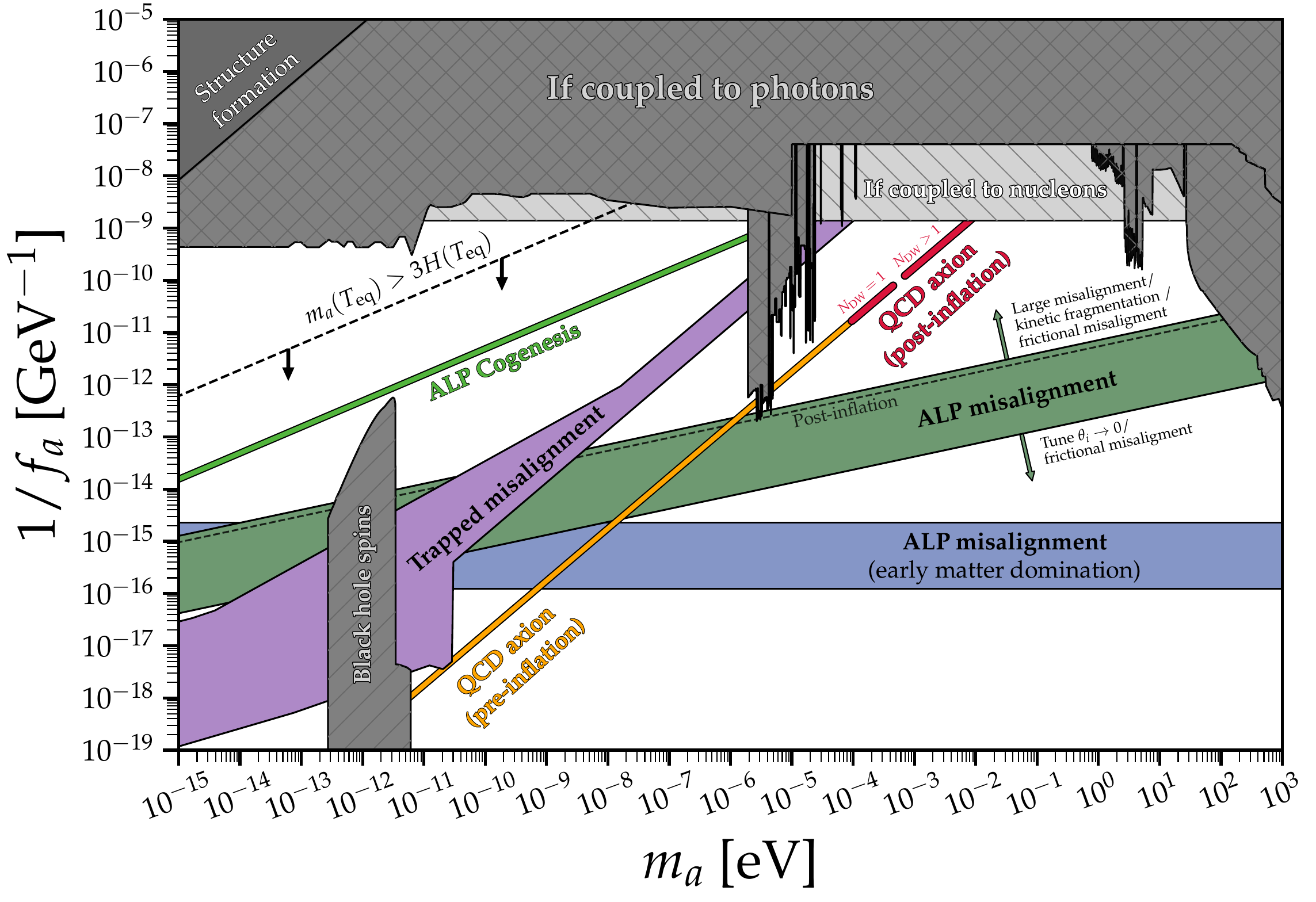}
    \caption{The parameter space of the cosmological ALP in terms of its mass $m_a$ and PQ scale, $f_a$, which are defined in analogy to the QCD axion but kept as free parameters. The QCD axion lives along a single line defined by Eq.(\ref{eq:axionmass})---the yellow and red segments of that line show the best current estimates of the viable mass windows where QCD axions can produce 100\% of the dark matter under the pre- and post-inflationary scenarios respectively. The various other coloured bands enclose the equivalent spaces for ALP models and/or alternative cosmological scenarios, discussed in Sec.~\ref{sec:ALPs}. For sake of comparison and familiarity, I have recast existing bounds on axion-like couplings to the photon and to nucleons into this space, assuming the form $g_{a\gamma} = C_{a\gamma}\alpha/2\pi f_a$ and $g_{an} = C_{an} m_n/f_a$ with $C_{a\gamma,\, an}=1$. The only bound that is strictly on this parameter space is the one arising from the superradiant spin-down of black holes, taken from Ref.~\cite{Baryakhtar:2020gao} (but see also Refs.~\cite{ Mehta:2021pwf,Stott:2020gjj,Baryakhtar:2022hbu}). All data and a Python notebook to reproduce this plot can be found at Ref.~\cite{AxionLimits}.}
    \label{fig:Axion_fa_ALPDM}
\end{figure*}

Throughout this section so far, I have presented a relatively minimal story for the production of QCD axions in the form of dark matter without too many additional model-building bells and whistles. Yet even in this relatively straightforwardly defined scenario, we encountered several wrinkles along the way---considerations like when inflation occurs, limiting the production of isocurvature, avoiding the domain wall problem, the challenges of performing accurate numerical simulations, and the emergence of small-scale axion substructure. All of these problems came about solely as a consequence of trying to explain the dark matter in our Universe using what was essentially (up to details) a one-parameter model. However, there is much we do not know about the earliest stages of the Universe, and all bets are off when it comes to thinking about particle physics beyond the Standard Model. So we are permitted to depart from this simple setup if we have a good reason to, and imagine ways in which axions as dark matter could be produced differently.

We now enter the colourful---and at times confusing---world of alternative axion models and cosmologies. I will not even attempt an exhaustive account of the literature on this, the number of variations people have conjured up is too vast. Instead, in this section, I will just mention a few of the main \textit{classes} of scenarios that I am aware of, particularly if they result in interesting departures from the baseline QCD axion dark matter cosmology. The main motivation for doing this is to set up what will come in the next two sections, which will be all about searches for axions across their wide parameter spaces---let us find out what kinds of models populate them.

A good place to start is to define the notion of an ``axion-like particle'' or ALP. An ALP arises when we relax the requirement of solving the strong CP problem but keep everything else, i.e.~all the phenomenology, qualitatively the same. For our purposes, the upshot of this is that we remove any connection between the ALP and QCD, and as a consequence, remove any connection between $m_a$ and $f_a$, keeping them as free parameters. The QCD axion is therefore a special case of the more broadly defined ALP, and sits on a precise line through the parameter space formed out of those two quantities.

In that spirit, Fig.~\ref{fig:Axion_fa_ALPDM} shows the 2-dimensional parameter space for an ALP. In grey are bounds on ALP properties under the assumption they have some coupling to either the photon or to nucleons (summarised in Ref.~\cite{AxionLimits}). The coloured bands indicate regions where specific cosmological scenarios or axion-like models generate the correct dark matter abundance. Since axion couplings have not come up yet in the discussion, for concreteness let us just take a moment to define what they are. An effective theory for an axion-like particle has the following general interaction Lagrangian:
\begin{equation}\label{eq:axionlagrangian}
    \mathcal{L}_a \supset  -\underbrace{\frac{\alpha_s}{8\pi} \frac{\phi}{f_a} G \widetilde{G}}_{\text{QCD axion only}} - \underbrace{\frac{1}{4} g_{a\gamma} \phi F \widetilde{F} 
    + \,\partial_\mu \phi \sum_{f} g_{af} \bar{f} \gamma^\mu \gamma^5 f - \phi\sum_{f} g_{sf} \bar{f}f}_{\text{QCD axion and ALPs}} \, .
\end{equation}
The first term describes the anomalous axion coupling to the gluons with strength $1/f_a$. The following is true about this term: 1) it is the one required to solve the strong CP problem, 2) it applies \textit{exclusively} to the QCD axion for that reason, and 3) it generates the axion's effective potential $V_{\rm QCD}(\theta)$ and therefore its mass, $m_a$. The second term describes the coupling to electromagnetism, and the third term sums over all possible couplings to fermions $f$. The axion is pseudo-Nambu-Goldstone boson, so its fermion couplings have the shift-symmetry-respecting derivative form. For completeness, I have also included the possibility of scalar, i.e.~CP-violating, couplings to the fermions, which may exist in addition to the latter CP-preserving ones but I won't mention these further (see e.g.~Refs.~\cite{Raffelt:2012sp, OHare:2020wah, Dekens:2022gha, Plakkot:2023pui, DiLuzio:2023lmd} for discussion on them).\footnote{You are also allowed to write down couplings that are quadratic in the field, i.e.~$(\phi/f_a)^2 F \tilde{F}$, which yield distinct phenomenology~\cite{Kim:2022ype,Beadle:2023flm, Kim:2023pvt, Bouley:2022eer, Banerjee:2022sqg, VanTilburg:2024tst}---I will not cover them either.}

The couplings to the photon and to fermions take the following form for the QCD axion:
\begin{equation}\label{eq:ALPcouplings}
    g_{a\gamma} = C_{a\gamma}\frac{\alpha}{2\pi f_a}, \quad g_{af} = \frac{C_{af}m_f}{f_a} \, ,
\end{equation}
where $\alpha \approx 1/137$ and $m_f$ is the mass of fermion $f$. The order-1 dimensionless coupling constants labelled $C_{a\gamma, af}$ encapsulate details about the UV model (e.g.~KSVZ or DFSZ models), and in the case of the photon coupling also low-energy phenomena like QCD-axion-pion mixing. Since all couplings are suppressed by the PQ scale: $g \sim 1/f_a$, this is how we imagine the ALP will couple too. The only difference is that we have more freedom because $f_a$ is not tied to $m_a$ like it is for the QCD axion. In Fig.~\ref{fig:Axion_fa_ALPDM} I have recast constraints on $g_{a\gamma}$ and $g_{an}$ (where $n$ is the neutron) into constraints on $1/f_a$. I have assumed representative values $C_{a\gamma}=1$ and $C_{an} =1$, but this choice is arbitrary and so the location of those recasted bounds is only indicative.

% QCD axion
The QCD axion lives along a single line in this parameter space defined by Eq.(\ref{eq:axionmass}). In reference to the discussion in earlier sections, I have put the ``predicted'' range for the QCD axion mass when produced via misalignment. The yellow line spans the pre-inflationary scenario, assuming $\theta_i$ can be tuned to anything between zero to $\pi$, whereas the red segments of the line correspond to the predictions under the post-inflationary scenario including the axions radiated from topological defects for models with one domain wall~\cite{Saikawa:2024bta}, and those with domain wall number greater than one assuming there is a mechanism for destroying them~\cite{Gorghetto:2020qws}. I am showing them here more for comparison's sake than anything, I will go through the various predictions at a more precise level and the calculations behind them in the next section. 

% ALP misalignment
The easiest way to imagine extending beyond QCD axion misalignment is to go back to the original simplified calculation presented in Sec.~\ref{sec:misalignment}, where we had both the mass and the initial field value left as free parameters. This can be straightforwardly mapped onto an axion-like particle with arbitrary $m_a$ and $f_a$ since we can just write $\phi_i = \theta_i f_a$ for the initial field value, and we are left with the same set of choices for what $\theta_i$ could be. For now, we assume that the ALP mass has no temperature dependence and define the temperature of oscillations to be when $c_{\rm osc}H(T_{\rm osc}) = m_a$, which means,
\begin{equation}
    T_{\rm osc} = \left( \frac{m_a M_{\rm Pl}}{c_{\rm osc}} \right)^{\frac{1}{2}} \left(\frac{90}{\pi^2 g_\star(T_{\rm osc})}\right)^{\frac{1}{4}} \, .
\end{equation}
Recalling Sec.~\ref{sec:misalignment}, the present-day energy density is,
\begin{equation}
\rho_\phi(T_0) = \rho_\phi(T_{\rm osc}) \frac{g_{\star,s}(T_0)}{g_{\star,s}(T_{\rm osc})} \left(\frac{T_0}{T_{\rm osc}} \right)^3 \, ,
\end{equation}
where $\rho_\phi(T_{\rm osc}) \approx \frac{1}{2} m_\phi^2 \theta_i^2 f_a^2$. This means we can evaluate $\Omega_{\rm ALP} h^2$ to find the parameter values that match the dark matter abundance
\begin{equation}
\Omega_{\rm ALP} h^2 \approx 0.12 \, \theta_i^2 \left(\frac{f_a}{1.3\times 10^{13}\,{\rm GeV}} \right)^2 \left(\frac{m_a}{1\,\upmu{\rm eV}} \right)^{\frac{1}{2}} \left(\frac{c_{\rm osc}}{3} \right)^{\frac{3}{2}} \left(\frac{g_\star(T_{\rm osc})}{106} \right)^{\frac{3}{4}}  \left(\frac{g_{\star,s}(T_{\rm osc})}{106} \right)^{-1} \, ,
\end{equation}
where we are again left with the choice of initial angle.

% ALP misalignment angle
The post-inflation ALP was explored in the landmark ``WISPy Cold Dark Matter'' paper by Arias et al.~\cite{Arias:2012az}. The topological defects in ALP models have also been simulated in Ref.~\cite{OHare:2021zrq}, finding an additional 25\% contribution to the axion number density. This can be parameterised by plugging in an ``effective'' value for $\theta^2_i = \theta^2_{\rm eff} \equiv n_a / m_a f_a^2 a(T_{\rm osc})^3\approx 2.33^2$, where $n_a$ is measured from the axion distribution within the simulation~\cite{OHare:2021zrq}. The specific post-inflation prediction, including these contributions from the defect network, is shown by the dashed line in Fig.~\ref{fig:Axion_fa_ALPDM}, whereas the dark-green band encapsulates pre-inflationary ALPS with $\theta_i$ in the range $(0.1,\pi-0.1)$. It is not too hard to imagine that our Universe happened to be given a value of $\theta_i$ that was tuned to small or large values, in which case the region above and below this dark green band respectively would be where an ALP lies. The former case could result from applying an anthropic argument as discussed in Sec.~\ref{sec:preinflation}, although a dynamical explanation is also easier to engineer in the case of an ALP where there is more parameter freedom~\cite{Co:2018phi}. The opposite scenario, in which the axion is tuned up to the maximum in its potential, could also result from additional dynamics in the model perhaps related to inflation, that might drive the axion field's minimum during inflation to be where its maximum would eventually end up afterwards~\cite{Co:2018mho, Takahashi:2019pqf, Huang:2020etx}. There are interesting observational consequences in this `large misalignment' scenario, like enhanced axion substructure, as explored in Ref.~\cite{Arvanitaki:2019rax}.

% Domain wall larger than 1
ALPs with $N_{\rm DW}>1$ have also been considered in Refs.~\cite{Gelmini:2021yzu,Gorghetto:2022ikz}, and are a major component of the
catastrogenesis model of Gelmini et al.~\cite{Gelmini:2022nim, Gelmini:2023ngs, Gelmini:2023kvo}, which ambitiously seeks to generate ALP dark matter alongside primordial black holes from collapsed domain walls, as well as observable gravitational wave signatures.

% Temperature-dependent ALP
Let us now consider some additional details we might be inclined to include in this model. Taking more inspiration from the QCD axion case, you might want to think about ALPs that acquire a temperature-dependent mass like $m^2_a(T)\sim T^{-n}$. Including these effects, the dark-matter-consistency band extends above $T$-independent ALP band shown in dark green and would have a slightly steeper gradient on the plot depending on how large $n$ is. Ultimately this scenario is constrained by the condition that the mass be larger than the Hubble scale at matter-radiation equality for the ALP to be behaving like a proper dark-matter candidate at the latest time in cosmology that we need it to be doing so---this constraint is shown as a black dashed line on Fig.~\ref{fig:Axion_fa_ALPDM}. This scenario has been studied in Refs.~\cite{Arias:2012az, Blinov:2019rhb}, and the additional dark matter contributed by the decay of topological defects for various $n$ scenarios was also quantified in Ref.~\cite{OHare:2021zrq}. While this is simply a phenomenological consideration, the origin of the temperature dependence on the ALP need not be so unnatural. It could come about from a coupling to some hidden but strongly-coupled gauge group~\cite{Arias:2012az}, or from some connection to the electroweak sector through a Higgs portal~\cite{Manna:2023zuq} for example.

% Axion rotations/kinetic misalignment
Another way to extend the logic of the ALP misalignment is to reconsider the various assumptions that were made in the original calculation. Calling back to the derivation presented in Sec.~\ref{sec:misalignment}, one of those assumptions was that the field began with no initial velocity, i.e.~$\dot{\theta}_i = 0$. Models that relax this assumption go by the name ``kinetic misalignment''~\cite{Chang:2019tvx, Co:2019jts}. If the axion field starts out with a large velocity, then it will be cycling around its potential over and over at early times, and potentially still be rolling around by the QCD phase transition. So in this case it is no longer $\theta_i$ that is important---the axion always gains the maximal initial angle at some point---but rather it is how large $\dot{\theta}_i$ is which sets when axion makes it over the potential barrier for the final time and therefore when it starts its oscillations. Because of this, it is easy to see how the abundance gets enhanced because everything in the story is delayed. So to keep the eventual numbers consistent, we have to \textit{lower} $f_a$ for a given mass. Kinetic misalignment populates the parameter space in Fig.~\ref{fig:Axion_fa_ALPDM} above the standard ALP misalignment band. An initial velocity is not an especially strange thing to have emerged in a model either. The example given in the original Refs.~\cite{Chang:2019tvx, Co:2019jts} is based on an Affleck-Dine-inspired model involving the explicit breaking of the PQ symmetry by higher dimensional, $M_{\rm Pl}$-suppressed operators. Explicit breaking of global symmetries is generally anticipated in quantum gravity anyway and is the origin of the so-called PQ quality problem.

% Fragmentation
The spectrum of axions from the kinetic misalignment mechanism turns out to be a little more complicated than in the standard cosmology. If the field is rolling many times over its potential, the precise number of times it does so depends on how much energy the field has. So if the axion field has perturbations then these will all finish their final roll over the top of the potential at slightly different times, resulting in the ``fragmentation'' of the field in which a previously-homogeneous energy density is distributed amongst a spectrum of higher-order modes, as discussed in Refs.~\cite{Fonseca:2019ypl, Morgante:2021bks, Eroncel:2022vjg}.

% Cogenesis
An elaboration even further beyond the kinetic misalignment mechanism is to imagine these initial axion rotations are wrapped up in the generation of the baryon asymmetry of the Universe---another outstanding problem in cosmology so far without a definitive resolution. The models in which both QCD axion dark matter and the baryon asymmetry are generated go by the name ``axiogenesis'' of which there are several variants~\cite{Co:2019wyp, Co:2020jtv, Barnes:2024jap}---they predict a fairly high QCD axion mass, but still not too far from the standard post-inflation case. The non-QCD case known as ``ALP Cogenesis''~\cite{Co:2020xlh,Co:2023mhe} sits on a line in the parameter space defined by $f_a \sim 2\times 10^9 \, {\rm GeV} \left(\upmu{\rm eV}/m_a\right)^{1/2}$ which is shown in light green in Fig.~\ref{fig:Axion_fa_ALPDM}. This construction is based on the idea that a large rotation of the axion field around its domain implies there will be asymmetry in the PQ charge across the Universe, which can then be converted into a baryon asymmetry through strong and/or electroweak sphaleron processes.

% Trapped misalignment
Next up, the purple band in Fig.~\ref{fig:Axion_fa_ALPDM}---this is not truly an ALP model but rather a kind of alternative to the QCD-axion, that happens to occupy part of the $(m_a,f_a)$ parameter space that is slightly shifted away from the conventional one. This model goes by the name ``$Z_N$ axion'', in which we imagine that there are $N$ mirror Standard Models which are unconnected with one another, except for the axion that acts as a portal between them. This slightly unusual-sounding scenario was originally motivated by an earlier proposal for a solution to the hierarchy problem~\cite{Hook:2018jle}. The main reason for mentioning this model here is that the phenomenology and predicted dark-matter parameter space for this modified QCD axion turn out to be quite different. One immediate difference is that the axion's mass is \textit{lighter} for the same value of $f_a$~\cite{DiLuzio:2021pxd}---thus shifting the QCD line to the left on the plot. The misalignment mechanism also proceeds with some differences because of the multiple minima in the axion's potential and its unique temperature dependence. The authors of Ref.~\cite{DiLuzio:2021gos} call it ``trapped misalignment'' because the axion oscillates for a time in the wrong minimum of its potential before finding the correct one (see also Ref.~\cite{Higaki:2016yqk} for an earlier instance of this). The onset of true oscillations is delayed because of this, and so the resulting dark matter density is enhanced. Similar effects are achieved in other models, like the bubble misalignment mechanism where the axion is produced alongside a first-order phase transition~\cite{Lee:2024oaz}, but the enhanced abundance is not necessarily a generic conclusion---Refs.~\cite{Nakagawa:2020zjr, Jeong:2022kdr, Li:2023uvt} show that under certain conditions, the trapping effect can suppress the abundance as well.

% Non-standard cosmology
It is also interesting to ponder how the physics of axion production might depend on the background cosmology beyond simply when inflation happens. As I've mentioned a few times, we have no idea what really took place prior to BBN---keeping the abundances consistent requires the reheating temperature was $\sim$5~MeV at least~\cite{Kawasaki:2000en}. We presume the Universe entered a period of radiation domination following inflationary reheating which persisted until matter-radiation equality 50,000 years later, but we have no evidence for that or anything else. Blinov et al.~\cite{Blinov:2019rhb} evaluated the consequences of a few alternative cosmological histories. Just as an example, I have taken their prediction for ALP dark matter when there is a period of early matter domination---a period where the energy is dominated by some form of matter until times just prior to BBN, after which we are sure the Universe must be radiation dominated. Since the axions are diluted by this period of matter domination, they end up dominating the energy density late, and so the values of $f_a$ consistent with the correct abundance of dark matter have to be very large. See also Ref.~\cite{Visinelli:2009kt, Arias:2021rer, Arias:2022qjt, Cheek:2023fht,Xu:2023lxw} for a discussion of other alternative cosmological scenarios in the context of ALP dark matter.

Another alternative cosmological period that may take place prior to BBN is called kination. This is when the energy density of the Universe is dominated by a scalar field---specifically, its kinetic energy $\frac{1}{2} m_\phi^2 \dot{\phi}^2$. The equation of state for scalar-kinetic-energy domination is $w=1$, and so the energy density scales as $\rho\sim a^{-6}$. Ref.~\cite{Blinov:2019rhb} find that the predicted band for ALPs where this period exists are generally pushed to smaller $f_a$. Axion rotations in the kinetic misalignment mechanism can also lead to a period of kination because $\dot{\theta}$ can be large~\cite{Co:2021lkc}.

% Parametric resonance
That now covers most of what appears in Fig.~\ref{fig:Axion_fa_ALPDM}, but I will highlight a few other possibilities very briefly just to really drive home the point that there is a lot of flexibility here. Another type of phenomenon that can be engineered is known as parametric resonance~\cite{Co:2017mop, Harigaya:2019qnl, Co:2020dya}. Parametric resonance is when a system is driven by some external source of energy at its natural resonant frequency. In the case of the axion, this source of energy could come from the other degree of freedom it is automatically connected to: the radial ``saxion'' mode. In the standard story, the saxion begins with a small value $|\Phi|\sim 0$, and then it rolls down the potential to where $|\Phi| = 0$ and is converted into axions. If, instead, the saxion began at a large value $|\Phi|\gg f_a$ (for example if it was deposited there during inflation), it can still roll down and convert to axions, but may do so much more efficiently because of the possibility for a resonant transfer of energy from one degree of freedom to the other.

% Frictional misalignment
Yet another modification to the standard ALP story goes by the name ``frictional misalignment''. Here, sphaleron-like transitions at high temperatures are invoked to create an additional friction term in the axion equation of motion beyond the standard Hubble friction~\cite{McLerran:1990de}. Depending on the nature and origin of this friction the correct dark matter abundance can be obtained above or below the standard ALP misalignment band as shown in Ref.~\cite{Papageorgiou:2022prc}.

% Multi-axion models
And finally, a general category of models that has a rich additional set of phenomena, are those with multiple axions. Multi-axion models have several sources of inspiration, for example if there are multiple independent sources of CP-violation contributing to the neutron electric dipole moment that need to be dealt with~\cite{Chen:2021jcb, Chen:2021hfq, Chen:2021wcf}, as well as more phenomenological constructions~\cite{Agrawal:2017cmd, Gavela:2023tzu, Hu:2020cga}--however the most common instance of proposed multi-axion scenarios come about as a result of the deluge of axions anticipated in the context of string theory, see e.g.~\cite{Witten:1984dg, Masso:1995tw, Masso:2002ip, Svrcek:2006yi, Conlon:2006tq, Arvanitaki:2009fg, Acharya:2010zx, Jaeckel:2010ni, Cicoli:2012sz, Ringwald:2012hr, Ringwald:2012cu, Demirtas:2018akl, Halverson:2017deq, Halverson:2019cmy, Demirtas:2021gsq, Cicoli:2021gss, Carta:2021uwv}. In the string-inspired scenario of the ``axiverse'', a large number of axions is suggested to emerge in the low-energy phenomenology of particle physics, and one linear combination of axions would be involved in solving the strong CP problem. If more than one of these axions contributes to the dark matter abundance, then the available parameter space that avoids overproduction must be considered carefully~\cite{Marsh:2019bjr}, and this will bring in added constraints if we still need there to be a viable QCD axion amongst them~\cite{Reig:2021ipa, Murai:2023xjn}. These multiple axions states are also expected to mix with each other, which will qualitatively change the formation and behaviour of solitons~\cite{Jain:2023ojg,Jain:2023tsr}, interference effects in the field~\cite{Amin:2022pzv}, topological defects~\cite{Benabou:2023npn}, and miniclusters~\cite{Hardy:2016mns} compared to single-axion cosmologies. To give one specific example, a fun possibility within the axiverse is for there to be pairs of so-called ``friendly'' axions~\cite{Cyncynates:2021xzw} occurring when two mass states happen to be nearly-degenerate and therefore undergo strong quantum mechanical mixing effects similar to neutrino oscillations~\cite{Kitajima:2014xla, Ho:2018qur, Chadha-Day:2023wub, Chadha-Day:2021uyt}. Deriving the expected couplings of these axion-like particles is non-trivial but it is clear that there will be distinct observational signatures in scenarios where the dark matter is made up of many particles across a broad spectrum of masses. The first comprehensive study of an axiverse coupled to electromagnetism was presented recently in Ref.~\cite{Gendler:2023kjt}, to which I refer you for an up-to-date discussion on this rich topic in general.

So, to summarise this whistle-stop tour of alternative axion cosmologies: there are many of them. The classes of scenarios that I've highlighted here should be taken as a representative sample only---a sample that I hope drives home the point that a dark matter ALP could have any set of properties you might dream of.

\section{Cosmological bounds on the QCD axion}
\begin{figure*}[t]
    \centering
    \includegraphics[width=0.99\textwidth]{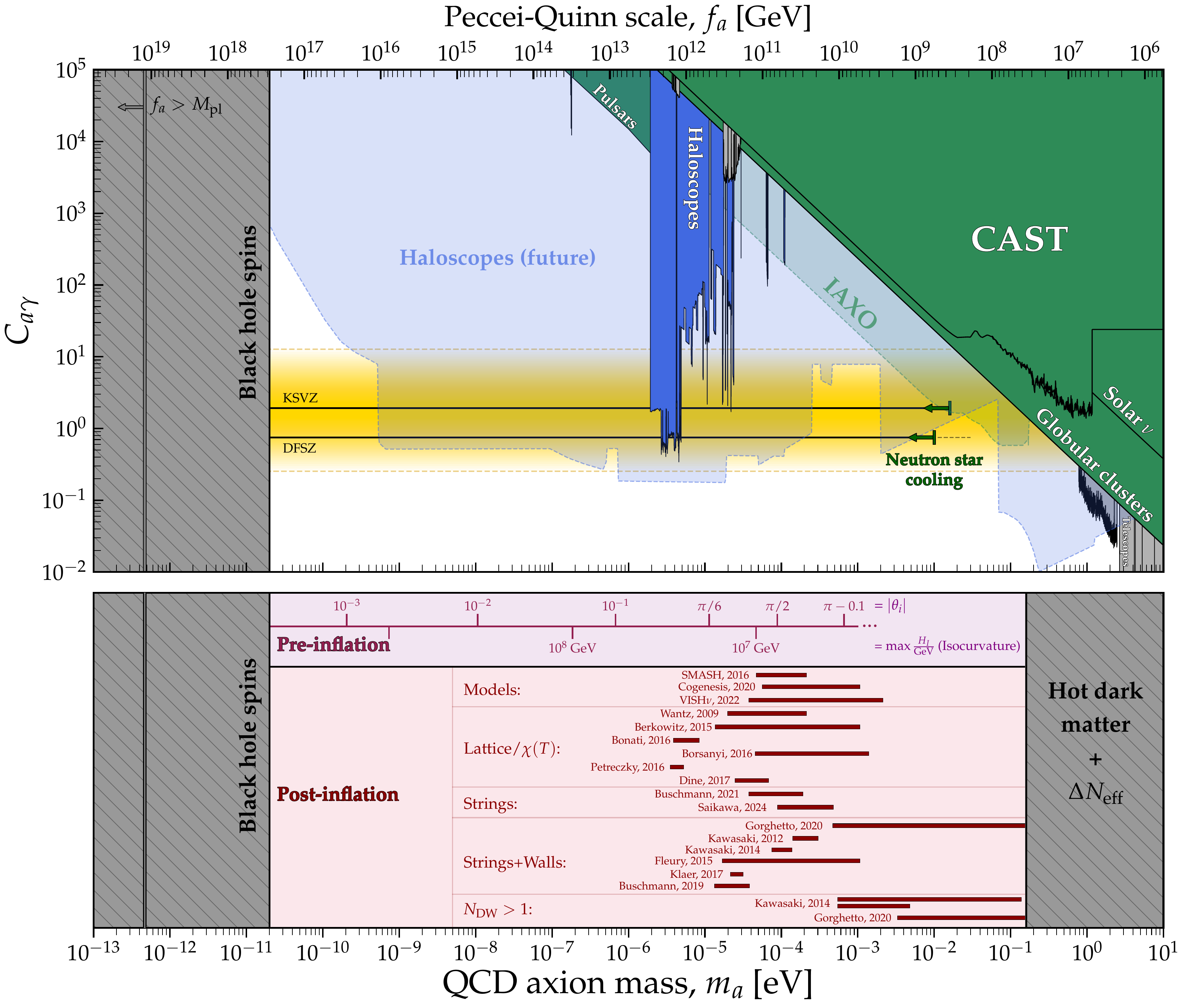}
    \caption{\textbf{Top panel:} constraints on the dimensionless coupling constant between the QCD axion and the photon, defined in Eq.(\ref{eq:C_agamma}). I have cut off the mass scale with a lower bound coming from the superradiant spin-down argument applied to stellar-mass black holes~\cite{Baryakhtar:2020gao, Mehta:2021pwf, Stott:2020gjj, Baryakhtar:2022hbu}, as well as an upper bound from thermally-produced relic axions (see the discussion in Sec.~\ref{sec:thermalaxions}). Bounds on axions produced astrophysically that do not rely on the assumption that they are dark matter are shown in green. Direct searches for axions as dark matter are shown in blue, and indirect searches for axions as dark matter are shown in light grey. \textbf{Bottom panel:} QCD axion dark-matter mass predictions under both the pre-inflation and post-inflation scenarios. In the pre-inflation case, the number-line shows the value of $\theta_i$ required for the axion to make up $\Omega_a h^2 = 0.12$, as well as the maximum Hubble scale of inflation ($H_I$) that avoids overproduction of primordial CDM isocurvature (see Sec.~\ref{sec:preinflation}). In the post-inflation scenario, I list a selection of predictions for the axion mass that can be found in the literature and are discussed in Sec.~\ref{sec:postinflation}. All data and a Python notebook to reproduce this plot can be found at Ref.~\cite{AxionLimits}.}
    \label{fig:AxionMass}
\end{figure*}

The previous section dealt with the many issues that must be understood as a consequence of postulating the axion as a dark matter candidate. This is all interesting enough, but ultimately we want to know if this is really how the Universe works or not. So what we need to do is confront these scenarios with data. Obviously what we would like is a positive signal of axion interactions so that we can measure its properties like its mass and coupling---these I will cover in the final section. However, in the absence of any signals, we can still use data to guide where to focus our search by instead constraining the properties the axion can and cannot have. In particular, what cosmology has been most powerful in telling us about the axion so far regards its \textit{mass}.

In this section, I will go through two types of bound on the QCD axion mass arising from cosmological arguments. But first I need to summarise some of the properties of the QCD axion, particularly its couplings and interactions, that we will need for later sections. Then, I will go through something of a recap of the last section which will involve me summarising the specific mass windows for QCD axions that are consistent with the measurement $\Omega_{\rm DM}h^2 = 0.12$---the predictions that I only alluded to in the previous section. The second part will explain how we can set an \textit{upper bound} on the mass by thinking about the effects of axions produced out of the thermal bath of the early Universe. In short, this section is all about understanding everything going into the summary plot shown in Fig.~\ref{fig:AxionMass}.

\subsection{Properties of the QCD axion}
Let me now introduce some of the important technical concepts associated with the QCD axion that I will make use of in the later discussion. I will mostly be stating facts here and not deriving anything, for detailed first-principles derivations of the axion's various UV-completions, interactions, and couplings, go to Ref.~\cite{DiLuzio:2020wdo}. 

Most constraints on the QCD axion so far arise from its coupling to the photon, so this is why I have chosen to show Fig.~\ref{fig:AxionMass} with $C_{a\gamma}$ as the vertical axis. The general expression for the dimensionless coupling constant in the case of the QCD axion is,
\begin{equation}\label{eq:C_agamma}
    C_{a\gamma} = \frac{E}{N} - 1.92 \, ,
\end{equation}
where $C_{a\gamma}$ appears in the expression for the dimensionful photon coupling as in Eq.(\ref{eq:ALPcouplings}). The first term in $C_{a\gamma}$ involves two parameters that result from the high-energy contents of a full UV-complete axion theory. They are known as the electromagnetic and colour anomaly coefficients. Anomalies are a class of particle interactions that show up quantum mechanically and violate the conservation of a symmetry that is preserved in the classical equation of motion. They can be quantified by finding the terms that violate the conservation of the symmetry's N\"other current. The anomaly \textit{coefficient} is found by summing over the different contributions to the anomaly from different particles, weighted by the charges of those particles under the symmetry in question. Here it's the $U(1)$ PQ symmetry of which the axion is the Goldstone boson. The Noether current $J^\mu_{\rm PQ}$ is conserved ($\partial_\mu J_{\mathrm{PQ}}^\mu = 0$) in the classical equation of motion, but not at the quantum level because of interactions involving triangular loops of whatever quarks are carrying the PQ charge.

We categorise the anomalies into two types: electromagnetic and colour/QCD, which have coefficients $E$ and $N$,
\begin{equation}
    \partial_\mu J_{\mathrm{PQ}}^\mu=N \frac{\alpha_s}{4 \pi} G \tilde{G}+E \frac{\alpha}{4 \pi} F \tilde{F} \, .
\end{equation}
The first anomaly involves only the QCD axion's defining interaction with the gluons, and this turns out to give us our guaranteed contribution to $C_{a\gamma}$. It comes about because we can connect the axion to two gluons which can then connect to a triangle of quarks, then to a pion, and then to two photons. The only axion coupling present there is the effective gluon coupling which is the only term required to solve the strong CP problem. So if the QCD axion exists, we really do expect it to couple to the photon via this pion interaction. Working through this calculation leads to a photon coupling constant of $C_{a\gamma} = -1.92(4)$~\cite{GrillidiCortona:2015jxo}. The second anomaly, the electromagnetic one, is more model-dependent. We include it if the axion also couples directly to electrically charged fermions. In some models, this is the case (e.g.~DFSZ), and in some, it isn't (e.g.~KSVZ).

In Fig.~\ref{fig:AxionMass} I also show the constraints on the space $(m_a,C_{a\gamma})$. Keep in mind that the actual photon coupling for the QCD axion is $g_{a\gamma} \propto C_{a\gamma}/f_a \propto m_a C_{a\gamma}$, so the QCD axion lives along a horizontal line in this plot, up to order-1 model uncertainty in the value of $C_{a\gamma}$. It is common to show a few benchmark models, e.g., 
\begin{equation}
C_{a\gamma} \equiv \frac{E}{N} - 1.92(4) = 
\begin{cases}
  -1.92, & {\rm KSVZ} \\
  0.75, & {\rm DFSZ\,I} \\
  -1.25, & {\rm DFSZ\,II} \, .
\end{cases}
\end{equation}
which are the The Kim-Shifmann-Vainshtein-Zakharov (KSVZ)~\cite{Kim:1979if, Shifman:1979if} and Dine-Fischler-Srednicki-Zhitnitsky (DFSZ) models~\cite{Dine:1981rt, Zhitnitsky:1980tq}. In the plot, I am showing DFSZ I, not II.

These models are essentially ways of extending the SM into a renomalisable theory that contains the QCD axion in the low-energy spectrum of particles. KSVZ and the DFSZ both do this by introducing a Standard Model singlet of which the axion is the pseudo-Nambu-Goldstone boson resulting from the breaking of the global symmetry $U(1)_{\rm PQ}$. The difference between the two is primarily that the KSVZ does this while introducing \textit{additional} non-Standard-Model but coloured quarks that are charged under this new symmetry, while the DFSZ is basically a two-Higgs-doublet model with the SM quarks charged under the PQ symmetry. At some level, quarks (Standard-Model ones or otherwise) have to be involved in a QCD axion model because we also need to explicitly break the $U(1)_{\rm PQ}$ symmetry with the colour anomaly to solve the strong CP problem (all axion models must have $N>0$).\footnote{The $N$ appears on the bottom of the first term in Eq.(\ref{eq:C_agamma}) because it is convention to absorb it into the definition of $f_a$ for convenience} There are two benchmark DFSZ models because the Yukawa couplings between the quarks and the Higgs can be with either of the two Higgs doublets. The KSVZ benchmark assumed above is for only one extra non-Standard-Model quark, but there is a set of possible representations that it could have under the Standard Model gauge group, and even more possibilities if you add more than one quark. Di Luzio et al.~\cite{DiLuzio:2016sbl,DiLuzio:2017pfr} put forward a helpful window for the possible KSVZ-like ``hadronic'' models that cover all possible representations of the quark, whilst excluding the models that have pathologies like Landau poles (strong coupling below the Planck scale), or cosmologically stable and strongly-interacting relics. This window, which spans $E/N \in [5/3,44/3]$, is what the fuzzy yellow band in Fig.~\ref{fig:AxionMass} represents, where I choose to give it a fuzziness to emphasise that this is still just an indication of plausible models---it is certainly possible for a QCD axion to lie outside of this window, for example if there is partial cancellation between the two contributions to $C_{a\gamma}$ in Eq.(\ref{eq:C_agamma}). Presumably, the further outside the band you go, the more enhancement or cancellation you are probably required to have, so the standard hadronic band is a decent benchmark.\footnote{Extended or non-minimal versions of the KSVZ or DFSZ models are more likely to encompass models that lie further away from the canonical QCD band (e.g.~\cite{Darme:2020gyx}). Exploring this possibility requires a kind of combinatorial survey of anomaly coefficients, which has been done for the KSVZ~\cite{Plakkot:2021xyx} and DFSZ~\cite{Diehl:2023uui} models recently. In the KSVZ case, these correspond to models where there are multiple heavy quarks rather than just one, and in the DFSZ case, it is when there is more than one additional Higgs doublet.}

There are a few important constraints on the QCD axion that are independent of cosmology and must be considered before we try and make additional ones. Most QCD axion models will also have couplings to nucleons, and these couplings lead to the strongest upper limits on the axion mass. In particular, the most stringent of these to date comes from making sure that the emission of axions from neutron stars does not cool them faster than we observe them to~\cite{Buschmann:2021juv}. This imposes an upper bound on $m_a$ between $10$ to $30$~meV for DFSZ models\footnote{A technical note: there is an uncertainty on the QCD axion nucleon couplings driven by the unknown Higgs vacuum expectation values in those models, the range here spans over a range of the parameter $\tan{\beta}$.}, and $m_a<16$~meV for KSVZ. Preventing there from being too much axion emission from the Sun~\cite{Vinyoles:2015aba}, or from stars in globular clusters~\cite{Dolan:2022kul, Capozzi:2020cbu}, also implies upper bounds on the QCD axion mass. For most models these are less competitive, but apply to any cases where the axion couples to the photon or the electron.\footnote{Another method for arriving at an upper limit on the axion mass is to consider the extremely stringent constraints on rare, flavour-violating processes like $K\rightarrow \pi +a$. These have been searched for in experiments like E949, NA62 and KOTO. Together they impose an upper limit on $m_a$ in the context of some KSVZ models which possess particular representations for the heavy quark~\cite{MartinCamalich:2020dfe, Alonso-Alvarez:2023wig}.} In Sec.~\ref{sec:thermalaxions} I will describe how similar emission mechanisms occurring in the thermal plasma of the early Universe also allow us to set a \textit{cosmological} upper bound on the QCD axion mass.

So far the only \textit{lower} bound on the QCD axion mass we can draw is from black hole spins. The general idea is that black holes are able to excite bound states of axions around them, which in the process causes them to lose their angular momentum via superradiance. The fact we observe black holes with nonzero spin therefore excludes the existence of axions that would have spun them down~\cite{Baryakhtar:2020gao, Mehta:2021pwf, Stott:2020gjj, Baryakhtar:2022hbu}. Technically this bound applies over a limited range of masses where the axion's Compton wavelength is of a similar size to the black holes we have observed spinning (in this case stellar mass ones). However for this discussion the superradiant argument functions as a lower bound because going beyond it leaves us in the region where we have super-Planckian values of the PQ scale: $f_a>M_{\rm Pl}$.

\subsection{Dark matter predictions for the QCD axion mass}\label{sec:predictions}
Attempts to directly detect axion dark matter lead to the bounds shown in blue in Fig.~\ref{fig:AxionMass}. Even without knowing any details of these experiments, it is easy to see that despite decades of searching, only narrow windows in axion mass have been excluded. This frustrating state of affairs is precisely where cosmology comes to the fore---we should be in a position to give experiments a location on this plot to target their searches.

In the lower panel of Fig.~\ref{fig:AxionMass} I included several categories of predictions for the QCD axion mass. As I hope I made clear in Sec.~\ref{sec:postinflation}, most of these predictions should not be taken as firm or definitive. The point of showing them is to give a general feeling for the areas of parameter space that we can loosely assign to different cosmological scenarios. Still, the end goal is to work towards something much more precise. This is especially the case for the post-inflationary scenario where a precise prediction does seem to be possible, at least in principle. Many of these predictions I have already alluded to in Sec.~\ref{sec:postinflation}, so the purpose of this subsection is to go through them in the order they appear on the plot from top to bottom.
\\~\\
\noindent {\bf Pre-inflation}: This number line is basically identical to the one described in the previous section. You can obtain the correct abundance of dark matter for any axion mass as long as you are happy to accept that the initial angle bestowed upon our observable Universe was a particular number. For low axion masses with $f_a$ around the scale of Grand Unification, for example, the angle has to be tuned close to zero, so you might also want to invoke an anthropic argument. Because we need to limit the amount of isocurvature pre-inflationary axions generate, certain values of the axion mass then also come with an associated upper limit on how large the scale of inflation can be. For example, if the axion were produced pre-inflation, and the mass was around an neV, then this means $H_I$ would have to be less than $\sim 10^9$ GeV.
\\~\\
\noindent {\bf Models:} Several interesting models of the QCD axion have been constructed that attempt to simultaneously solve other problems in physics, or fill other gaps in our knowledge. These include providing a mechanism for generating neutrino masses, a field to drive inflation or an origin for the matter-antimatter asymmetry of the Universe. Two examples involving neutrino masses are the SMASH model of Ballesteros et al.~\cite{Ballesteros:2016euj, Ringwald:2023anj, Berbig:2022pye} which has its basis in the KSVZ axion (see also~\cite{Salvio:2015cja,Salvio:2018rv,Salvio:2021puw}), as well as the DFSZ-inspired ``VISH$\nu$''model of Sopov and Volkas~\cite{Sopov:2022bog}. The ``Cogenesis''~\cite{Co:2020xlh} scenario is designed to solve the baryon asymmetry, as described in Sec.~\ref{sec:ALPs}. Here it is the rotations in the \textit{QCD} axion field that generate the initial charge asymmetry which is then fed into a chiral asymmetry in the quarks, and then the baryon asymmetry.
\\~\\
\noindent {\bf Lattice QCD}/$\chi(T)$: Several of the earliest predictions of the QCD axion mass in the post-inflationary scenario were refined using calculations of the topological susceptibility (the axion-mass temperature dependence) using lattice QCD. This allowed them to account for one of the uncertainties that ultimately influences the dark matter abundance. Note that these calculations still do not consider the effects of axion field gradients, and therefore the contribution to the axion abundance from topological defects is not included---the takeaway from these predictions should be that their lower edges are robust, but their upper edges are not because there could be additional axions from strings. The corresponding references are the following: Bonati et al.~\cite{Bonati:2015vqz}, Buchoff et al.~\cite{Buchoff:2013nra}, Borsanyi et al.~\cite{Borsanyi:2016ksw}, Berkowitz et al.~\cite{Berkowitz:2015aua}, Dine et al.~\cite{Dine:2017swf}, and Petreczky et al.~\cite{Petreczky:2016vrs}.
\\~\\
\noindent {\bf Strings}: To include the axions radiated by cosmic strings, extensive cosmological lattice simulations evolving the axion field numerically through cosmic time are needed. These are challenging, and currently physical values for the parameters nor simulations lasting physical durations are possible. Extending the early simulations from over a decade ago by Hiramatsu et al.~\cite{Hiramatsu:2010yu}, three groups have recently published major results on axion string radiation and have given out mass predictions: Gorghetto et al.~\cite{Gorghetto:2018myk, Gorghetto:2020qws}, Buschmann et al.~\cite{Buschmann:2021sdq} and Saikawa et al.~\cite{Saikawa:2024bta}.\footnote{A comparison of several string simulation results was made in Ref.~\cite{Hoof:2021jft}.} All groups see evidence for a logarithmic increase to the number of strings per Hubble volume, as well as the growth of the string-radiation spectral index $q$ over time---both initially observed by Gorghetto et al.~\cite{Gorghetto:2018myk, Gorghetto:2020qws}. Buschmann et al.~\cite{Buschmann:2021sdq} successfully employed adaptive mesh refinement to extend the duration of their simulation by enhancing the grid resolution around the string cores. Despite there being several sources of quantitative disagreements, it is safe to say that the ultimate mass prediction is most heavily influenced by the many-orders-of-magnitude extrapolation of the axion emission spectrum beyond the final times of the simulation. Saikawa et al.~\cite{Saikawa:2024bta} performed the most recent and largest set of simulations, and has made a concerted effort to quantify numerical and extrapolation uncertainties.
\\~\\
{\bf Strings+Walls}: In this set of predictions, the axion field is evolved not only during the string-dominated period after the PQ phase transition but also through the QCD phase transition during which domain walls appear and act to collapse the network. All of these assume one species of the domain wall, i.e.~$N_{\rm DW} = 1$, as in, for example, the one-quark KSVZ model. Some of the first serious estimates of the axions emitted from the combined string-wall system were by Hiramatsu et al.~\cite{Hiramatsu:2012gg} and Kawasaki et al.~\cite{Kawasaki:2014sqa}, pointing towards a relatively wide range. A few years later a more precise prediction was made by Fleury \& Moore~\cite{Fleury:2015aca} and then Klaer \& Moore~\cite{Klaer:2017qhr, Klaer:2017ond}. The most recent and largest simulations of the string-wall decay were performed after that by Vaquero et al.~\cite{Vaquero:2018tib} and Buschmann et al.~\cite{Buschmann:2019icd}. The simulations of Vaquero et al.~are so far the largest in this classification in terms of numbers of lattice sites at $\sim 8000^3$, but they did not quote a mass prediction. To simulate the destruction of the strings by walls, these simulations must bring about the QCD phase transition at unphysical values of the string tension, so although they allow the distribution of the fluctuations that would eventually become miniclusters to be understood, predicting an axion mass still requires the same extrapolation to high values of the string tension. The eventual mass predictions are, therefore, only as reliable as whatever is learned from the string-only simulations, hence why most current efforts are focused there.
\\~\\
\noindent {\bf Domain wall number larger than one}: This last set of predictions follows the same principle as the previous ones but for the multi-domain wall scenarios, like the DFSZ model. As discussed earlier, this results in a higher abundance of axions from the collapse of the extra walls, and hence the consistent values of the axion mass must be larger in order to suppress the overall energy density down to the cosmologically observed value. The group of Hiramatsu, Kawasaki et al.~have performed numerical simulations of these scenarios~\cite{Hiramatsu:2012sc, Kawasaki:2014sqa}, and the $N_{\rm DW}=6$ case was also extrapolated using a dimensional scaling argument from an $N_{\rm DW}=1$ simulations by Gorghetto et al.~\cite{Gorghetto:2020qws}. I emphasise again that simply taking $N_{\rm DW}>1$ leads to a cosmology that is not compatible with our Universe, and some additional explicit breaking of the PQ symmetry is needed to break the degeneracy and destroy the network. Many of these scenarios result in additional strong CP-violation that must be then confronted with observational constraints like the neutron electric dipole moment~\cite{Hiramatsu:2012sc, Kawasaki:2014sqa, Beyer:2022ywc}.

\subsection{Thermal relic axions}\label{sec:thermalaxions}

\begin{figure}
    \centering
    \includegraphics[width=0.99\linewidth]{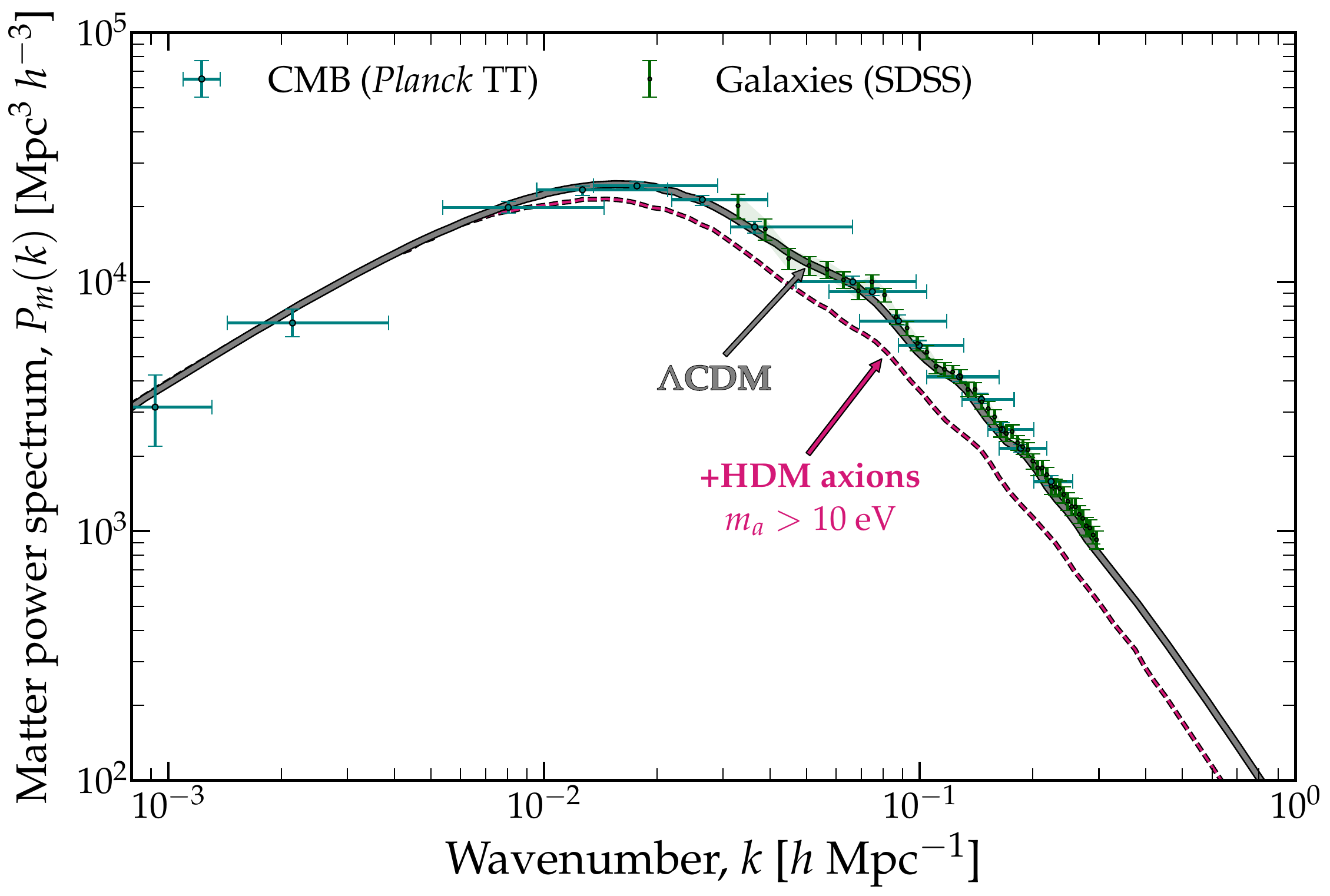}
    \caption{The suppression in the matter power spectrum expected if there is a form of hot dark matter in the Universe. For the sake of illustration, I have chosen a case where the suppression is unphysically large, corresponding to an axion mass in excess of $m_a = 10$~eV. The effect of hot dark matter is to hinder gravitational clustering towards small scales (large $k$) which is why the power spectrum is suppressed. The power spectrum nowadays is very precisely measured, so an axion this heavy is massively ruled out, as is easily seen by eye. Masses at the current upper limit allowed for the QCD axion $\mathcal{O}(0.1~{\rm eV})$ would leave the matter power spectrum essentially unchanged at the level of this plot.}
    \label{fig:Pk_HDM}
\end{figure}

Aside from explaining the cold dark matter abundance in the Universe, there is another cosmological bound that we may draw on the QCD axion that is both robust and independent of any assumptions about what particles ultimately make up the dark matter. This bound results from considering their production in thermal processes occurring in the hot plasma of the early universe. Since the QCD axion must couple to particles that existed in the thermal bath at high temperatures---gluons, quarks, pions etc.---this population is guaranteed at some level, but if the interactions with those particles are strong enough then it is also possible that the axions were actually in thermal equilibrium with them~\cite{Turner:1986tb, Berezhiani:1992rk}. Inevitably, once the interaction rate $\Gamma$ drops below the Hubble expansion rate $H$, these processes cannot maintain equilibrium, and the axions decouple from the bath---however, their presence may still be felt by the particles they decoupled from.

The early thermal history of the Universe---tested through cosmological data like the CMB, BBN, and large-scale structure---is consistent with all the degrees of freedom available in the Standard Model and none extra. So any effect of a population axions generated and kept in equilibrium at early times must be extremely subtle if it is present at all. This then implies we can place bounds on the properties of the QCD axion. Specifically, it will give us an upper limit on the mass---heavier axions being more strongly coupled. Although we only currently have a constraint, this idea may get much more interesting in the future thanks to upcoming high-resolution probes of the CMB. The statement that the Universe's thermal history is consistent with the Standard Model is only true as far as we have been able to measure it; future data may reveal the presence of additional as-yet-unaccounted-for degrees of freedom. Like the C$\gamma$B (aka CMB) and the C$\nu$B, it may well be there is a C$a$B---a cosmic axion background---as well.

This thermally-produced C$a$B will certainly have to be out of equilibrium by recombination, $T>0.26$~eV, which means our thermal axions will be decoupling from the thermal bath while relativistic. If the axion mass is on the larger side though, they may end up redshifting down to non-relativistic speeds given enough time. This is the case if the axion mass is $m_a\gtrsim 1$~eV, where we would consider the C$a$B to be a form of ``hot dark matter''---a component that behaves like matter at late times, but was initially produced while relativistic and so free-streamed during early structure formation. Hot dark matter is imprinted on the matter power spectrum because it is able to escape small-scale structures that have begun gravitational collapse while it is still free-streaming relativistically. A hot dark matter component is tightly bound to be a tiny fraction of the cosmic energy budget thanks to probes of the distribution of matter~\cite{Baumann:2017gkg}. The C$a$B doesn't necessarily have to be hot dark matter though. If the mass of the axion is negligible compared to its thermal energy, then for much of cosmic history the C$a$B constitutes a form of ``dark radiation'', whose effects are imprinted much more subtly in the Universe's thermal history, as I will come to soon.

Axions are now firmly ruled out as a hot dark matter candidate---which is to say that the mass is bounded to be so light that the C$a$B never constitutes a matter component. Fig.~\ref{fig:Pk_HDM} shows the matter power spectrum for $\Lambda$CDM which fits the data perfectly, and an illustrative example of the case when there is a population of thermally produced axions that were massive enough to become non-relativistic on cosmological timescales. Since this component of matter free-streams over much larger scales than the highly non-relativistic cold dark matter, a Universe with even a small amount of hot dark matter in it has noticeably less small-scale structure than one with none. Since the effect of a hot-dark-matter population of axions is identical to the cosmic neutrino background, if the neutrino masses are large enough, analyses that seek to set bounds on $m_a$ must do so by assuming a model that includes both $m_a$ and $\sum m_\nu$ as parameters. Studies doing this yield constraints at the level $m_a\lesssim 0.5$~eV, see e.g.~Ref.~\cite{Hannestad:2010yi, Cadamuro:2010cz, Archidiacono:2013cha}. So axions are ruled out as hot dark matter, and therefore, a C$a$B, if it exists, is a form of dark radiation. This pushes us into thinking instead about the effects of a thermally-produced axion as an additional relativistic degree of freedom in play during the pre-CMB universe. As such, the rest of this section is devoted to showing how we can derive this constraint.

The effects of dark radiation are quantified and constrained via a cosmological parameter called $N_{\rm eff}$---the ``effective'' number of neutrino flavours. The way this parameter comes about is by writing down the overall energy density in radiation. In the Standard Model, the only forms of radiation we care about are our beloved photons and our $N_\nu = 3$ neutrinos, so this should be:
\begin{equation}
    \rho_r = \rho_\gamma + \rho_\nu = \left[1+ \frac{\rho_\nu}{\rho_\gamma} \right]\rho_\gamma = \left[1+\frac{7}{8}\left(\frac{T_\nu}{T_\gamma}\right)^{4} N_{\mathrm{eff}}\right] \rho_\gamma
\end{equation}
where I use Eqs.(\ref{eq:rho_vs_T}) to convert energy densities into temperatures for bosons and fermions--- $N_\nu$ neutrinos contribute $\rho_\nu \propto 2\times N_\nu \times \frac{7}{8} T^4$ compared to photons that contribute $\rho_\gamma \propto 2\times T^4$. The factor $(T_\nu/T_\gamma) = (4/11)^{1/3}$ is the famous ratio between the cosmic neutrino background and cosmic photon background temperatures derived in Sec.~\ref{sec:thermodynamics} which arises because the photons get an extra injection from $e^+ e^-$ annihilation after the neutrinos have already decoupled. Except: notice now that in the formula I didn't put in $N_\nu$ for the number of neutrino flavours, but rather some ``effective'' number, $N_{\rm eff}$ which is $\neq N_\nu$. Even though there are indeed only three flavours of neutrino in the Standard Model, the value we need for this calculation is not exactly three because (among several other corrections) the neutrinos haven't totally fallen out of equilibrium with the rest of the bath by the time the temperature scale is roughly at the electron mass $T \sim 0.5$~MeV---for example, there is some amount of $e^+ e^- \rightarrow \bar{\nu} \nu$ still going on. State-of-the-art calculations yield $N_{\text {eff }}=3.0440 \pm 0.0002$ for the Standard Model, see e.g.~Refs.~\cite{Gnedin:1997vn,Mangano:2001iu,deSalas:2016ztq,Bennett:2020zkv,Akita:2020szl,Froustey:2020mcq,Drewes:2024wbw}.

If we now add some relativistic axions into the mix, we will need to include a third type of radiation:
\begin{equation}
    \rho_r = \rho_\gamma + \rho_\nu + \rho_a = \left[1+\frac{7}{8}\left(\frac{T_\nu}{T_\gamma}\right)^{4} N_{\mathrm{eff}} + \frac{1}{2}\left(\frac{T_a}{T_\gamma}\right)^4\right] \rho_\gamma \, .
\end{equation}
where the factor of $1/2$ is because axions only contribute one extra degree of freedom compared to the photon's two. Since there could be any number of additional new particles contributing to $\rho_r$, in practice these are lumped into one, and early-Universe data like the CMB and BBN constrain a generic parameter called $\Delta N_{\rm eff}$ which is how many extra degrees of freedom there are in units of a single neutrino. So since we are only adding one new degree of freedom described by some temperature $T_a$, then the parameter $\Delta N_{\rm eff}$ is,
\begin{equation}
    \Delta N_{\rm eff} = N_{\rm eff}-3.044 =\frac{4}{7}\left(\frac{T_a}{T_\nu}\right)^4 = \frac{8}{7}\left(\frac{11}{4}\right)^{4 / 3} \frac{\rho_a}{\rho_\gamma} \,  .
\end{equation}

I won't get into exactly how the CMB gives us access to $\Delta N_{\rm eff}$ here\footnote{I'd recommend lectures by Green~\cite{Green:2022bre} as an excellent pedagogical introduction to this physics, and Ref.~\cite{Dvorkin:2022jyg} for a timely discussion of recent progress in the physics of dark radiation.}, but in a nutshell these are general implications of a value of $\Delta N_{\rm eff}>0$. The main effect is that more radiation enhances the expansion rate during radiation domination and also extends its duration, lowering the redshift of matter-radiation equality in the process. An enhanced expansion rate acts to shrink the size of the sound horizon projected on the sky, shifting the CMB's peaks to smaller scales. It also increases the amount of diffusion of radiation out of the smaller fluctuations, and so suppresses the amplitude of the power spectrum further at high $\ell$ through Silk damping~\cite{Hou:2011ec}. The latter two effects are somewhat coupled together---because the $\ell$ of the first acoustic peak is so well measured, values of $N_{\rm eff}$ are best tested through the latter effect on the damping tail. In doing so, the inferred value of $H_0$ will necessarily have to shift slightly to keep the peaks in the same position. Ultimately all of these effects must be very mild though (if they weren't, we would have seen them already), so we really need precise measurements of the CMB power spectra at the very small scales/high-$\ell$ where these effects are apparent to push the frontier of our knowledge forward~\cite{CMB-S4:2016ple, Abazajian:2019eic}.

The \textit{Planck}+BAO 2018 analysis measures $N_{\rm eff} =2.99\pm0.17$~\cite{Planck:2018vyg}, which is in agreement with a slightly less-precise value inferred from getting the abundance of the primordial elements correct within the framework of BBN: $N_{\rm eff} = 2.89 \pm 0.28$~\cite{Cyburt:2015mya}. Moreover, both values are consistent with the Standard Model prediction within their uncertainties. So this means that anything like $\Delta N_{\rm eff}>0.5$ or so would have been spotted already in these analyses as a 2$\sigma$ discrepancy with theory. We must make sure our thermal cosmic axion background doesn't give us a $\Delta N_{\rm eff}$ larger than this. As we will see, imposing this constraint will give us an upper bound on the axion mass.

So going back to axions now, what we want to get out is a prediction for $\Delta N_{\rm eff}$. For a particle that was once in thermal equilibrium and then decoupled, it turns out that getting this answer requires nothing more than the temperature at which this decoupling occurred. Let's say axions decouple very early on, before various expected phases of the Universe's evolution like the electroweak or QCD phase transitions---in this case, the ratio $T_a/T_\gamma$ will be small because the photons get all that extra injection from those eras while the axions miss out. On the other hand, let's say the axions decoupled at a similar temperature to the neutrinos, then the only entropy injection they miss out on is from $e^+ e^-$ annihilation, and so the ratio $T_a/T_\gamma$ will be bit larger, comparable to that of the neutrinos. In this scenario we observe $N_{\rm eff}\sim 3.044 + 1$---or we would, if this was not already wildly inconsistent with the \textit{Planck} measurement of $N_{\rm eff}\sim 3$. So we conclude already that axions must decouple before the neutrinos (if they were ever even in thermal equilibrium to begin with).

So if we want $\Delta N_{\rm eff}$ then the question we should be asking is: what temperature did the axions decouple from the thermal bath? Let us define decoupling first of all. Axions (or any particles) decouple when the rate of production and annihilation processes that keep them in equilibrium drops below the expansion rate of the Universe: $\Gamma(T_d) < H(T_d)$. Plugging in the temperature dependence of the Hubble parameter during radiation domination, Eq.(\ref{eq:Hubble_raddom}), we get:
\begin{equation}
T_d \approx \left(M^2_{\rm Pl} \Gamma(T)^2 \frac{90}{\pi^2 g_\star(T_d)} \right)^\frac{1}{4} \, .
\end{equation}
only this is still an equation we must solve since the production rate $\Gamma$ is probably going to be temperature-dependent.

So what is $\Gamma$ for the QCD axion? The axion's defining interaction is a vertex connecting the axion to two gluons. So we know that there will be $gg \leftrightarrow g\phi$ going on, as well as processes involving quarks in the initial and final state like $q \phi \leftrightarrow g q$ and $ag \leftrightarrow q\bar{q}$. In fact, these processes will dominate at temperatures above the confinement scale $T>100$~MeV where we have free quarks and gluons flying around. However, if we go towards temperatures around or below this scale, then axion interactions are better described using the axion-pion effective Lagrangian and we have $\pi^{ \pm} \pi^0 \leftrightarrow \pi^{ \pm} a$, but which also emerges from the axion-gluon interaction. This era actually turns out to be the one that is relevant for deriving the current bound on the axion mass because $T_d\sim 100$~MeV, but for the purposes of doing a simple calculation let us stay at high temperatures.

The general formula for the interaction rate is just the number density multiplied by the velocity-averaged cross-section:
\begin{equation}
    \Gamma = n_g \langle \sigma v\rangle \, ,
\end{equation}
where we put in $n_g$ as the gluon number density for now, and just the $gg \leftrightarrow g\phi$ process for simplicity. To get $n$ for a boson in thermal equilibrium we just do a similar integral to Eq.(\ref{eq:energydensity_gas}) but without the factor of $E$, which will give us the number density rather than the energy density. The result for 8 gluons is:
\begin{equation}
    n_g(T) = 2\times 8 \times \frac{\zeta(3)}{\pi^2} T^3 \, ,
\end{equation}
where $\zeta(3) \approx 1.202$ is the Riemann-zeta function. The number density of fermions is the same up to an extra factor of 3/4. The process in question comes about from the interaction $\mathcal{L} = \frac{\alpha_s}{8\pi} \frac{\phi}{f_a}G\tilde{G}$, and the cross-section has the form\footnote{In fact the cross section for the processes involving quarks will have essentially the same parametric dependence.},
\begin{equation}
    \langle \sigma v\rangle \sim \frac{\alpha^3_s}{8\pi^2 f_a^2} \, ,
\end{equation}
which yields the decoupling temperature:
\begin{equation}\label{eq:Td}
    T_d \lesssim \frac{\sqrt{g_\star(T_d)}\pi^5}{2\sqrt{90}\zeta(3)\alpha_s^3}\frac{f_a^2}{M_{\rm Pl}} \approx  15\times 10^{11} \,{\rm GeV} \left(\frac{f_a}{10^{12}\,{\rm GeV}}\right)^2 \approx 7 \times 10^{9} \,{\rm GeV} \left(\frac{m_a}{100\,\upmu{\rm eV}}\right)^{-2} \, .
\end{equation}
I put a ``$\lesssim$'' because this overestimates $T_d$ on account of the fact we have considered only a single production/annihilation process. Adding more will increase the factor on the denominator and thus decrease $T_d$. For example, including the processes involving quarks, we would instead get a value of $T_d \approx 3\times 10^{11}$~GeV for $f_a = 10^{12}$~GeV. To get the numerical value written above I have taken $g_\star(T_d) = 107.75$ (+1 compared to the usual value at temperatures above the electroweak scale because we have added the axion) and $\alpha^{-1}_s(10^{11}\,{\rm GeV}) \approx 30$ from the running of the strong gauge coupling constant. 

Even though this answer is fairly close to accurate estimates, if we want to do things properly we must account for all possible processes that could have kept axions in equilibrium with the Standard Model bath out of the primordial plasma. For some studies of the production rates arising from processes involving certain couplings see e.g.:~Refs.~\cite{Brust:2013ova, Salvio:2013iaa,Ferreira:2018vjj, Green:2021hjh} for the Standard Model fermions; Refs.~\cite{Masso:2002np, Graf:2010tv} for the gluon; and Ref.~\cite{Bolz:2000fu} for photons. In general, however, concrete axion models are expected to possess whole sets of couplings to several particles at once, as studied in Refs.~\cite{Chang:1993gm, DEramo:2021usm,DEramo:2021lgb} for example. The bounds I will quote towards the end of the section will be the ones derived in this spirit and apply to specific named UV-complete axion models like KSVZ or DFSZ. Nonetheless, it is still informative to compute bounds where only one coupling is ``switched on'' at a time, as in e.g.~Refs.~\cite{Ferreira:2018vjj, DEramo:2018vss, Arias-Aragon:2020shv, Green:2021hjh, DEramo:2021psx,Caloni:2022uya, Giare:2020vzo}.

Another interesting thing to point out about Eq.(\ref{eq:Td}), is that it leaves open the possibility that the decoupling temperature could be \textit{higher} than $f_a$. In this case, the axions are born decoupled---the thermal bath is never hot enough to maintain a thermal population, and so none of what I'm talking about ever applies to our Universe. We can see that $T_d>f_a$ will occur if $f_a \gtrsim 10^{12}$~GeV or $m_a\lesssim 6\,\upmu$eV, in which case there is no thermal background of axions. Similarly, if inflation occurs but only reheats the Universe to a temperature $T_{\rm RH}<T_d$ then we also have no thermal axions.

Back to the matter at hand, since we know $T_d$ we can now go and relate this to $\Delta N_{\rm eff}$. We arrive at this by writing down the formula for $\Delta N_{\rm eff}$ in terms of a temperature ratio between the axion and photon,
\begin{equation}\label{eq:dNeff_step}
    \Delta N_{\rm eff} = \frac{4}{7} \left( \frac{11}{4}\right)^\frac{4}{3} \left(\frac{T_a}{T_\gamma} \right) \, .
\end{equation}
To get $T_a$, we can enforce the conservation of comoving entropy density between two temperatures, e.g.~between $T_d$ and the temperature of the photon bath today $T_0$. Recall Eq.(\ref{eq:entropydensity}), where I stated that the entropy density of the Universe is $s(T) = (2\pi^2/45) g_{\star,s}(T)T^3$. Since we have axions in the mix, we will add $+1$ to $g_{\star,s}$ when working it out at temperatures $T>T_d$, but at temperatures $T<T_d$ we only add $+(T_a/T)^3$ since the axions have fallen out of equilibrium. The conservation of comoving entropy density between $T_d$ and $T_0$ therefore states:
\begin{equation}
\begin{aligned}
    s(T_d) a(T_d)^3 &= s(T_0)a(T_0)^3\, , \\
    \Rightarrow \left(g_{\star,s}(T_d)+1\right) T_d^3 a(T_d)^3 &= \left(g_{\star,s}(T_0) + \left(\frac{T_a}{T_0}\right)^3\right) T_0^3 \, ,
\end{aligned}
\end{equation}
where in the second step we set $a(T_0) = 1$. Since axions are decoupled for temperatures below $T_d$, this means their temperature today is just redshifted from their decoupling temperature: \mbox{$T_a = a(T_d) T_d$}. Writing $T_0 = T_\gamma$ as the temperature of the photon bath today we can rearrange to get:
\begin{equation}
    \frac{T_a}{T_\gamma} = \left(\frac{g_{\star,s}(T_0)}{g_{\star,s}(T_d)}\right)^\frac{1}{3} \, ,
\end{equation}
where the $g_{\star,s}(T)$ refers specifically to the entropic degrees of freedom \textit{in the Standard Model}. Plugging this back into Eq.(\ref{eq:dNeff_step}), we find that axions decoupling at $T_d$ will contribute the following amount to $\Delta N_{\rm eff}$:
\begin{equation}
\Delta N_{\mathrm{eff}} = \frac{4}{7}\left(\frac{11}{4} \frac{g_{\star s}\left(T_0\right)}{g_{\star s}\left(T_d\right)}\right)^\frac{4}{3} \approx 0.027 \left( \frac{106.75}{g_{\star,s}(T_d)}\right)^\frac{4}{3} \, .
\end{equation}
Since $g_{\star,s}$ in the Standard Model is a constant for temperatures above the top-quark mass, $T_d\gtrsim100$~GeV, we see that axions are actually guaranteed to contribute \textit{at least} 0.027 to $N_{\rm eff}$, if they were ever in thermal equilibrium.

\begin{figure}
    \centering
    \includegraphics[width=0.99\linewidth]{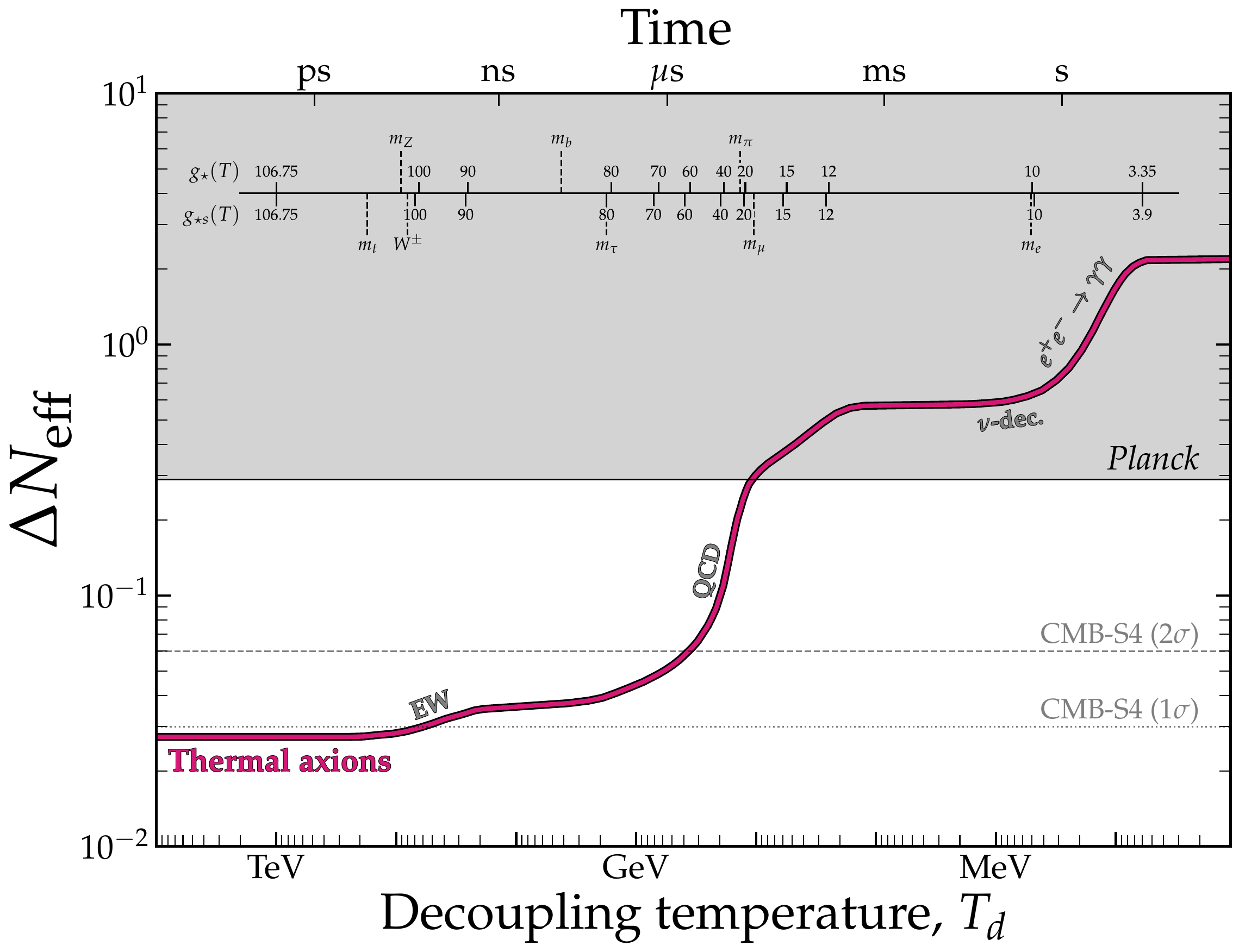}
    \caption{The number of relativistic degrees of freedom added to the effective number of neutrinos active in the Universe when an axion (or any particle with one internal degree of freedom) decouples at some temperature, $T_d$. The longer the axions can remain in thermal equilibrium the more they will contribute to the radiation density of the Universe. If they can last as long as neutrinos do (down to $\sim1$~MeV) then they contribute $\Delta N_{\rm eff} \sim 1$, which is what this number is measured relative to. This case is confidently ruled out by \textit{Planck}~\cite{Planck:2018vyg}, and future CMB probes~\cite{CMB-S4:2016ple, Abazajian:2019eic} aim to be sensitive to \textit{any} new particles freezing out in this manner.}
    \label{fig:dNeff}
\end{figure}

The $\Delta N_{\rm eff}$ contribution from thermally-produced axions as a function of the decoupling temperature is shown in Fig.~\ref{fig:dNeff}. For comparison, I have shown the existing \emph{Planck} bound at the 68\% level, which rules out axions decoupling at temperatures lower than $\lesssim$0.1 GeV---almost but not quite up to the QCD phase transition. I have also shown the forecasted 68\% and 95\% CL sensitivity expected from the upcoming CMB-Stage 4 program, which is the collective name for a series of ground-based CMB experiments that will map the high-$\ell$ CMB sky to even greater precision~\cite{CMB-S4:2016ple, Abazajian:2019eic}. Tantalisingly, the minimal contribution to $\Delta N_{\rm eff}$ expected from a thermal axion background in the Universe is just about within reach.\footnote{This, of course, is by design---0.027 corresponds to the smallest increase to $N_{\rm eff}$ for a 1-degree-of-freedom particle that was ever in thermal equilibrium.} The ultimate target is, therefore, to push down the precision to $\Delta N_{\rm eff}\lesssim 0.027$ so that $T_d$ can be bounded to above the electroweak phase transition, which is the earliest point in time where we believe new degrees of freedom ought to have been created in the Standard Model\footnote{This statement is really resting on some quite big assumptions about cosmology, namely that the Universe was radiation dominated. Strictly we have no concrete handle on anything happening before BBN. It is quite possible for the Universe to have reheated to a temperature just above the minimal temperature allowed by BBN of $T_{\rm RH}>5$~MeV~\cite{Kawasaki:2000en}, or for there to have been a period of early matter domination. In other cosmological histories the axion mass bounds can be significantly relaxed~\cite{Grin:2007yg, Carenza:2021ebx, Arias:2023wyg}.}.

So now let me briefly discuss a bit about the actual bounds. Increasingly stringent upper bounds on the axion mass under various model configurations have been derived over the years, beginning with Refs.~\cite{Hannestad:2003ye, Hannestad:2005df, Hannestad:2007dd} at the level of $m_a\lesssim 1$--$3$~eV. Sequential improvements in the extent and quality of cosmological datasets brought the bound down to the sub-eV level during the WMAP~\cite{Melchiorri:2007cd, Hannestad:2010yi, Hannestad:2008js} and \textit{Planck} eras~\cite{Archidiacono:2013cha, Giusarma:2014zza, DiValentino:2015zta, DiValentino:2015wba}.\footnote{That is for models with the usual KSVZ-like couplings. In models with suppressed couplings, like the `astrophobic' models~\cite{DiLuzio:2017ogq,Bjorkeroth:2019jtx,Badziak:2021apn,Badziak:2023fsc}, the mass limits are weaker than this~\cite{Badziak:2024szg}.} Beyond this, advancements in this area have been driven thanks to people endeavouring increasingly accurate computations of $T_d$ for realistic models. Some examples of recent sophistications include going beyond the assumption that decoupling happens instantaneously, and correctly computing the production rates while in the midst of a phase transition~\cite{Arias-Aragon:2020shv, DEramo:2021psx, DEramo:2021lgb, Notari:2022ffe, DEramo:2022nvb}. Another interesting case from recently is Ref.~\cite{DiValentino:2022edq} who performed an analysis that also marginalised over alternative cosmological histories.

The frontier where the bound lies is now around the QCD phase transition, which means the simplifying assumption I made of axion production via gluons alone is not a good one. Relevant calculations of the production rate are those that are accurate at 100~MeV temperatures. While this fact has been appreciated for a long time, it was pointed out only relatively recently by Di Luzio et al.~\cite{DiLuzio:2021vjd}, that the existing calculations based on an extrapolation of the lowest-order chiral perturbation expansion of the axion-pion effective Lagrangian were not valid at temperatures above $\sim$60 MeV. In the few years since then, people have been trying to remedy this situation. Two approaches have emerged: one that involves smoothly interpolating over the QCD crossover up to the regime where axion-gluon (rather than axion-pion) scattering dominates~\cite{DEramo:2021psx, DEramo:2021lgb, DEramo:2022nvb}, and the other which is based on an empirical derivation of the $a \pi \rightarrow \pi \pi$ amplitude from experimental pion scattering data~\cite{DiLuzio:2022gsc,Notari:2022ffe}. The two methods were compared in Ref.~\cite{Bianchini:2023ubu} and shown to disagree by $4$\% at most, in terms of the value of $\Delta N_{\rm eff}$. However the interpolation approach has been questioned in Ref.~\cite{Notari:2022ffe} which pointed out the importance of including strong sphaleron contributions at these temperatures. A full exploration of this regime within the context of lattice QCD is probably needed to finally settle this issue~\cite{Bonanno:2023thi}.

The state-of-the-art cosmological extraction of the thermal axion mass bound including all available data from BBN down to the CMB, and incorporating correct computations of the axion production rate around the QCD phase transition, results in mass bounds of $m_a<0.28$~eV for the KSVZ axion and $m_a\lesssim0.2$~eV at 95\% CL~\cite{DEramo:2022nvb}---the DFSZ case being marginally stronger due to the additional production channels open in that model~\cite{Ferreira:2020bpb}. Other bounds also based on the now-corrected production rates show general consistency with this~\cite{Notari:2022ffe, Bianchini:2023ubu}, with~\cite{Bianchini:2023ubu} quoting the strongest bound to date of $m_a<0.16$~eV.

As expressed in Fig.~\ref{fig:dNeff}, the situation stands to improve in the future~\cite{Archidiacono:2015mda, Baumann:2016wac, Green:2019glg, Green:2021hjh} thanks to CMB-Stage 4~\cite{CMB-S4:2016ple, Abazajian:2019eic} which has as one of its central goals to make a sub-percent level measurement of the abundance of relativistic species in the universe. At this level of precision, even those axions decoupling at temperatures prior to the electroweak phase transition leave a hint of their existence in the data.

Let's say instead that we pessimistically assume $N_{\rm eff}$ remains consistent with the SM to that level of precision, mapping this constraint onto axion parameters leads to a potential bound of $m_a<10^{-4}$~eV below which the couplings were simply not large enough for the axions to have ever thermalised.\footnote{See the projections in Ref.~\cite{Baumann:2016wac} but note the strong dependence on the unknown reheating temperature of the Universe. As discussed earlier, if inflation ended and reheated the universe at a temperature below the temperature at which the axions would decouple then there is no population of thermal axions around to leave any imprints in cosmological data.} Thinking optimistically though, it is quite possible that we may reveal that there really are additional forms of dark radiation in the Universe! The one downside is the fact that this discovery is essentially encapsulated in a single number---a positive detection of $\Delta N_{\rm eff}>0$ does not necessarily tell us that the \textit{axion} exists, only that there is some form of beyond-Standard-Model dark radiation. It would nevertheless be a very promising nudge in the right direction, so these near-future constraints on $N_{\rm eff}$ represent one of the most exciting tests of the axion we will have gotten for many years.

% Cosmic axion background
Having covered most of what there is to say about the axion mass bounds derived from thermal production, there is one more tangent it is worth going off on that is related to this subject. If things conspire in such a way that the early universe did indeed create a large population of thermally produced axions, then even if we can't see their imprints in data, they should still be flying all around us in the Universe right now, just like the cosmic background of photons and neutrinos. In fact, with all the numbers we have worked out already, it is very simple to see how substantial this C$a$B will be at the present day. We just use the formula for the number density of bosons with 1 degree of freedom as a function of their temperature:
\begin{equation}
    n_{\rm C{\it a}B} = \frac{\zeta(3)}{\pi^2} T_a^3 = \frac{1}{2} n_\gamma \left(\frac{g_{\star,s}(T_0)}{g_{\star,s}(T_d)}\right) = 7.5 \, {\rm cm}^{-3} \left(\frac{106.75}{g_{\star,s}(T_d)}\right) \, .
\end{equation}
In the second step, I have brought in $n_\gamma = 411$~cm$^{-3}$, which is the present-day number density of CMB photons. Just like the CMB, the C$a$B will be distributed in energy with a redshifted Black Body spectrum:
\begin{equation}
    \frac{\textrm{d}\rho_{\rm C{\it a}B}}{\textrm{d}\omega} = \frac{\omega^3}{\pi^2(e^{\omega/T_a}-1)} \, ,
\end{equation}
The peak frequency of a Blackbody at temperature $T$ is $\omega_{\rm peak} = 2.8 T$, which for the cosmic axion background is:
\begin{equation}
    \omega_{\rm peak} = 2.8 T_a \approx 2.8 T_\gamma \left( \frac{g_{\star,s}(T_0)}{g_{\star,s}(T_d)}\right)^\frac{1}{3} = 218 \, \upmu {\rm eV} \, \left(\frac{106.75}{g_{\star,s}(T_d)}\right)^\frac{1}{3} \, ,
\end{equation}
whilst this turns out to be within the frequency range of a few proposed searches for axions as galactic dark matter, unfortunately, the energy density is shockingly low. Even if the whole spectrum could be somehow collected in a single instrument, we still have to contend with the fact that $\rho_{\rm C{\it a}B} = \frac{\pi^2}{30} T_a^4 = 10^{-12}$~GeV~cm$^{-3}$---11 orders of magnitude smaller than the galactic dark matter density. At the moment it is not clear to me if our chances of ever detecting the C$a$B directly are as dim as they are for the C$\nu$B~\cite{PTOLEMY:2018jst, Bauer:2022lri}. 

% Other axion backgrounds
Keep in mind that the only parameter controlling $\rho_{\rm C{\it a}B}$ here is $T_d$ so there is little room for ALP models with enhanced couplings to make up the numbers and enable detection. The only way a detectable C$a$B might exist is if there are processes that can generate a relic population of axions other than thermal contact with the Standard Model, for example the decays of some heavier parent particle at early times. There have been discussions about this, e.g.~Refs.~\cite{Higaki:2013qka, Conlon:2013txa, Conlon:2013isa, Evoli:2016zhj, Dror:2021nyr}, and even attempts at direct detection~\cite{ADMX:2023rsk}, however the readily-detectable models at the moment typically require that the heavier particle also makes up the dark matter and its subsequent decays into axions over the age of the Universe cause a background of them to build up~\cite{Cui:2017ytb, Dror:2021nyr, Kar:2022ngx}. Many other examples of diffuse axion backgrounds were catalogued recently in Ref.~\cite{Eby:2024mhd}.

% Freeze-in axions
Another class of cosmological background for axion-like particles was explored in Ref.~\cite{Balazs:2022tjl, Langhoff:2022bij}---the so-called ``irreducible axion background''. Here the axions are also thermal relics, but of a different type---ones which are said to ``freeze-in'' to the Universe as opposed to the ``freeze-out'' mechanism we have been working through. Axions freeze in when the processes coupling them to the Standard Model bath are so feeble that they just accumulate in the Universe gradually and out of thermal equilibrium. Within certain parts of the parameter space of a keV--MeV scale ALP coupled to the photon, such an abundance could build up in sufficient amounts through processes like $\gamma e \to a e$, $
\gamma \gamma \to a$ or, $e^+ e^- \to \gamma a$. It is then possible the subsequent decays of this population \textit{back} into photons over longer timescales could be observable in the form of excess energy injection, or as an anomalous X-ray emission line in the local universe.\footnote{In fact this argument represents one of the strongest constraint on the axion-photon coupling for keV-MeV mass ALPs that does not rest on any assumption about the nature of dark matter. This is the part of the green bound in Fig.~\ref{fig:AxionPhoton_UltraSimple_FullParameterSpace} that stretches downwards diagonally.} As with all of the other thermal axion bounds I have been discussing, the critical question is: what is the maximum temperature the Universe reached prior to BBN? However, Ref.~\cite{Langhoff:2022bij} argue that even in the minimal case where the reheating occurs just above the temperature allowed by BBN ($T_{\rm RH} = 5$~MeV)~\cite{Kawasaki:2000en}, there is still a residual abundance of freeze-in axions that would have been detected in some parts of parameter space---hence the ``\textit{irreducible}'' axion background.

\section{Cosmological bounds on ultralight axions}\label{sec:ultralight}
Having now discussed the upper limit on the axion mass that we can derive using cosmology, we will now head down to the complete opposite end of the spectrum. We are back to discussing the dark matter of the Universe, but when we enter the ultralight regime we really have to confront the concepts that I introduced at the beginning of Sec.~\ref{sec:axionCDM}---the concept of \textit{wave-like dark matter}.

\subsection{Wave-like dark matter}
To refresh your memory, wave-like dark matter is when we model the dark matter in terms of a macroscopically occupied state of some large population of bosons, which mathematically can be expressed straightforwardly in terms of a classical field. We can write down the equation,
\begin{equation}\label{eq:singlesine}
    \phi(x,t) \approx \frac{\sqrt{2 \rho_{\rm DM,local}}}{m_\phi} \sin( m_\phi t + m_a \frac{1}{2} v^2t + m_a\mathbf{v}\cdot \mathbf{x}) \, .
\end{equation}
if we want to know the value of the field at some point in space and time. Since a dark matter population will be described with a phase space distribution in terms of position and velocity, this single sine can only be accurate for describing the field on short timescales and distances. If we imagine the dark matter has some spread in velocities $\sigma_v$, then its wave-like nature implies that when we compare field values between two spatially or temporally separated points, those two values will be out of phase with each other if those separations are larger than,
\begin{equation}
    \tau_{\rm coh} = \frac{2\pi}{m_a v \sigma_v} = 3.2 \, {\rm Myr} \left( \frac{10^{-22}\,{\rm eV}}{m_a} \right) \left(\frac{167 \, {\rm km/s}}{\sigma_v} \right) \left(\frac{220 \, {\rm km/s}}{v} \right) \, ,
\end{equation}
and,
\begin{equation}\label{eq:coherencelength}
    \lambda_{\rm coh} = \frac{2\pi}{m_a \sigma_v} = 0.72 \, {\rm kpc} \, \left(\frac{10^{-22}\,{\rm eV}}{m_a} \right) \left( \frac{167 \,{\rm km/s}}{\sigma_v}\right) \, .
\end{equation}
The numerical value evaluated for $\lambda_{\rm coh}$ is intended to be thought-provoking in the context of dark matter. 

When we start thinking about the concept of low-mass wave-like dark matter in general, we will naturally be led down the path of thinking about \textit{ultralight} dark matter. How small can the dark matter mass be? The limiting factor in answering this is precisely $\lambda_{\rm coh}$---which can be thought of as the physical size of the eponymous waves of wave-like dark matter.\footnote{However see Ref.~\cite{Amin:2022nlh} for an interesting alternative viewpoint on deriving a minimum dark-matter mass.} In the equation above I plugged in $m_a = 10^{-22}$~eV: dark matter candidates with masses this light are often classed as ``fuzzy dark matter'' to distinguish them from CDM---at this scale, the wave-like characteristics start to exhibit themselves on scales comparable in size to the cosmic structures they form~\cite{Hu:2000ke, Guzman:2003kt, Schive:2014dra, Hui:2016ltb}. Fuzzy dark matter and CDM universes look very different, so we must understand which picture is correct.\footnote{There are some historically noteworthy papers from decades ago introducing this idea~\cite{Baldeschi:1983mq, Sin:1992bg, Ji:1994xh}, predating the recent resurgence in interest towards wave dark matter by several decades.}

Because the lighter you go, the larger the dark matter waves get, it is clear that dark matter cannot be a particle with an arbitrarily small mass. We invented dark matter because we needed a way to explain why structures like galaxies form and look the way they do. In fact, our entire conception of why the large-scale structure of the cosmos even exists is a direct result of the initial gravitational collapse and subsequent hierarchical merging of dark matter halos. So a very minimal requirement we can enforce on our fuzzy dark-matter model is that these waves aren't bigger than the smallest structures we need them to fit inside. This puts a relatively strict bound on dark matter models having masses less than around $10^{-22}$~eV, where its wavelength is already bigger than things like dwarf galaxies which have $\mathcal{O}({\rm kpc})$ radii and contain lots of dark matter.

In spirit, this is how we can set a lower bound on the mass of ultralight axions. To set a robust exclusion limit though, we need to understand more concretely what fuzzy dark matter halos look like and compare that with data. Specifically, we want to know whether or not the structures and abundances of fuzzy halos are distinct in any way from the ones formed by plain-old collisionless CDM. Recall the discussion at the end of  Sec.~\ref{sec:cosmologicalobservations}: standard cosmology really does fit practically all of the cosmological data very well and with no sign that we require any additional parameters beyond the 6 of $\Lambda$CDM. Dark matter may differ from CDM in some interesting way, but at the very least it should reproduce what CDM already succeeds in explaining.

There has been a lot of recent progress in understanding the nature of fuzzy dark matter and there are several important lower bounds on $m_a$ that people have now drawn. I think many would argue that an extremely fuzzy dark matter cosmology is probably at the cusp of being ruled out---which is to say that the dark matter \textit{could} be a very light particle, but not so light that this fuzziness is made apparent on the scale of galaxies. As you can already see from Eq.(\ref{eq:coherencelength}), as we increase the dark matter mass, these waves will shrink, which means the fuzziness becomes too small to see. At this point fuzzy dark matter converges on cold dark matter from the perspective of any realistic astronomical observation.\footnote{In fact, this hand-wavy statement has a solid mathematical underpinning. It can be shown that the limiting behaviour of the equations describing fuzzy dark matter as $1/m_a \to 0$ is to transform into the Vlasov-Poisson equation for a self-gravitating gas of collisionless particles. This is referred to as the Schr\"odinger-Vlasov correspondence~\cite{Mocz:2018ium}.}

\subsection{The Schr\"odinger-Poisson system}
\begin{figure}
    \centering
    \includegraphics[width=0.99\linewidth]{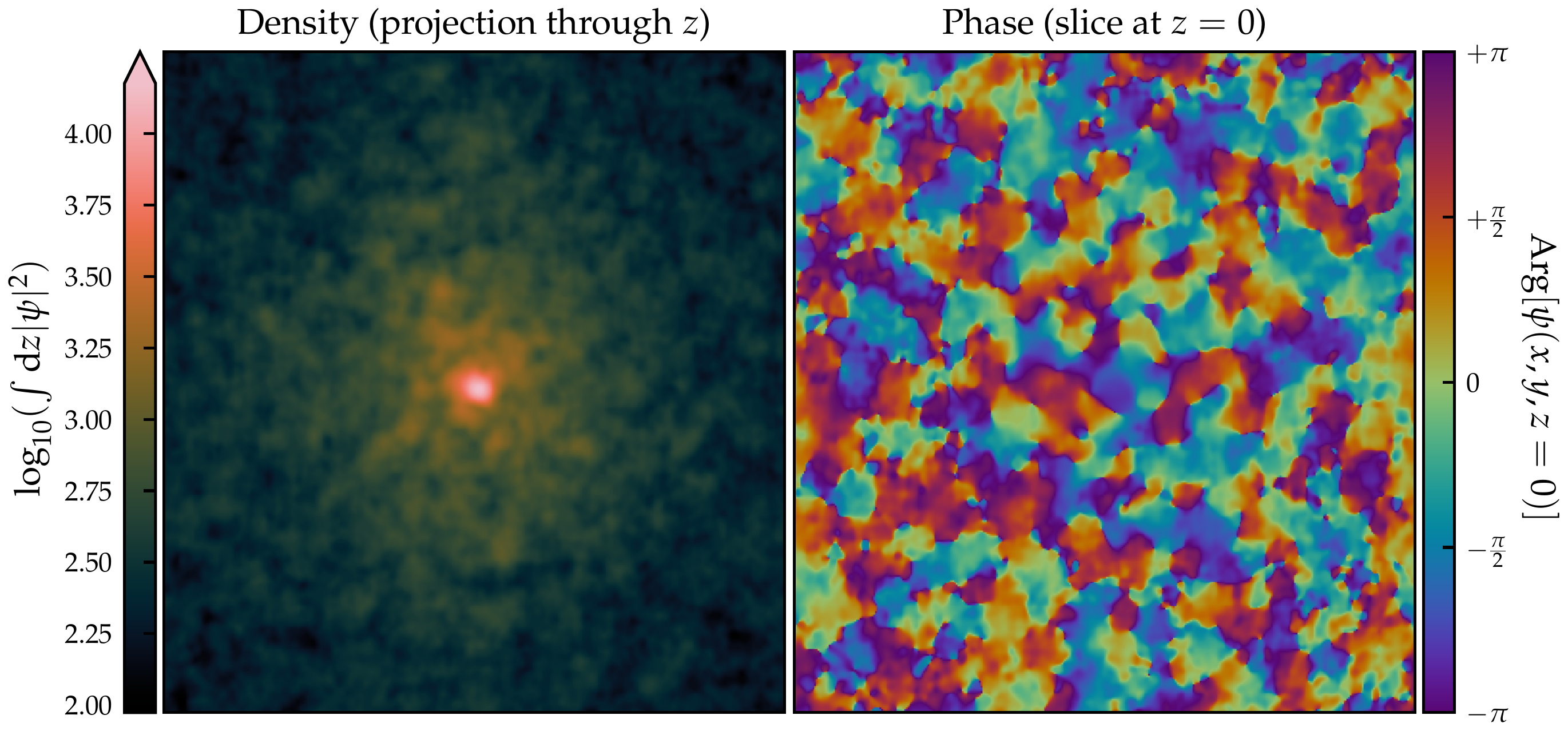}
    \caption{A toy-model simulation of a clump of ultralight axions evolved under the Schr\"odinger-Poisson equations. The Schr\"odinger-Poisson equations govern the evolution of the envelope of the field's oscillations, $\psi$, rather than the value of the oscillating field itself, see Eq.(\ref{eq:WKB}). The left-hand panel shows the density, which is related to the amplitude of this complex envelope, while the right panel shows the phase $\vartheta$, where $\nabla \vartheta$ is interpreted as the dark-matter velocity. The density is shown as a projection through the box, whereas the phase is only shown as a slice through the middle of the box.}
    \label{fig:fuzzy_halo}
\end{figure}

Now we delve a bit further into the physics of fuzzy dark matter. The starting point is to understand how an ultralight bosonic field evolves under its own gravity. The relevant equation for this is aptly named the ``Schr\"odinger-Poisson'' equation, something I mentioned when discussing miniclusters in Sec.(\ref{sec:miniclusters}) but did not write down. Let us see now where this comes from.

First of all, we need a better description of the field for this new context. Equation~(\ref{eq:singlesine}) above might be fine for describing the field over short distances\footnote{and indeed it will show up once again when we talk about direct detection which is a small-scale situation.} but for most situations in astrophysics and cosmology, we don't really care about the microscopic oscillations $\sim \sin(m_a t)$, but the large-scale behaviour of a classical field composed of many interfering modes. So we roll out the usual WKB trick of factoring out the fastly-oscillating parts and focus our attention towards some envelope, $\psi$ that we assume is slowly varying compared to the oscillations: $\dot{\psi} \ll m_a \psi$ and $\ddot{\psi} \ll m_a \dot{\psi}$. The real classical field $\phi$ in terms of this new one $\psi$ is written as,
\begin{equation}\label{eq:WKB}
\phi=\frac{1}{\sqrt{2 m_a}}\left(\psi e^{-i m_a t}+\psi^* e^{i m_a t}\right) \, ,
\end{equation}
where $\psi(\mathbf{x},t)$ is now a complex scalar field whose spatial and temporal variations will encapsulate fluctuations and motions of the dark matter on scales much larger than the field's Compton wavelength, $1/m_a$. We can get the energy density of the field by taking the amplitude: $\rho = m_a |\psi|^2 $, and this energy density will source a Newtonian gravitational potential $\Psi$ on sub-horizon scales. This gravitational potential then also feeds back into the equation of the motion for the waves in $\psi$, and this is encapsulated in a coupled system of equations of motion---the Schr\"odinger-Poisson equations.

To get these equations, we first extract the equation of motion for $\phi$ in a similar way to the beginning of Sec.~\ref{sec:misalignment}, except now we implicitly include a gravitational coupling by giving our underlying metric some scalar perturbations, $\Phi$ and $\Psi$, which in the Newtonian gauge gives us:
\begin{equation}
    g_{\mu \nu} = {\rm diag}[-(1+2\Psi),\, 1 + 2\Phi,\, 1 + 2\Phi,\, 1 + 2\Phi] \, .
\end{equation}
For brevity I will ignore the $a(t)$ that would usually go in front of the spatial diagonal when working in the cosmological setting. As we will only be dealing with a non-relativistic scalar field with no anisotropic stress, the energy-momentum tensor tells us that $\Phi = -\Psi$, and so we can just use the Newtonian potential, $\Psi$.

It takes a little extra work, but solving the Euler-Lagrange equation for a massive scalar field with this metric will result in the same Klein-Gordon equation as before,
\begin{equation}\label{eq:KG_pert}
    \square \phi - m_a^2 \phi = 0 \, ,
\end{equation}
but where all of the details about the gravitational coupling of the field are stored in the d'Alembertian which is now,
\begin{equation}
    \square =\frac{1}{\sqrt{-g}} \partial_\mu\left(\sqrt{-g} g^{\mu \nu} \partial_\nu\right) = -(1-2 \Psi)\frac{\partial^2}{\partial t^2}+ 4 \dot{\Psi} \frac{\partial}{\partial t} + (1+2 \Psi) \nabla^2 \, .
\end{equation}
where $\sqrt{-g} \approx 1-2\Psi$, and $g^{\mu \nu} \approx \operatorname{diag}[-(1-2 \Psi), 1+2 \Psi, 1+2 \Psi, 1+2 \Psi]$, and we work to linear order in the metric perturbation, ignoring any $\Psi^2$ and $\dot{\Psi}\Psi$. We then rewrite Eq.(\ref{eq:KG_pert}) in terms of our complex envelope field $\psi$, and enforce our assumption that $\dot{\psi} \ll m_a \psi$ and $\ddot{\psi} \ll m_a \dot{\psi}$. After some more work, this yields something that looks very much like a Schr\"odinger equation:
\begin{equation}
\begin{aligned}
i \partial_t \psi & =-\frac{1}{2 m_a} \nabla^2 \psi+m_a \Psi \psi \, ,\\
\nabla^2 \Psi & = 4 \pi G_N (\rho-\bar{\rho}) \, .
\end{aligned}
\end{equation}
where in the second line I have just written out Poisson's equation for a scalar perturbation over a cosmological background density $\bar{\rho}$. Because $\rho = m_a|\psi|^2$ these two equations are coupled, and together they make up the Schr\"odinger-Poisson system. 

In Fig.~\ref{fig:fuzzy_halo} I show a visual illustration of the behaviour of an initially smooth-ish and Gaussian-ish clump of axions evolved under the Schr\"odinger-Poisson equations. The clump collapses under its own gravitational attraction as expected from the Poisson part of the system, however on top of this clump are imprinted distinct ``granules'' which arise because of wave-interference governed by the Schr\"odinger part of the system. The left-hand panel shows the density $\rho \propto |\psi|^2$, whereas the right-hand panel shows the phase of the complex field, which, as we will see in a moment, gives us information about the velocity of the dark matter at that position. In the centre of the image, there is a small and very dense region, which is yet another interesting feature of fuzzy dark matter halos known as a soliton.

A further manipulation that people often do is to rewrite this complex envelope in terms of some more physically interesting variables: the energy density $\rho$ and a velocity $\mathbf{v}$~\cite{Widrow_and_Kaiser,Suarez:2015fga,Chavanis2012_growth}. They are related to the original field via,
\begin{equation}
    \psi=\sqrt{\frac{\rho}{m_a}} e^{i m_a \vartheta} \, ,
\end{equation}
where the gradient of this new phase gives the velocity,
\begin{equation}
    \mathbf{v} = \nabla \vartheta \, .
\end{equation}
This is called the Madelung transformation. We can plug this into the Schr\"odinger equation, and separate the imaginary and real parts to get two new equations, which are respectively\footnote{To get the second line out of the equation for the real component, you also need to take its spatial derivative.}: 
\begin{equation}
    \begin{aligned}
 \dot{\rho}+\nabla \cdot (\rho \mathbf{v}) &= 0  \, ,\\
 \dot{\mathbf{v}}+(\nabla \cdot \mathbf{v}) \mathbf{v} + \nabla 
 \bigg(\Psi-\frac{\nabla^2 \sqrt{\rho}}{2 m_a^2\sqrt{\rho}}\bigg)&= 0 \, . \\
\end{aligned}
\end{equation}
These are reminiscent of equations that appear in the context of fluid dynamics---the first one expresses the conservation of mass, and the second is the conservation of momentum.

The only conspicuous term in the Schr\"odinger-Poisson system in the fluid description is the last one: $\nabla(\nabla^2\sqrt{\rho}/\sqrt{\rho})$. This is one of the defining features of wave-like dark matter and is often referred to in the literature by a slightly unusual bit of terminology: ``quantum pressure''. While the quantumness and pressureness of this term are debatable, it \textit{is} nonetheless related to wave dynamics, and does act a bit like a pressure in the sense that it appears in the equation above with the opposite sign to the Newtonian potential and so counteracts gravity. It also enters proportional to $1/m_a^2$, meaning it will be increasingly relevant in the dynamics for ultralight masses while being unimportant for heavier dark matter on the same spatial scale. Because it counteracts gravity when the spatial gradients are large, the role played by the quantum pressure is to set the scale below which perturbations in the dark matter cannot grow.

To see this explicitly we must go through a bit of linear perturbation theory rigmarole that is outlined in more detail in Ref.~\cite{Marsh:2015daa}. We write out the density in terms of a perturbation around the average background density: $\rho = \bar{\rho}(1+\delta)$, and see how the fluid equations written above will dictate the evolution of those perturbations. In terms of $\delta$, they are,
\begin{equation}
    \begin{aligned}
\dot{\delta}+ \mathbf{v} \cdot \nabla \delta + \left(1+\delta\right) \nabla \cdot \mathbf{v} & = 0 \, , \\
\dot{\mathbf{v}}+\left(\mathbf{v} \cdot \nabla\right) \mathbf{v} + \nabla\left(\Psi-\frac{\nabla^2 \sqrt{1+\delta}}{2 m_a^2\sqrt{1+\delta}}\right)& = 0 \, .
\end{aligned}
\end{equation}
The trick to proceed from here is to go into Fourier space, which essentially just entails converting all the derivatives into $ik$. After doing this and keeping everything at first order in $\delta$ and $v$, the equations become:
\begin{equation}
    \begin{aligned}
\dot{\delta}_k+ i k v & =0 \, ,\\
\dot{v}+ i k \Psi + \frac{i k^3 \delta_k}{4 m_a^2} & =0 \, .
\end{aligned}
\end{equation}
Now we can use the fact that in Fourier space Poisson's equation can be written $-k^2 \Psi = 4\pi G_N \bar{\rho}\delta_k$. Plugging this in for $\Psi$, and taking the time derivative of the first equation, we can combine them both into one 2nd-order differential equation for $\delta_k$:
\begin{equation}
    \ddot{\delta}_k+\left(\frac{k^4}{4 m_a^2}-4 \pi G_N \bar{\rho} \right) \delta_k=0 
    \, .
\end{equation}
If you have read Sec.(\ref{sec:miniclusters}), this is (almost) exactly the same equation of motion for an axion density perturbation that I was quoting in the context of miniclusters. In fact the only difference from this previous instance is that here we have not incorporated the background cosmological expansion for simplicity. It is a useful exercise to go back and repeat the derivation with the $a(t)$'s in the metric. If you do that you will get one extra term which corresponds to the Hubble friction\footnote{You will also need to assume $m_a\gg H$.}, and every $\nabla$ will become a $\nabla/a$:
\begin{equation}
     \ddot{\delta}_k +2 H \dot{\delta}_k +\left(\frac{k^4}{4 m_a^2 a^2}-4 \pi G_N \bar{\rho}\right) \delta_k=0 \, .
\end{equation}
where $k$ is now comoving momentum. It is a direct result of the scaling symmetry of the Schr\"odinger-Poisson system that we can use this same equation to describe the behaviour of $\upmu$eV-scale axions on AU--pc scales as we can $10^{-22}$~eV-scale axions on kpc--Mpc scales.

The key feature of this equation is the final term proportional to $\delta_k$, which changes sign for wavenumbers below the Jeans wavenumber, $k_J$, defined by,
\begin{equation}
    k_J=\left(16 \pi G a \bar{\rho}\right)^{1 / 4} m_a^{1 / 2}=66.5 \, \mathrm{Mpc}^{-1} \, a^{1 / 4}\,\left(\frac{m_a}{10^{-22} \,\mathrm{eV}}\right)^{1 / 2} = 8.7 \left(\frac{1+z_{\rm eq}}{1+z}\right)^\frac{1}{4}\left(\frac{m}{10^{-22} \,\mathrm{eV}}\right)^{1 / 2} \mathrm{Mpc}^{-1} \, .
\end{equation}
Interestingly, this is the exact same numerical value as we got for the typical Jeans mass for the QCD axion---except for the unit, which is now inverse Mpc as opposed to inverse mpc.

Again, the idea with the axion Jeans scale is that if you're a mode with a wavelength \emph{larger} than the Jeans length ($L>2\pi/k_J$), gravity dominates over the gradient pressure term, but when you're \emph{below} it ($L<2\pi/k_J$), the pressure dominates and halts any further gravitational collapse. You can think of it like Heisenberg's uncertainty principle. If we tried to confine our dark matter waves into too small a space, then Heisenberg tells us the spread of momenta has to grow. But if it grows to the point of the dark matter exceeding the local escape speed, then that dark matter is no longer gravitationally bound, and we don't have a halo anymore. 

Solving the equations properly, beginning from the adiabatic initial perturbations of matter, this heuristic understanding is demonstrated. Fuzzy dark matter has suppressed levels of structure relative to cold dark matter at scales close to a cut-off that is proportional to $m_a^{-1/2}$~\cite{Hu:2000ke, Hui:2016ltb}. Reference~\cite{Hu:2000ke} provides a handy formula that captures the level of suppression in the present-day linear matter power spectrum in a fuzzy dark matter universe compared to a CDM one:
\begin{equation}
P_{\mathrm{FDM}}(k)=T_{\mathrm{FDM}}^2(k) P_{\mathrm{CDM}}(k) \, ,
\end{equation}
where the `transfer function' for fuzzy DM with respect to CDM is parameterised as,
\begin{equation}
    T_{\mathrm{FDM}} \simeq \frac{\cos x^3}{1+x^8} \quad\text{with}, \quad x=1.61\,\left(\frac{m}{10^{-22}\, \mathrm{eV}}\right)^\frac{1}{18}\left(\frac{k}{k_{J}(z_{\rm eq})}\right) .
\end{equation}
This is what was used to make the dotted line in Fig.~\ref{fig:DimensionlessPowerSpectrum}.

A test of the fuzzy dark matter hypothesis then involves hunting for any evidence of suppressed structure growth below some scale~\cite{Hu:2000ke}. The smaller scales you can probe down to, the \textit{larger} the value of $m_a$ you can test. If things all look normal down to the limits of your observations, then you set a bound instead, and so far that's what people have done, as I will discuss in the next section.

Before that though, one final matter to discuss is about axion stars. Just like on small scales in the QCD axion context, on very large scales we have the possibility for solitonic objects to form---stable configurations where the gradient energy pressure and gravity are precisely balanced. An example of one can be seen in the centre of the toy halo shown in Fig.~\ref{fig:fuzzy_halo}. In this context, people don't call them axion stars but rather refer to them as ``cores'' because the places they tend to end up are in the centres of galactic halos. The typical size of one of these solitons for a halo mass $M_h$ at the present day is~\cite{Schive:2014dra, Schive:2014hza},
\begin{equation}\label{eq:axionstarradius_ultralight}
    R_s = 0.16 \, {\rm kpc} \left( \frac{10^{-22}\,{\rm eV}}{m_a} \right) \left(\frac{M_h}{10^{12}\, M_{\odot}}\right)^{-1 / 3} \, ,
\end{equation}
which, as expected, is comparable to the coherence length, Eq.(\ref{eq:coherencelength}), for a virialised halo around the Milky Way mass. This formula makes use of the fact that the mass of the solitonic core $M_c$ is proportional to the size of the halo they are embedded inside like $M_c \propto M_h^{1/3}$, but as discussed in Sec.(\ref{sec:axionstars}) there is debate about this exact scaling.\footnote{In fact, I have told something of an inverted history here---these relations were discovered first in the context of fuzzy dark matter cosmology, not in the context of miniclusters.} We will see next that the formation of dense kpc-scale cores turns out to be one of the reasons to take the idea of fuzzy dark matter seriously.

\subsection{Lower bounds on the axion mass}
% Halo mass function
I am not going to go through every single lower bound people have drawn on the fuzzy dark matter mass, but I'll mention a few. A simple one to start with is the masses of halos. Structures come in a range of sizes, from clusters, to galaxies, to dwarf galaxies. One way to implement the general idea I described above is to build a distribution of the halo masses of these objects and look for the way fuzzy dark matter would suppress the abundance of the smaller ones. While the matter power spectrum receives a suppression from fuzzy dark matter above some value of $k$, in terms of a halo mass function this maps onto a suppression in the numbers of halos below a certain mass. This could be observed on galactic scales using surveys~\cite{Bozek:2014uqa, Kulkarni:2020pnb, Dentler:2021zij,Lague:2021frh,Winch:2024mrt}---the ``halo mass function''---or using the dwarf galaxies that orbit around a larger host like the Milky Way---the ``subhalo mass function''~\cite{DES:2020fxi, Schutz:2020jox}. The halo and subhalo mass functions have been used to set bounds at the level of a few $\times 10^{-22}$ to $\times 10^{-21}$ eV, respectively.

% CMB
On the other hand, constraints on much larger scales like using the CMB typically reach $10^{-24}$~eV~\cite{Hlozek:2014lca, Hlozek_2018, Farren:2021jcd,Rogers:2023ezo} but have the advantage that they apply exclusively to scales where linear perturbation theory works and so you only have to solve the Boltzmann equation as opposed to running N-body simulations. CMB constraints are also the most powerful for axions with masses even lighter than this which could serve as a candidate for dark energy.

% Simulations
However, linear perturbation can only get you so far, and the scales that are at the frontier of our measurements right now are already those where $\Delta^2(k)>1$, and so perturbations are indeed evolving non-linearly, which requires simulations. Simulations of fuzzy dark matter have become increasingly sophisticated in a relatively short time (compare the earliest instance, Ref.~\cite{Woo:2008nn} with the state-of-the-art like Refs.~\cite{Schwabe:2021jne, May:2022gus, Shen:2023lsf}), but it is still safe to say they are less well-developed than the exquisite hydrodynamic simulations that have underlined the success of the cold dark matter paradigm~\cite{Vogelsberger:2019ynw}. In particular, the role of baryons in influencing the dynamics of galaxies, especially in their inner regions, is not a minor one.

% Lyman-alpha
One of the most powerful techniques for mapping the small-scale distribution of matter on non-linear scales is via the so-called ``Lyman-$\alpha$ forest''~\cite{Croft:1997jf, Chabanier:2019eai}. The ``forest'' here is actually a forest of hydrogen absorption lines in the spectrum of some bright quasar shining at us from far off in the universe. Hydrogen only has one set of absorption lines, but when we look at quasars we see a whole forest of lines because the light has had to pass through the clouds of neutral hydrogen whilst continuously getting redshifted by the expansion of the universe. This moves the lines taken out earlier in their journey down the spectrum to lower wavelengths so there is always a fresh piece of the spectrum for the next cloud of hydrogen to eat. Eventually, the spectrum we receive has the spatial distribution of neutral hydrogen clouds along the line-of-sight imprinted onto it. So once armed with some understanding of how neutral hydrogen clouds trace the underlying matter distribution, we can use the Lyman-$\alpha$ forest to measure a 1-dimensional projection of the power spectrum. Importantly, the Lyman-$\alpha$ forest is nice because the resolution in terms of scale is governed by the resolution with which a spectrum can be measured---which is to say, very well.

Several groups have drawn constraints on the fuzzy dark matter mass using the Lyman-$\alpha$ forest, taking into account all of the various uncertainties coming from our limited knowledge of the intergalactic medium and reionisation, and using N-body simulations to predict the non-linear power spectrum. The results are at the level of 
\begin{equation}
    m_a > 2 \times 10^{-20}~{\rm eV} \, (95\% \, \text{CL})
\end{equation}
for the most recent study~\cite{Rogers:2020ltq}---see also Refs.~\cite{Kobayashi:2017jcf,Armengaud:2017nkf,Irsic:2017yje} for earlier ones. This represents one of the strongest bounds on ultralight dark matter that I am aware of---but it could be argued this is still subject to astrophysical uncertainties. Reference~\cite{Rogers:2020ltq} also made use of the results from hydrodynamic simulations with a range of assumptions about the fuzzy dark matter transfer function---this is a computationally demanding procedure but is enabled here through the use of a technique called emulation that allows a more efficient scan over the various parameter dependencies.

% Small-scale problems
Although there are now strong bounds, I cannot move on without mentioning one of the major reasons why so many people are interested in fuzzy dark matter. Although CDM works well almost everywhere, it turns out that fuzzy dark matter has the promise of explaining the handful of cases where the CDM seems to fall short~\cite{Matos:2023usa}. These are the so-called ``small-scale problems'' of cold dark matter---problems that are primarily those of inconsistency between simulations and observations~\cite{Bullock:2017xww,Salucci:2018hqu}, with names like the `core-cusp', `too-big-to-fail', and `missing-satellites'. The status of these problems is still under some debate~\cite{Kim:2017iwr}. At least part of the problem was with the fact that early simulations did not include baryonic physics (supernovae, gas, space dust, even crazier space dust etc.), and once baryons were accounted for, some of the tensions were alleviated. Advancements were made on the observational side too: more of the faintest dwarf galaxies that lay just below the detection thresholds of earlier observations were eventually found.

The small-scale crisis that seems most naturally resolved by fuzzy dark matter is that of the core-cusp problem: CDM-only simulations of galaxies formed halos with central cusps to their density profiles, whereas observed ones had much shallower profiles and sometimes flat cores~\cite{Salucci2000,DeLaurentis:2022nrv}. As discussed above, one of the hallmarks of fuzzy dark matter halos is the formation of solitonic cores. Unfortunately, the physical size of the core is inversely proportional to the dark matter mass, so there is a conflict between the core-cusp issue wanting for a light mass to explain the sizes of cores, while the Lyman-$\alpha$ forest tightly constraining the axion mass to be above $10^{-21}$~eV in order to not underproduce small-scale structure~\cite{Marsh:2015wka}.\footnote{One way to circumvent this tension is to add more degrees of freedom to the model, for example by incorporating self-interactions~\cite{Mocz:2023adf, Painter:2024rnc}. Another possibility could be within an axion model with a large initial misalignment angle which can enhance rather than suppress structure around the Jeans scale~\cite{Winch:2023qzl}}

% Further phenomenology
Nonetheless, the phenomenology of the fuzzy dark matter scenario remains intriguing, and there are several other implications of the system that, if they can be understood properly, will allow much more precise tests of the model. For instance, one such phenomenon is the existence of fluctuating ``quasiparticles'' due to wave interference~\cite{Chavanis:2020upb} as seen in Fig.~\ref{fig:fuzzy_halo}. They have a physical size given by the coherence length and although they are transient in nature, they do correspond to real, physical enhancements in the local density of matter, and as a result, it is plausible they could jostle other forms of matter around. These wave interference effects could potentially heat up star clusters~\cite{Bar-Or:2018pxz, Marsh:2018zyw, Lancaster:2019mde, Buehler:2022tmr}, modify the geometries of stellar streams~\cite{Dalal:2020mjw}, as well as influence the motions of heavy objects like black holes~\cite{Wang:2021udl, Boey:2024dks}, and even the central solitonic cores themselves~\cite{Chowdhury:2021zik}. Some bounds have been drawn based on these effects that are claimed to be slightly stronger than the Lyman-$\alpha$ mass bound~\cite{Dalal:2022rmp} and may reach up to a few $\times 10^{-19}$ eV, but it seems likely that a better understanding of the combined baryon+fuzzy-dark-matter system is needed before bounds based on physics inside baryon-rich environments can be fully trusted.\footnote{See e.g.~Ref.~\cite{Robles:2023xtw} for a recent study on the impact of a supernova on the structure of a central soliton.}

There are more interesting probes of the effects of fuzzy dark matter being suggested all the time. To list a few without going into much detail: using gravitational lensing to pick up small halos or see the granules~\cite{Laroche2022, Powell:2023jns}\footnote{A recent study suggested that a $\sim10^{-22}$~eV fuzzy dark matter halo was a better fit compared to cold dark matter for reconstructing a quasar imaged multiple times by a galaxy cluster~\cite{Amruth:2023ehw}. A precise mass was not given but unfortunately the range considered is firmly ruled out by other bounds.}; high-precision mapping the halo mass function of neutral hydrogen clouds~\cite{Garland:2024csi}; using galactic rotation curves~\cite{Bar:2021kti}, or measuring the heating up of the galactic disk~\cite{Yang:2024hvb}; tracking the orbits of stars around the supermassive black hole of the Milky Way~\cite{DellaMonica:2023dcw}; or lastly, by measuring the local oscillations in the potential using low to mid-frequency gravitational waves~\cite{Khmelnitsky:2013lxt, Porayko:2018sfa, Kato:2019bqz, Sun:2021yra, Unal:2022ooa, Kim:2023pkx, Kim:2023kyy}.
Fuzzy dark matter has also been proposed as a possible solution to the so-called ``final parsec problem'' in simulations of supermassive black hole mergers~\cite{Begelman:1980vb, Milosavljevic:2002ht}, where inspirals stall when they around a parsec apart due to the insufficient gravitational wave emission for orbits on those scales. The fact they cannot merge fast enough is a problem given the number of very massive black holes seen in the centres of modern galaxies. However, orbiting in the vicinity of a galactic solitonic core~\cite{Koo:2023gfm} or interactions with the ultralight dark-matter granules~\cite{Bromley:2023yfi} could provide mechanisms to facilitate orbital decay.

What is promising in this field is that there are datasets forthcoming from a range of very fancy new astronomical facilities. Relevant for probing ultralight dark matter include the nascent JWST, the upcoming Euclid for surveying galaxies, future high-precision measurements of the cosmic microwave background~\cite{Grin:2019mub, Hlozek:2016lzm, Farren:2021jcd}, as well as large radio observatories like the giant Square Kilometre Array (SKA), currently being built in Australia~\cite{Weltman:2018zrl}. The SKA will be able to uncover the universe's history all the way up to the period of cosmic dawn when the first round of star formation was triggered~\cite{Flitter:2022pzf, Hotinli:2021vxg}. Projections for these promise to push the lower bound on the dark matter mass up to around $10^{-18}$~eV~\cite{Vanzan:2023gui}, but if we imagine masses much heavier than this then the fuzzy dark matter and cold dark matter universes do start to converge.

Perhaps then only a bit of the dark matter is fuzzy\footnote{On that subject, there was a very recent hint of a tension in the linear matter power spectrum measured using CMB+BAO+Supernovae compared with the Lyman-$\alpha$ forest could be explained using a percent-level amount of ultralight axion dark matter~\cite{Rogers:2023upm}.}, or maybe the fuzzy dark matter people have been modelling is too simplistic. Interestingly, the axiverse scenario that I mentioned a few times already would explain why our Universe contains ultralight axions. String theorists suggest that we should anticipate an entire spectrum of such states, including ultralight ones and ones with heavier masses. So the recent simulations of situations consisting of a mixture of fuzzy+cold dark matter~\cite{Marsh:2013ywa, Schwabe:2020eac, Vogt:2022bwy, Lague:2023wes} or multi-field ultralight axion scenarios~\cite{Luu:2018afg, Guo:2020tla, Eby:2020eas, Huang:2022ffc, Gosenca:2023yjc, Glennon:2023jsp} are very much warranted.

\section{Searches for axion dark matter}
Over the preceding sections, most of what I have been talking about are ideas for how we can engineer a particle like an axion to explain the dark matter we see across the Universe. Addressing this question first was essential because cosmological data is exquisite and makes it hard to argue with the list of known properties of dark matter, as small as that list may be. So it seems we can achieve that in a wide parameter space: an axion is a satisfactory candidate that, (1) minimally explains why $\Omega_{\rm DM}h^2 = 0.12$; (2) explains why dark matter halos form across the wide range of scales they do; and (3) has signatures with which we might be able to distinguish it from other hypotheses about dark matter. But is just the bare minimum---we want to know if we have the right answer, so we must go out and try to observe the innumerable quantities of axions we suppose are filling up the halos of galaxies across the universe, including the one in which we live.

So in this final section, I plan to highlight the ways people are using the ideas I introduced in previous sections to come up with tests of the axionic dark matter hypothesis. The main theme here is on looking for interactions that do not solely rely on the axion's gravitational effects in the Universe, which was the focus of the previous sections. In the end, it is only a non-gravitational signature of the axion's particle identity that will tell us we have got something right, and lead us into the next era of particle physics and cosmology.

\begin{figure*}[t]
    \centering
    \includegraphics[width=0.99\textwidth]{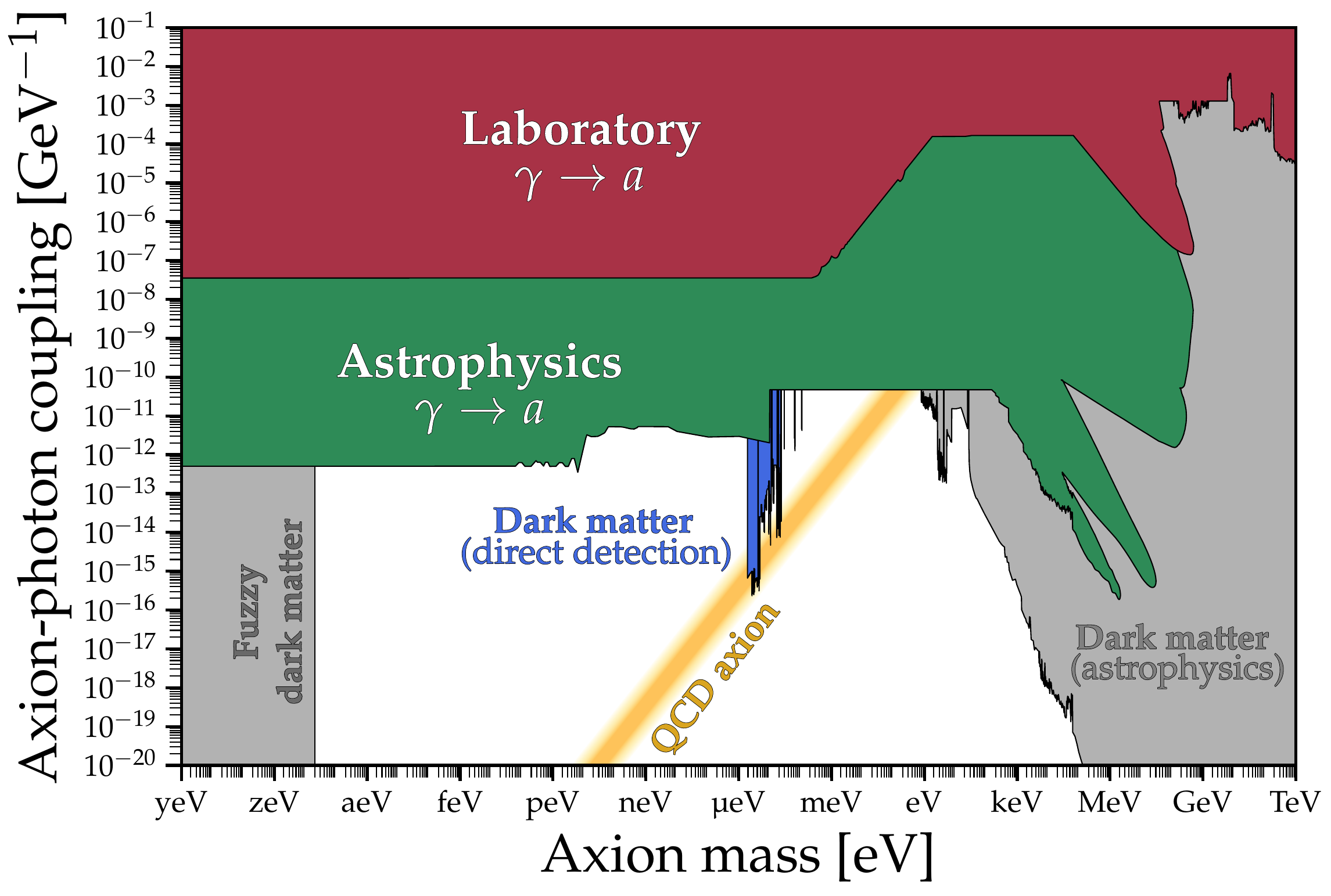}
    \caption{A simplified panorama of the axion-photon coupling parameter space. Constraints are grouped together according to the level of assumption they work under: pure laboratory searches (red), searches for axions produced in astrophysical environments (green), \textit{direct} searches for axions as dark matter around the Earth (blue) and \textit{indirect} searches for axions as dark matter elsewhere in the Universe (grey). The canonical QCD axion falls within the yellow band. The different bounds are layered on top of one another in an intentional way---red supersedes green, which supersedes blue and grey. This is because of the levels of assumption being made in each case: the red and green bounds assume axions are produced from photons that we already know exist, whereas the blue and grey bounds require us to posit that axions make up 100\% of the dark matter in the Universe. All data and a Python notebook to reproduce this plot can be found at Ref.~\cite{AxionLimits}.}
    \label{fig:AxionPhoton_UltraSimple_FullParameterSpace}
\end{figure*}

I will cover many different ideas in this section, but the only unifying idea will be that we are using the axion's coupling to the photon to try and test it.\footnote{There are good reasons to want to look for axion dark matter through its other couplings, like those with the electrons or nucleons, but in the context of dark matter searches there are surprisingly few instances apart from a few direct detection experiments where a dark-matter bound can surpass the more stringent and more generic astrophysical bounds in terms of sensitivity (see e.g.~Refs.~\cite{AxionLimits, Berlin:2023ubt, Bloch:2019lcy, ParticleDataGroup:2022pth}) for summaries.} So we we will go back to the more phenomenological definition of the axion, in terms of the effective theory Eq.(\ref{eq:axionlagrangian}), and pick out the term involving the electromagnetic field strength, $F^{\mu \nu}$. Including this term in the Lagrangian for electromagnetism we get ``\textit{axion electrodynamics}'':
\begin{equation}
    \mathcal{L} = -\frac{1}{4} F_{\mu \nu} F^{\mu \nu} - J^\mu A_\mu + \frac{1}{2} \partial_\mu \phi \partial^\mu \phi - \frac{1}{2} m_a^2 \phi^2 - \frac{1}{4} g_{a\gamma} \phi F_{\mu \nu} \tilde{F}^{\mu \nu} \, .
\end{equation}
The interaction $-\frac{1}{4} g_{a\gamma} \phi F_{\mu \nu} \tilde{F}^{\mu \nu}$ can be equivalently written in terms of electric and magnetic fields as $ g_{a\gamma} \phi  \mathbf{E}\cdot \mathbf{B}$. To get this you put in the definition of $F$ in terms of electric and magnetic fields, and contract it with its dual, defined as $\tilde{F}^{\mu \nu} = 1/2 \epsilon^{\mu \nu \sigma \rho}F_{\sigma \rho}$. The dual tensor is essentially the same object except you swap the electric and magnetic fields with a minus sign.

Working out the Euler-Lagrange equations gives us two equations of motion for the axion and the photon, which are:
\begin{equation}\label{eq:axionphoton_eom}
    \begin{aligned}
    \square \phi+m_\phi^2 \phi&= -\frac{1}{4} g_{a\gamma} \phi F_{\mu \nu} \tilde{F}^{\mu \nu} \, , \\
        \partial_\mu F^{\mu \nu}&=J^\nu-g_{a \gamma} \tilde{F}^{\mu \nu} \partial_\mu \phi \, .
    \end{aligned}
\end{equation}
There are several interesting processes that this axion-two photon vertex can give us. One could be that axions decay into two photons, or alternatively, axions and photons could convert into one another in the presence of the virtual second photon provided by a pre-existing electromagnetic field. The former is often only useful in astrophysical settings because the rate at which decays happen is going to be extremely small. Whereas the latter is often more practical for a laboratory setting since electromagnetic fields are things we can make.

The combination of electric and magnetic fields that governs this interaction, $\mathbf{E}\cdot \mathbf{B}$, is quite peculiar. It violates $CP$, and nowhere does this appear in Maxwell's equations---electric and magnetic fields are usually supposed to be perpendicular to each other. We can derive two of these Maxwell's equations by writing out the photon's equation of motion in terms of electric and magnetic fields, whereas the other two can be derived from the Bianchi identities: $\partial_\gamma F_{\alpha \beta}+\partial_\alpha F_{\beta \gamma}+\partial_\beta F_{\gamma \alpha}=0$. Together they reveal the four \textit{axion-modified} Maxwell's equations,
\begin{equation}\label{eq:maxwell}
\begin{aligned}
\nabla \cdot \mathbf{E} & =\rho_q-g_{a \gamma} \mathbf{B} \cdot \nabla \phi \, ,\\
\nabla \times \mathbf{B}-\dot{\mathbf{E}} & =\mathbf{J}+g_{a \gamma}(\mathbf{B} \dot{\phi}-\mathbf{E} \times \nabla \phi)  \, ,\\
\nabla \cdot \mathbf{B} & =0 \, , \\
\nabla \times \mathbf{E}+\dot{\mathbf{B}} & =0\, .
\end{aligned}
\end{equation}
From inspecting where the axion shows up here we see that it plays the role of an effective current and charge density: $  J_{\rm eff}^\mu=g_{a \gamma}\left(-\mathbf{B} \cdot \nabla \phi,\mathbf{B}\dot{\phi} -\mathbf{E} \times \nabla \phi\right)$.

Solving these equations in different settings where we have electromagnetic fields along with the axion results in some quite interesting electrodynamics that are unique to the axion, and indeed many experiments looking for the axion directly exploit this violation of classical electromagnetism.\footnote{In fact axion electrodynamics shows up in scenarios completely unrelated to particle physics, for example in a certain class of materials called topological insulators~\cite{Wilczek:1987mv, Qi:2008ew}.}

In Fig.~\ref{fig:AxionPhoton_UltraSimple_FullParameterSpace}, I have displayed the panorama of all constraints on the axion-photon coupling over 36 orders of magnitude in mass. It is important to emphasise the way that these constraints have been grouped and layered. Red encodes bounds set by laboratory experiments where there are no assumptions being made other than the axion exists as a particle in Nature. Most of these experiments are classed as light-shining-through-walls experiments (e.g.~\cite{DellaValle:2015xxa, SAPPHIRES:2021vkz, Ehret:2010mh, Betz:2013dza, OSQAR:2015qdv, NOMAD:2000usb}), but at high-enough masses, axions can undergo detectable decays after production in collider~\cite{Bauer:2017ris, ATLAS:2020hii, CMS:2018erd, Knapen:2016moh, Jaeckel:2015jla, BESIII:2022rzz} and beam dump experiments~\cite{Dobrich:2015jyk, Dolan:2017osp, NA64:2020qwq}. In green, I show astrophysical bounds that arise because axions should be produced in plasmas like the Sun, white dwarfs or neutron stars, or would be visible undergoing mixing with photons as they propagate through astrophysical B-fields. These bounds are subject to astrophysical uncertainties, but there are a great many bounds here, and most are trustworthy to the level of an order of magnitude at the very least (if not much more for some of the stellar cooling arguments). Finally, in blue and grey, I show direct and indirect constraints on axions as a dark matter candidate. This is the largest assumption one can make here, and as a consequence, you can only self-consistently search for dark matter in the parameter space not excluded by the other dark-matter-independent bounds. This point is not a particularly subtle one in my opinion, yet it is one that is ritually ignored or brushed under the carpet in the literature. 

The goal of this section is to go through the bounds filled in grey, where we are interested in the processes whereby dark matter axions can interact with photons in space in a way that makes them observable. I will group these different processes and searches primarily by the mass ranges that they are most effective at searching over, starting from the heaviest masses in Sec.~\ref{sec:axiondecay} and ending with the lightest in  Sec.~\ref{sec:axionbirefringence}. Then in the final Secs.~\ref{sec:minicluster_signatures} and Sec.~\ref{sec:directdetection} I will discuss the signatures associated with axion substructure and direct detection in the solar neighbourhood respectively.

\subsection{Axion decay}\label{sec:axiondecay}
\begin{figure*}[t]
    \centering
    \includegraphics[width=0.99\textwidth]{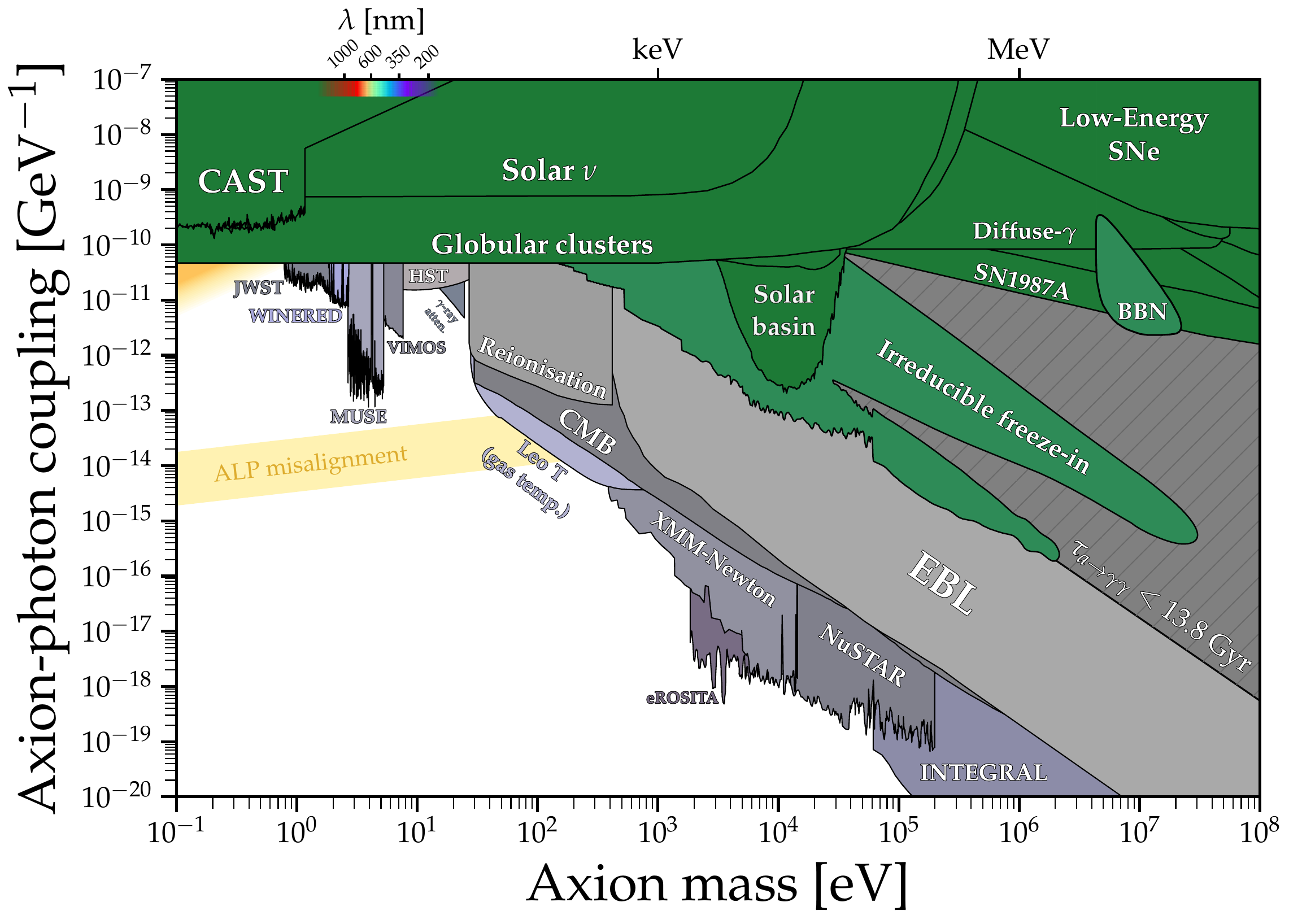}
    \caption{A closeup of the ALP parameter space for heavy axions, relevant for searches based around axion decay to two photons as covered in Sec.~\ref{sec:axiondecay}. As in other plots of this parameter space, green colours are used for bounds that do not assume anything about dark matter, and grey is used for indirect detection constraints that assume axions make up 100\% of galactic and cosmological dark matter. The constraints under the former category are from CAST~\cite{CAST:2007jps, CAST:2017uph}, the solar luminosity/neutrino bound~\cite{Vinyoles2015}, emission by globular cluster stars~\cite{Dolan:2022kul}, solar axions trapped in the Sun's gravitational basin~\cite{DeRocco:2022jyq, Beaufort:2023zuj}, gamma/X-rays from axions produced in supernovae and compact object mergers~\cite{Lucente:2020whw, Caputo:2021rux, Caputo:2022mah, Diamond:2023scc, Diamond:2023cto, Dev:2023hax}, and finally the irreducible cosmological background of axions produced under the minimal assumption about pre-BBN cosmology~\cite{Langhoff:2022bij, Depta:2020wmr}. The dark-matter decay bounds in order from infrared to gamma-rays are: JWST/blank sky~\cite{Janish:2023kvi}, WINERED/dwarf galaxies~\cite{Yin:2024lla}, MUSE/dwarf galaxies~\cite{Todarello:2023hdk}, VIMOS/galaxy clusters~\cite{Grin:2006aw}, HST/cosmic optical background~\cite{Nakayama:2022jza, Carenza:2023qxh}, gamma-ray attenuation by background photons from axion decay~\cite{Bernal:2022xyi}; axion energy injection affecting reionisation~\cite{Cadamuro:2011fd}, the CMB~\cite{Capozzi:2023xie, Liu:2023nct}, and the gas temperature of Leo T~\cite{Wadekar:2021qae}; and decay lines visible in spectra of the EBL~\cite{Cadamuro:2011fd}, XMM-\textit{Newton}~\cite{Foster:2021ngm}, \textit{NuSTAR}~\cite{Perez:2016tcq, Ng:2019gch, Roach:2022lgo}, eROSITA~\cite{Fong:2024qeq} and INTEGRAL~\cite{Calore:2022pks}. All data and a Python notebook to reproduce this plot can be found at Ref.~\cite{AxionLimits}.}
    \label{fig:AxionPhoton_HeavyAxionDecayCloseUp}
\end{figure*}

The axion to two-photon coupling permits populations of axion dark matter residing in galaxies to occasionally decay into pairs of photons with energies $\omega \approx \frac{1}{2}m_a + \mathcal{O}(v^2)$ where $v$ will usually be $\sim 10^{-3}$ because cold dark matter is non-relativistic. The decay rate for this process is given by the formula, 
\begin{equation}\label{eq:axiondecayrate}
    \Gamma_{a\to \gamma\gamma} =\frac{g_{a \gamma \gamma}^2 m_a^3}{64 \pi} \, .
\end{equation}
The timescale associated with axion decay is therefore,
\begin{equation}\label{eq:axionlifetime}
    \tau_{a\to \gamma\gamma} = \Gamma^{-1}_{a\to \gamma\gamma} = t_U \bigg(\frac{5.5 \times 10^{-7} \,{\rm GeV}^{-1}}{g_{a\gamma}}\bigg)^2 \bigg(\frac{1\,{\rm eV}}{m_a}\bigg)^3 \, ,
\end{equation}
where $t_U = 13.8$~Gyr is the age of the Universe. The steep $1/m^3_a$ dependence is why spontaneous axion decay is only a practical signal to look for in the case of axions at the heavier end of their parameter space.

A zoom-in on the part of the parameter space containing axion decay bounds from infrared to gamma-ray energies is shown in Fig.~\ref{fig:AxionPhoton_HeavyAxionDecayCloseUp}. The colour scheme used here is the same as in the panorama in Fig.~\ref{fig:AxionPhoton_UltraSimple_FullParameterSpace}: green is for bounds that do not assume axions are dark matter, and grey is for bounds that do. Axions with $m_a$ or $g_{a\gamma}$ lying above the line defined by $\tau_{a\to \gamma\gamma}=t_U$ are trivially ruled out as a dark matter candidate since they wouldn't survive until the present day in sufficient numbers. The various bounds displayed here can be broadly divided into two types that depend on whether we are searching for axion dark matter decaying at early times (i.e.~around recombination, reionisation etc.) or at late times (in nearby galaxies or the Milky Way).

\subsubsection{Early decays}
Before discussing some of the early-decay bounds, it is worth mentioning that cosmological bounds can be drawn on ALPs decaying to photons that do not require us to enforce they make up all of the dark matter. This works because, at some level, a population of ALPs is guaranteed to be produced even if we only switch on the coupling to the photon. They would be made from processes like the inverse decay $\gamma \gamma \to \phi$, or Primakoff production $\gamma q \to \phi q$ in the hot bath of photons and quarks in the early Universe. The energy injection from the subsequent decays of those $\phi$ back into photons may then interfere with BBN, recombination, or reionisation if there was too many of them. These decay photons could also contribute an observable excess of photons on top of the CMB's Blackbody spectrum (known as CMB spectral distortions) or to the integrated photon spectrum of the Universe (known as the extragalactic background light or EBL). 

All of these considerations lead to tight bounds ALPs that decay during the various phases of the Universe's evolution. The first comprehensive map of these constraints was presented in Ref.~\cite{Cadamuro:2011fd}. The only major uncertainty in drawing such constraints is related to the unknown temperature that the Universe was reheated to after inflation, as this dictates the duration over which these ALP populations could be produced via either the freeze-out~\cite{Masso:1995tw, Masso:1997ru, Cadamuro:2010cz, Millea:2015qra, Depta:2020wmr} or freeze-in mechanisms~\cite{Balazs:2022tjl, Langhoff:2022bij}. As discussed at the end of Sec.~\ref{sec:thermalaxions} there is an irreducible population of ALPs that arise when the reheating temperature is set to its minimal allowed value just prior to BBN, $T_{\rm RH}\sim 5$~MeV~\cite{Langhoff:2022bij, Balazs:2022tjl, Depta:2020wmr}. The bounds on this irreducible population are shown in Fig.~\ref{fig:AxionPhoton_HeavyAxionDecayCloseUp} in slightly brighter colours of green to distinguish them from the stellar/supernova axion bounds. It is possible to set tighter dark-matter-independent constraints on heavy ALPs coupled to photons than this, but only at the expense of additional model-dependence, see e.g.~Ref.~\cite{Depta:2020wmr}.

In the context of dark matter, our attention is drawn towards the part of Fig.~\ref{fig:AxionPhoton_HeavyAxionDecayCloseUp} that lies below the line $\tau_{a\to \gamma\gamma} = t_U$ where axion dark matter halos wouldn't expire before the present day. Even though these axions would be cosmologically stable, the bounds on them do not necessarily have to be weak. The act of enforcing that axions make up 100\% of dark matter across the Universe implies that there is also a huge population of them out there which we can exploit to overcome their feeble decay rates. 

If axions make up the dark matter and decay into photons they will inject energy into the Universe at a rate,
\begin{equation}
    \left(\frac{\textrm{d} E(z)}{\textrm{d} t \textrm{d} V}\right)_{\text {inj }}=\rho_a(1+z)^3 \Gamma_{a\to\gamma\gamma} \, .
\end{equation}
The most powerful indirect constraints on axion decays come from considering environments that are especially sensitive to this injected energy. For example, one period of cosmic history that might be particularly sensitive to this is reionisation, especially when the decay photons are in the UV to soft-X-ray part of the spectrum. The fraction of ionised hydrogen as a function of cosmic time that is inferred via the optical depth felt by CMB photons passing through the era of reionisation from $z_{\rm re}\approx 1100$ until today,
\begin{equation}
    \tau= \sigma_T \int_0^{z_{\rm re}} \textrm{d}z \, n_e(z) \frac{\textrm{d} r}{\textrm{d} z} \, ,
\end{equation}
where $\sigma_T$ is the Thomson cross section and $\textrm{d} r / \textrm{d} z= (1+z)^{-1} H(z)$ is the redshift dependence of proper distance along the line-of-sight. Since axions will inject photons that will ionise neutral hydrogen, they will shift the redshift dependence of the electron density $n_e$ away from standard cosmology. Keeping things consistent with the CMB requires a strong upper bound on the photon coupling for $m_a>30$~eV where the decay photons are highly ionising to neutral hydrogen, as originally done in Ref.~\cite{Cadamuro:2011fd}. A later study that incorporated an analysis of spectral distortions in the cosmic microwave background measured by the FIRAS instrument on the COBE satellite improved upon this bound slightly~\cite{Bolliet:2020ofj}.

A pair of works also came out very recently that have derived a more refined cosmological constraint on early axion dark matter decays~\cite{Capozzi:2023xie, Liu:2023nct}, including the generation of ionisation measurable through the CMB, as well as the heating of the intergalactic medium using the Lyman-$\alpha$ forest. Both groups made use of the recently-developed \texttt{DarkHistory} code package~\cite{Liu:2019bbm} which calculates how the cosmic ionisation history is modified in the presence of exotic sources of energy injection. I have shown the bound from Ref.~\cite{Liu:2023nct} in Fig.~\ref{fig:AxionPhoton_HeavyAxionDecayCloseUp} labelled as ``CMB'', which made use of the most recent version of the code, but the two are in broad agreement. In the near future, the cosmological energy injection bound stands to improve by another order of magnitude by mapping the distribution of neutral hydrogen during reionisation via its 21~cm emission~\cite{Facchinetti:2023slb,Sun:2023acy}.

\subsubsection{Late decays}
If axions have lifetimes not too much longer than the age of the Universe and do make up the dark matter, then what we should see is all halos across the universe glowing very dimly with some as-yet-undiscovered line emission at a frequency given by half the axion mass. This is potentially a much more striking signal than searching indirectly for some source of extra heat, and it is one that astronomers already have the tools to look for. However, unlike the case of early decays where the entire Universe was our sample, we will necessarily be looking at a much smaller population of axions here. So we must play the statistical game---find environments where we have high densities of axions in one place and try to hunt for their emission lines very carefully in precisely measured, background-subtracted spectra.

A nice simple example to demonstrate how this works in practice is with dwarf galaxies. The Milky Way has many well-studied dwarf galaxies which have masses $M\sim 10^{6}$--$10^{9}\,M_\odot$ and half-light radii $r \sim \mathcal{O}(10$--$100)$~pc. Those dwarf galaxies that are especially faint---so-called ultra-faint dwarfs---are some of the favourite objects of dark-matter hunters for several reasons. The primary one in this context is that they have huge amounts of dark matter relative to their size and don't contain a very substantial luminous baryonic component whose emission might overpower a feeble dark-matter decay signal. This fact is encapsulated in the so-called mass-to-light ratio, measured in units of solar masses per solar luminosity. Ultra-faint dwarf galaxies have mass-to-light ratios in the ballpark of $M/L \sim \mathcal{O}(10^2$--$10^3) \, M_\odot/L_\odot$---they are dark-matter-rich environments with low astrophysical backgrounds. 

We can get an initial estimate of the flux of photons arriving at Earth from a nearby dwarf by doing,
\begin{equation}
    F \sim \frac{M_{\rm DM}}{m_a} \frac{\Gamma_{a\to\gamma\gamma}}{4 \pi d^2} \approx 8.8\times 10^{-5} {\rm cm}^{-2} \, {\rm s}^{-1} \, \left( \frac{M_{\rm DM}}{10^7\,M_\odot} \right) \left( \frac{409\,{\rm kpc}}{d} \right)^2 \left( \frac{m_a}{3\,{\rm eV}} \right)^2 \left( \frac{g_{a\gamma}}{10^{-12}\,{\rm GeV}^{-1}} \right)^2 \, ,
\end{equation}
where I have put in the total mass of dark matter, $M_{\rm DM}$, and the distance, $d$, to the Leo T ultra-faint dwarf, which has a mass-to-light ratio of $\sim 100$~\cite{Clementini2012}. The photons are emitted at energies $E_\gamma = m_a/2$ which corresponds to wavelengths of,
\begin{equation}
    \lambda_a = \frac{4\pi}{m_a} = 2.48 \,\upmu {\rm m}\, \left( \frac{1\,{\rm eV}}{m_a} \right) \, .
\end{equation}
To estimate the expected sensitivity of some axion emission line hunt in a dwarf galaxy, we need to compare this to the background level in the region of interest. An interesting and simple comparison to make is with just the upper limit set by Ref.~\cite{MUSEFaint_LeoT} on the surface brightness of the Leo T dwarf, obtained through observations using the MUSE infrared integral field spectrograph on the Very Large Telescope (VLT) in Chile. Their limit is $<1 \times 10^{-20}$~erg/s/cm$^2$/arcsec$^2$ for extended line emission across the dwarf. We work out the expected surface brightness from axion line emission by multiplying the flux by the photon energy and dividing by the solid angle subtended by the dwarf: $\Delta\Omega \sim 2\pi \theta^2 = 2\pi(r/d)^2$ where $r\sim 120$~pc is its radius. Doing that, we get,
\begin{equation}
    I \sim \frac{E_\gamma F}{\Delta \Omega} \sim 9\times 10^{-21}\,{\rm erg}\,{\rm cm}^{-2}\,{\rm s}^{-1}\,{\rm arcsec}^{-2} \left( \frac{m_a}{3\,{\rm eV}} \right)^3 \, \left(\frac{g_{a\gamma}}{10^{-12}\,{\rm GeV}^{-1}} \right)^2\,\left( \frac{M_{\rm DM}}{10^7 \, M_\odot} \right) \left( \frac{120\,{\rm pc}}{r} \right)^2  \, ,
\end{equation}
which implies that couplings below $g_{a\gamma}<10^{-12}$~GeV$^{-1}$ should be constrainable by this observation of Leo T, and by comparing with Fig.~\ref{fig:AxionPhoton_HeavyAxionDecayCloseUp} we see this roughly matches the level of the MUSE bound derived from a detailed analysis~\cite{Regis:2020fhw, Todarello:2023hdk}.

% %Note also that the specific intensity of the atmosphere is also at the level of $10^{-17}$ erg/s/cm$^2$/\AA, so this axion line at 8000\AA of this strength will be several million times fainter than the night sky so real observations like this require careful subtraction of the background. 
% Note that Leo T is an extremely faint source and so extracting these limits requires a careful subtraction of the background from the night sky.
% Fortunately, the signal of the axion is quite striking: an emission line that probably will not line up with any known atomic or molecular line observed in astronomy. 

So how is this done beyond a back-of-the-envelope estimate? Firstly we need to rethink how we are adding up the dark matter. How do we know the value of $M_{\rm DM}$ to put in? This can be inferred from the kinematics of stars in the dwarf, but what we really get out of doing that procedure is how the dark matter is distributed throughout the object. This is a more useful thing to quantify anyway because in doing so we can make sure we are pointing our telescope at where the dark matter is. To calculate the flux coming at us from some spatial distribution of dark matter parameterised by a density $\rho(\mathbf{x})$ we consider some direction  $\hat{\mathbf{n}}$ that we have pointed our telescope, and add up all of the dark matter we have along that line of sight. The more correct formula for the flux then becomes,
\begin{equation}\label{eq:decayflux}
\frac{\textrm{d}^2 F}{\textrm{d}E_\gamma \textrm{d}\Omega} = \frac{\Gamma_{a\to \gamma \gamma}}{4 \pi m_a} \frac{\textrm{d} N_\gamma}{\textrm{d} E_\gamma} D(\hat{\mathbf{n}}) \, .
\end{equation}
where $\textrm{d}N_\gamma/\textrm{d}E_\gamma$ is the photon emission spectrum. The dark matter density is encapsulated in the so-called $D$-factor,
\begin{equation}\label{eq:Dfactor}
    D = \int_0^\infty \textrm{d}s \, \rho\big(s,\hat{\mathbf{n}}\big) \, ,
\end{equation}
where $s$ is a coordinate running from our location out to infinity in the direction of $\hat{\mathbf{n}}$.

The $D$-factors for Milky-Way dwarf galaxies are typically in the range $\log_{10}(D/{\rm GeV\,cm}^{-2}) = 15$--$17$, see Ref.~\cite{PaceStrigari} for a list. But notice that this formula doesn't actually contain any information about specific objects, it is just the cumulative density of dark matter between us and infinity. This makes sense because even when we point a telescope at a dwarf galaxy. there is nothing to stop us from simultaneously seeing the photons coming from all the dark matter in the Milky Way halo that happens to be decaying along that same line of sight.\footnote{In fact, we should also see the dark matter decaying across the entire Universe as well. However, this is usually not dominant, and there is also the redshifting of the photons, which turns the flux from a narrow line into a spread-out distribution towards longer wavelengths.} We can account for this straightforwardly by adding another component to the density in the $D$-factor, e.g by assuming a spherically symmetric NFW or cored profile for the Milky Way halo $\rho_{\rm MW}(r)$, and using $r = \left(R_{\odot}^2+s^2-2 R_{\odot} s \cos \theta\right)^{1 / 2}$ for the galactocentric radius, where $R_\odot\approx 8$~kpc is the solar orbital radius and $\theta$ is the angle between $\hat{\mathbf{n}}$ and the galactic centre.

Depending on the nature of the observation---e.g.~the type of instrument being used, its spatial resolution, field of view, available integration time etc.---the best signal to noise may not come from looking at single objects like dwarfs, but from integrating over a large fields of view where there happens to be low backgrounds, and focusing instead on the expected diffuse line emission from the Milky Way halo that surrounds us. The $D$-factor for the Milky Way can reach values between $\log_{10}(D/{\rm GeV\,cm}^{-2}) = 21$--$22$, peaking towards the galactic centre, so in many situations in fact this emission is expected to dominate the decay signal. In a few paragraphs when I will discuss a few bounds, several of these will have been obtained from this way through a ``blank sky'' approach, as opposed to targeting specific objects.

In Eq.(\ref{eq:decayflux}) I wrote the spectrum of photons emitted by the dark matter in terms of a generic function $\textrm{d}N_\gamma/\textrm{d}E_\gamma$. It may seem like we have the answer for this already: $\textrm{d}N_\gamma/\textrm{d}E_\gamma = \delta(E_\gamma - m_a/2)$. However, if we adopted this assumption we would miss several effects that, depending on the instrument being used and the source of the dark matter, could be important. 

One obvious effect a delta-function emission spectrum misses is the Doppler shifting and broadening of the line---the axions will not be precisely at rest and nor are we as observers. So we must account for the spread in dark-matter velocities---$\sigma_v \sim \mathcal{O}(10\,{\rm km/s})$ for dwarfs and $\mathcal{O}(100\, {\rm km/s})$ for dark matter in galaxies and in the Milky Way---as well as the fact that we are moving with respect to the galactic centre at a velocity $|\mathbf{v}_\odot| \approx 248$~km~s$^{-1}$. The former Doppler effect will broaden the emission line while the latter will shift it up or down in wavelength slightly depending on the sign of $\mathbf{v}_\odot \cdot \hat{\mathbf{n}}$. On top of that, there will be the finite spectral resolution of the instrument, which will cause photons to be slightly mis-measured at different wavelengths, usually with some Gaussian spread around the true value. In most searches that I will mention in a few paragraphs, this spectral resolution dominates over the Doppler broadening. However, some new instruments, like the infrared spectrographs onboard JWST, have 0.1\%-level resolution, which is sufficient to resolve the axion decay line in the case of the Milky Way emission~\cite{Roy:2023omw}. 

The Doppler effect will also make $\textrm{d}N_\gamma/\textrm{d}E_\gamma$ direction and position-dependent in general. Another position-dependent effect is the presence of dust lying along the line of sight that acts to attenuate photons arriving from certain directions, especially those along the plane of the Milky Way. See Ref.~\cite{Roy:2023omw} for how all of these effects can be combined together into an effective energy and direction-dependent photon emission spectrum for dark-matter decay.

I will now summarise the rest of the bounds appearing in Fig.~\ref{fig:AxionPhoton_HeavyAxionDecayCloseUp}, moving from the infrared to the hard X-ray. Several other searches for spontaneous axion decay lines at longer wavelengths can be found in the literature---in the microwave~\cite{Blout:2000uc} and radio~\cite{Chan:2021gjl, Keller:2021zbl} bands, for example---but unsurprisingly, these fall short of the stellar bounds on axions and so do not probe any self-consistent parts of parameter space.
\\~\\
\noindent {\bf Infrared}: The first telescope bounds on axion decay at masses below 3 eV were set only very recently. Observations of the Leo V and Tucana II dwarfs using the WINERED infrared spectrograph on the Clay Telescope in Chile were used by Ref.~\cite{Yin:2024lla} to cover the gap between 1.8--2.7~eV. As mentioned above, one of the most powerful space telescopes ever launched is operating right now in the infrared: JWST. An early bound on the Milky Way decay signal using 2000 seconds of blank-sky data was drawn in Ref.~\cite{Janish:2023kvi}, whereas the full end-of-mission sensitivity was forecast in Ref.~\cite{Roy:2023omw}. Projections for JWST observations of the dark-matter-rich Segue~I dwarf have also been forecast in Ref.~\cite{Bessho:2022yyu}.
\\~\\
\noindent {\bf Optical}: One of the earliest axion decay bounds, to my knowledge, was set in the 1990s by Refs.~\cite{Bershady:1990sw, Ressell:1991zv} who observed several galaxy clusters in the optical using the Kitt Peak Observatory and managed to rule out the QCD axion in the 3--8 eV window (at the time this was much less constrained than it is now). A much more sensitive search for axions decaying in galaxy clusters was done 16 years later using the VIMOS multi-object spectrograph on the VLT in Ref.~\cite{Grin:2006aw}.\footnote{I would like to point out that the data for this bound listed on my GitHub repository was scaled incorrectly for a long time, causing the error to propagate into several subsequent publications. I thank the authors of Ref.~\cite{Carenza:2023qxh} for bringing this to my attention.} Much more recently, Refs.~\cite{Regis:2020fhw, Todarello:2023hdk} used data from a survey of ultra-faint dwarfs using the MUSE integral-field spectrograph (also on the VLT). Slightly higher masses than this fall within the sensitivity of the famous Hubble Space Telescope (HST), but while there do not seem to be any dedicated axion line hunts using HST data, Refs.~\cite{Nakayama:2022jza, Carenza:2023qxh} set bounds in the 5--20 eV window by excluding any anomalous contributions to the cosmic optical background light~\cite{Overduin:1993kz}. In particular, they use the fact that axion decay across the Universe should generate anisotropies that are distinct from those of astrophysical foregrounds like the Zodiacal light. A semi-related idea employed by Ref.~\cite{Bernal:2022xyi} involves \textit{indirectly} searching for the axion contribution to the cosmic optical background via the way it would attenuate fluxes of very high-energy gamma-rays through $\gamma \gamma \to e^+ e^-$ pair-production. This attenuation can be extracted from the redshift-dependence of the flux of powerful gamma-ray emitters like blazars, and this can be used to set bounds on any extra contributions to the EBL.\footnote{Reference~\cite{Bernal:2022xyi} also point out that their inferred optical depth to gamma-rays was larger than expected, which could be interpreted as a hint of additional photons produced by an axion with a coupling and mass lying just below the quoted $\gamma$-ray attenuation bound in Fig.~\ref{fig:AxionPhoton_HeavyAxionDecayCloseUp}.}
\\~\\
\noindent {\bf X-ray}: One of the first examples of a constraint on keV-mass axion decay is from Cadamuro and Redondo's 2011 catalogue~\cite{Cadamuro:2011fd}, where they estimated a bound from published X-ray background data available at the time. Since then, there have been more dedicated searches for dark-matter emission lines using data from the handful of X-ray space telescopes that are available\footnote{This activity, it should be said, was primarily driven by interest in sterile neutrino decays rather than the less well-motivated keV-mass ALPs}. In particular, XMM-\textit{Newton}~\cite{Foster:2021ngm}, and \textit{NuSTAR}~\cite{Perez:2016tcq, Ng:2019gch, Roach:2022lgo} are sensitive across 1--100 keV energies and constraints have been derived using observations of a variety of different objects, including blank sky observations~\cite{Foster:2021ngm}.\footnote{These observations cover the mass range where a decaying ALP could have been involved in one of the explanations for the anomalous 3.5 keV X-ray emission line observed by several groups in objects like Andromeda and the Perseus cluster~\cite{Boyarsky:2014jta, Bulbul:2014sua, Cappelluti:2017ywp}. While standard dark matter ALP decays would not be able to create a line this strong without falling foul of non-observations elsewhere, models invoking some dark matter$\to$ALP$\to \gamma$ process were proposed as possible alternatives~\cite{Conlon:2014xsa, Conlon:2014wna, Alvarez:2014gua}. Unfortunately, the line is highly unlikely to be from dark matter decay due to it not being present at any measurable level in the Milky Way~\cite{Gewering-Peine:2016yoj, Dessert:2018qih} (and see also~\cite{Jeltema:2015mee, Hitomi:2016mun}). Worse still, Ref.~\cite{Dessert:2023fen} have recently argued that the original claims about the existence of the line may have been the result of flawed analysis procedures or background mis-modelling.} The INTEGRAL space telescope on the other hand covers the hard-X-ray gap between \textit{NuSTAR} and the gamma-ray observatory \textit{Fermi}. The highest-mass axion-decay bound appearing on the plot was derived from INTEGRAL data in a $95^\circ \times 95^\circ$ region around the galactic centre by Ref.~\cite{Calore:2022pks}. Some forecasts have been made for in-progress, forthcoming, or concept X-ray telescopes like XRISM~\cite{Dessert:2023vyl}, the Line-Emission Mapper~\cite{Krnjaic:2023odw}, eROSITA~\cite{Caputo:2019djj, Dekker:2021bos} and THESEUS~\cite{Thorpe-Morgan:2020rwc}. In fact, a constraint has already been derived from an early release of data from the eROSITA all-sky X-ray survey by Ref.~\cite{Fong:2024qeq}, which appears in Fig.~\ref{fig:AxionPhoton_HeavyAxionDecayCloseUp}.
\\~\\
% Gas heating
There is one final late decay bound appearing in Fig.~\ref{fig:AxionPhoton_HeavyAxionDecayCloseUp} that operates slightly differently. Like Refs.~\cite{Regis:2020fhw, Todarello:2023hdk}, it also makes use of the Leo T dwarf, but is not a direct search for a decay line. Leo T lies far enough away from the Milky Way for its gas to have not yet been stripped away from it by the bulk drag force exerted by the gas found deeper in the Milky Way halo---a process known as Ram pressure stripping. The cooling rate of gas inside isolated, metal-poor dwarf galaxies like Leo T is expected to be extremely slow, and so its measured temperature rules out excessive sources of energy injection that would heat it up. Reference~\cite{Wadekar:2021qae} employed this argument to set a bound on axions in the sub-keV mass window. For similar reasons, it is challenging to imagine any more direct searches for axion emission lines in this mass range, because photons with UV-X-ray energies will rapidly photo-ionise any neutral hydrogen they come across. Indirect arguments like gas cooling (as well as early decays which follow a similar logic) are therefore likely to be the only way to probe this window of masses. 

\subsection{Axion stimulated decay}
Before moving on to the non-decay searches, I will briefly comment on some proposals to look for axion decay that purport to be able to cover lower masses. As you can see from the tiny segment of the yellow band that appears in Fig.~\ref{fig:AxionPhoton_HeavyAxionDecayCloseUp}, none of the axion decay searches I mentioned above are in any way relevant to our favourite target: the QCD axion. Nonetheless, axion decay is a good signature to look for because emission lines are fairly unambiguous once they are detected with confidence above the background. So it would be nice if it were possible to see decay signals from much lighter axion masses. The 100 MHz--100~GHz radio-frequency range ($\mu$eV--100$\mu$eV) would be an especially attractive window to explore because it is within this window that the misalignment mechanism most naturally generates the correct dark matter abundance. Moreover, there are plenty of big radio telescopes around, as well as even bigger ones on the horizon that could go and look for such a signal. Unfortunately, the big problem here is that you have to fight against the punishing $m_a^{3}$ dependence in the decay rate which makes searches for spontaneous axion decay totally impractical below infrared wavelengths. There is a possible route out of this however---the photon rate via decay can be enhanced by instead considering \textit{stimulated} axion decay~\cite{Tkachev:1987cd,Kephart:1994uy}, as opposed to \textit{spontaneous} decay.

Stimulated decay involves the same axion$\rightarrow$ 2 photon diagram, but instead assumes that the process happens in the presence of a background radiation field. If some of that background radiation exists underneath the axion dark matter at exactly the right frequency ($m_a/2$), then this can stimulate the emission of two photons at a much higher rate than spontaneous decay. This is the same idea as how a laser works---the transition rate of an atomic species from some excited state down to its ground state is enhanced when there is a substantial background of photons with energies that are the same as the transition energy. Stimulated axion decay can happen, therefore, because the axion can be thought of as the ``excited state'' of the photon that is an energy of $m_a/2$ above its ground state.

The flux density (flux divided by the signal bandwidth, $\Delta \nu$) including the possibility for stimulated decay is given by~\cite{Caputo:2018ljp, Caputo:2018vmy},
\begin{equation}
    S =\frac{\Gamma_{a\to\gamma\gamma}}{4 \pi \Delta \nu} \int \textrm{d} \Omega\, \textrm{d} s \,\rho_a(s, \Omega) e^{-\tau\left(s, \Omega\right)}\left[1+2 f_\gamma\left(s, \hat{\mathbf{n}}\right)\right] \, ,
\end{equation}
where,
\begin{equation}
 f_\gamma(s,\hat{\mathbf{n}}) = \frac{(2\pi)^3}{E_\gamma^3} \frac{\textrm{d}\rho_\gamma(s,\hat{\mathbf{n}})}{\textrm{d}\omega}\bigg|_{\omega = E_\gamma} \, ,
\end{equation}
is the occupation number of photons of energy $E_\gamma = m_a/2$ at the location along the line-of-sight $(s,\hat{\mathbf{n}})$. The $e^{-\tau}$ accounts for the attenuation of the photon flux parameterised by the optical depth, $\tau$. Note that this effect will only dominate if the number density of axions is much larger than the number density of photons; otherwise, the photon re-absorption will dominate and suppress the signal.

This idea was proposed by Caputo et al.~\cite{Caputo:2018ljp, Caputo:2018vmy} in the context of radio observations of nearby dwarf galaxies. The radiation background could be provided by whichever ambient photons happened to be flying around the dark matter halo, for example, the ever-present CMB or the extragalactic radio background, as well as synchrotron emission from the galaxy itself. Because $f_\gamma$ decreases with increasing photon energy, it is only photon backgrounds in the microwave to radio bands that will enhance emission by any substantial amount. Unfortunately, even with the SKA, it is challenging to find the right combination of numbers where the sensitivity can brought down much below the stellar axion bounds. 

Some other ideas have been put forward to look for the stimulated decay of axions in our own dark matter halo, with the photons provided either by some bright radio source~\cite{Ghosh:2020hgd} or from a semi-recent (within last few hundred years) supernova~\cite{Sun:2021oqp, Buen-Abad:2021qvj}. One neat aspect of this latter kind of search is that because the axions are (almost) at rest the two stimulated photons are emitted (almost) back to back. This means that one of those photons will return back in the direction of the source. People have referred to this as ``axion gegenschein\footnote{Gegenschein being an astronomical phenomenon where a faint glow appears in the night sky opposite to the position of the sun due to sunlight backscattered by solar-system dust.}'' or ``axion echo''. For example, you might be able to see the light from some historical supernova reflected back at you if you pointed a radio telescope towards the opposing direction in the sky.

Since those initial papers on the axion echo, several more recent studies have forecasted the full sensitivity of existing and upcoming radio facilities to the axion-stimulated decay signal. These calculations now include both the gegenschein (backwards emitted) and vorwärdsschein (forward emitted) photons that could be stimulated by radio sources from across the whole Milky Way, including supernovae, pulsars, and galactic synchrotron emission. See Refs.~\cite{Sun:2023gic, Dev:2023ijb, Todarello:2023xuf} for further details. Unfortunately, it seems that even accounting for all possible emissions and forecasting for a transformative facility like the SKA, it is still challenging to surpass other non-dark matter bounds, particularly the one derived from axion emission from pulsar cap regions~\cite{Noordhuis:2022ljw} which rules out axions in a broad window around $m_a \sim 1\,\upmu$eV with $g_{a\gamma}>3 \times 10^{-12}$~GeV$^{-1}$ (see Fig.~\ref{fig:AxionPhoton_Closeup_AltColours}).

But then there is another even more radical iteration on this idea. Instead of looking for axions stimulated into decaying by astrophysical photons, you could instead fire a powerful beam of your own photons into the sky and look for the axion dark matter around the Earth gegenscheining it back at you in real time~\cite{Arza:2019nta, Arza:2021nec, Arza:2021zqc, Arza:2022dng}. A pathfinder search is currently being pursued within the Chinese project, the 21 CentiMeter Array~\cite{Arza:2023rcs} between 50--200 MHz (0.41--1.6~$\upmu$eV) using a 1 MW transmitter fired into the sky. But again, a much more powerful search than this will be required to surpass the non-dark matter pulsar bound.

\subsection{Axion conversion}\label{sec:axionconversion}
To make up sensitivity to axions in the radio frequency band we must move away from decay and start thinking about other interactions they might undergo with photons. Axion \textit{conversion} refers to the process whereby an axion can turn into a photon with energy $\omega = m_a + \frac{1}{2} m_a v^2$ (for non-relativistic axions) with a little help from a virtual photon supplied by a background electromagnetic field. In almost all cases relevant to dark matter, strong \textit{magnetic} fields are the most fruitful for this, primarily because the electromagnetic energy stored in an astrophysical B-field over astrophysical volumes is much larger than any electric fields.

The probability for an axion to convert into a photon after propagating a length $L$ through a vacuum containing a magnetic field of strength $B$ is~\cite{Raffelt:1987im},
\begin{equation}\label{eq:conversionprob}
    P_{a \to \gamma}=\left(\frac{g_{a \gamma} B}{q}\right)^2 \sin ^2\left(\frac{q L}{2}\right) \, ,
\end{equation}
where $q = k_\gamma - k_a$ is the momentum transfer between the massive axion and massless photon (this fact will be important later). Let us first appreciate that the value of this probability is likely going to be very small in most places---axions are feebly coupled particles after all. So we have to try to figure out how to make up gains somehow if we want to see the photons they are supposed to convert into. The obvious way to enhance $P_{a\to \gamma}$ is to find a B-field that is as large as possible; another is to just get lots of axions so there are many possible instances of conversion. In other words, we need a big magnet with lots of dark matter around it---it turns out that \textit{neutron stars} fit the bill on both counts.

In fact, all of the searches I will discuss in this subsection make use of neutron stars in some way\footnote{Although I don't discuss them here, there are a couple of other ideas for indirect detection of axion dark matter through its conversion inside astrophysical magnetic fields. These include the examples given in Ref.~\cite{Sigl:2017sew} as well as ideas involving the magnetic field in the Sun's corona~\cite{An:2023wij} or its sunspots which can have B-fields of a few thousand Gauss~\cite{Todarello:2023ptf}.}. The classic kind of neutron stars are those that spin such their radio-emitting jets periodically align with the Earth, in which case they are called pulsars, but nearby, radio-quiet neutron stars where this isn't the case have also been observed and studied---for example the group of seven neutron stars within a few hundred parsecs of Earth known as the ``Magnificent Seven''. To give a sense of the numbers involved, neutron stars possess staggering magnetic fields of up to $10^{14}$~Gauss\footnote{$1 \mathrm{Gauss} \approx 1.95 \times 10^{-2} \mathrm{eV}^2$.}, and have masses comparable to the Sun crammed inside radii of only $r_{\rm NS}\sim10$~km or so. 

Before we go plugging these numbers into Eq.(\ref{eq:conversionprob}), we need to think more carefully about the physics of axion-photon conversion. The axion-photon system in this context is best thought of as a two-particle mixing phenomenon similar to neutrino oscillations. As stated above, in a vacuum there is a momentum mismatch $q = k_\gamma-k_a$ between the massive and massless components of the axion+photon superposition that propagates. This means that the axion and the photon can only stay in phase with each other over a length scale given by $\sim 1/q$. It is only in the limit that $q L \ll 1$ that we get,
\begin{equation}\label{eq:conversionprob2}
     P_{a\to \gamma} \approx \left(\frac{g_{a \gamma} B L}{2}\right)^2 \, ,
\end{equation}
i.e.~a conversion probability that is parametrically enhanced by how long the B-field is. However, if we find ourselves in the opposite regime, $L>1/q$, then the axion and photon are already destructively interfering over part of the B-field's total length. Ultimately, this means the probability of seeing a photon at the other end of the B-field is suppressed, and in the extreme case where $L\gg 1/q$ it can be \textit{extremely} suppressed. Naively it would seem that non-relativistic axion dark matter conversion in a neutron-star B-field is in this latter regime because $k_a \sim m_a v$ and $k_\gamma = \omega_\gamma \approx m_a + \frac{1}{2}m_a v^2$. The length-scale \mbox{$1/q \sim m_a^{-1} \sim 0.2 \,{\rm metres}\,(1\,\upmu{\rm eV}/m_a)$} is much more smaller than $r_{\rm NS}$ unless we go to very small masses.

However there is something we are missing about neutron star environments: their rotating magnetic fields are so strong that the subsequently generated electric fields can rip charges straight off of their surfaces. So it is expected that neutron star magnetospheres will be dressed in a dense \textit{plasma} of electrons, positrons and ions. This fact is of critical importance for the problem at hand because, inside a plasma, the photon's dispersion relation is modified in a way that gives it an effective mass. This completely changes the physics of axion-photon conversion. If we just so happen to have a medium in which the effective mass of the photon is equal to the axion mass, there is no associated momentum mismatch and $1/q \to \infty$---in this case the conversion is \textit{resonant}. The photon signal we get out of axions propagating through the B-field when $q = 0$ solely depends on how long that propagation length is, like in Eq.(\ref{eq:conversionprob2}).\footnote{This effect is used by the solar axion experiment CAST to obtain sensitivity to higher-mass axions where the momentum mismatch would cause $1/q$ to be shorter than the length of their magnet. They pump in a buffer gas at a certain pressure that gives the photon a tunable mass, thereby facilitating resonant conversion~\cite{CAST:2011rjr}.}. It turns out that this phenomenon will occur around neutron star magnetospheres.

\begin{figure}
        \centering
        \includegraphics[width=0.7\linewidth]{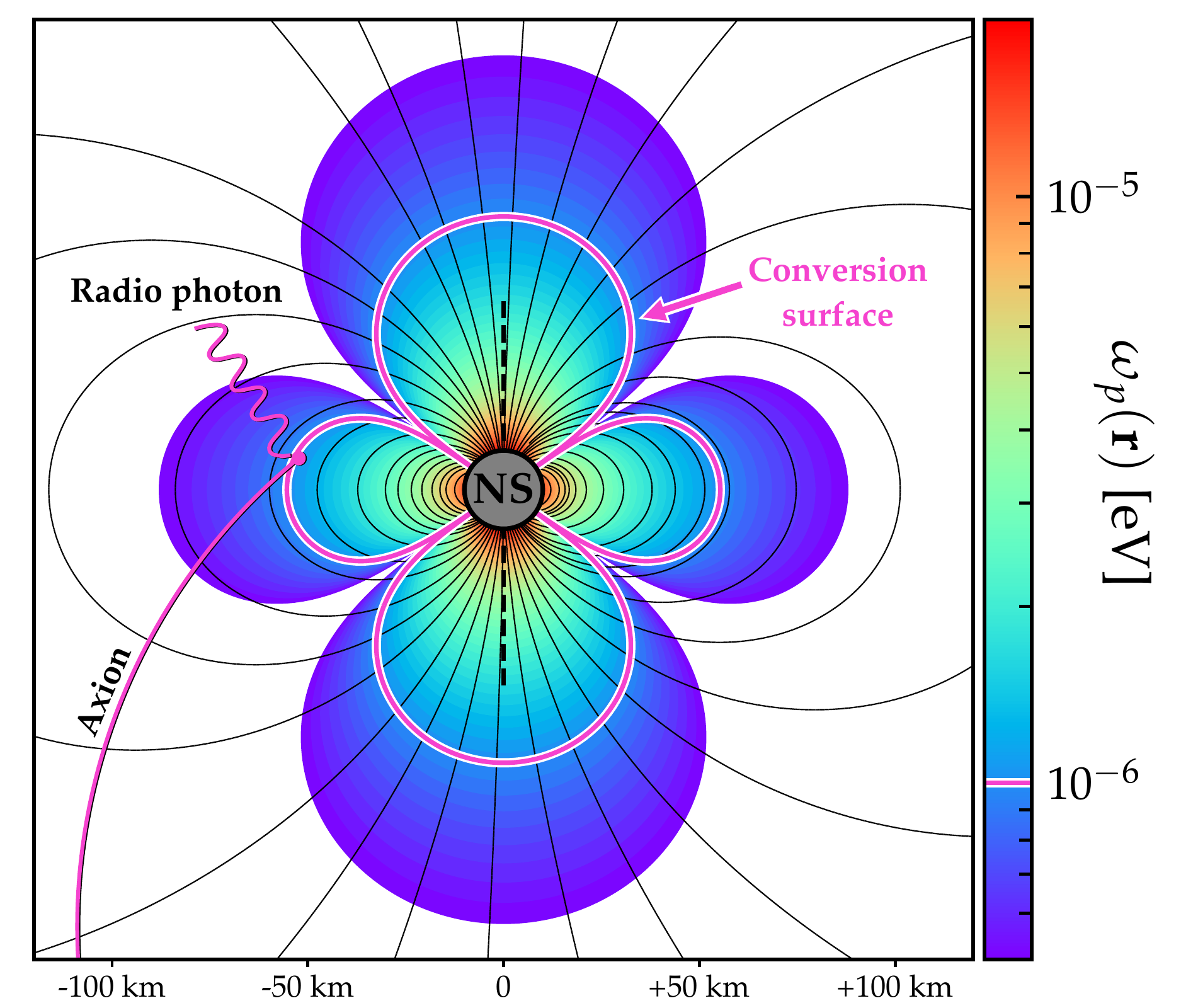}
        \caption{A visualisation of the neutron star magnetosphere and plasma frequency within the Goldreich-Julian model. An axion falls onto the neutron star from infinity, and when it crosses the surface defined by $\omega_p(\mathbf{r}) = m_a$, it resonantly converts into radio-frequency photons, which may then leave the neutron star and be observed in a telescope.}
        \label{fig:NS}
    \end{figure}

Let's start by writing down the effective mass for the photon in a plasma, also called the plasma frequency---all it depends on is the number density of electric charges. In this case, we'll write it more simply in terms of just the number density of electrons $n_e(\mathbf{r})$,
\begin{equation}
    \omega_p(\mathbf{r})=\sqrt{\frac{4 \pi \alpha n_e(\mathbf{r})}{m_e}} \, ,
\end{equation}
where $e = \sqrt{4\pi \alpha}$ and the electron mass is $m_e = 511$~keV. So what we need is a model for the electron density around a neutron star. The assumption used in the majority of studies is a model from a 1969 paper by Goldreich \& Julian~\cite{GoldreichJulian}, which takes the form,
\begin{equation}
n_{\mathrm{GJ}}\left(\mathbf{r}_{\mathrm{NS}}\right)=\frac{2 \boldsymbol{\Omega} \cdot \mathbf{B}_{\mathrm{NS}}}{e} \frac{1}{1-\Omega^2 r^2 \sin ^2 \theta} \, ,
\end{equation}
where $\boldsymbol{\Omega} = (2\pi/P) \hat{\mathbf{z}}$ is the rotation vector of the magnetic field which is taken to align with the $\hat{\mathbf{z}}$ direction (and is generally assumed to be misaligned from the axis of symmetry for the $B$-field), and $\theta$ is the polar angle away from the $z$-axis. The rotation period is $P\sim \mathcal{O}({\rm secs})$ for typical neutron stars.

The Goldreich-Julian model is the simplest stable solution for a plasma surrounding a neutron star. It is derived by finding a co-rotating charge density generated by electric fields at the neutron star surface, where the electric field that is then induced parallel to the neutron star surface by the subsequent rotation and rearrangement of the charge density is exactly cancelled. It is the minimal co-rotating charge density needed to screen the combination $\mathbf{E}\cdot \mathbf{B}$ to zero at the neutron star surface. Although the Goldreich-Julian model only sets the absolute charge density and not specifically the electron density, it is common to assume $n_e = |n_{\rm GJ}|$. 

The magnetic field on the other hand is usually modelled as a misaligned dipole: $B \sim B_0(r_{\rm NS}/r)^3$, where the field strength at the surface is on the order of $B_0 \sim 10^{13}$--$10^{14}$~Gauss. In a dipolar field, the plasma density will decrease moving radially away from the neutron star surface. Thinking about this gradient in the plasma frequency is key to seeing how we get our axions to convert into photons. Depending on how the numbers work out, it is possible for there to be some ``conversion surface'' defined by the set of spatial points where $\omega_p(\mathbf{r}) = m_a$.\footnote{Strictly speaking this resonant conversion region is not a surface but has a thickness depending on the local photon dispersion relation and the orientation of the B-field, as I discuss later. But for this simplified discussion it is enough to just think of it as a surface.} It is precisely at this surface where we expect the resonant conversion to take place because this is where the photon and axion dispersion relations match. Within the Goldreich-Julian model, we can work this out to be,
\begin{equation}\label{eq:NS_omega_p}
\begin{aligned}
\omega_p\left(\mathbf{r}\right)&=\sqrt{e \mathbf{B} \cdot \hat{\mathbf{z}} \frac{4 \pi}{m_e P} \frac{1}{1-\Omega^2 r^2 \sin ^2 \theta}} \\ 
& \approx 69 \, \upmu \mathrm{eV} \, \left(\frac{r_{\mathrm{NS}}}{r}\right)^\frac{3}{2} \,\left(\frac{B_0}{10^{14}  \, \mathrm{G}}\right)^\frac{1}{2} \left(\frac{1 \mathrm{~s}}{P} \right)^\frac{1}{2} \, ,
    \end{aligned}
\end{equation}
where in the second line I have suppressed all dependence on the coordinates other than the radius, which also ignores the dependence on the angle between the magnetic field and the spin axis. Given the typical range of neutron stars that are out there, axion masses within the range $\upmu{\rm eV} \lesssim m_a \lesssim {\rm meV}$ naively look within reach to this sort of probe. The only hard boundary on the mass window is at the very high-mass end, where the conversion surface for a typical neutron star would end up \textit{inside} of it, $r_c<r_{\rm NS}$. 

This mass range we can cover using neutron stars spans photon frequencies in the radio to microwave bands, which makes it an attractive probe considering the masses predicted from the misalignment mechanism. What's more, the signal of axions converting around neutron stars could be rather striking if detected with statistical significance. When some axion dark matter flows past the neutron star and encounters this particular region they will convert into radio photons at a frequency somewhere close to $f = m_a/2\pi$, and with a very narrow spectral width given by the narrow non-relativistic spread of dark matter velocities.

The plasma frequency and magnetic field for this simplified neutron star model are visualised in Fig.~\ref{fig:NS}. The conversion surface for a $\upmu$eV axion is shown as a purple line, which is where axions of that mass could convert resonantly into photons. As is quite apparent from this image, the conversion surface is very much not a sphere, and this peculiar geometry has been shown to strongly influence the predicted photon flux. Simulations show that the outgoing photons propagated to infinity end up actually concentrated towards points of origin around the `throat' regions----the narrow gap between the polar lobes and equatorial torus of the conversion surface. This is due to the fact that the throats lie closest to the neutron star surface and that photons emitted from other locations tend to get focused down to those regions~\cite{Witte:2021arp}. Because neutron stars rotate on second timescales, this direction dependence to the outgoing photon trajectories also implies there will be a strong time dependence in the signal too. For example, we expect stronger signals for neutron stars having observer and magnetosphere alignments such that the throats pass over our line of sight.

The question then becomes whether or not the radio flux from some nearby neutron star is large enough to be detected in existing or near-future telescopes. The first calculations of the flux expected from axions around neutron stars were presented in Refs.~\cite{Pshirkov:2007st, Huang:2018lxq, Hook:2018iia, Safdi:2018oeu}, and I will now briefly go through their logic, but keep in mind there are several assumptions here that turn out to be overly crude. I will discuss a few of these caveats once we've got a first estimate of the signal strength.

We'll start by stating what we want to give to our radio observer: the expected \textit{flux density}, $S$, defined as the power per unit area over the signal bandwidth:
\begin{equation}
    S = \frac{1}{(\text{BW})\,d^2}\frac{\textrm{d}\mathcal{P}_{\rm NS\to \gamma}}{\textrm{d}\Omega}
\end{equation}
where ${\rm BW} \approx v^2 (m_a/2\pi) \approx 2.6 \, {\rm kHz} (m_a/20\,\upmu {\rm eV})$ is the expected frequency bandwidth of the axion signal coming from its spread of incoming velocities. In other words, the radio line will be about a million times narrower than the central frequency because $v^2 \sim 10^{-6}$. The power output per unit solid angle is very approximately set by,
\begin{equation}
    \frac{\textrm{d}\mathcal{P}_{{\rm NS}\to\gamma}}{\textrm{d}\Omega} \sim v_c P_{a\to \gamma} \, \rho_{\rm DM}(r_c) r_c^2
\end{equation}
where $r_c \propto B_0^{1/3} P^{-1/3} m_a^{2/3}$ is the radius of the conversion surface that satisfies $\omega_p(r_c) = m_a$ as in Eq.(\ref{eq:NS_omega_p}), and $v_c$ is the axion velocity at $r_c$. So as would be naively expected, the power radiated out is proportional to 1) the axion-photon conversion probability because that's how many photons you get out; 2) the dark matter density at the conversion radius $\rho_{\rm DM}(r_c)$ because that's how many axions go in; and 3) a factor of $r_c^2$ because the power is radiated out of a surface with area $\sim 4 \pi r_c^2$. 

Next, recall that the conversion probability on dimensional grounds must be some combination that looks like $P_{a\to\gamma} \sim (g_{a\gamma} B L)^2$. It turns out (see e.g.~Refs.~\cite{Leroy:2019ghm, Witte:2021arp, Millar:2021gzs}) that since the conversion takes place over a thin region where the plasma frequency is changing, the probability ends up looking like\footnote{Note that this assumes the axion takes a radial trajectory, which is one of the several bad assumptions we're making.}: 
\begin{equation}
    P_{a \rightarrow \gamma} \sim \frac{\pi}{2} g^2_{a \gamma} B^2(r_c) \frac{1}{v_c\left|\omega_p^{\prime}\right|} \, .
\end{equation}
So the relevant length scale for conversion turns out to be related to the gradient of the plasma frequency,
\begin{equation}
    L \sim 1/\sqrt{\omega_p^{\prime}} \sim \sqrt{2 r_c/3 m_a} \, .
\end{equation}
The last quantities we need before we can work out $S$ are the axion velocity and the DM density at the conversion surface. As opposed to simply hitting the neutron star like a projectile, axions are actually focused down towards the magnetospheres from over a much wider cross-sectional area because of the intense gravitational fields. They can get accelerated up to semi-relativistic velocities $v_c \sim \sqrt{2 G_N M_{\rm NS}/r_c}$ and the DM density is also correspondingly enhanced by this focusing effect to \mbox{$\rho_{\rm DM}(r_c) \sim \rho_{\rm DM}^{\infty} (v_c/v_0)$}. See e.g.~Refs.~\cite{Alenazi:2006wu, Hook:2018iia, Millar:2021gzs} for how to calculate this and for the full expressions beyond the tilde level.

Plugging all this in we can get a rough feel of the amplitude of the signal and its parameter dependencies,
\begin{equation}\label{eq:NS_fluxdensity}
\begin{aligned}
    S \sim 4 \times 10^{-5} \, & {\rm Jy} \underbrace{\left(\frac{g_{a\gamma}}{10^{-12}\,{\rm GeV}^{-1}}\right)^2 \left(\frac{m_a}{20\,\upmu {\rm eV}}\right)^{4/3}}_{\text{Axion parameters}} \times \underbrace{\left(\frac{\rho_{\rm DM}}{0.3\,{\rm GeV}\,{\rm cm}^{-3}} \right) \left(\frac{d}{100\,{\rm pc}} \right)^{-2}}_{\text{Neutron star location}}\\
    &\times \underbrace{\left( \frac{M_{\rm NS}}{1\,M_\odot} \right)^\frac{1}{2} \left( \frac{r_{\rm NS}}{10\,{\rm km}} \right)^\frac{5}{2}  \left( \frac{P}{1\,{\rm s}} \right)^\frac{7}{6}\,\left( \frac{B_0}{10^{14}\,{\rm G}} \right)^\frac{5}{6}}_{\text{Neutron star properties}}  \, , \\
\end{aligned}
\end{equation}
This is expressed in units of \mbox{$1\,{\rm Jy} = 10^{-23}\,{\rm erg}\,{\rm s}^{-1}\,{\rm cm}^{-2}\,{\rm Hz}^{-1}$}, which is the preferred radio-astronomy unit of flux density called the Jansky. 

To compare this number with the real world, consider Ref.~\cite{Foster:2020pgt} where $\sim$20 minutes on the Green Bank Telescope observing two nearby neutron stars ($d\sim \mathcal{O}(100\,\text{pc}))$ yielded a flux-density upper limit of $\sim 0.01$~Jy on lines with $\sim$~kHz width. If we assume every other number plugged into the formula above is reasonable, this suggests that coupling strengths down to $g_{a\gamma} = \text{few}\, \times 10^{-11}$~GeV$^{-1}$ can be reached using existing facilities and without extensive observational campaigns.

I have to take a moment now to emphasise that this is a very simplistic calculation that turns out to be missing numerous effects whose importance goes beyond mere corrections to the various formulae I have written down. A fairly large literature has now built up, with many groups tackling different aspects of the problem. Steadily, a much more sophisticated picture of axion-photon conversion around neutron star magnetospheres has been building up. I direct you to Refs.~\cite{Battye:2019aco, Leroy:2019ghm, Witte:2021arp, Millar:2021gzs, Battye:2021xvt, McDonald:2023ohd} for a sample, because I will only be able to give a taste to what some of these sophistications involve here.

One of the issues that needs immediate attention is the modelling of the trajectories of both the incoming axions and the outgoing photons. We know the axions are gravitationally focused down towards the neutron star, but as well as enhancing their density and speeding them up, this process also affects their phase space distribution and makes their trajectories non-radial. This issue is relatively straightforwardly worked out by starting from some assumption about the infinite-distance phase space distribution and then applying Liouville's theorem---indeed that procedure leads to the formula for $\rho_{\rm DM}(r_c)$ at first order.

The issue of the photon trajectories, on the other hand, is much more involved. When the axion encounters the conversion surface we should assume it can do so with an arbitrary angle of attack as opposed to enforcing that its trajectory be radial. This angle will also be arbitrary with respect to the inherently anisotropic plasma that surrounds it, turning the problem into a fully three-dimensional one. The upshot is that we should expect some kind of direction and momentum-dependent modifications going into $P_{a\to\gamma}$. But in fact, even the usual technique for calculating what is called the ``axion-conversion probability'' turns out to be oversimplified too. In a vacuum, the quantum mechanical notion of a probability, $P_{a\to\gamma}$, only comes about by drawing an analogy with the classical equation of motion for the axion and the two-photon polarisations transverse to the direction of propagation. The final squared amplitude of the photon divided by the axion ends up giving the same answer as would a quantum mechanical calculation for the conversion probability. In an anisotropic plasma, this is not the case. The axion can also mix with a third \textit{longitudinal} polarisation, and so when doing a 3-dimensional calculation it is more appropriate to calculate instead a `flux transfer' to account for the energy density in the photon field in the longitudinal mode~\cite{Millar:2021gzs}.

Continuing this theme, we should expect that the subsequently emitted photon will also not follow a straight trajectory out to infinity. Rather it will refract and reflect by amounts that depend on the surrounding density, anisotropy, and rotation of the plasma. Recall that the conversion takes place over the length-scale $L\sim \sqrt{r/m_a}$ within which the plasma gives the photon a mass similar to $m_a$ and lets it mix with the axion in phase. But if this propagation isn't perpendicular to the conversion surface, the photon will suffer refraction in the medium whereas the axion will not. As a result, there has been the suggestion that a dephasing will occur between the two, incurring yet another position/direction/time-dependent modification to the outgoing signal\footnote{Although see Ref.~\cite{McDonald:2023ohd} whose treatment did not reveal any dephasing effect at leading order}. The state-of-the-art for dealing with all of this photon propagation business is to solve it all in one go numerically using a technique called ray-tracing~\cite{Leroy:2019ghm, Witte:2021arp, McDonald:2023shx, Tjemsland:2023vvc}.

As well as modifying the size of the expected flux, these subtleties impact the spectral shape of the signal too~\cite{Battye:2021xvt, Xue:2023ejt}. The baseline expectation is for axions to generate a narrow spectral line of width ${\rm BW} \sim \mathcal{O}(\text{kHz})$, But when accounting for all the effects mentioned above, as well as the fact that the neutron star has a relative velocity with respect to us, any particular line from a given neutron star will be broadened as well as shifted away from $f = m_a/2\pi$.

% Adiabatic conversion
When it comes to setting constraints, yet another important theoretical issue arises, as pointed out in Ref.~\cite{Foster:2022fxn}: in a certain regime, when the axion-photon conversion probability is sufficiently large, the \textit{back-conversion} of photons into axions again cannot be ignored. The picture described originally is one in which the infalling axions encounter the conversion surface and convert into photons of energy $E_\gamma = m_a + \mathcal{O}(v_c^2)$. But if those photons then continue to propagate inwards---the direction in which the plasma frequency is increasing---a short distance later they will encounter another surface where the plasma frequency crosses $\omega_p(r)=E_\gamma$. The photons then can't propagate further beyond this second surface, so they reflect off of it and head back outwards. In the small-coupling ``\textit{non-adiabatic}'' limit, the axion-to-photon conversion is only assumed to happen once. However, if parameters like $B(r_c)$ or $g_{a\gamma}$ are large enough, then it is possible $P_{a\to\gamma}$ could take on values approaching $\sim 1$, in which case it is very likely that the photons reflected outwards will convert back into axions as they traverse the conversion surface again. So in this regime, the overall photon flux getting out of the neutron must be suppressed. Unfortunately, current observations turn out to be in the awkward intermediate regime where these effects do impact the flux non-negligibly, but not enough that the conversion can be modelled as fully adiabatic with $P_{a\gamma}\sim 1$.\footnote{An approach for dealing this is to adopt a probability that is somewhat closer to the adiabatic limit where multiple level-crossings occur, i.e.~approaching the Landau-Zener formula: $P_{a \rightarrow \gamma}^{\mathrm{ad}}=1-e^{-\gamma}$ where $\gamma = P_{a \rightarrow \gamma}^{\text {non-ad }}$~\cite{Battye:2019aco}. However, I should point out the recent Ref.~\cite{Tjemsland:2023vvc} which suggests that Ref.~\cite{Foster:2022fxn}'s approach to dealing with the transition between the adiabatic and non-adiabatic regimes by multiplying the axion$\to$photon and photon$\to$axion probabilities together: $P = e^{-\gamma}(1-e^{-\gamma})$ is overly conservative. This could mean some of the latest bounds discussed below could stand to improve (although erring on the side of being conservative is never a bad thing when setting limits).}

\begin{figure}
        \centering
        \includegraphics[width=0.99\linewidth]{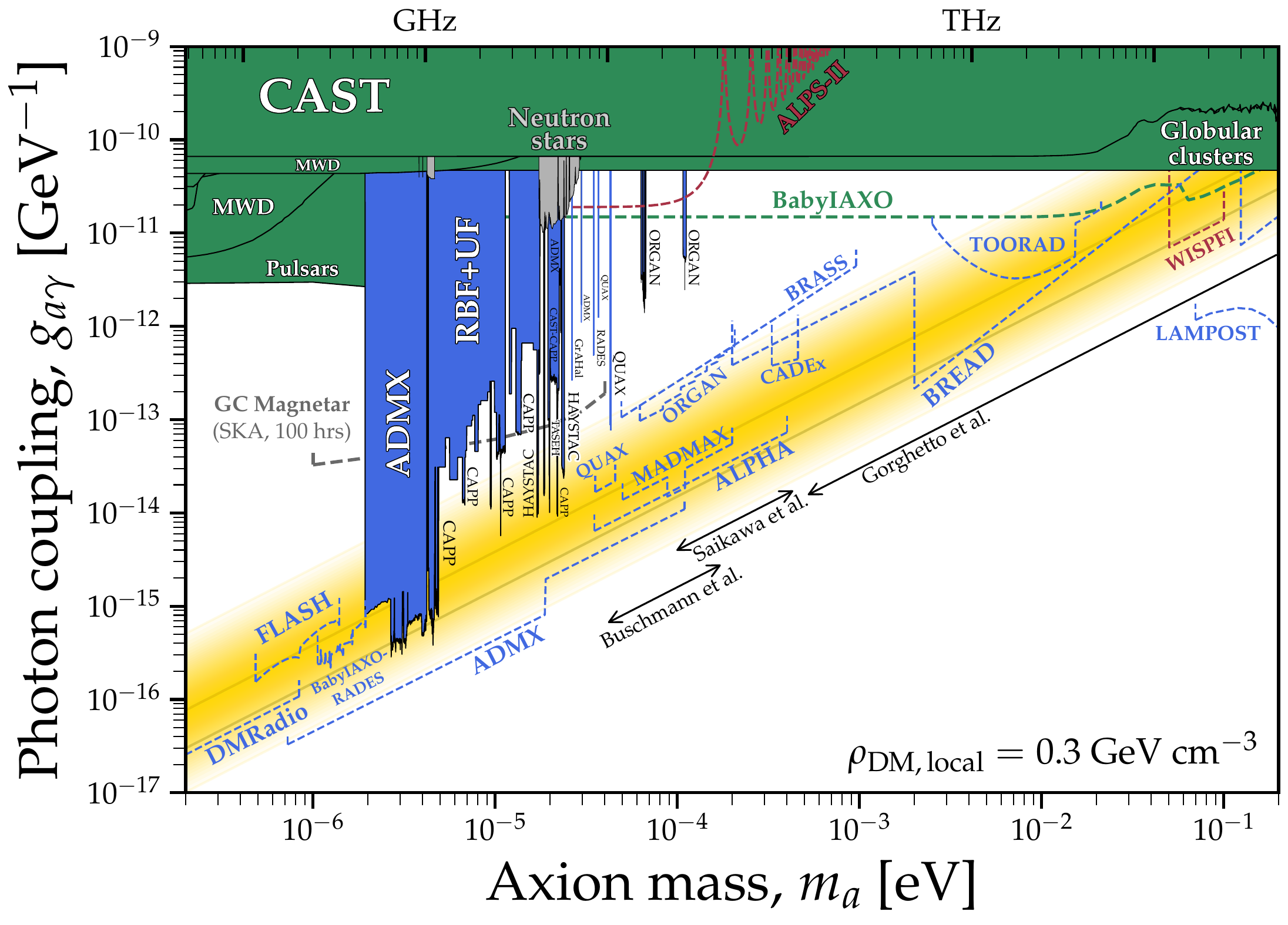}
        \caption{Constraints on the axion-photon coupling in the MHz-THz frequency range. The colour scheme is the same as other constraint plots and based around the level of assumption: green for bounds that do not assume axions are dark matter~\cite{CAST:2007jps,CAST:2017uph,Dolan:2022kul,Dessert:2021bkv,Dessert:2022yqq,Noordhuis:2022ljw}, blue for direct detection bounds~\cite{ADMX:2018gho,ADMX:2019uok,ADMX:2021nhd,ADMX:2018ogs,Bartram:2021ysp,Lee:2020cfj,Jeong:2020cwz,CAPP:2020utb,CAPP:2024dtx,Lee:2022mnc,Kim:2022hmg,Yi:2022fmn,Yang:2023yry,Kim:2023vpo,Adair:2022rtw,Grenet:2021vbb,HAYSTAC:2018rwy,HAYSTAC:2020kwv,HAYSTAC:2023cam,McAllister:2017lkb,McAllister:2022ibe,Quiskamp:2022pks,Quiskamp:2023ehr,Alesini:2019ajt,Alesini:2020vny,Alesini:2022lnp,QUAX:2023gop,CAST:2020rlf,DePanfilis,TASEH:2022vvu,Hagmann} (rescaled to a common conservative assumption about the local density of dark matter $\rho_0 = 0.3$~GeV~cm$^{-3}$), and grey for indirect/astrophysical searches for axions as dark matter which have much larger systematic uncertainties. In this plot, the only two examples of the latter are searches for axion dark matter converting in the magnetospheres of neutron stars~\cite{Foster:2022fxn, Battye:2023oac}. Red dashed lines show two proposed solely laboratory-based searches for axions~\cite{Bahre:2013ywa, Batllori:2023pap} and the green-dashed line shows the reach of the upcoming helioscope BabyIAXO~\cite{IAXO:2020wwp}---these projections are independent of any assumption about dark matter. The reach of some proposed haloscopes~\cite{Stern:2016bbw,Lawson:2019brd,ALPHA:2022rxj,Ahyoune:2023gfw,Liu:2021pei,Aja:2022csb,DeMiguel:2023nmz,Baryakhtar:2018doz,Beurthey:2020yuq,Alesini:2023qed,Schutte-Engel:2021bqm,Marsh:2018dlj} are shown with blue dashed lines. I have also shown one possible projection for a search for axion dark matter conversion around a neutron-star taken from Ref.~\cite{Witte:2021arp}, which assumes 100 hours of SKA time on the galactic centre magnetar, PSR J1745-2900.}
        \label{fig:AxionPhoton_Closeup_AltColours}
    \end{figure}
    
% Individual NSs
Having highlighted just a handful of the many subtle issues that arise when people attempt to do serious computations of the axion signal from neutron stars, let us now move to the observational side. From inspecting Eq.(\ref{eq:NS_fluxdensity}), we see that the ideal target neutron star will be nearby, with a high B-field and lower spin rates\footnote{Naively, a lower spin rate would seem to imply that older neutron stars would be better. However it has been suggested that neutron star magnetic fields will decay over their lives~\cite{Popov_2010}, so there is a trade-off between these two parameters.}. In conjunction, it would be useful also for the neutron star to be in a radio-quiet environment and be radio-quiet itself. ``Magnetars''---neutron stars with astonishing magnetic field strengths up to $\sim 10^{15}$~Gauss---might sound like the ideal targets, however the detailed magnetic field structures of these extreme objects are not well understood. So far, most of the calculations have relied upon the simple Goldreich-Julian model which will be a poor reflection of a magnetar. So instead, nearby and isolated neutron stars have been typically chosen for individual observations.

% Population NSs
The only problem with looking at nearby objects is that the dark matter density may not be as high as we can get it to be. One way to boost the signal then could be to search for entire \textit{populations} of faraway neutron stars that happen to be situated in regions with a high dark matter density~\cite{Safdi:2018oeu}. The obvious example is the galactic centre, which is expected to host a population of around a million neutron stars within the central parsec where the dark matter density is expected to be orders of magnitude higher than in the solar neighbourhood.\footnote{Globular clusters have also been suggested~\cite{Wang:2021hfb}, but while they do host vast numbers of old stellar remnants, even anomalous cases like $\omega$Cen are not expected to be particularly rich in dark matter.} The signal in a population study would not be a single radio line but an entire forest of them, where each line is shifted and broadened by an amount specific to a given neutron star. In a neutron star population, many unknown quantities---like the observer-neutron star alignment and magnetosphere parameters--are straightforwardly averaged away. Unfortunately, this is counterbalanced by the \textit{additional} uncertainty coming from synthesising the population of neutron stars, as well as the highly uncertain dark matter density profile at the galactic centre, e.g.~is it cusped or cored? Is there a dark matter spike? etc.

% Early studies
Foster et al.~\cite{Foster:2020pgt} was the first group to attempt an axion search using real neutron star data. They took dedicated radio observations of two isolated neutron stars using the Green Bank Telescope, whilst also observing the galactic centre using the Effelsberg telescope. One of these same neutron stars was also followed up for another axion search a few years later but a lower frequency band by the MeerKAT telescope~\cite{Zhou:2022yxp}, and there were also a handful of other limits set using archival data from the Jansky Very Large Array (VLA) on the galactic centre magnetar \mbox{PSR~J1745-2900}~\cite{Darling:2020uyo, Darling:2020plz, Battye:2021yue}. All of these limits unfortunately suffered from some of the overly simplistic assumptions in the theoretical modelling of the signal, and likely do not achieve their claimed sensitivity.

% Breakthrough listen
In 2022 however, Foster et al.~\cite{Foster:2022fxn} performed another search towards the galactic centre that accounted for many of the additional problems mentioned above, e.g.~doing ray-tracing to model the photons leaving the magnetosphere, as well as dealing with the fact that the relevant values of the axion-photon conversion probability for their assumed neutron star populations left them just shy of the non-adiabatic regime. Interestingly, they did not require dedicated time on radio facilities for this, but rather their data came about due to a serendipitous alignment with the needs of the \textit{Breakthrough Listen} project~\cite{Gajjar:2021ifn}. This is a program in which extensive radio telescope time is devoted to searching for ultra-narrow spectral lines that might be being transmitted into space by aliens\footnote{It's possible of course that aliens have already detected the axion and are attempting to prove their scientific credentials by beaming out a radio signal at $f=m_a/2\pi$ towards their galactic neighbours. As far as I know, no analysis has yet included the irreducible alien background in their noise model.}. In Ref.~\cite{Foster:2022fxn} they were able to set limits using galactic-centre data that had already been collected at a spectral resolution far exceeding the requirement for looking for an axion line. The resulting bound can be seen in Fig.~\ref{fig:AxionPhoton_Closeup_AltColours} spanning 15--35~$\upmu$eV, and corresponds to their most conservative neutron star population model.

% Battye et al.
The only other neutron-star bound I am showing in Fig.~\ref{fig:AxionPhoton_Closeup_AltColours} is from Battye et al.'s most recent constraint in the $m_a=3.9$--$4.7\,\upmu \mathrm{eV}$ region, derived using MeerKAT data on the pulsar \mbox{PSR J2144-3933}~\cite{Battye:2023oac}. This pulsar happens to have a small enough B-field that the axion conversion is safely in the non-adiabatic conversion regime. This analysis also incorporated the expected time dependence in the signal.

% Future
Note that the non-adiabatic regime holds again for small enough $g_{a\gamma}<10^{-11}$~GeV$^{-1}$ for the neutron star populations considered in the Breakthrough Listen study~\cite{Foster:2022fxn}. This was unfortunate in that study since they were in the regime where the measured flux scaled very slowly with the value of $g_{a\gamma}$. However, future searches that will probe more deeply will eventually break past this regime, thereby providing the potential to achieve much better sensitivity than a naive extrapolation of existing limits might suggest. The case is therefore quite compelling for embarking on a campaign of galactic-centre neutron-star line hunts with the SKA (see e.g.~\cite{Witte:2021arp, Foster:2022fxn}).

As well as population searches, the recent Ref.~\cite{McDonald:2023ohd} has revisited the possibility of using the magnetar PSR J1745-2900. This object has an estimated field of $10^{14}$~G and is within 0.2 parsecs of the galactic centre---a promising combination of features. However magnetars are significantly more complex objects, with twisted and non-dipolar magnetic fields, and charge densities that are driven to much higher values than the minimal co-rotating Goldreich-Julian configuration. Nonetheless, by scanning over possible models and forecasting SKA-level observational sensitivity, it is plausible that QCD-axion sensitivity will eventually be within reach of neutron star observations. In Fig.~\ref{fig:AxionPhoton_Closeup_AltColours} I have shown one possible projection for 100 hours of observation on the galactic centre magnetar~\cite{Witte:2021arp}.

\subsection{Axion birefringence}\label{sec:axionbirefringence}
Having gone already down to $\upmu$eV masses, let us now think about the possible signatures of much lighter dark matter axions coupled to the photon. The most important overarching concept I will discuss here is that of \textit{birefringence}. There are several types of birefringence that are relevant for axions, but the one we will look at here is the polarisation rotation effect that a background axion field has on photons propagating through it\footnote{A different type of polarisation-dependent effect that I am not mentioning here is the one caused by the fact that the axion only mixes with the component of the photon transverse to whatever external magnetic field it is propagating through~\cite{Maiani:1986md, Raffelt:1987im}. This effect is present even without any pre-existing axion abundance in the form of dark matter, and so it is the basis of some laboratory tests of the axion~\cite{DellaValle:2015xxa}, and some methods of detecting non-dark-matter astrophysical ALPs~\cite{Dessert:2022yqq, Yao:2022col, Song:2024rru,Fortin:2018aom,Fortin:2018ehg,Fortin:2021sst}.}.

I will follow the derivation presented in Ref.~\cite{Harari:1992ea} and consider a situation in which we have some linearly polarised electromagnetic wave that is propagating through a background axion field, $\phi(t,\mathbf{x})$. If we assume that the axion field is varying over spatial and temporal scales much slower than the photon's frequency and wavelength, we can then combine the photon and axion equations of motion Eqs.(\ref{eq:maxwell}) together yielding
\begin{equation}\label{eq:maxwell_biref}
    \begin{aligned}
& \square\left(\boldsymbol{E}+\frac{1}{2} g_{a\gamma} \phi \, \boldsymbol{B}\right)=\frac{1}{2} g_{a\gamma} \phi \square \boldsymbol{B} \, , \\
& \square\left(\boldsymbol{B}-\frac{1}{2} g_{a\gamma} \phi \boldsymbol{E}\right)=-\frac{1}{2} g_{a\gamma} \phi \, \square \boldsymbol{E} \, .
\end{aligned}
\end{equation}
If we assume $\phi\sim \text{const.}$, then at lowest order we find $\square \mathbf{E} = \square \mathbf{B} \approx 0$, and reveal that it is the mixed-up combinations $\boldsymbol{E}+\frac{1}{2} g_{a\gamma} \phi \boldsymbol{B}$ and $\boldsymbol{B}-\frac{1}{2} g_{a\gamma} \phi \boldsymbol{E}$ that approximately obey a wave equation, as opposed to $\mathbf{E}$ and $\mathbf{B}$ independently as in standard electromagnetism. If we now go and assume the value of $\phi$ is changing along the photon's propagation, albeit very slowly so the approximation remains valid, then the $\mathbf{E}$ field of some initially polarised wave will not propagate as a wave but will appear to rotate by some angle into $B$.

This polarisation rotation is known as birefringence, and it comes about because the left and right-handed circular polarisations have slightly different dispersion relations in the presence of a slowly-varying axion field~\cite{Carroll:1989vb, Carroll:1991zs}:
\begin{equation}
    \omega_{\pm} \approx k \pm \frac{1}{2} g_{a\gamma}\left(\dot{\phi}+\nabla \phi \cdot \hat{\mathbf{k}}\right) \, .
\end{equation}
Let's say we have some linearly polarised photon with a known polarisation angle at the point of emission $(t_i,\mathbf{x}_i)$, and then we detect that photon at some other spacetime point $(t_f,\mathbf{x}_f)$. Recall that any linear polarisation can be thought of as the superposition of the left and right-handed circular polarisations. So because one part of this superposition lags behind the other, when we measure the linear polarisation at $(t_f,\mathbf{x}_f)$ the angle that it will have rotated by, $\Delta \vartheta$, can be calculated by how much this lag accumulates over the photon's path, i.e.~\cite{Carroll:1989vb, Carroll:1991zs, Harari:1992ea},
\begin{equation}
    \Delta \vartheta = \frac{1}{2} \int_{(t_i,\mathbf{x}_i)}^{(t_f,\mathbf{x}_f)}\left(\omega_{+}-\omega_{-}\right) \mathrm{d} s = \frac{1}{2} g_{a\gamma}\big(\phi(t_f,\mathbf{x}_f) - \phi(t_i,\mathbf{x}_i) \big)\, .
\end{equation}
By evaluating this integral we see it is only the difference in the values at the points of emission and detection that count towards the rotation angle. The intuition for why the effect only cares about the endpoints is essentially because the photon is coupled to the \textit{derivative} of the axion field\footnote{This fact can be observed directly in Eqs.(\ref{eq:maxwell_biref}), but derivative interactions are ultimately a feature of the axion being a Goldstone boson.}. Integrating a derivative along a path yields only the difference between the values at each end of the path. See Ref.~\cite{Fedderke:2019ajk} for a more detailed discussion and an alternative derivation of this in the cosmological setting that will be relevant for a few of the examples mentioned below. The underlying reason for why the axion creates specifically a \textit{birefringent} effect on the photon on the other hand, is that the axion-photon interaction is parity-violating.\footnote{Note however that parity is also broken when photons propagate through a magnetic field, even in the absence of an axion---this also causes a rotation of the polarisation but here it is a well-known effect called Faraday rotation.}

This is an interesting effect that is also very nice observationally because there are many possible sources of photons that are already linearly polarised in some fashion.\footnote{This same birefringent effect is also the basis of a handful of direct detection experiments that are looking for axion dark matter in the low-mass regime, e.g.~\cite{Melissinos:2008vn, DeRocco:2018jwe, Obata:2018vvr, Liu:2018icu, Martynov:2019azm, Michimura:2019qxr, Oshima:2021irp, Oshima:2021ear, Fujimoto:2021mft, Martynov:2019azm, Heinze:2023nfb}.} If these photons are passing through axion dark matter then we expect there to be some $\Delta\phi \neq 0$ between the points of emission and detection and therefore a polarisation rotation. Additionally, because the axion dark matter field is oscillating at the frequency of the axion mass, so too would we expect this polarisation rotation angle to also oscillate at the same frequency if we were to measure it at multiple instances in time. Recall that we model axion dark matter as a classical field oscillating with amplitude $\phi \sim \sqrt{2\rho_a}/m_a$. It is because of this inverse mass scaling, that the best prospects are to use this effect to search for axions in the ultralight regime. 

\begin{figure*}[t]
    \centering
    \includegraphics[width=0.99\textwidth]{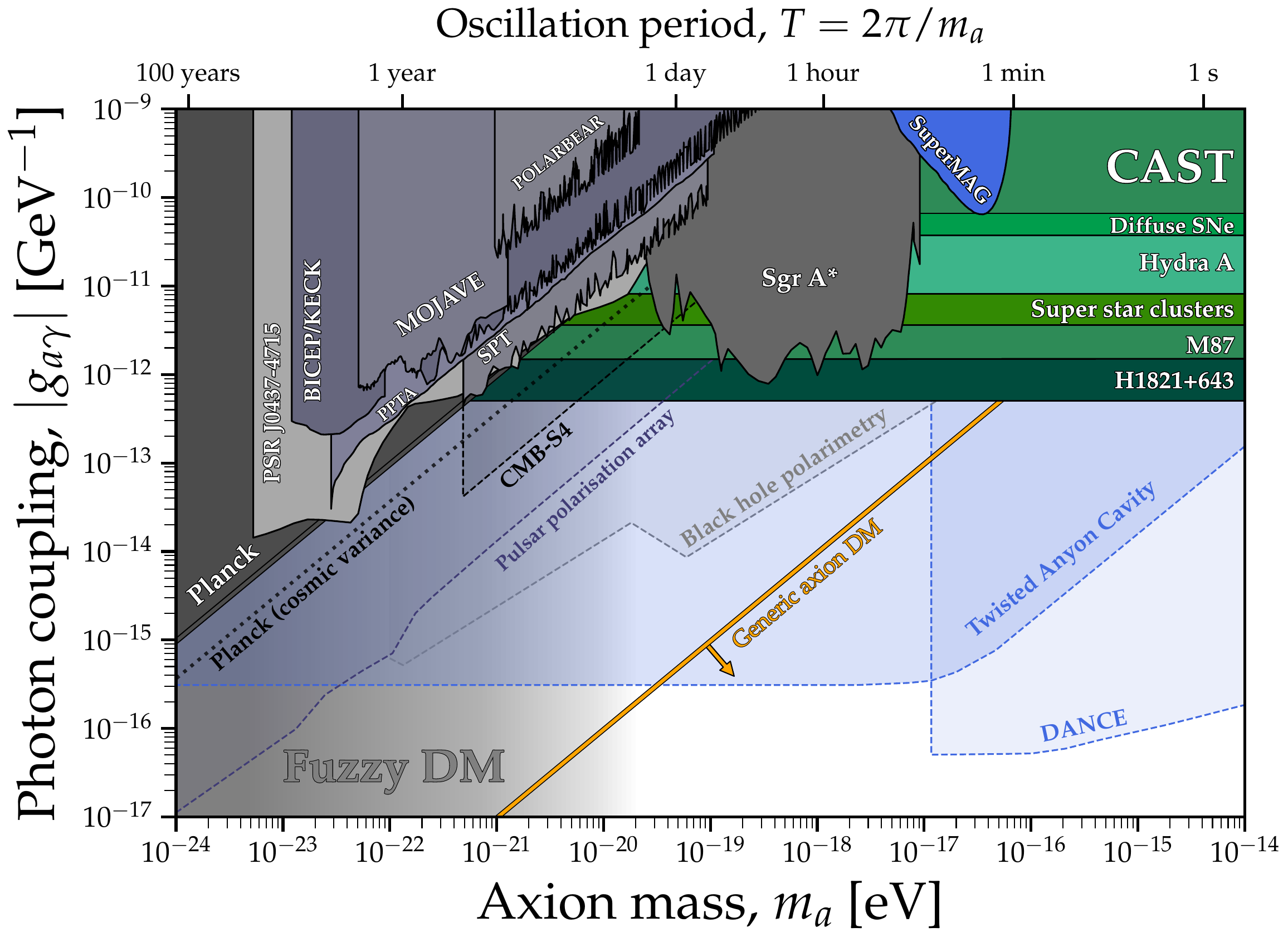}
    \caption{Closeup of the `ultralight' axion regime. The grey regions are all bounds based on the axion dark-matter birefringence effect discussed in Sec.~\ref{sec:axionbirefringence}. The green bounds come from searches that do not assume axions are dark matter---I am showing the bound from CAST~\cite{CAST:2007jps, CAST:2017uph} as well as various bounds on ALP-photon mixing in astrophysical B-fields~\cite{Calore:2021hhn, Calore:2020tjw, Wouters:2013hua, Dessert:2020lil, Marsh:2017yvc, Reynes:2021bpe}. SuperMAG is a direct search for dark matter around the Earth~\cite{Arza:2021ekq}, and I am also shown two projected limits for other proposed direct detection experiments with dashed blue lines~\cite{Bourhill:2022alm, Oshima:2023csb}. A ``generic axion dark matter'' model without an excessively enhanced photon coupling should fall below the orange line, as defined in Ref.~\cite{Dror:2020zru} and discussed at the end of Sec.~\ref{sec:axionbirefringence}.}
    \label{fig:AxionPhoton_Ultralight_with_Projections}
\end{figure*}

To search for axion dark matter using this effect we want to have polarised photons that are emitted and absorbed at points where we have some knowledge about the typical dark matter densities. Since we already know the density at the point of detection, i.e.~here, we get that one for free. I will now list a few examples of different searches that have been done using this effect in different contexts. All of these appear on the plot Fig.~\ref{fig:AxionPhoton_Ultralight_with_Projections} which zooms in on the ultralight regime. 

Several groups have considered pulsar timing arrays (PTAs) for this purpose. PTAs are large networks of continuously monitored nearby pulsars in the Milky Way. The plane of linear polarisation of the radio emission from these objects is observed to be essentially constant, meaning a bound can be placed on any extra oscillations in the angle caused by the changing axion field value between us and the various pulsars in the network~\cite{Caputo:2019tms, Liu:2019brz}. The bound from Ref.\cite{Castillo:2022zfl} is labelled PPTA---they used twenty pulsars from the Parkes Pulsar Timing Array of the Murriyang radio telescope in Australia, in conjunction with data on the Crab pulsar from QUIJOTE, a CMB experiment on the Canary Islands. The ultimate reach of the PTA approach is to think about pulsar \textit{polarisation} arrays, in which the polarisation information on different pulsars in the network is spatially and temporally correlated rather than considered separately as a measure of Faraday rotation~\cite{Liu:2021zlt}.

In Ref.~\cite{Yuan:2020xui} on the other hand, the authors used polarisation data from the Event Horizon Telescope project around the central supermassive black hole of the Milky Way called Sgr A$^\star$. Note that this bound requires knowledge of the dark matter density around the galactic centre. In this case they assumed the central halo consisted of a solitonic core surrounded by an NFW profile, which is the natural choice to make for this ultralight mass regime. The potential reach of this kind of probe is shown as ``black hole polarimetry'', taken from Ref.~\cite{Gan:2023swl}, which assumes that axion stars are able to grow around supermassive black holes.\footnote{Other works discussing polarisation data around supermassive black holes include Refs.~\cite{Wang:2023eip, Chen:2021lvo,Chen:2022oad,Chen:2019fsq} who consider the birefringence signal due to a cloud of axions generated around the black hole via superradiance.}

Many more constraints have been set using e.g.~the UV polarisation of radio galaxies~\cite{diSeregoAlighieri:2010ejm}, polarimetry of the starlight reflected off of protoplanetary disks~\cite{Fujita:2018zaj, Davydov:2023hgw} as well as an active fast radio burst~\cite{Wang:2024sdz}; and there are several further potential sources of polarisation have also been suggested like supernova remnants~\cite{Chigusa:2019rra}. I have not included these less-sensitive constraints or projections due to the already excessive clutter in Fig.~\ref{fig:AxionPhoton_Ultralight_with_Projections}. 

An issue with performing measurements of the birefringence angle for single objects is that there is the possibility for confusion with Faraday rotation as well as a fundamental systematic uncertainty arising from the fact that polarisation angles in radio astronomy have to be calibrated against something. A clever workaround to both of these issues was suggested by Basu et al.~\cite{Basu:2020gsy} which involves measuring the polarisation of multiple images in a gravitational lens system. Doing a \textit{differential} birefringence measurement (i.e.~the difference in the angle between two images) would cancel the unwanted contributions from Faraday rotation in the source and local environment as well as the calibration uncertainty. What would be left behind then are just the contributions from the axion-induced rotation, which---because the two spacetime paths taken by photons in the two images are slightly different---would not be the same as each other. A bound was set in Ref.~\cite{Basu:2020gsy} using a multiply imaged quasar, but sadly it is above the CAST limit. Note that the use of two images of a quasar means the contribution to the birefringence angle from the local dark matter is cancelled, leaving only a dependence on the axion field at the source. This would be different in something like a multiply-lensed fast radio burst which would be observed at several distinct instances in time, as suggested by Ref.~\cite{Gao:2023xbi}. Another bound set over extragalactic scales was the one using radio data collected for the MOJAVE project whose aim was to monitor the polarisation coming from the jets of some active galactic nuclei using the Very Long Baseline Array in Hawaii~\cite{Ivanov:2018byi}. 

Some of the best bounds set exploiting the ultralight axion birefringence technique are from CMB experiments. To explain, first let's recap the fact that the CMB has a well-studied polarisation signal due to Thomson scattering of photons on electrons when surrounded by the primordial temperature fluctuations. The various orientations of fluctuations in the polarisation direction on the sky are referred to as $E$ and $B$ modes, which are parity-even and parity-odd patterns respectively.\footnote{Note there is also a feature of the $EB$ correlation that is of broader interest to CMB cosmologists called ``cosmic birefringence''~\cite{Lue:1998mq}. This is related to, but not the same as, the effect I'm discussing here. Cosmic birefringence is a hypothetical all-sky static polarisation rotation of E-modes into B-modes. Although axion dark matter does not generate this, if the cosmic birefringence angle is nonzero that does still hint at the existence of a parity-violating interaction with photons, which could perhaps be generated by some configuration of an axion-like model. Nonzero cosmic birefringence in the \textit{Planck} data was actually claimed recently~\cite{Minami:2020odp}, but the hint is still somewhat up in the air~\cite{Diego-Palazuelos:2022dsq}. Dark-energy-like axions with masses between $10^{-33}$--$10^{-28}$~eV, if coupled to the photon, could provide a natural explanation of this observation, see e.g.~Refs.~\cite{Fujita:2020ecn, Gasparotto:2023psh, Gendler:2023kjt}.} The formalism for searching for axion dark matter using the CMB polarisation was detailed in Ref.~\cite{Fedderke:2019ajk}, and the two lines drawn on Fig.~\ref{fig:AxionPhoton_Ultralight_with_Projections} labelled ``washout'' and ``cosmic variance'' are from that work.\footnote{See also Refs.~\cite{Finelli:2008jv, Liu:2016dcg, Sigl:2018fba} for earlier papers on broadly the same idea but which suffered from some confusion about the nature of the axion birefringence effect (the differences are explained fully in Sec.~VI of Ref.~\cite{Fedderke:2019ajk}).} 

There are two ways the CMB polarisation can be used to constrain axion dark matter birefringence. They can be thought of as an early-time and a late-time effect. Both are expressed in the formula for the Stokes $Q$ and $U$ parameters of a polarised electromagnetic wave. These quantities measure the polarisation along two sets of axes offset by 45$^\circ$. The Stokes $Q$ and $U$ of the polarised CMB radiation measured in some direction are modified to~\cite{Fedderke:2019ajk},
\begin{equation}\label{eq:stokes}
    (Q \pm i U)(\hat{\boldsymbol{n}}) \to J_0\left(g_{a \gamma}\bar{\phi}(z_{\rm re})\right) \exp \left[ \pm 2 i\left(\frac{g_{a \gamma}}{2} \phi_0 \cos \left(m_a t+\beta\right)\right)\right](Q \pm i U)(\hat{\boldsymbol{n}}) \, ,
\end{equation}
where $J_0$ is a Bessel function of the 1st kind. The early and late time effects are encapsulated in the first and second quantities in this expression respectively.

The early-time effect comes from $J_0$, which takes as its argument $g_{a \gamma}\bar{\phi}(z_{\rm re})$ where $\bar{\phi}(z_{\rm re})$ is averaged axion field amplitude around recombination. Because it is a Bessel function, this piece will suppress the level of polarisation when $g_{a\gamma}\bar{\phi}$ is large. The reason behind this suppression is that the axion field will oscillate many times within the era of recombination when the CMB photons we observe coming from some direction on the sky were first emitted. Although in Sec.~\ref{sec:cosmologicalobservations} I talked about photon decoupling as though it happens at a single moment in time, this is not exactly true. The CMB photons are emitted from a surface that is not infinitesimally thin but has a thickness given by how long the era of recombination lasted. Although $z_{\rm re}$ is centred around a redshift of 1100, we observe CMB photons that have been emitted from a range of redshifts with a full-width at half maximum between 900 and 1200---a period lasting around 120,000 years. Once we average over all the possible birefringence angles experienced by different photons released at slightly different times, the net polarisation for that direction will end up suppressed ever so slightly from what we would expect in the absence of the axion.

We \textit{do} observe a polarisation signal clearly on the sky in the \textit{Planck} polarisation maps, so from that a bound can be drawn on axion couplings that are too large. This was done in Ref.~\cite{Fedderke:2019ajk}, by comparing how the E-mode polarisation and temperature correlation coefficients are modified due to Eq.(\ref{eq:stokes}) above:
\begin{equation}
C^{T E}_{\ell} \rightarrow J_0\big(g_{a \gamma}\bar{\phi}(z_{\rm re})\big)\,C^{T E}_{\ell} \, .
\end{equation}
So the washout is dependent on the quantity $(g_{a \gamma} \bar{\phi}(z_{\rm re}))^2$ alone---the dependence on the local dark matter field value cancels for this effect because we are integrating many photons emitted at different positions but all detected at the same position. A dependence on the axion mass is also present but only comes in because $\phi(z) \approx \sqrt{2\rho_a(z)}/m_a$. Fedderke et al.~\cite{Fedderke:2019ajk}'s analysis of the \emph{Planck} polarisation power spectrum under a minimal $\Lambda$CDM$+$1 parameter model, resulted in an upper limit of $g_{a \gamma} \bar{\phi}(z_{\rm re}) \lesssim 0.15$ (95\% CL). Converting this into an upper bound on the coupling as a function of the axion mass we get,
\begin{equation}
\begin{aligned}
    g_{a\gamma} \lesssim & \, 0.15 \frac{m_a}{\sqrt{2\rho_a(z_{\rm re})}} \, ,\\
    \lesssim & \, 9\times 10^{-14}\, {\rm GeV}^{-1} \left( \frac{m_a}{10^{-22}\,{\rm eV}} \right) \left( \frac{0.12}{\Omega_{\rm DM}h^2}\right)^\frac{1}{2} \left( \frac{1101}{1+z_{\rm re}} \right)^\frac{3}{2} \, .
\end{aligned}
\end{equation}
where the field value at recombination is obtained by unredshifting the present-day dark matter density back to $z_{\rm re} = 1100$.\footnote{This makes the usual assumption that the dark matter is described as a homogeneous oscillating classical field. However, if there are objects in the field like topological defects then we expect the impact on the CMB polarisation to be qualitatively different~\cite{Agrawal:2019lkr}.} Ultimately this bound is limited by the finite statistics available to us because there is only one CMB we can observe. This limit is called the ``cosmic variance''. The ultimate cosmic-variance-limited reach of the washout effect is shown as a dotted black line in Fig.~\ref{fig:AxionPhoton_Ultralight_with_Projections}. 

The second, late-time effect of axion dark matter on the CMB polarisation was pointed out also in Ref.~\cite{Fedderke:2019ajk} and then pursued in dedicated analyses by CMB experiments. In this case, the signal is an all-sky oscillation in the polarisation angle generated by the oscillation in the dark matter at the location of the Earth---the second piece in Eq.(\ref{eq:stokes}). The signal here appears not in the CMB power spectrum but rather in timeseries data, which is why it required dedicated analyses within the collaborations. Three CMB experiments have taken up this idea since it was proposed: BICEP/KECK~\cite{BICEPKeck:2020hhe, BICEPKeck:2021sbt}, the South Pole Telescope (SPT)~\cite{SPT-3G:2022ods}, and POLARBEAR~\cite{POLARBEAR:2023ric}. The latter has also very recently improved their best constraint by looking at polarisation oscillations towards the direction of the Crab nebula~\cite{POLARBEAR:2024vel}. Currently, the bounds set by CMB experiments do not improve much upon the \textit{Planck} washout limit, but unlike \textit{Planck}, there is room for improvement here with more data because the local effect is not cosmic-variance limited. A projection for the potential ultimate reach of the next-generation ground-based CMB surveys, CMB-S4, is also shown---taken from SPT's study~\cite{SPT-3G:2022ods}.

% Natural models
To round off this section on a more negative note: I will point out that you should expect to run into difficulty when coming up with models that populate this ultralight mass regime. To do it, we need to have an axion whose mass is suppressed by some huge scale but has a coupling to the photon that is not nearly as suppressed. It has been shown that many of the bounds discussed here are really constraining a part of parameter space that is hard to explain in a minimal setup~\cite{Dror:2020zru}. Reference~\cite{Poulin:2018dzj} quantified structure formation in a cosmology where a very light scalar evolves under an axion-like potential, $V \sim (1-\cos{\phi/f_a})^n$. By demanding that the CMB and matter power spectrum remain consistent with the levels of structure in $\Lambda$CDM on large scales, they find that there is a restriction on how late this axion can start oscillating ($z_{\rm osc}>9\times 10^4$), and therefore a lower bound on the scale $f_a$ for a given mass in the ultralight regime. The bound is:
\begin{equation}
    \begin{aligned}
f_a & \gtrsim \, \sqrt{\frac{2 \rho_{\mathrm{DM}}\left({\rm today}\right)}{m_a^2}}\left(1+z_{\rm osc}\right)^{3 / 2}, \\
 & \gtrsim \, 1.2 \times 10^{13} \, \mathrm{GeV}\left(\frac{10^{-20} \mathrm{eV}}{m_a}\right) .
\end{aligned}
\end{equation}
If we then assume that the photon coupling should also take on an axion-like form $g_{a\gamma} = C_{a\gamma} \alpha/2 \pi f_a$ where the dimensionless coupling constant $C_{a\gamma}$ is not permitted to be excessively enhanced beyond order-1 numbers, this rules out much of Fig.~\ref{fig:AxionPhoton_Ultralight_with_Projections} above the orange line as explainable with a generic model. There are workarounds, if one is willing to put in the model-building effort~\cite{Dror:2020zru,Ismail:2024zbq}, but they are far from generic.

\subsection{Signatures of miniclusters and axion stars}\label{sec:minicluster_signatures}
Here's where things get a little wacky. The preceding sections went through most of the standard ways to attempt an indirect detection of axion interactions in space. By `standard' what I mean is that you make no other assumption than there being some dark matter present in the object or environment that you have pointed your telescope at. Much of these assumptions are driven by wisdom gained within the context of the CDM paradigm. However, as discussed in Sec.~\ref{sec:miniclusters} in the context of the QCD axion, in Sec.~\ref{sec:ALPs} in the context of axion-like particles, and in Sec.~\ref{sec:ultralight} in the context of the ultralight axion, the {distribution of axion dark matter is not always expected to mimic CDM. 

When setting bounds on dark matter, there is usually the implicit assumption that its distribution on sub-galactic scales is smooth and homogeneous. But this is an assumption that is only backed up by simulations of vanilla and exactly collisionless CDM beginning from purely adiabatic perturbations. Axions, however, are not collisionless, and in the post-inflationary scenario, the field does not start from smooth initial conditions either. We did not account for the possibility that the axion has small-scale structures like miniclusters and axion stars in most of the earlier discussions. And yes, for many of those probes, this issue was of little consequence, but if we take the possibility that axion dark matter could possess beyond-CDM substructure seriously, then there is a longer list of additional signals that we could go and look for, as I will now discuss. Since we're adding a layer of speculation here, my aim in this brief section is to mention as many of these ideas as I can without going into excessive detail.

% Miniclusters
Let us first begin with miniclusters. If we imagine a situation in which the axion dark matter distribution on the AU--mpc scale is fundamentally granular---clumped into many small asteroid-to-planetary mass clumps---the consequences for present-day observables should be quite different. However, at least for direct detection the situation at first glance seems rather pessimistic. If we assume that a large fraction of the Milky Way's dark matter is clumped up inside miniclusters, the rate at which we would encounter one would be less than once every 100,000 years. So there is little hope that we will observe one directly in an experiment~\cite{Eggemeier:2022hqa} (although such encounters would be very interesting~\cite{Dandoy:2023zbi}, and I will talk about a caveat to this statement in the next section).

So what about indirectly detecting one? There have been many suggestions for how to do it. One extension of an idea introduced in Sec.~\ref{sec:axionconversion} is to look for axion miniclusters encountering neutron star magnetospheres~\cite{Tkachev:2014dpa, Edwards:2020afl, Witte:2022cjj}. Most of the same logic from the earlier discussion comes through here\footnote{Although of course when digging into the details there are differences in the way a compact clump of axions with a low-velocity dispersion would fall onto a neutron star and convert into photons compared to the much more dispersed and evenly distributed halo dark matter usually considered. I point you to Ref.~\cite{Witte:2022cjj} which looked at the axion-photon conversion in a neutron-star-minicluster encounter specifically.}, only now because miniclusters are spatially concentrated our signal becomes a \textit{transient} radio emission line around the axion mass, as opposed to a persistent one. Although the encounter rate of any one minicluster with any one neutron star is as low as it would be for the Earth, we can make up the statistics by observing a large population at once towards the galactic centre. The calculations of Ref.~\cite{Edwards:2020afl} suggest a rate of radio transient events of around 1-100 per day, each one lasting days--months at a time. 

% microlensing
Another proposed indirect detection technique is gravitational microlensing~\cite{Kolb:1995bu, Fairbairn:2017dmf, Fairbairn:2017sil}, wherein an axion minicluster in the halo would pass along the line-of-sight between us and a star, temporarily magnifying its brightness in a distinct pattern. This was originally thought to be a promising concept because there are many pre-existing astronomical time-domain surveys looking for such events caused by compact objects in the galactic halo like MACHOs or primordial black holes. Unfortunately, recent calculations incorporating the internal structures of axion miniclusters found that the majority of them may be too diffuse to generate a detectable magnification, except for inside a narrow window of parameter space~\cite{Ellis:2022grh}. This result was obtained for some N-body miniclusters that were well-fit by an NFW profile. For the majority of miniclusters in current simulations, however, the characteristic scale radius can not be resolved, so simulations of higher spatial resolution are required for a more in-depth assessment. I will remark, however, that results from the most recent N-body simulations suggest that a population of the miniclusters with masses above $10^{-12}~M_\odot$ develop central profiles that are steeper than NFW~\cite{Eggemeier:2024fzs}---so it may be too soon to totally dismiss microlensing as a way to search for miniclusters.

% tidal disruption
A word of caution is in order regarding all proposed indirect probes of axion miniclusters. There are a fair number of uncertainties that remain to be worked through and studied in more detail. For one, the fluffiness of the miniclusters makes them susceptible to tidal disruption~\cite{Tinyakov:2015cgg, Dokuchaev:2017psd, Kavanagh:2020gcy, Shen:2022ltx, Dandoy:2022prp, OHare:2023rtm, DSouza:2024flu}, which contributes another source of theoretical uncertainty\footnote{In contrast, this effect improves the prospects for \textit{direct} detection on Earth~\cite{OHare:2023rtm} as I will mention again in Sec.~\ref{sec:directdetection}.}, beyond the already large theoretical uncertainties inherited from the limited nature of both the early universe string simulations and N-body simulations that follow them. At the current time, further investigation into the structure growth and evolution in the post-inflationary scenario is critical if we want to have confidence in any future detection efforts.

% Axion stars
The other class of substructure are axion stars. These objects are distinct from miniclusters because they emerge from different physics. Axion stars, or solitons, are balanced against gravity by an outward pressure created by the field's gradient energy. These objects are simply one possible solution for any form of dark matter describable in terms of a classical field, and as a consequence, they are somewhat more generic than miniclusters, which emerge only in a specific cosmological setting. However, while they are generic solutions, the question of whether they form is less certain. For example, the seeding of axion stars inside QCD axion miniclusters and minicluster halos has been demonstrated in bespoke simulations~\cite{Eggemeier:2019jsu}, but it is not yet definitive if every minicluster will host one. On the other hand, in the context of ultralight dark matter, the formation of solitonic cores at the centres of galaxies is one of the model's primary features. However, because baryons dominate the potential in the central regions of galaxies, no firm conclusions can be drawn about them until extensive ultralight dark matter simulations include hydrodynamical physics.

Assuming for now that these objects \textit{are} abundant throughout the Universe, there are many possible intriguing signatures we could consider. Many of them are spiritually the same as those considered for miniclusters: gravitational lensing~\cite{Prabhu:2020pzm} and interactions with neutron star magnetospheres~\cite{Witte:2022cjj, Iwazaki:2014wka, Raby:2016deh, Dietrich:2018jov, Bai:2021nrs, Buckley:2020fmh, Kouvaris:2022guf, Kyriazis:2022gvw}---collisions with regular stars have also been considered~\cite{Eby:2017xaw}. There have also been numerous attempts to link them to fast radio bursts~\cite{Tkachev:2014dpa, Pshirkov:2016bjr, Clough:2018exo, Buckley:2020fmh, Iwazaki:2022bur, Di:2023nnb}. Of particular note are signatures arising from axion star ``explosions''. These can occur if stars exceed a critical mass (e.g.~through a merger) such that a runaway cascade of axion decay processes is triggered. This leads to the exponential dissipation of the star's energy into relativistic axions~\cite{Eby:2021ece, Fox:2023xgx}, photons~\cite{Amin:2020vja,Amin:2021tnq, Hertzberg:2020dbk,Chung-Jukko:2023cow}, or gravitational waves~\cite{Chung-Jukko:2024hod}, depending on which couplings are active.

There are a handful of model-dependent constraints people have drawn recently using these kinds of arguments. One is a constraint on axion self-interactions derived by assuming axion stars explode into relativistic axions~\cite{Fox:2023xgx}, thereby draining the cold dark matter abundance over time. Another is based on axion star explosions into photons~\cite{Escudero:2023vgv}. If these explosions go off at early times, they could trigger early reionisation of the Universe, which would have been spotted already in the optical depth to the CMB. This leads to a constraint on $10^{-14}$--$10^{-8}$~eV mass axions because the explosion into photons requires the axion mass to be above the plasma frequency of the surrounding medium or otherwise the decay of the axions inside the star is blocked.

\subsection{Direct detection}\label{sec:directdetection}

I want to finish a bit closer to home. Everything else I have written about here has to do with the imprints and signals of axion dark matter in cosmology and astrophysics, but after all, we live in a dark matter halo of our own---there is dark matter around us right now. If it turns out that we are correct and the dark matter is made of axions then we may even have a decent chance of directly detecting some on Earth inside a type of experiment called a \textit{haloscope}~\cite{Sikivie:1983ip}.

I will not touch on anything about the nature of those experiments--you can find out more about them here:~\cite{Irastorza:2018dyq}---but I will come very close. In any case, the general idea behind essentially all these experiments is the same: they are devices that aim to tap into the local oscillations in the axion field in some manner. So my goal with this final section is to use some of the concepts introduced in earlier sections and come up with a list of generic and largely experiment-independent behaviours we should expect this oscillating axion field to exhibit inside our Solar System. This is where the worlds of axion cosmology and axion experiments collide. 

\subsubsection{Dark matter in the solar neighbourhood}\label{sec:directdetection_DM}
Before focusing on the behaviour of the axion specifically, let us discuss how we expect dark matter to behave in our Solar System, independent of its particle nature.

Here are some things we know for sure: we live in a spiral galaxy around 8 kpc away from its centre. The Sun's orbit is, to good approximation, a circle, with a current orbital velocity of $248\pm 1$~km/s with respect to the central supermassive black hole~\cite{GRAVITY2019}. The Milky Way is surrounded by a dark matter halo which has a total mass somewhere between a ${\rm few}\times 10^{11}$ and a ${\rm few} \times 10^{12}$~$M_\odot$ (e.g.~\cite{Patel_MWmass, Eilers_rotationcurve, Fritz_MWmass, Koppelman_vesc, Roche_vesc, Monari_vesc}). The dark matter density at our location can be estimated to be in the range \mbox{$\rho_{\rm DM,\,local} = 0.3$--$0.5$~GeV~cm$^{-3}$}~\cite{deSalas:2020hbh}.

The parameter $\rho_{\rm DM,\,local}$ is of high priority for terrestrial searches for dark matter, so a lot of effort has been devoted over the last few decades to trying to nail down its value as precisely as possible. The local density can be extracted using several different techniques, but the principle behind most of them is to use the positions and kinematics of some tracer population of stars to infer the gravitational potential they are feeling. Then, by subtracting off models for the baryonic contribution to the potential from other stars, as well as the gas and dust, whatever remains ought to be the dark matter density. This latter point is important to emphasise because the gravitational potential at our position in the inner galaxy is dominated not by dark matter, but by baryons. 

In more detail, there are two general categories of technique for measuring $\rho_{\rm DM,\, local}$. One approach is to look at the problem globally, in which a large population of stars spread over kiloparsecs are used to construct a spatially-averaged rotation curve and mass model for the Milky Way (see e.g.~\cite{Ou:2023adg, Jiao:2023aci, Zhou:2022lar} for a few recent examples).\footnote{It has been confirmed recently that the Milky Way rotation curve is actually slowly declining towards large radii as opposed to remaining flat as expected for canonical dark matter halos~\cite{Jiao:2023aci}. This result impacts the inference of the total mass of the Milky Way halo but does not appreciably affect the local dark matter density.} Another approach is more local---consider just nearby stellar tracers and infer the local gravitational potential by, for example, modelling their vertical harmonic oscillations in and out of the disk plane. This latter technique provides the answer we need for direct detection but suffers larger systematic uncertainties due to the smaller volume of the sample and the requirement of an accurate subtraction of the baryonic gravitational potential. On the other hand, the rotation curve approach uses more data and is less influenced by these systematics, but only gives us an indirect idea of what the local dark matter density could be at our exact position. Reassuringly though, both techniques, when applied to the best stellar kinematic data available, yield similar answers within their quoted uncertainties: values in the range 0.3 to 0.5~GeV~cm$^{-3}$, see e.g.~Ref.~\cite{deSalas:2020hbh} for a summary and discussion of a few other techniques that have been developed.

With the recent map of nearly 2 billion stars surveyed by the \textit{Gaia} mission~\cite{GaiaDR2}, measurements of the dark matter density are much less limited by statistics than in the decades prior. The effects limiting improvements in $\rho_{\rm DM,\,local}$ are now much more subtle. For example, the role of systematic uncertainties in the baryonic density model is now central, and any departures from the usual simplifying assumptions like axisymmetry and equilibrium---due to, for example, the influence of the Milky Way's stellar bar---are simultaneously important while being challenging to incorporate into models.

I emphasise that $\rho_{\rm DM,\,local}$ is a fundamental systematic uncertainty that all searches for dark matter inside our Solar System are subject to. Although the convention in the field is to simply adopt a benchmark value for this parameter, that is a potentially fatal practice. Particularly in the case of the QCD axion where there is a defined point inside a vast parameter space that we are targeting. Given how painstaking it is to make progress through this parameter space, it is highly unlikely, once any range of masses and couplings is claimed to be excluded, that this region will ever be probed again by another experiment. It is painful to imagine a situation in which an axion was within reach of an experiment but simply missed it because the true dark matter density lay closer to the lower edge of its expected window than the upper edge. This is the reason why I rescaled all haloscope exclusion limits in Fig.~\ref{fig:AxionPhoton_Closeup_AltColours} from the convention in the field of $0.45$~GeV~cm$^{-3}$, to the more conservative $0.3$~GeV~cm$^{-3}$. To re-do this yourself, use the fact that experimental signals in the case of the axion-photon coupling are sensitive to the combination $g_{a\gamma}\sqrt{\rho_{\rm DM,\,local}}$.

As well as the local density, another ingredient to any dark matter signal in the Solar System will be the distribution of velocities. Speaking more generally though, in astrophysics, the two are combined into a full 6-dimensional phase space distribution function, $f(\mathbf{x},\mathbf{v})$. This object obeys the non-relativistic and collisionless Boltzmann equation:
\begin{equation}
\frac{\partial f}{\partial t}+\dot{\mathbf{x}} \frac{\partial f}{\partial \mathbf{x}}+\dot{\mathbf{v}} \frac{\partial f}{\partial \mathbf{v}}=0 \, .
\end{equation}
The density of particles at some point $\mathbf{x}$, e.g.~in the solar system, is defined by integrating out the velocity dependence,
\begin{equation}
    \rho(\mathbf{x})=\int \mathrm{d}^3 v \,f(\mathbf{x}, \mathbf{v})  \, ,
\end{equation}
whereas the distribution of velocities is found by integrating out the spatial dependence,
\begin{equation}\label{eq:distributionfunction_fv}
    f(\mathbf{v})=\frac{1}{M} \int \textrm{d}^3 x \,f(\mathbf{x}, \mathbf{v})  \, .
\end{equation}
We are assuming that the distribution function is normalised such that integrating over phase space gives back the total mass in particles, $M$,
\begin{equation}
    \int \textrm{d}^3 x \,\textrm{d}^3 v \,f(\mathbf{x}, \mathbf{v}) = M \, .
\end{equation}
To simplify things we deal only with steady-state solutions $\partial f / \partial t = 0$, which is appropriate for describing the Milky Way halo on the fleeting timescales that human beings will ever be able to observe it for. As we know, Poisson's equation connects the Newtonian gravitational potential to the density: \mbox{$\nabla^2 \Psi = - 4\pi G_N \rho$}, so if we have a model for the gravitational potential then we can figure out both the density and the velocity distribution in the Solar System at the same time.

In practice, we can only find analytic solutions to the Boltzmann equation under some fairly sweeping assumptions and symmetries. A good starting point, and a widespread baseline for dark matter searches, is the Standard Halo Model. We assume the dark matter halo is spherical, isotropic, and \textit{isothermal}---described by one spatially-independent temperature or equivalently, one velocity dispersion, $\sigma_v$. The Standard Halo Model is just about the simplest model we could have come up with that performs the basic job of a dark matter halo. We may depart from its core assumptions if we want, but it is nonetheless a good example to try out first.

It can be shown that the distribution function for a steady-state spherically symmetric system need only depend on two integrals of motion: the total energy and total angular momentum: $f(E, L)$. The assumption of isotropy, on the other hand, states that: $\langle v_i v_j \rangle = 0$ for all $i\neq j$, which also implies in this case that we have no net angular momentum, and so we are down to just $f(E)$.

The next step is to define a useful measure of energy, which is taken to be the relative energy per unit particle mass $\mathcal{E} = \Psi - \frac{1}{2} v^2$. The Standard Halo Model takes the ansatz,
\begin{equation}
    f(r,v) = \frac{\rho_0}{\left(2 \pi \sigma_v^2\right)^{3 / 2}} \exp \left(\frac{\Psi-\frac{1}{2} v^2}{\sigma_v^2}\right) \, ,
\end{equation}
where $\rho_0$ and $\sigma_v$ are constants that we may fix. Integrating the distribution function over velocity to obtain $\rho\propto e^\Psi$, we can then insert this into Poisson's equation to get,
\begin{equation}
    \frac{\mathrm{d}}{\mathrm{d} r}\left(r^2 \frac{\mathrm{d} \ln \rho}{\mathrm{d} r}\right)=-\frac{4 \pi G_N}{\sigma_v^2} r^2 \rho \, ,
\end{equation}
which sees solutions of the form,
\begin{equation}\label{eq:rhor_isothermal}
   \rho(r)=\frac{\sigma_v^2}{2 \pi G_N r^2} \, .
\end{equation}
Now integrating the distribution function over $r$ gives us the velocity distribution at a point,
\begin{equation}\label{eq:velocitydistribution_galactic}
    f(\mathbf{v}) = \frac{1}{\sqrt{2\pi \sigma^2}} \exp \left(-\frac{|\mathbf{v}|^2}{2 \sigma_v^2}\right) \, .
\end{equation}
You can work out the mass contained within some radius by integrating Eq.(\ref{eq:rhor_isothermal}), and then use that to work out the rotation speed of a circular orbit at that radius, in which case you will confirm that this halo generates a flat rotation curve, $v_c=\sqrt{2}\sigma_v = $~constant.

A few more modifications are needed in order to get the velocity distribution we need. The halo model is in the rest frame of the galactic centre, so to get the distribution we observe, we boost Eq.(\ref{eq:velocitydistribution_galactic}) into our frame of reference.
\begin{equation}
       f(\mathbf{v}) = \frac{1}{(2\pi \sigma^2)^{3/2}} \exp \left(-\frac{|\mathbf{v}+\mathbf{v}_{\rm lab}|^2}{2 \sigma^2}\right) \, ,
\end{equation}
where $\mathbf{v}_{\rm lab}$ is the velocity of the observer in their laboratory with respect to the galactic centre. In most (but not all!) situations, however, we don't have any sensitivity to the direction of the direction of the axion waves given by $\mathbf{v}$, but just the frequency of them, which depends on $v = |\mathbf{v}|$. So we require the speed distribution, which we can get by integrating out the angular dependence from the velocity distribution. For the boosted Gaussian this takes a Maxwell-Boltzmann-like form:
\begin{equation}
    f_{\text {lab }}(v, t)=\frac{\sqrt{2} v}{\pi \sqrt{\pi} \sigma_v v_{\mathrm{lab}}(t)} \exp\left(-\frac{v^2+v_{\mathrm{lab}}(t)^2}{2\sigma_v^2}\right) \sinh \left(\frac{v_{\mathrm{lab}}(t) v}{\sigma_v^2}\right) \, .
\end{equation}
In many other contexts in direct dark matter detection it is common to also truncate the speed distribution at the local escape speed. This in turn is related to the fact we must regulate the divergent total mass of the Standard Halo Model by cutting off the density profile at some radius. Since the escape speed can be measured by looking at the high-speed tail of some velocity distribution of stars, we take an empirically motivated value of something like $v_{\rm esc}\approx 490$-$580$~km~s$^{-1}$~(see e.g.~Refs.~\cite{Koppelman_vesc, Roche_vesc, Monari_vesc}). In any case, this is something of a tangent for this discussion since searches for the axion are not at all sensitive to this number.

As for the velocity $\mathbf{v}_{\rm lab}$, while we expect that this will primarily be on the order of the circular rotation speed, \mbox{$v_c \sim \mathcal{O}(200\,{\rm km\,s}^{-1})$}, the Solar System does not move on a precisely circular orbit around the Milky Way. Moreover, we as observers are not moving with the same velocity that the Sun is moving around the galaxy. In general, it is best to decompose $\mathbf{v}_{\rm lab}$ into a sum of several velocities that are all known to, at worst, $\mathcal{O}(1\,{\rm km\,s}^{-1})$ accuracy. They are:
\begin{equation}\label{eq:labvelocity}
    \mathbf{v}_{\text {lab}}(t)=\mathbf{v}_{\text {LSR }}+\mathbf{v}_{\text {pec }}+\mathbf{v}_{\oplus,\text{rev}}+\mathbf{v}_{\oplus,\,\text {rot }} \, .
\end{equation}
This velocity is time-dependent, but which components take on that time dependence depends on the coordinate system and the reference frame being used. When I give numerical values I assume they are written in the rest-frame of the galactic centre.

The first component, $|\mathbf{v}_{\rm LSR}| \approx 233$~km~s$^{-1}$, is the velocity of the Local Standard of Rest (LSR) and measures the bulk rotational motion of the disk in our neighbourhood, whereas $|\mathbf{v}_{\rm pec}| \sim 18$~km~s$^{-1}$ measures the peculiar motion of the Sun with respect to the LSR~\cite{Schoenrich}. The third and fourth velocities represent the velocities of the Earth revolving around the Sun, $|\mathbf{v}_{\oplus,\text{rev}}|\sim 30$~km~s$^{-1}$, and the observer rotating around the Earth's axis, $|\mathbf{v}_{\oplus,\text {rot }}|\approx 0.46$~km~s$^{-1}$. They will generate an annual and daily variation in $f_{\rm lab}(v)$ respectively. See Ref.~\cite{Mayet:2016zxu} for detailed formulae for these velocities in various coordinate systems.

The boost by $\mathbf{v}_{\rm lab}$ substantially enhances the flux of the dark matter particles arriving at Earth---a phenomenon known as the ``dark matter wind''. The dark matter wind is expected to be one of the most unambiguous signatures of any particle that originates from the \textit{halo} and not from the baryonic part of the galaxy, or the Solar System, or the Earth, or something else. So a measurement of some feature of this wind will be required before any claimed detection of a signal can be taken seriously as a discovery of dark matter. With that in mind, let me now summarise a few of the expected features of the dark matter wind from the perspective of a terrestrial observer.
\begin{itemize}
    \item \textbf{Annual modulation}: because $\mathbf{v}_{\rm lab}$ contains the velocity of the Earth around the Sun we expect the flux of incoming dark matter to vary (approximately) sinusoidally over the year, peaking in June.
    \item \textbf{Directionality}: $|\mathbf{v}_{\rm LSR}|$ is about an order of magnitude larger than any other velocity in Eq.(\ref{eq:labvelocity}), which means the flux of incoming dark matter peaks towards a direction that is tangent to our circular galactic orbit. This direction is a $90^\circ$ anti-clockwise rotation away from the galactic centre in the disk plane. If you look up a map of the constellations in galactic coordinates you will discover that this direction points towards Cygnus~\cite{Spergel:1987kx}.
    \item \textbf{Sidereal modulation}: in the rest-frame of an Earthbound laboratory, the velocity $\mathbf{v}_{\rm lab}$, will rotate around the sky with a period of one \textit{sidereal} day, which is about 3.9 minutes shorter than the 24-hour Solar day. The dark matter flux also undergoes a tiny daily oscillation at the level of $v_{\oplus,\,{\rm rot}} \sim 0.46$~km/s due to the relative motion of the Earth's surface with respect to the galaxy.
    \item \textbf{Gravitational focusing}: when the Earth is down-wind of Cygnus, the gravitational field of the Sun will bend the trajectories of dark matter with speeds close to the solar-system escape velocity at the Earth's location: $v\sim\sqrt{2G_N M_\odot/{\rm 1AU}}\sim 42.3$~km/s~\cite{Griest:1987vc, Sikivie:2002bj}. This causes a small shift in the speed distribution at these low speeds, as well as a few-percent enhancement in the dark matter density peaking around March~\cite{Alenazi:2006wu, Lee:2013wza, Bozorgnia:2014dqa, DelNobile:2015nua}.
\end{itemize}
To be clear, these features apply to \textit{any} particle dark matter candidate. I will extend this list with additional features specific to wave-like dark matter in the next section.

I am claiming that these features are generic, however as I have written them here, it may seem like many of them arise from what was an overly simplistic model for the Milky Way's dark matter halo. This may worry you because, unlike the dark matter's local density, we have no direct methods of measuring its local velocity distribution. So far we have only been able to make educated guesses about the expected properties of $f(\mathbf{v})$ using simulations and the Milky Way's \emph{stellar} halo as guides. The spherically symmetric and isothermal assumptions that form the basis of the Standard Halo Model are reasonable starting points for devising a benchmark model, but are unlikely to be accurate in detail, as summarised in Ref.~\cite{Evans:2018bqy}. In fact, there are several well-motivated ways in which the true dark matter distribution function in the solar neighbourhood could be more complex than this.

The halos of Milky Way-like galaxies emerging in simulations of $\Lambda$CDM are observed to be very smooth and largely featureless in their inner regions---well-modelled using standard density profiles like the NFW (keep in mind that we are reasonably far away from the galactic centre which is the region where we should be cautious when making statements about density profiles). However, the Standard Halo Model is spherical, whereas the Milky Way halo is \textit{known} to be triaxial~\cite{Iorio:2019}. Triaxial halos do emerge ubiquitously in simulations of $\Lambda$CDM halo formation, and this is important in our context because triaxial halos are also expected to exhibit some degree of overall figure rotation. This would counteract my strong statement above about directionality as a generic feature of the dark matter wind, as this implicitly relies on the assumption that the halo does not co-rotate in the same direction as the disk. 

Fortunately, while we expect the Milky Way halo to spin, the speed of this figure rotation inferred from simulations may only be as high as a few tens of degrees over Gyr timescales, assuming there haven't been large numbers high-angular-momentum merger events in the galaxy's recent past~\cite{Bailin:2004rz, Bryan:2007ap, Vogelsberger:2014dza, Ash:2023fwc}. The slow rotation of the Milky Way halo is also backed up by observational tests of rotation of the Milky Way's stellar halo, which spins at $\sim 20$~km/s~\cite{Deason2017_spin}; as well as attempts to constrain anomalously fast rotation in the dark matter halo itself using the geometry of stellar streams~\cite{Valluri:2020lsq}. Put together, this evidence suggests that the assumption that the dark matter flux points back towards Cygnus is likely very robust.%, after all this is a feature of our motion through the halo more so than anything about the halo itself, and our galactocentric velocity vector is known very precisely. A daily sidereal modulation on the other hand is in some ways even more robust than this for the same reason, in fact any signal with a period of a solar day must be related to the solar rest-frame by definition. 

Whether or not the velocity distribution is well-modelled as a Gaussian is more up for debate. There has been a long history of extracting simulation-inspired local dark matter velocity distributions, see e.g.~\cite{Lacroix:2020lhn, Lentz:2017aay, Necib2019_FIRE, Poole-McKenzie:2020dbo} for a very small sample. To highlight one recent example, Ref.~\cite{Staudt:2024tdq} has taken Milky Way analogues in the FIRE suite of zoom-in hydrodynamic simulations to infer a local speed distribution that is roughly Maxwellian but with a dispersion of $\sigma_v = 161$~km~s$^{-1}$, a peak speed of $250$~km~s$^{-1}$, and a high-speed cutoff of $v_{\rm esc} = 470$~km~s$^{-1}$.

Of course, a simulated galaxy is not \textit{the} Milky Way, and so many of the features peculiar to the exact merger history our own galaxy has undergone in its recent or not-so-recent history will not be reflected. It turns out that a somewhat blurry view of this history is precisely what the recent \textit{Gaia} survey, and the field of `galactic archaeology' is able to provide us with. In particular, there is a population of stars that themselves orbit within the halo of our galaxy called the \textit{stellar} halo. While this halo population makes up less than 1\% of all the stars in the galaxy, their distribution is nonetheless flush with evidence about the Milky Way's assembly history.

One of the most important revelations of the \textit{Gaia} era is that the inner halo of the Milky Way is rife with phase space substructure, see e.g.~\cite{HelmiReview, Naidu2020, KoppelmanHelmiStreams, MyeongSequoia}. Satellites in the process of being tidally disrupted have been visible stretching across the sky for many years ~\cite{Belokurov:2006ms}, but only thanks to the quantity and quality of the \textit{Gaia} data can we map this structure throughout the 6-dimensional phase, allowing even objects that are invisible when simply counting stars to be picked up. But while dark matter halos are built up from a large number of mergers---including with many small subhalos that never hosted substantial numbers of baryons to begin with~\cite{Purcell2007}---the bulk of the \textit{stellar} halos of Milky Way-like galaxies on the other hand are built from a relatively small handful of significant merger events~\cite{Cooper2010, Monachesi2019, Fattahi2020}. Moreover, and thanks chiefly to \textit{Gaia}, the evidence is now overwhelming that the inner stellar halo of the galaxy is dominated by one merger specifically: the \textit{Gaia Sausage-Enceladus} (GSE)~\cite{HelmiEnceladus, Be18}. 

The debris of the collision with the progenitor dwarf galaxy that formed the GSE is distributed all across the Milky Way and is marked by its peculiar kinematics and chemistry. We see it today as a population of halo stars in a triaxial figure~\cite{Iorio:2019} out to around 20 kpc from the galactic centre~\cite{DeasonApocenterPileup} and on highly radial (i.e.~eccentric) orbits. The inferred stellar mass of this primordial dwarf galaxy is around $\sim 10^8~M_\odot$~\cite{Naidu:2021snw, Lane2023}, massive enough to host a decent number of its own globular clusters~\cite{MyeongwCen, Callingham2022}, which means a substantial dark matter halo was likely also deposited into the halo as well due to known correlation between the number of globular clusters and the mass of the halo they orbit~\cite{Dornan2023}. See Ref.~\cite{Deason_review} for a recent historical account of how evidence for this structure built up over the years from the pre-\textit{Gaia} era until now.

Because of the reasons mentioned above, the Gaia Sausage-Enceladus will most likely be less noticeable in the dark matter distribution compared to the stellar halo. However, the kinematics of this accreted material are so distinct from the rest of the Milky Way's halo that there is a strong implication that our assumption of an isotropic velocity distribution will not be correct. A simple way to model it would be to apply an anisotropy in the galactic radial direction (i.e.~a wider distribution along an axis perpendicular to the DM wind)~\cite{Necib2019_SDSSGaia, Evans:2018bqy}. Evidence from simulations~\cite{Naidu:2021snw, Bozorgnia:2019mjk, Necib2019_FIRE}, as well as halo-shape arguments~\cite{Evans:2018bqy}, both hint that something at the level of 20\% of the Milky Way's DM was brought in by the GSE at most. I will also point out that the overall shape of the stellar halo is observed to be triaxial and tilted with respect to the galactic disk~\cite{Iorio:2019}, which is a configuration that can only have remained stable for so many Gyrs if the dark matter halo is also tilted and triaxial~\cite{Han22_tilt, Han23_tilt, Han23_tilt2}, potentially as a response to the impact with the GSE~\cite{Dodge_tilt}. This fact would be expected to influence the relative density of dark matter at the solar position as well as generate anisotropy--- effects that push us further towards reconsidering the basic assumptions of the Standard Halo Model.

Fortunately though, unlike many other types of dark matter candidates, the consequences of the various forms of kinematic substructure or departures from spherical symmetry are expected to be rather mild for axion experiments~\cite{OHare:2018trr, OHare:2019qxc}. Anisotropies would likely only show up if the velocity distribution or the annual modulation signal were measured at a relatively precise level~\cite{Knirck:2018knd, Foster:2020fln}. So while the \textit{Gaia} era may have completely transformed our view of our galaxy's structure and evolution, for the time being, a model for the velocity distribution in the form of a simplified isotropic Gaussian, remains a sensible enough assumption until the axion is discovered~\cite{Evans:2018bqy}.

\subsubsection{Axion dark matter in the solar neighbourhood}
So far this is all fairly standard stuff for dark matter detection, but when it comes to axions we are sort of talking the wrong language. We want to model the axion as a classical field made up of lots of waves with dispersion relation: $\omega^2=|\mathbf{p}|^2+m_a^2$. So we ought to somehow make the connection between the particle description of a dark matter distribution and some equivalent statement about the Fourier modes of an oscillating classical field. A suitable way to do this is to first write the value of the real axion field at a spacetime point as a sum over waves with complex Fourier amplitudes $\phi(\mathbf{p})$:
\begin{equation}\label{eq:fourierdecomp}
\phi(\mathbf{x}, t)=\sqrt{V_{\odot}} \int \frac{\mathrm{d}^3 \mathbf{p}}{(2 \pi)^3} \frac{1}{2}\left[\phi(\mathbf{p}) e^{i\left(\mathbf{p} \cdot \mathbf{x}-\omega t + \beta_{\mathbf{p}}\right)}+\phi^*(\mathbf{p}) e^{-i\left(\mathbf{p} \cdot \mathbf{x}-\omega t + \beta_{\mathbf{p}}\right)}\right] \, ,
\end{equation}
where $\beta_{\mathbf{p}}$ is some arbitrary phase which is assigned to each momentum. The energy density in a scalar field is $\rho = \frac{1}{2}\dot{\phi}^2 + \frac{1}{2}(\nabla \phi)^2 + \frac{1}{2} m_a^2 \phi^2$, which can be used to work out the spatially averaged density:
\begin{equation}
    \bar{\rho}_a=\frac{1}{V_{\odot}} \int_{V_{\odot}} \mathrm{d}^3 \mathbf{x}\, \rho_a(\mathbf{x})\approx\int \frac{\mathrm{d}^3 \mathbf{p}}{(2 \pi)^3} \frac{1}{2} \omega^2|\phi(\mathbf{p})|^2 \, .
\end{equation}
By comparing this with the other definition of the spatially averaged density from a distribution function, Eq.(\ref{eq:distributionfunction_fv}), we can make the connection between the velocity distribution and the amplitudes of a Fourier decomposition,
\begin{equation}
    |\phi(\mathbf{p})|^2 \approx (2\pi)^3 \frac{2\rho}{m^2_a} f(v) \, .
\end{equation}

\begin{figure*}[t]
    \centering
    \includegraphics[width=0.99\textwidth]{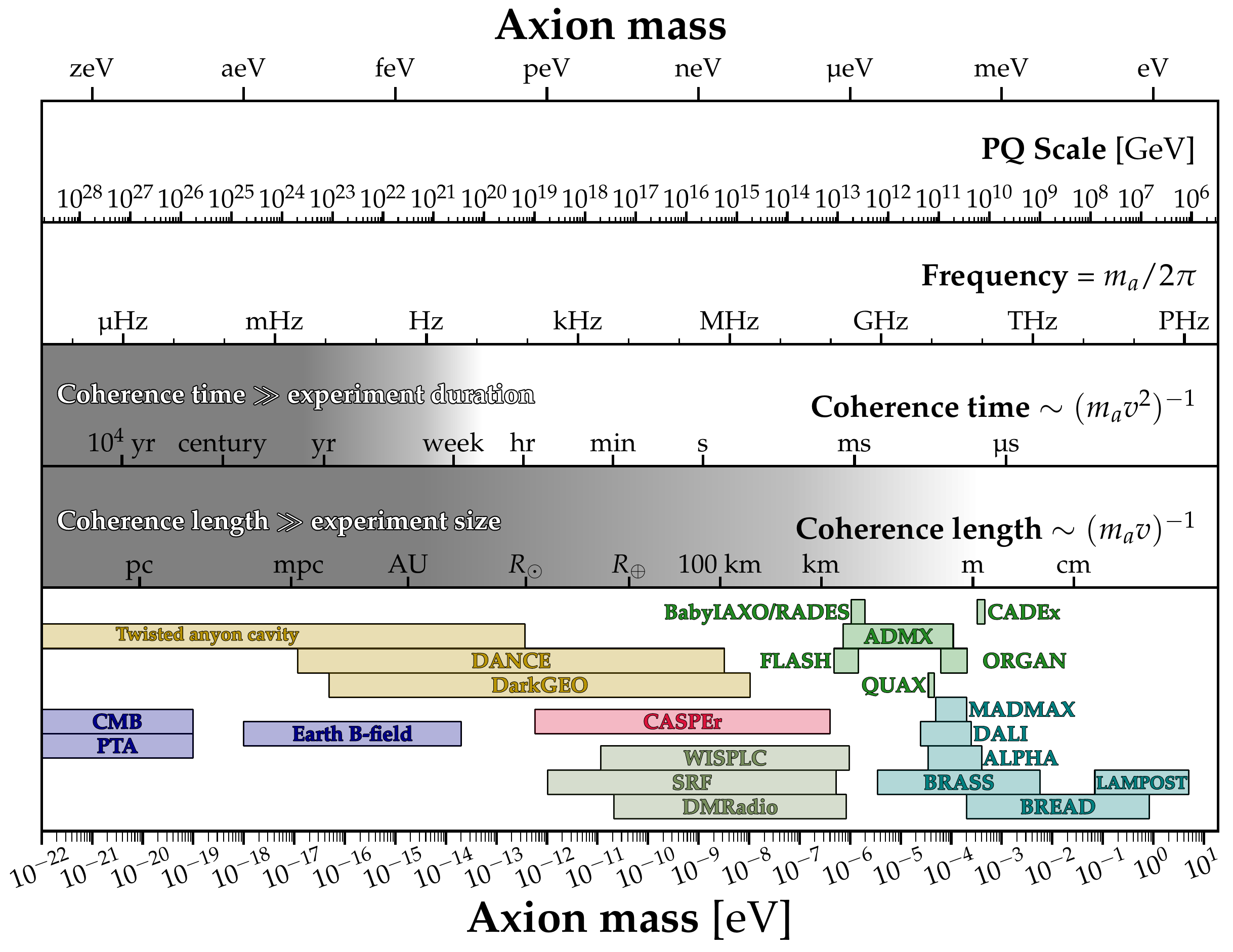}
\caption{A cheatsheet for comparing a dark matter mass with its oscillation frequency, coherence length, and coherence time, assuming $v\sim300$~km/s. Note that the latter two quantities are not rigorously defined and should be viewed as the approximate length and timescales where the loss of coherence due to the distribution of dark-matter velocities will be important. For comparison, I have shown the axion mass windows targeted by the following haloscopes: Twisted Anyon Cavity~\cite{Bourhill:2022alm}, DANCE~\cite{Michimura:2019qxr}, DarkGEO~\cite{Heinze:2024bdc}, CASPEr~\cite{JacksonKimball:2017elr}, WISPLC~\cite{Zhang:2021bpa}, SRF~\cite{Berlin:2020vrk}, DMRadio~\cite{DMRadio:2022pkf}, BabyIAXO-RADES~\cite{Ahyoune:2023gfw}, AMDX~\cite{Stern:2016bbw}, FLASH~\cite{Alesini:2023qed}, CADEx~\cite{Aja:2022csb}, ORGAN~\cite{McAllister:2017lkb}, MADMAX~\cite{Beurthey:2020yuq}, DALI~\cite{DeMiguel:2023nmz}, ALPHA~\cite{ALPHA:2022rxj}, BREAD~\cite{Liu:2021pei} and LAMPOST~\cite{Baryakhtar:2018doz}. See Ref.~\cite{AxionLimits} for data and code to reproduce this figure.}
    \label{fig:DM_cheatsheet}
\end{figure*}

To make progress from here we have to think about a measurement of the field $\phi(\mathbf{x},t)$, which will necessarily take place over some finite temporal and spatial scale. Typically, this will be done by measuring the field value over a volume as a function of time and then taking a discrete Fourier transform of that timeseries. The frequency resolution in $\omega$ of this discrete Fourier transform will be related to the duration of that temporal sample, i.e.~$\delta \omega = 2\pi/T$. 

An important fact that emerges when thinking about the expected spread in axion momenta $\phi(\mathbf{p})$ is the idea that when these samples are short enough in time or small enough in distance, the oscillations will appear coherent. For example, if we measure the field over some short distance $\Delta x$, then any two axion momenta separated by anything smaller than $\Delta p<2\pi/{\Delta x}$ will fall within the same momentum bin. Similarly, if we measure the field over a timescale $\Delta t$, modes separated by values smaller than the $\Delta \omega < 2\pi/\Delta t$ will also end up in the same bin. An equivalent statement is that two modes spaced by \textit{larger} momentum or frequency separations than this will go out of phase with each other by more than $2\pi$ across the measurement, and so we can tell them apart.

So we can now ask, given that the dark matter velocity distribution \textit{does} have a spread in momentum and frequency given by \mbox{$\Delta p = m_a \Delta v \sim m_a \sigma_v$} and \mbox{$\Delta \omega = m_a v \Delta v \sim m_a v \sigma_v$}, over what length or time is this spread apparent? Equating these expressions with the ones in the previous paragraph we see this is simply when,
\begin{equation}
   \Delta x \gtrsim \frac{2 \pi}{m v} \, , \quad \Delta t \gtrsim \frac{2 \pi}{m v \sigma_v}
\end{equation}
which, as mentioned several times already, are the coherence time and coherence length for the field. Here, $v$ should just be thought of as a `typical' velocity, e.g.~the mean speed coming from $f(v)$, whereas $\sigma_v$ is the velocity spread. In the Standard Halo Model, these are of the same order of magnitude, but there are instances where this may not be the case. 

The notion of the coherence time and coherence length has been important in many different contexts so far, and in the context of direct detection, they encapsulate another important idea. What they tell us is the typical length and time over which the axion field can be modelled as a single, monochromatic and spatially coherent oscillation:
\begin{equation}
    \phi=\phi_0 \cos (\omega t-\mathbf{p} \cdot \mathbf{x}+\beta)
\end{equation}
For comparison, I show Fig.~\ref{fig:DM_cheatsheet}, which compares the size of the coherence length and coherence times across the axion mass range with some experimentally ``reasonable'' scales, i.e.~experiments on the scale of metres in size and data-taking durations not exceeding several weeks. For masses $m_a<10^{-4}$~eV the field within a metre-scale experiment will be absolutely spatially coherent, whereas for masses $m_a<10^{-15}$~eV, it can be modelled as monochromatic as well. For higher masses than these limits, a typical integration time will span many coherence times and more modes from across the velocity distribution will be sampled. For masses above $m_a>10^{-4}$~eV, the axion field may also not be coherent across the entire experimental volume, leading generally to a mild direction-dependent suppression to the signal integrated over that volume~\cite{Irastorza:2012jq, Knirck:2018knd}.

Another important issue that arises from Eq.(\ref{eq:fourierdecomp}) relates to the measured \textit{amplitude} of the field. We know that on average $\langle \phi \rangle = \sqrt{2\rho_{\rm DM,\,local}}/m_a$, because this is the value that we engineered the field's energy density to have. However, within any one coherence volume, the amplitude will not take on this averaged value. The reason for this lies in $\beta_{\mathbf{p}}$. The field at a single point and time is still the sum over many different modes, however in the situation we are interested in, all of these modes should be assumed to have unrelated \textit{phases}---i.e.~$\beta_{\mathbf{p}}$ for each mode is completely random. The reason for assuming random phases is linked to the implicit assumption that we make in direct detection that the dark matter around us is spatially homogeneous. If there was any strong phase coherence in $\beta_{\mathbf{p}}$, this would enhance the amplitude of the field and hence constitute an overdensity. So, when we sum up all of the modes under the assumption of homogeneity, strictly we must do so with randomly chosen phases for every mode. This may still lead to the possibility of over or underdense patches, but these are purely statistical in nature and occur due to constructive or destructive interference---similar to the ``granules' discussed in the context of fuzzy dark matter in Sec.~\ref{sec:ultralight}.\footnote{In fact if the density of dark matter happened to be larger than anticipated in the solar system, these density fluctuations could be detected purely gravitationally (i.e.~even if all other couplings were suppressed) using gravitational wave interferometers~\cite{Kim:2023pkx}.}

Let us now see what the distribution of amplitudes will be. I will sketch the basic idea, but you can go to Ref.~\cite{Foster:2017hbq} if you want a mathematical proof. First, imagine we are measuring the field for periods well within the coherence time. In this case, we are summing up many modes $\sum \cos{(\omega_i t + \beta_i)},$ all of which appear to have practically identical frequencies within the measurement time: $\omega_i \approx m_a$; but where each phase $\beta_i$ is random. So while the frequency of this sum will still be $m_a$, because of the incoherent sum over many waves with random phases, the final amplitude we measure will be randomly drawn and suppressed from down from the extreme upper limit of $N\times A$, where $A$ is the amplitude of one of the waves and $N$ the number of them. It can be shown that the measured amplitude arising from this incoherent sum will follow the statistics of a Rayleigh distribution: $p(A) = (2A/N)e^{-A^2/N}$. 

This logic can be applied when the measurement extends longer than the coherence time too. In this case, the Fourier spectrum is split across more finely spaced bins, but the modes going into each bin still add up incoherently and so each one gets a random amplitude drawn from a Rayleigh distribution. The only difference in this case is that the value of $N$ for each bin is different because the modes take on some distribution of frequencies given by $f(v)$ where $v \approx \sqrt{2(\omega/m_a-1)}$. At the level of the Fourier power spectrum, which is what is usually measured, the distribution of each bin has the statistics of a \textit{squared} Rayleigh variable, which is the exponential distribution. 

\begin{figure*}[t]
    \centering
    \includegraphics[width=0.99\textwidth]{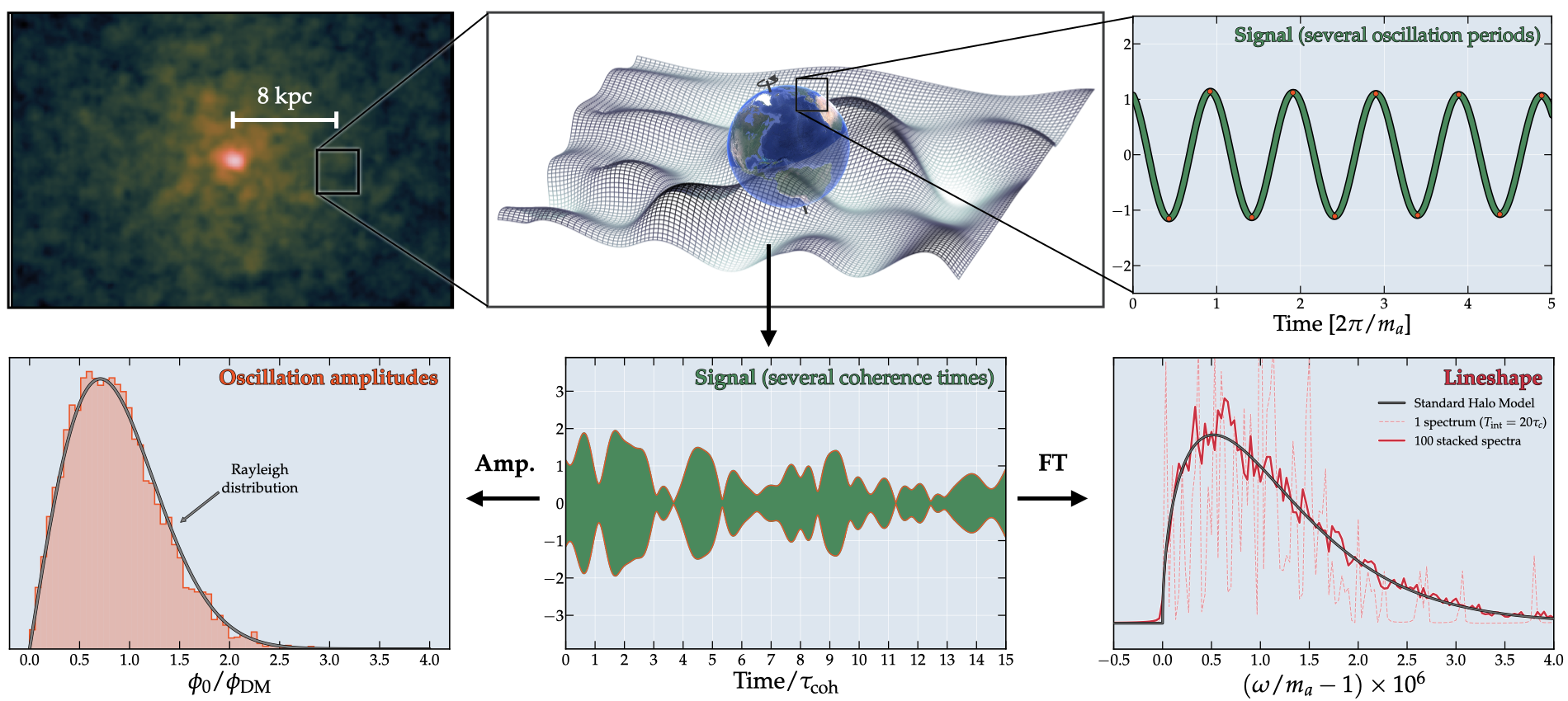}
    \caption{Illustration of the stochastic properties of a wave-like dark matter signal observed on Earth. These are obtained via a Monte Carlo simulation of the classical field assuming random phases and the dark matter's Fourier mode distribution implied by the Standard Halo Model's velocity distribution.}
    \label{fig:stochastic}
\end{figure*}

In the limit of a measurement lasting much longer than the coherence time, the full power spectrum has the following general form~\cite{Derevianko:2016vpm, Foster:2017hbq}:
\begin{equation}\label{eq:lineshape}
    P(\omega) = P_0 \, \varepsilon_\omega f(\omega) =  P_0 \,\varepsilon_\omega  f(v) \frac{\mathrm{d}v}{\mathrm{d}\omega} =  P_0 \,\varepsilon_\omega \left. \frac{f(v)}{m_a v}\right|_{v=\sqrt{2(\omega / m_a-1)}} \, ,
\end{equation}
where $P_0$ is some arbitrary signal strength that will depend on the specifics of the experiment as well as the dark matter density and axion coupling. The values of $\varepsilon_\omega$ are the randomly drawn amplitudes. Assuming there are no phase correlations between Fourier modes, i.e.~there are no spatial overdensities in the field nearby, then at every value of $\omega$ in the discrete Fourier transform of some sample of the field, a different $\varepsilon_\omega$ is drawn from an exponential distribution. 

Although this intrinsic stochasticity suppresses the power in each bin, if the measurement times are not too long then it can be suppressed by stacking multiple power spectra together. Unfortunately, for very low axion masses this procedure won't be possible over any realistic timescale, and so these random amplitudes can severely impact the expected signal statistics~\cite{Centers:2019dyn}. In fact, for experiments operating in the ultralight regime, entire measurement campaigns may fit inside a single coherence time. The full power spectrum is then only contained in a single frequency bin with one randomly drawn amplitude, and it is possible to get unlucky where, by random chance, you draw a very small number. 

These various issues are expressed in Fig.~\ref{fig:stochastic}, which shows how the signal on the timescale of the axion period $2\pi/m_a$ looks like a single wave, whereas over many coherence times, there is a slow variation in the amplitude and frequency. The Fourier transform reveals the frequency dependence $f(\omega)$, called the \textit{lineshape}, whereas measuring the values of the amplitude within each oscillation period (shown in orange), pulls out the Rayleigh distribution.

\begin{figure}
        \centering
        \includegraphics[width=0.99\linewidth]{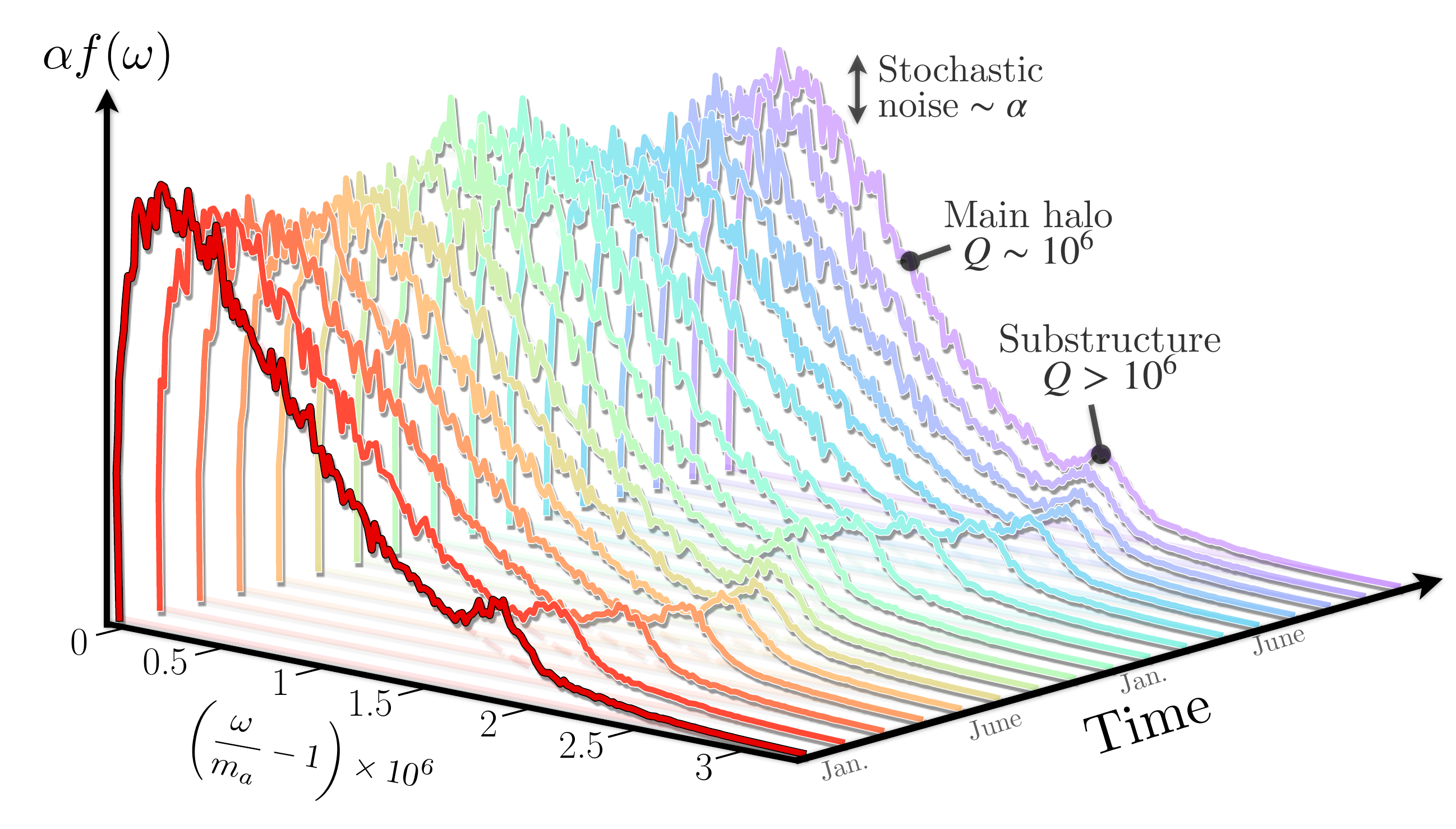}
        \caption{The axion lineshape $f(\omega)$ as it would be observed over a duration of two years. The typical width of the lineshape for a Maxwellian speed distribution is around $Q = m_a/\Delta \omega \sim 10^6$. The annual modulation of the Earth's velocity acts to vary the width of the lineshape over the course of the year. Any substructure in the local velocity distribution would appear as a distinct peak in the lineshape, with a higher value of $Q$, and with its own phase of annual modulation. This plot is based on a figure I originally made for Ref.~\cite{Boutan:2024sxz}.}
        \label{fig:lineshape}
    \end{figure}
    
We see that all of the characteristics of the dark matter wind mentioned in Sec.~\ref{sec:directdetection_DM} are encapsulated in the lineshape $f(\omega)$~\cite{Krauss:1985ub, Turner:1990qx, OHare:2017yze, Lentz:2017aay}. In Fig.~\ref{fig:lineshape} I show how the lineshape measured at multiple instances in time would be expected to vary. From the variation in $\mathbf{v}_{\rm lab}(t)$, we expect the lineshape to get slightly narrower in December when we move with the dark matter wind, and broader in June when we move against it~\cite{Ling:2004aj}.

The lineshape in this simple example is simply related to the dark-matter speed distribution transformed into a distribution of frequencies. This form of the lineshape is relevant for the vast majority of experiments looking for axions, but not all of them. One notable exception applies to any experiment that operates over length scales larger than the axion's coherence length~\cite{Irastorza:2012jq, Knirck:2018knd}, or equivalently if multiple axion signals are measured and correlated over a spatial extent that extends outside the axion's coherence length~\cite{Foster:2020fln}. Another exception applies to experiments that don't couple directly to $\phi$ but to $\nabla \phi$, which pulls out a factor of $\mathbf{v}$~\cite{Gramolin:2021mqv, Lisanti:2021vij, Foster:2023bxl}. The latter is the case for experiments probing axion-fermion couplings, e.g.~Refs.~\cite{JacksonKimball:2017elr, Garcon:2019inh, Lee:2022vvb, Gao:2022nuq}. In all of these exceptions, the fact that there are effects that depend on the vector $\mathbf{v}$ implies that they are sensitive to the directionality of the dark matter wind, and so will be subject to a strong modulation over a period of a sidereal day due to the rotation of the experiment with respect to Cygnus.

So what about the various assumptions we made in Sec.~\ref{sec:directdetection_DM} about the dark matter halo? Do we expect these to apply to wave-like dark matter? Let us first address the issue of homogeneity. While it is generally believed that there should not be any fine-grained fluctuations in the dark matter density on the mpc scales probed by experimental campaigns, it is worth recalling why we believe this to be the case. Dark matter halos are formed from a complicated folding procedure of an initially smooth 3-dimensional sheet embedded in 6-dimensional phase space. The density at a point in some inner part of a halo must, therefore, be set related to some huge number of folds of this sheet, which physically corresponds to innumerable streams of dark matter particles with identical velocities. All of these streams sum up to make the virialised halo that, for most practical purposes, would appear perfectly smooth both spatially and kinematically~\cite{Vogelsberger2011}. These objects are referred to as fundamental streams (to distinguish them from the streams originating from the tidal disruption of accreted subhalos), and they are a generic---albeit probably unobservable---consequence of halo formation under pure cold dark matter.

Before I return to that idea, we must also think about the ways that axions differ from purely cold dark matter. For example, in post-inflationary axion and ALP scenarios we have the expectation that the primordial density sheet is not as smooth as in cold dark matter. In particular we expect there to be AU-scale clumping of axions into miniclusters whose subsequent mergers will boost the level of substructure in axion dark matter beyond that expected from the adiabatic perturbations. In the case of the QCD axion, this enhancement emerges around halo masses from the Earth mass down to asteroid masses~\cite{Vaquero:2018tib, Eggemeier:2019khm}. This implies yet another modification to the picture of a smooth DM distribution, to the point that in the worst-case scenario in which the vast majority of axions end up inside miniclusters, there may be no dark matter around us to detect. However, this pessimistic conclusion is moderated by several effects. Firstly, it has been shown in several N-body simulations that around 10-20\% of the axions never end up bound inside miniclusters, and instead fill the pc-scale minivoids that separate them in the galaxy. Additionally, while axion miniclusters are compact, they are not so tightly bound for them to survive tidal disruption~\cite{Tinyakov:2015cgg, Dokuchaev:2017psd, Kavanagh:2020gcy}. As any given minicluster orbits the galaxy, it will come within a few parsecs of many millions of stars. And for a subset of those encounters, the tidal energy injection they will experience can be sufficient to strip a substantial amount of their mass from them. For the heavy and more loosely bound miniclusters that are well described by NFW profiles, they may be disrupted almost in their entirety, with a hefty fraction of the total dark matter mass spilling back into the minivoids~\cite{OHare:2023rtm}. This acts to refill the phase space distribution, but miniclusters have internal velocity dispersions many orders of magnitude narrower than the halo, which means that these form another type of stream on top of the fundamental streams from the halo infall of the dark matter that was never inside miniclusters, or got stripped early on.

A simple model for a distribution made up of these kinds of fine-grained streams is to simply sum up many Gaussians~\cite{OHare:2018trr, OHare:2019qxc},
\begin{equation}
    f_{\mathrm{str}}(\mathbf{v})= \sum_{N_{\rm str}} \frac{\zeta^i_{\rm str}}{\left(2 \pi \sigma_{\mathrm{str}}^2\right)^{3 / 2}} \exp \left(-\frac{\left(\mathbf{v} + \mathbf{v}_{\rm lab} - \mathbf{v}^i_{\mathrm{str}}\right)^2}{2 \sigma_{\mathrm{str}}^2}\right)
\end{equation}
where we impose the normalisation condition $\sum \zeta_{\rm str}^i = 1$. Here the distribution in general is poorly reflected by a Gaussian, however we should still draw $v_{\rm str}$ from the halo's large-scale multivariate Gaussian. Only now, at the level of the solar neighbourhood the distribution is intrinsically granular. For a halo entirely made of fundamental streams, we expect something like $N_{\rm str} = 10^{14}$ of them~\cite{Vogelsberger2011}, each one having a velocity dispersion $\sigma_{\rm str}/v_{\rm str}\sim 10^{-10}$. Realistically, this distribution may as well be perfectly smooth. However, at least one of those streams is expected to contribute around 0.1\% of the local density alone~\cite{Vogelsberger2011}, which could be observed in ultra-high-frequency resolution measurements of the axion field oscillations, as attempted recently in Ref.~\cite{ADMX:2023ctd} for example.

Before finishing, I will mention that there are a handful of other characteristic behaviours of the axion dark matter inside the Solar System that we can add to the list of possible effects at the end of Sec.~\ref{sec:directdetection_DM}. For instance, we may wish to account for the wave-like nature of dark matter when calculating gravitational focusing~\cite{Kim:2021yyo}. The presence of the Sun as a massive body influencing the wave dynamics of the incoming dark matter leads to additional patterns in the density and spectrum when the Earth is downwind of Cygnus. The additional wave-like signatures are most pronounced when the coherence length is on the order of the Earth's orbital radius, $10^{-15}$--$10^{-14}$~eV---although even in this regime it is still a percent-level effect. 

Another possibility is that there could be some amount of gravitational condensation of the dark matter inside the Solar System similar to how axion stars form. If the formation of a bound state of axions gets accelerated by the presence of a massive body like the Sun or the Earth, this could lead to small miniature versions of halos that could substantially boost the detectable dark matter density~\cite{Banerjee:2019xuy}. The estimated radius of these halos would be proportional to $1/m_a^2$ and so having a solar axion halo with a radius that reaches the Earth's orbit requires $m_a \lesssim 10^{-14}$---but this also cannot extend to arbitrarily low masses because the bound state should not be permitted to be too large or too dense or it would disrupt the orbits of the planets. Having an Earth-bound halo that extends up to the Earth's surface on the other hand requires slightly higher masses, $m_a\lesssim 10^{9}$~eV. Quartic self-interactions (controlled by the parameter $1/f_a$) may be able to accelerate the formation and saturation of the bound state through scattering processes amongst axions~\cite{Budker:2023sex}.

Finally, a feature of wave-like dark matter that has been discussed in the literature, but perhaps not investigated in enough detail, is the idea that the local distribution should contain an abundance of vortices---regions where the dark matter velocity field has nonzero curl. Vortex rings are a natural consequence of wave interference in a non-relativistic field obeying the Schr\"odinger-Poisson equation. I refer you to Ref.~\cite{Hui:2021tkt, Hui:2020hbq} for more detail and Ref.~\cite{Liu:2022rss} for visualisations of these in an ultralight dark matter halo. In the centre of one of these vortex rings, the field will destructively interfere with itself, causing the amplitude to vanish, although it will still have a nonzero gradient with the density scaling like $1/r^2$ in its vicinity. These could lead to interesting implications for experiments, especially those in which signals are temporally or spatially correlated. To model one of these objects individually we would need to depart from the assumption about $\beta_{\mathbf{p}}$ being totally uncorrelated. The phase would circulate around the vortex and so we would expect to exhibit that specific pattern in physical space through some, $\beta(t,\mathbf{x})$.

So let us put together these ideas and extend the list of expected signatures of dark matter in the solar neighbourhood to a few more that are specific to wave-like dark matter. Note that while many of these emerge as consequences of specific models or regimes in parameter space. They are, as a consequence, necessarily model-dependent, and so not all of them would be expected to be present at the same time. Nonetheless, we can conjecture the following additional signatures: 
\begin{itemize}
    \item \textbf{Fundamental streams}: if the signal can be measured at a frequency resolution at the level of $\omega/\Delta \omega>10^{10}$ or better, the fundamental streams expected as a result of the phase-space folding of the primordial dark matter sheet into a virialised halo could be made visible. 
    \item \textbf{Fossils of post-inflationary substructure}: in post-inflation axion misalignment scenarios we expect enhanced small-scale structure---the growth and merger of miniclusters into minicluster halos. The effective lineshape of these objects would be sufficiently narrow-band that even if all of them were tidally disrupted, their remnants may still persist when resolving the axion signal in frequency.
    \item \textbf{Wave-like gravitational focusing}: for wave-like dark matter masses with a coherence length that are of a similar order of magnitude to an astronomical unit, $m_a \sim 10^{-14}$~eV, the gravitational focusing of the waves can cause additional interference effects beyond the standard particle gravitational focusing effect mentioned in the previous list.
    \item \textbf{Solar halos}: for light particle masses which undergo self-interactions, a gravitationally bound halo of dark matter may build up around the Sun or the Earth, enhancing the local density by several orders of magnitude.
    \item \textbf{Vortices}: resulting from wave dynamics alone, within every coherence volume, there should be a region around which the velocity circulates with nonzero curl and in the centre, the field value vanishes.
\end{itemize}

Clearly, there is much that could be going on in the dark matter inside our solar system---it is now in the hands of experimentalists to determine the nature of the dark matter and reveal which, if any, of these peculiar phenomena are present. It is interesting to think about this in the broader context. As discussed earlier, the \textit{Gaia} satellite is right now revolutionising our understanding of the Milky Way's halo on a much larger scale than the solar system. There are many exciting features that are being unravelled all the time, but so far we can only make educated guesses about what they might imply for dark matter. There is the tantalising possibility that one day, after a positive detection of the axion is made, it will become our messenger from the dark side of our galaxy, and contribute as much to a revolution in astronomy as it will to particle physics...

\section{Conclusion}
\noindent But then again, maybe the axion doesn't exist.

\section{Acknowledgments}

\noindent I thank the organisers of the 1st COSMIC WISPers Training School for inviting me to write these lecture notes and for their patience while I finished them. I must thank those with whom I have had discussions over the years who have helped clarify parts of the material presented here, including Andrea Caputo, David Marsh, Alex Millar, Giovanni Pierobon, Georg Raffelt, Javier Redondo, Andreas Ringwald, Fuminobu Takahashi, Edoardo Vitagliano and Sam Witte. I also thank those who sent me comments on an early draft of this paper: Luca Di Luzio, Jan Sch\"utte-Engel, Yong Xu, Bingrong Yu, Kuver Sinha, Paolo Salucci, Mario Reig, Marcin Badziak, Yifan Chen, Shuailiang Ge, Marco Gorghetto, Pierre-Henri Chavanis, Eoin O Colgain, Mustafa Amin, Keir Rogers and Alberto Salvio. This article is based upon work from COST Action COSMIC WISPers CA21106, supported by COST (European Cooperation in Science and Technology). The Australian Research Council and The University of Sydney fund me through an ARC Discovery Early Career Research Award, grant number DE220100225.

\bibliographystyle{JHEP}
\bibliography{biblio}

\end{document}